\documentclass[12pt,a4paper,twoside]{report}
\newcommand{\intk}{\int\frac{d^4k}{(2 \pi)^4}}
\newcommand{\intp}{\int\frac{d^4p}{(2 \pi)^4}}
\newcommand{\intpt}{T \sum_n\int\frac{d^3p}{(2 \pi)^3}}
\newcommand{\intkt}{T \sum_n\int\frac{d^3k}{(2 \pi)^3}}
\def\roughly#1{\mathrel{\raise.3ex\hbox{$#1$\kern-.75em%
\lower1ex\hbox{$\sim$}}}}
\def\lsim{\roughly<}
\def\gsim{\roughly>}

\newcommand{\ave}[1]{\langle {#1} \rangle}

\newcommand{\pslash}{p\!\!\!/}
\newcommand{\qslash}{q\!\!\!/}
\newcommand{\dslash}{\partial\!\!\!/}
\newcommand{\pb}{\bar\psi}

\newcommand{\qq}{\ave{\pb\psi}}
\newcommand{\dg}{\delta g_{\pi qq}^{-2}}
\newcommand{\fig}[1]{Fig.~\ref{#1}}
\newcommand{\intf}[1]{\int d^4 x_{#1} {\mathrm e}^{-i q_{#1}\cdot x_{#1}}}
\newcommand{\Sec}[1]{section~\ref{#1}}
\def\rrr{\longrightarrow}
\newcommand{\eq}[1]{Eq.~(\ref{#1})}
\newcommand{\eqs}[1]{Eqs.~(\ref{#1})}
\newcommand{\nce}{$1/N_c$-expansion scheme}
\newcommand{\bce}{\begin{center}}
\newcommand{\ece}{\end{center}}
\newcommand{\bmi}{\begin{minipage}[t]{12.0cm}}
\newcommand{\emi}{\end{minipage}}
\newcommand{\beq}{\begin{equation}}
\newcommand{\eeq}{\end{equation}}
\newcommand{\bea}{\vspace{0.25cm}\begin{eqnarray}}
\newcommand{\eea}{\end{eqnarray}}
\newcommand{\eps}{{\epsilon}}
\newcommand{\mn}{{\mu\nu}}

\newcommand{\Tr}{{\mbox{\rm Tr}}}

\newcommand{\ba}{\begin{array}}
\newcommand{\ea}{\end{array}}

\newcommand{\komma}{\mbox{ ,}}

\usepackage{epsfig}
\usepackage{feynmp}
\begin{document}
\newcommand{\clearemptydoublepage}
  {\newpage{\pagestyle{empty}\cleardoublepage}}
\pagestyle{plain}
\begin{titlepage}
\begin{center}
\vspace*{3.5cm}
{\bfseries \Large Investigation of meson loop effects in the 
  Nambu--Jona-Lasinio model}\\
\vspace{2cm} 
{\large Vom Fachbereich Physik\\
der Technischen Universit{\"at} Darmstadt\\
\vspace{2cm}
zur Erlangung des Grades\\
eines Doktors der Naturwissenschaften\\
(Dr.~rer.~nat.)\\
\vspace{2cm}
genehmigte Dissertation von \\
Dipl.-Phys. Micaela Oertel\\
aus Langen (Hessen)\\
\vspace{5cm}
Darmstadt 2000 \\
D17}
\end{center}
\end{titlepage}
\newpage
\begin{abstract}
The influence of mesonic fluctuations on quantities in the Nambu--Jona-Lasinio
model is examined. To that end different approximation schemes are introduced
which guarantee the consistency with several relations following from chiral
symmetry. In particular this refers to the Goldstone theorem, which states the
existence of massless bosons, in our case pions, if the symmetry of the
Lagrangian is spontaneously broken. The Gell-Mann--Oakes--Renner
relation describes the behavior of the pion mass for small current quark
masses, which explicitly break chiral symmetry. Three different schemes are
presented. These are the inclusion of the ring sum in a so-called
``$\Phi$-derivable-method'', an expansion in powers of $1/N_c$ and an
expansion up to one-meson loop in the effective action formalism. Since the
two latter schemes are used for explicit calculations, it is explicitely
proved that the
Goldstone theorem as well as the Gell-Mann--Oakes--Renner relation hold within
those schemes.  

The influence of meson-loop effects on the quark condensate $\qq$, the pion
mass, the pion decay constant $f_\pi$ and properties of $\rho$- and
$\sigma$-meson are investigated. First we focus on the determination of a
consistent set of parameters. In the {\nce} it is possible to find a set of
parameters which allows to simultaneously describe the quantities in the pion
sector and those related to the $\rho$-meson, whereas this turns out to be not
possible within the expansion of the effective action. Results for
the $\sigma$-meson are also discussed. Here, similarly to the $\rho$-meson,
mesonic intermediate states are essential. Besides, the relation of our model
to hadronic models is discussed.

In the last part of this thesis the behavior of the quark condensate at
nonzero temperature is studied. In the low-temperature region agreement with
the model independent chiral perturbation theory result in lowest order can be
obtained. The perturbative {\nce} does not allow for an examination of the
chiral phase transition at higher temperatures, whereas this is possible
within the meson-loop approximation scheme. A first order phase transition is
found.  
\end{abstract}
\newpage
\pagestyle{plain}
\pagenumbering{arabic}

\tableofcontents
\begin{fmffile}{feyngph}
\chapter*{Introduction}
\addcontentsline{toc}{chapter}{Introduction}
\label{Introduction}
Since the substructure of the nucleus was discovered at the beginning of the
century one tries to understand in which way its constituents, the nucleons,
interact. Already in the thirties Yukawa suggested~\cite{yukawa} that this
interaction could be described by an exchange of a massive meson which was
later identified with the pion. Presently this idea is still the basis for
most phenomenological models of the interaction between nucleons which mainly
rely on the exchange of mesons with various quantum numbers~\cite{bonn,paris}.

Not only our present notion of the interaction between nucleons, but almost
all phenomenological models which try to describe strongly interacting
particles, mesons and baryons, in the low-energy region, are essentially
influenced by the idea of Yukawa. These models incorporate mesons and baryons
as the principal degrees of freedom in the low-energy region.  In many cases
reasonable descriptions of hadronic spectra, decays and scattering processes
have been obtained within these phenomenological models.  For instance, the
pion electromagnetic form factor in the time-like region can be reproduced
rather well within a simple vector dominance model with a dressed $\rho$-meson
which is constructed by coupling a bare $\rho$-meson to a two-pion
intermediate state~\cite{brown,herrmann}.

Recently much experimental and theoretical effort has been undertaken to
understand the properties of hadrons not only in vacuum but also in a strongly
interacting medium and, in in the same context, the phase structure of
strongly interacting matter. Besides for nuclear physics, these questions are
of importance mainly for astrophysical applications: In the evolution of the
early universe one presumes hot strongly interacting matter with small baryon
densities having existed (temporarily) shortly after the ``big bang'', and in
the interior of neutron stars one expects to find densities of a few times
nuclear matter density at very low temperatures.  Experimentally one tries to
gain information about the state of matter under various conditions with the
help of heavy-ion experiments.  Since experimentally only secondary hadrons
are measurable, it is extremely difficult to obtain direct information about
the state of matter in the early stages of the collision. Dilepton spectra
($e^+e^-$ or $\mu^+\mu^-$) offer a possibility to learn, at least indirectly,
something about the early stages, because leptons do not interact strongly. In
this context, the vector mesons, especially the $\rho$-meson, play an
important role. Here phenomenological hadronic models have been successfully
extended to investigate medium modifications of vector mesons and to calculate
dilepton production rates in hot and dense hadronic matter \cite{rapp}.

However, hadrons do not represent the fundamental degrees of freedom.
Nowadays quantum chromodynamics (QCD) is generally accepted as the theory of
strong interaction, which contains quarks and gluons as fundamental degrees of
freedom. QCD is constructed in the sense of a Yang-Mills gauge
theory~\cite{yang}. It is based on the color $SU(3)$-gauge group with the
gluons representing the gauge bosons. From the non-abelian structure of the
gauge group it follows that the gluons themselves carry color charge and
therefore can interact with each other. Quarks exist in six different flavors,
``up'', ``down'', ``strange'', ``charm'', ``bottom'' and ``top'', whereof the
two first are almost massless compared with the other ones and typical
hadronic mass scales.

Two prominent features of QCD are ``confinement'', which describes the fact
that in nature only colorless objects are observed, and the so-called
``asymptotic freedom''. The former
means in particular that no free quarks and gluons can be observed, they are
``confined'' to hadrons. Up to know the origin of this phenomenon in QCD has
not been understood. This is in contrast to asymptotic freedom, which reflects the fact
that the coupling constant decreases with increasing energies.
As a consequence QCD can be treated perturbatively at sufficiently high
energies, whereas in the low-energy region, which is the relevant one for
our hadronic world, perturbative methods are not applicable. 

In principle lattice calculations (see e.g.\@ Ref.~\cite{rothe}) provide us with the possibility to obtain
results directly from QCD also at lower energies.  Among the successes of
these ab-initio lattice calculations are the simulation of the linear rising
(confining) potential between two heavy quarks as well as the description of
the mass spectrum of hadrons, and in particular calculations at nonzero
temperatures. However, up to now these calculations mainly have dealt with
thermodynamic observables. In the sector of light quarks, up and down, on the
contrary, one is less successful, because of fundamental problems with
treating light fermions on the lattice. In addition, till now no reliable
results have been obtained at nonzero density.

According to the Goldstone theorem~\cite{goldstone} a symmetry of the
Lagrangian which is spontaneously broken in the ground state enforces
the existence of massless bosons. In the present case this symmetry is
the chiral $SU(2)_L\otimes SU(2)_R$-symmetry in flavor space which is
exhibited by QCD in the sector of up- and down-quarks.  The small mass
of the pions as compared with other hadronic masses is thought to
reflect their nature as Goldstone bosons.  The nonzero mass is thereby
attributed to the fact that this symmetry is explicitly broken because
of the nonzero masses of up- and down-quarks. The spontaneous breaking
of chiral symmetry in the physical vacuum is usually attributed to
instantons~\cite{schaefer}, semi-classical gluon field
configurations. A method which takes advantage of the small mass of
the pions and the fact that Goldstone bosons interact only weakly is
``Chiral perturbation theory''~\cite{CPT}, which corresponds to an
expansion in powers of momenta and pion masses. However, this method
also has its shortcomings: It is e.g.\@ not suited to treat
resonances.  Thus, besides the small numbers of observables which can
be studied on the lattice one has to rely here on calculations in
effective models.

Though hadronic calculations prove very successful in describing the
properties of light hadrons, one might certainly ask in which way they
can be understood from the underlying quark substructure. Since this
question cannot be answered from first principles it has to be
addressed within quark models. The main objection one can raise
against most of the known quark models is the lack of
confinement. However, at least for light hadrons chiral symmetry and
its spontaneous breaking in the physical vacuum seems to play the
decisive role in describing their properties with confinement being
much less important~\cite{schaefer}. Two of the most prominent
examples of quark models incorporating chiral symmetry were already
invented in the sixties: The Gell-Mann--L\'evy
model~\cite{gellmann}, dealing originally with nucleons, pions and $\sigma$-mesons,
and the Nambu--Jona-Lasinio (NJL) model~\cite{njl}, which originally was a
model which exclusively contained nucleons. Nowadays these models are
reinterpreted with quarks instead of nucleons.  

The NJL model has been applied by many authors to study properties of hadrons.
Baryons can be constructed either by directly building a bound state of three
quarks by solving Fadeev equations~\cite{fadeev} or by considering chiral
solitons~\cite{solitons}. Mesons of various quantum numbers have already been
investigated by Nambu and Jona-Lasinio~\cite{njl} and by many authors
thereafter (for reviews see~\cite{vogl,klevansky,hatsuda}). In most of these
works one starts by calculating quarks in mean-field approximation (Hartree or
Hartree-Fock) and then constructs mesons as correlated
quark-antiquark states in ``Random Phase Approximation'' (RPA). With a
suitable choice of the model parameters, chiral symmetry is spontaneously
broken in the vacuum and pions emerge as massless Goldstone bosons. The
spontaneously broken symmetry is also reflected by the nonzero value of the
quark condensate, being closely related to the nonzero mass the constituent
quarks acquire within this approximation. 

The breaking of chiral symmetry in the vacuum via dynamical effects and the
consistent description of the pion is certainly one of the successes of the
model. In contrast, the description of other mesons is more problematic, mainly
because of the missing confinement mechanism in the NJL model. This means that
a meson can decay into free constituent quarks if its energy exceeds twice
the constituent quark mass $m$. A typical parameter set, which is chosen in
such a way that the values for the pion decay constant $f_\pi$, the pion mass
$m_\pi$, and the quark condensate are in agreement with the
empirical values, leads to a quark mass of about $m\sim 300$~MeV. Thus, for
instance the $\rho$-meson with a mass of 770~MeV would be unstable against
decay into quarks. This is of course an unphysical feature. Besides, the
physically most important decay channel of the $\rho$-meson into two pions is
not included in those calculations. The same argument could be put forward for
other mesons like e.g.\@ the $\sigma$-meson which emerges in the standard
approximation scheme as (almost) sharp particle just above the threshold for
the decay into a quark-antiquark pair, although in nature it --if it exists at all--
has a very large width which can be mainly attributed to the decay into two
pions. The appearance of unphysical effects on the one hand and the
missing of physical ones on the other hand obviously results in a poor
description of the properties of those mesons and related quantities, e.g.\@
the pion electromagnetic form factor which is mainly determined by the
$\rho$-meson.

The phase structure of strongly interacting matter in the NJL model within a 
mean-field calculation has also been investigated by many authors (see
e.g.~\cite{klevansky,hatsuda,sklimt,mlutz}). In this way many of the
features induced by chiral symmetry, among others the expected restoration of
this symmetry at nonzero temperatures and densities, can be 
illustrated nicely. 

At this point we should mention that an examination of the chiral phase
transition is not the only thermodynamical application which has been
considered up to know in the NJL model within the standard approximation
scheme. For instance, in Ref.~\cite{BO} a hypothesis raised by Farhi and
Jaffe~\cite{farhi} that strange quark matter could be the absolute ground
state of matter is critically examined with the use of the $SU(3)$ version of the NJL
model, which includes in addition to up- and down quarks also strange quarks.
In the same
context one could mention that a few years ago the existence of a ``new''
state of strongly interacting matter was proposed~\cite{velkovsky,alford}: In
cold dense matter a ``color superconducting'' phase is likely to exist. In
that phase Cooper pairs of quarks and antiquarks are built similarly to
electronic Cooper pairs in metallic superconductors (see e.g.
Ref.~\cite{rajagopal} and references therein). Most of the corresponding
calculations have so far been performed either in instanton models or in
models of the same type as the NJL model.

However, the same objections which can be raised against the description of 
mesons concern also the investigations of properties below the phase 
transition. 
These calculations suffer from the fact that the thermodynamics is entirely
driven by unphysical unconfined quarks even at low temperatures and densities.
On the contrary the physical degrees of freedom, in particular the pions, are
missing.

Though there are some applications where the presence of unphysical free
quarks and the missing of relevant degrees of freedom seem to be less
important, e.g.\@ the study of bulk quark matter, for most calculations the standard approximation scheme proves
insufficient. Therefore several authors attempted to extend the standard
scheme. For instance, in Ref.~\cite{krewald} the coupling of a quark-antiquark
$\rho$-meson to a two-pion state via a quark triangle was considered. 
Also higher-order corrections to the quark self-energy \cite{quack}
and to the quark condensate \cite{blaschke} were investigated.
However, these attempts are insufficient in another sense: Since one of the
most important features of the NJL model is its chiral symmetry, any
approximation scheme which is applied should conserve properties connected to
that symmetry. Especially the existence of Goldstone bosons should be ensured
if the symmetry is spontaneously broken. 

To our knowledge Dmitra\v{s}inovi\'c et al.~\cite{dmitrasinovic} for the first
time discussed a symmetry conserving approximation scheme in the NJL model
which goes beyond the standard Hartree (-Fock) + RPA scheme. The authors
included a local correction term to the quark self-energy and, by an
appropriate choice of diagrams, succeeded to construct meson propagators
consistently. Among other things this guarantees that the Goldstone theorem
holds. In addition to the explicit proof of the existence of massless
Goldstone bosons, the authors of Ref.~\cite{dmitrasinovic} showed the validity
of the Goldberger-Treiman relation. A more systematic
derivation was presented by Nikolov et al.~\cite{nikolov}, who
used a one-meson-loop approximation to the effective action in a bosonized NJL
model.

An appealing feature of that scheme is that it incorporates mesonic
fluctuations. An examination of the effect of mesonic fluctuations on various
quantities, for instance on the pion electromagnetic form factor~\cite{lemmer}
or on $\pi$-$\pi$-scattering in the vector~\cite{he} and scalar
channel~\cite{huang}, has been performed, based on that approximation scheme.
However, because of the difficulties which arise in connection with the
numerical evaluation of the occurring multi-loop integrals, the authors of
these references approximated the exact expressions by low-momentum
expansions.
 
Are there other possibilities to construct symmetry conserving approximation
schemes? We could show that the most direct extension of the usual RPA in
terms of techniques used in many-body-theory, the so-called ``Second RPA''
(SRPA)~\cite{wambach}, does not preserve the necessary
symmetries~\cite{diplm}. On the other hand, in the literature other symmetry
conserving approximation schemes are known. In addition to the one-loop
expansion of the effective action~\cite{weinberg} mentioned above, these are
the ``$\Phi$-derivable''-method~\cite{luttinger,baym} and an expansion in
powers of $1/N_c$, the inverse number of colors. The mean-field (Hartree)
approximation in combination with the RPA corresponds to the leading order in
such an expansion. In Ref.~\cite{dmitrasinovic} it was shown that in the NJL
model the Goldstone theorem also remains valid in an expansion up to
next-to-leading order in $1/N_c$. A fundamental difference of this
approximation scheme to that mentioned above is the perturbative character of
the correction terms taken into account. More detailed investigations of quark
and meson properties within the {\nce} were performed in
Refs.~\cite{oertel,OBW}. Recently such an expansion has been discussed also in
the framework of a non-local generalization of the NJL model~\cite{plant}.

In this paper we will calculate quark and meson properties as well in the
non-perturbative scheme introduced in Refs.~\cite{dmitrasinovic,nikolov} as in
the {\nce}, including the full momentum dependence of all expressions. A
comparison of the results obtained in both schemes, focusing on the
pion and the $\rho$-meson in vacuum, can also be found in Ref.~\cite{OBW2}. The
investigation of pion properties was mainly motivated by recent papers by
Kleinert and van den Bossche~\cite{kleinert} who claim that in the NJL model
chiral symmetry is restored in vacuum due to strong mesonic fluctuations. One
argument which can be raised against this conjecture~\cite{oertel} is the
non-renormalizability of the NJL model which causes new divergencies to emerge
if further loops, in this case meson loops, are included. And, new divergencies
require new cutoff parameters. Following Refs.~\cite{dmitrasinovic}
and~\cite{nikolov} we introduce a cutoff $\Lambda_M$ for the meson-loops which
is independent of the regularization of the quark loops. For small values of
that cutoff the results in the pion sector change only quantitatively as
compared with those obtained in the Hartree + RPA scheme, whereas for large
values of the cutoff within both extended schemes instabilities in the pion
propagator are encountered which might be a hint for an unstable ground
state~\cite{oertel,OBW2}. However, in the non-perturbative scheme a closer
analysis revealed that these instabilities are not the consequence of
restoration of chiral symmetry.  

Encountering those instabilities one might fear that thereby a description of
meson properties will be made impossible. But since this is an effect which
strongly depends on the choice of the model parameters, especially the
meson-loop cutoff $\Lambda_M$, it is not ruled out that mesons can be
reasonably described with a parameter set far away from the region where
instabilities appear. In fact, in the {\nce} we determined such a parameter
set~\cite{OBW} by fitting the values of $m_\pi, f_\pi$, the quark condensate
$\qq$, and the pion electromagnetic form factor. The inclusion of meson loops,
in particular pion loops, is absolutely crucial to obtain a realistic
description of the latter, which is therefore well suited to fix our
parameters. 

We mentioned above that one disadvantage of hitherto
performed calculations in the NJL model is the existence of quark-antiquark
decay channels for all mesons if their energy is larger than twice the
constituent quark mass. Of course, our calculations cannot cure this
problem, but it can be by-passed: The analysis in
Ref.~\cite{OBW} shows that a realistic description of the above listed
observables is possible with a relatively large constituent quark mass of
about 450 MeV, such that the threshold for the decay into a $q\bar{q}$-pair
lies above the peak of the $\rho$-meson spectral function. This is an
important result since the constituent quark mass is not an independent input
parameter. The same analysis was performed in
Ref.~\cite{OBW2} for the selfconsistent scheme. It turned out that no fit can
be achieved with a constituent quark mass large enough to shift the threshold
above the $\rho$-meson peak.  

At low temperatures and low densities pionic degrees of freedom are expected
to be the dominant ones. Thus not only the description of mesons but also the
thermodynamics of the model should be considerably improved by including meson
loop effects. This becomes, for instance, obvious from the temperature
dependence of the quark condensate. For its low-temperature behavior chiral
perturbation theory provides us with model independent results relying mainly
on thermally excited pions~\cite{gasser}. In the standard approximation to 
the NJL model any attempt to describe this
behavior fails whereas it could be shown that as well in the
{\nce}~\cite{OBW2} as in the non-perturbative scheme~\cite{florkowski} it can
be reproduced. In the latter scheme also an examination of the chiral phase
phase transition is possible~\cite{florkowski,OBW2}.

In the first chapter we establish the different approximation schemes to the
NJL model. To that end we begin by briefly reviewing the standard
approximation, Hartree + RPA, and in which way quark and meson properties are
described within that scheme. Afterwards we present three different
possibilities to extend the standard scheme, first an expansion up to
next-to-leading order in $1/N_c$, then a ``$\Phi$-derivable method'', and
finally an one-meson-loop expansion of the effective action. The two latter
schemes have nonperturbative character in contrast to the former
scheme.  
For the {\nce} and the one-meson-loop approximation, the schemes which we will
later apply to study properties of quarks and mesons, we will explicitly show
the consistency with the Goldstone theorem and the Gell-Mann--Oakes--Renner
relation as well as the transversality of the polarization function in the
vector channel. 
The chapter will close with a discussion of a particular approximation to the
two extended schemes which points out the relation of our calculations to
hadronic models. 

The numerical results at zero temperature will be presented in
Chapter~\ref{numerics}. We begin by explaining the regularization procedure
before we come to investigate the influence of mesonic fluctuations on the
quark condensate and pion properties. Then we try to determine a set of model
parameters which is consistent with quantities as well in the pion sector as
in the $\rho$-meson sector. An examination of further quantities related to
the $\rho$-meson and a comparison with hadronic calculations follows. Finally
we briefly discuss properties of the $\sigma$-meson. 

Chapter~\ref{temperatur} will be devoted to a discussion of results at nonzero
temperature. We will begin by establishing the formalism for calculations at
nonzero temperature (and density) which will first be applied to illustrate
some basic results in the standard approximation scheme. Then we will discuss
results for the quark condensate obtained in the two extended approximation
schemes at nonzero temperature.

Finally we will summarize our main results and present some possible
objectives. 

\chapter{The Nambu--Jona-Lasinio model}
\label{NJLModell}
\section{Basics of the model}

The Nambu--Jona-Lasinio model was originally introduced by Nambu and
Jona-Lasinio~\cite{njl} in 1961 to describe an effective nucleon-nucleon
interaction. In its original form it contained an isospin-doublet built of
proton and neutron. Nowadays it is reinterpreted, incorporating quarks instead
of nucleons. In analogy to the nucleonic isospin-doublet one now deals with
up- and down quarks. If one extends the model to flavor-$SU(3)$ also strange
quarks can be described~\cite{hatkuni,bernardjaffe,takizawa,klimt}. We will
restrict our investigations to the model with two flavors. 
Thus, we consider the following Lagrangian:
\beq
   {\cal L} \;=\; \pb ( i \partial{\hskip-2.0mm}/ - m_0) \psi
            \;+\; g_s\,[(\pb\psi)^2 + (\pb i\gamma_5{\vec\tau}\psi)^2]  
            \;-\; g_v\,[(\pb\gamma^\mu{\vec\tau}\psi)^2 + 
                        (\pb\gamma^\mu\gamma_5{\vec\tau}\psi)^2]   
   \,.
\label{lagrange}
\eeq 
$\psi$ is a quark field with $N_f$~=~2 flavors and $N_c$~=~3 colors.
$g_s$ and $g_v$ are coupling constants of the dimension energy$^{-2}$. This
Lagrangian exhibits the same global symmetries as QCD if in flavor space only
the $SU(2)$-isospin-sector is taken into account. Via the Noether theorem
these global symmetries are related to conserved currents. The global
symmetries of the above Lagrangian are:
\begin{itemize}
\item{Invariance under $U_{V}(1)$ transformations 
\[ \psi\to \exp [-i\alpha]\psi~.\]  
The corresponding current is:
\[ j_{\mu}=\bar{\psi}\gamma_{\mu}\psi~.\]
This symmetry is related to
conservation of baryon number.} 
\item{Invariance under $SU_{V}(2)$ transformations 
\[
\psi\to 
\exp[-i\vec{\tau}\cdot\vec{\varphi}/2]\psi~.
\] 
This can be seen by recalling
that $\bar{\psi}i\gamma_5\vec{\tau}\psi$ is a vector in isospin space and the
transformation corresponds to a rotation in that space. The scalar product of
a vector with itself then obviously transforms like a scalar. The same
argumentation leads to the conclusion that as well $(\pb\gamma^\mu\vec{\tau}\psi)^2$ and
$(\pb\gamma^\mu\gamma_5{\vec\tau}\psi)^2$ transform like scalars. It is
obvious that also $(\pb\psi)^2$ transforms like a scalar.    
The corresponding current is:
\[  
   J_{\mu}^{k}=\bar{\psi}\gamma_{\mu}\tau^{k}\psi~. \]}
\item{In the chiral limit, i.e.\@ for $m_0=0$, in addition 
invariance under $SU_{A}(2)$ transformations  
\[\psi\to\exp[-i\gamma_5\vec{\tau}\cdot\vec{\theta}/2]\psi~\] is found.
One can most easily convince oneself of this invariance by considering the
following transformation properties:
\begin{eqnarray*}
(\bar{\psi}\psi)&\to& (\bar{\psi}\psi)\cos\theta-(\bar{\psi}i\gamma_5
\vec{\tau}\cdot\hat{\theta}\psi)\sin\theta\komma\nonumber\\
(\bar{\psi}i\gamma_5\vec{\tau}\psi)&\to&(\bar{\psi}i\gamma_5
\vec{\tau}\psi)+(\bar{\psi}\psi)\hat{\theta}\sin\theta
-(\bar{\psi}i\gamma_5\vec{\tau}\cdot\hat{\theta}\psi)\hat{\theta}
(1-\cos\theta)~.\nonumber\\
\end{eqnarray*}
Here $\hat{\theta}$ denotes a unit vector in $\vec{\theta}$-direction,  
${\vec{\theta}\over \theta}$.
The corresponding current is:
\[ 
   J_{\mu}^{k}=\bar{\psi}\gamma_{\mu}\gamma_5\tau^{k}\psi. \]}
\end{itemize}
We can conclude that in the chiral limit the Lagrangian, \eq{lagrange}, is
invariant under global $SU_{V}(2)\otimes SU_{A}(2)\otimes U_{V}(1)$
transformations.
The invariance under a $SU_{V}(2)\otimes SU_{A}(2)$ transformation is called
``chiral symmetry''. The missing of degenerate ``chiral partners'' in the
hadronic spectrum suggests that chiral symmetry is spontaneously broken in the
QCD vacuum. This has important consequences, among others this requires the
existence of massless Goldstone bosons. Because of the small mass of the pions
compared with other hadronic scales these are interpreted as Goldstone
bosons. The nonzero mass is thereby attributed to
the explicit breaking of chiral symmetry due to the nonzero current masses of
up- and down-quarks. The NJL model has been used by many authors (see e.g.\@ the
reviews~\cite{vogl, klevansky,hatsuda}) to study the
spontaneous breakdown of chiral symmetry in the vacuum and its restoration at
nonzero temperature and density. 

However, a principal difference to QCD is the treatment of color degrees of
freedom: In QCD those are related to an invariance under local transformations
of the gauge group $SU(N_c)$, whereas in the present model $N_c$ only counts
the number of identical copies of quark fields. Usually one chooses the coupling constants
$g_s$ and $g_v$ to be of order $1/N_c$~\cite{dmitrasinovic,nikolov} to
reproduce the large-$N_c$ behavior of QCD. Treating $N_c$ as an expansion
parameter will enable us to construct a scheme to describe mesons
which is consistent with the requirements of chiral symmetry. All calculations
will, however, be performed with the physical value $N_c=3$. 

\section{Standard approximation: Hartree-(Fock)+RPA}
\label{hartree} 
Detailed analyses of the results one obtains in the NJL model using the
standard approximation scheme can be found among others in the
reviews~\cite{vogl,klevansky, hatsuda}.  Before we will come to discuss
possible extensions of the standard approximation scheme in the following sections
we will briefly review the approximation usually applied to the NJL model. On
the quark level this corresponds to a (Bogoliubov-) Hartree approximation.
The local four-fermion interaction within the NJL model allows to cast
exchange terms, i.e.\@ Fock terms, in the form of direct terms with the help of
a Fierz transformation. Therfore a Hartree-Fock approximation is similar to a
Hartree one with a redefined coupling constant. Since the contributions from
Fock terms are suppressed by one order in $1/N_c$ as compared with the Hartree
terms, we will restrict ourselves here to the Hartree approximation.

\begin{figure}[b!]
\begin{center}
  \parbox{10cm}{ \epsfig{file=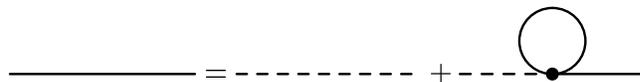}}
\end{center}
\caption{\it The Dyson-equation for the quark-propagator in  
         Hartree-approximation (solid). Dashed lines denote propagators of
         bare quarks.}
\label{fig1} 
\end{figure}
The Dyson-equation for the quark-propagator in Hartree approximation is shown
diagrammatically in \fig{fig1}. Note that we have renounced here to show the
arrows on the quark propagators which would be standard convention for fermion
lines. However, since later on most of the diagrams exist in two versions
which only differ by the orientation of the quark loop, it seems convenient
to display only one version which has then to be understood to contain all
orientations which lead to topological different diagrams. 

Via selfconsistently solving the Dyson-equation the quarks acquire a
momentum independent self-energy which leads to a nonzero ``constituent''
quark mass $m$ 
\beq 
m = m_0 + \Sigma_H = m_0 + \sum_M \; \Gamma_M 2i g_M\ 
\intp \;\Tr\,[\,\Gamma_M\,S(p)\,]~.
\label{gap} 
\eeq
$S(p) = (\pslash - m + i\eps)^{-1}$ denotes here a quark propagator in Hartree
approximation and the symbol ``$\Tr$'' stands for a trace over color-, flavor-
and Dirac-indices. In principle the sum over $M$ contains all interaction
channels, i.e.\@
$M = \sigma, \pi, \rho, a_1$ with
\begin{eqnarray} 
\Gamma_\sigma &=& 1\!\!1~,\quad \Gamma_\pi^a = i\gamma_5\tau^a~, \nonumber\\
\Gamma_\rho^{\mu\,a} &=& \gamma^\mu\tau^a\quad {\rm and}\quad 
\Gamma_{a_1}^{\mu\,a} = \gamma^\mu\gamma_5\tau^a~.
\label{gammas}
\end{eqnarray}
The corresponding coupling constants are $g_M = g_s$ for $M = \sigma$ or  
$M = \pi$ and $g_M = g_v$ for $M = \rho$ or $M = a_1$. 
It can easily be seen, however, that in vacuum only the scalar channel,
$M=\sigma$, contributes. Consequently the quark self-energy is proportional to
the identity matrix. One obtains 
\beq
m = m_0 + 2i g_s\ 4 N_c N_f \intp {m\over{p^2-m^2+i\epsilon}}~.
\label{gapexp}
\eeq
Since $g_s$ is of order $1/N_c$, the constituent quark mass $m$ and
consequently the quark-propagator are of ${\mathcal O}(1)$  in
Hartree approximation. This corresponds to the leading order in a
$1/N_c$-expansion for the quark self-energy.

In the chiral limit, i.e.\@ for vanishing current quark mass $m_0$, it is
obvious that always a
``trivial'' solution of \eq{gapexp} with $m=0$ exists. 
If the scalar coupling constant $g_s$ exceeds a certain critical value, in
addition there exists a solution of \eq{gapexp} with a nonzero constituent
quark mass $m$. Because of the resulting ``gap'' in
the spectrum, \eq{gapexp} is called, in imitation of BCS theory for
superconductors, ``gap equation''. A nonzero constituent quark mass reflects
the spontaneously broken symmetry of the underlying ground state. This can
e.g.\@ be
seen if one considers the relation between the constituent quark mass and the
order parameter of chiral symmetry, the quark condensate.
Generally the quark condensate is given by
\beq
    \qq = -i \intp \; \Tr\,S(p)~.
\label{qbq} 
\eeq
In the Hartree approximation it is directly related to the constituent quark
mass, 
\beq
    \qq^{(0)} = - \frac{m-m_0}{2g_s} ~.
\label{qbq0}
\eeq     
We have introduced here the superscript $(0)$ to indicate that we deal with a
quantity in Hartree approximation. 

\begin{figure}[t!]
\begin{center}
\parbox{10cm}{
     \epsfig{file=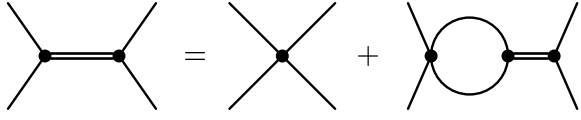}}\end{center}
\caption{\it The Bethe-Salpeter equation for the meson-propagators 
             in RPA (double line). The solid lines indicate quark propagators.}
\label{fig2} 
\end{figure}

Mesons can be described via a Bethe-Salpeter equation. The leading order in
$1/N_c$ is here given by a so-called ``Random Phase approximation'' (RPA)
without exchange terms. Diagrammatically this Bethe-Salpeter equation is
displayed in \fig{fig2}.

The propagators of the mesons can be extracted from the quark-antiquark
scattering matrix $T$, which satisfies the following equation
\beq
T_{M,ijkl}(q) = K_{M,ijkl} + K_{M,ijab}
J_{bcda}(q) T_{M,cdkl}(q)~,
\label{tmatrix}
\eeq
with $J_{bcda}(q) = -i \intp S_{bc}(p+q/2) S_{da}(p-q/2)$. The indices $a$ to
$d$ and $i$ to $l$ are multi-indices, indicating the components of the various
quantities in color-, flavor- and Dirac-space.
$K_{M}$ denotes the scattering kernel, which can be written in the following
way: 
\beq
K_{M,ijkl} = 2 g_{M}\Gamma_{M,ij} \Gamma_{M,kl}~.
\label{kkern}
\eeq
The vertices $\Gamma_M$, $M = \sigma, \pi, \rho, a_1$ have already been
defined in \eq{gammas} together with the corresponding coupling constants
$g_M$. If we make the ansatz
\beq
T_{M,ijkl}(q) = -D_{M}(q) \Gamma_{M,ij}\Gamma_{M,kl}~,
\eeq
for the $T$-matrix, \eq{tmatrix} can be transformed into an
equation for the meson propagators $D_M$, 
\beq
D_{M}(q) = -2 g_{M} + 2 g_{M}
\Pi_{M}(q) D_{M}(q)~,
\eeq
with the polarization functions $\Pi_M$. These consist of a
quark-antiquark-loop 
\beq
   \Pi_{M}(q)= \Gamma_{M,ab}J_{dabc}(q)\Gamma_{M,cd}=-i\intp \;\Tr\,[\,\Gamma_{M} \, iS(p+{q\over2})
                  \,\Gamma_{M} \, iS(p-{q\over2})\,] \;.
\label{pol0}
\eeq
Here again $\Tr$ denotes a trace over color-, flavor- and Dirac-indices.
In the scalar and the pseudoscalar channel, i.e.\@ for the $\sigma$-meson and
the pion, we obtain
\bea
D_{\sigma}(q)& =& \frac{-2 g_s}{1-2g_s\Pi_{\sigma}(q)}~, \nonumber\\
D_{\pi}^{ab}(q) &\equiv& D_{\pi}(q)\,\delta_{ab} 
= \frac{-2 g_s}{1-2g_s\Pi_{\pi}(q)}\,\delta_{ab}\;.
\label{dpisigma}
\eea
Here $a$ and $b$ are isospin indices. We have used the following notation:
$\Pi_{\pi}^{ab}(q) \equiv \Pi_{\pi}(q)\,\delta_{ab}$.

In the vector channel we proceed in a similar way. The only difference is the
the more complicated Lorentz structure of the polarization function and the
resulting propagator. Since the polarization function can depend only on one
four-momentum, in our notation $q$, its tensor structure is generally given by 
\beq
\Pi_\rho^{\mn,ab}(q) = \delta_{ab}(g^\mn f_1(q^2) + \frac{q^\mu q^\nu}{q^2}
f_2(q^2))~,
\eeq
with two scalar functions $f_1$ and $f_2$. This can be separated into a
tranverse and a longitudinal part using the corresponding projectors,
\beq
T^\mn = -g^\mn + \frac{q^\mu q^\nu}{q^2}\quad{\rm and} \quad L^\mn =
\frac{q^\mu q^\nu}{q^2}~.
\eeq
However, because of vector current conservation, the polarization function in
the vector channel has to be transverse, i.e.\@
\beq
    q_\mu\,{\Pi}_\rho^{\mu\nu, ab}(q) =
    q_\nu\,{\Pi}_\rho^{\mu\nu, ab}(q) = 0 \;.
\label{trans}
\eeq
The tranverse structure, together with the assumption of Lorentz invariance,  
not only simplifies the determination of the tensor structure of the
corresponding propagator but also the
evaluation of the polarization function itself: With the ansatz 
\beq
\Pi^{\mn,ab}_{\rho}(q) = T^\mn(q) \delta_{ab}\Pi_{\rho}(q) 
\label{tensortrans}
\eeq
we only need to evaluate the (Lorentz) scalar function 
\beq
\Pi^{ab}_{\rho}(q) = -\frac{1}{3} g_\mn\Pi^{\mn,ab}_\rho(q)~.
\label{scalartrans}
\eeq
Finally 
one arrives at the following expression for the propagator of the $\rho$-Meson
\beq
D_{\rho}^{\mu\nu,ab}(q) \;\equiv\; (D_{\rho}(q)\;T^{\mu\nu}+2 g_v\;L^{\mn})\;\delta_{ab}
= (\frac{-2 g_v}{1-2g_v\Pi_{\rho}(q)} \,T^{\mu\nu}+2
g_v\;L^{\mn})\;\delta_{ab}~.
\label{drho}
\eeq
In \Sec{transverse} we will explicitly show that the polarization function in
the vector channel is indeed transverse, provided that a suitable
regularization procedure is used. 
In \Sec{regularization} we will discuss a further consequence of
vector current conservation, namely that 
$\Pi_\rho(q)$ should vanish for $q^2 = 0$, in the context of the regularization
procedure we will apply.

In the same way the propagator of the $a_1$-meson can be obtained from the
transverse part of the axial polarization function 
$\Pi_{a_1}^T(q)\delta_{ab}=-\frac{1}{3} T_{\mn} \Pi_{a_1}^{\mn, ab}(q)$.
As discussed e.g.\@ in Ref.~\cite{klimt}, $\Pi_{a_1}^{\mu\nu}$ in addition
contains a longitudinal part which contributes to the pion propagator together
with the pseudoscalar polarization function $\Pi_{\pi}^{ab}$ and the mixed
ones, which contain one pseudoscalar and one axial vertex. There is no
conceptual difficulty in dealing with this so-called
``$\pi$-$a_1$-mixing''. We will neglect it in the following discussion, and 
only consider
the pseudoscalar part of the pion propagator, in order to keep
the formalism as simple as possible. The pion propagator including
$\pi$-$a_1$-mixing is discussed in App.~\ref{pia1rpa}.

It follows from \eqs{pol0} to (\ref{drho}) that the functions $D_M(q)$ are of
order $1/N_c$. The explicit form of these functions can be found in
App.~\ref{correlators}. We will call them ``propagators'' although strictly
speaking they have to be interpreted as the product of a meson propagator with
a meson-quark coupling constant squared.  
The latter is given by the inverse of the residue of the function $D_M(q)$ at
the pole, whereas the mass of the corresponding meson is determined by the
location of the pole,
\beq
   D_{M}^{-1}(q)|_{q^2 = m_M^{2 (0)}} = 0 \;,\qquad
    g_{Mqq}^{-2 (0)} = \frac{d\Pi_{M}(q)}{dq^2}|_{q^2 = m_M^{2 (0)}} \;.
\label{mesonmass0}
\eeq
The superscript $(0)$ here serves to indicate that $m_M^{2 (0)}$
and $g_{Mqq}^{(0)}$ are RPA quantities. In a $1/N_c$-expansion these
quantities are of leading order. One can easily convince oneself that they are
of order ${\mathcal O}(1)$ and $1/\sqrt{N_c}$, respectively. Dealing with
$\pi$-$a_1$-mixing, we have
only to keep in mind that in addition to a pseudoscalar pion-quark coupling
constant an axial one exists. This will be discussed in App.~\ref{pia1rpa}.
\section{Extensions of the standard approximation}
The Hartree + RPA (only direct contributions) scheme, which was discussed in
the previous section is consistent with chiral symmetry. This means, for
instance, that a spontaneously broken symmetry enforces the existence of
massless Goldstone bosons. Within that approximation scheme also the validity
of the Goldberger-Treiman and the Gell-Mann--Oakes--Renner relation, which
determines the behavior of the pion mass if the symmetry is explicitly broken
due to a small current quark mass $m_0$, can be
shown~\cite{vogl,klevansky,hatsuda}. We will present a proof for these
relations in \Sec{pionrpa}.  Since chiral symmetry is one of the main features
of the NJL model, any approximation should be consistent with chiral symmetry.
Within this section we will discuss different approximation schemes which have
in common that on the one hand they fulfill the requirements of chiral
symmetry and on the other hand they go beyond the standard Hartree + RPA
scheme. We will begin with a strict $1/N_c$-expansion up to next-to-leading
order, proceed by discussing the ``$\Phi$-derivable method'' and then consider
a one-loop expansion of the effective action. The latter corresponds to a
selfconsistent extension of the gap equation including a meson-loop term.
These methods enable us to include mesonic fluctuations in our investigations.
\subsection{$1/N_c$-expansion}
\label{1/N_c}
Within this section we will consider the quark self-energy and the
polarization functions of the mesons in a $1/N_c$-expansion up to
next-to-leading order. To determine the corresponding correction terms let us
first recall the $1/N_c$-counting rules for the leading-order quantities: Quark propagators
are of order unity whereas the meson propagators are of order $1/N_c$. 
In addition one has to keep in mind that, because of the trace in color space,
every quark loop contributes a factor $N_c$ and every four-fermion vertex a
factor $1/N_c$ due
to the order of the coupling constant.  

\begin{figure}[t!]
\begin{center}
\parbox{6cm}{
     \epsfig{file=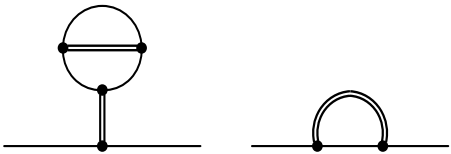,
     height=2.cm, width=6.0cm}}\end{center}
\caption{\it The $1/N_c$-correction terms $\delta\Sigma^{(a)}$ (left) and 
$\delta\Sigma^{(b)}$ (right) to the quark self-energy.}
\label{fig3} 
\end{figure}

For the quark self-energy we find two correction terms,
\beq
    \delta \Sigma(p) = \delta\Sigma^{(a)} + \delta\Sigma^{(b)}(p) \;.
\eeq
These terms are graphically shown in \fig{fig3}. A solid line corresponds here
to a quark propagator in Hartree approximation (${\cal O}(1)$) and a double line
to an RPA meson propagator (${\cal O}(1/N_c)$). With the help of the above
mentioned counting rules it is then easy to convince oneself that these are
indeed the only correction terms to the quark self-energy in next-to-leading
order. Any further contribution inevitably contains in addition at least one quark
loop and two coupling constants or one meson propagator, which leads altogether to
an additional factor $1/N_c$. 

The correction terms to the quark self-energy determine those to the quark
propagator to order $1/N_c$,
\beq 
    \delta S(p) = S(p)\,\delta\Sigma(p)\,S(p) ~.
\label{deltaS}
\eeq
$S(p)$ here denotes the quark propagator in Hartree approximation. In leading
order the self-energy has been iterated in order to obtain the corresponding
propagator. This has been done via the selfconsistent solution of the gap
equation, \eq{gap}. However, in next-to-leading order an iteration of the
self-energy would lead to terms of arbitrary order in $1/N_c$ in the
propagator. Following \eq{qbq} the
correction to the quark condensate is given by
\beq
    \delta\qq = -i \intp \; \Tr\,\delta S(p) \;.
\label{deltaqbq}
\eeq

\begin{figure}[t!]
\begin{center}
\parbox{14cm}{
     \epsfig{file=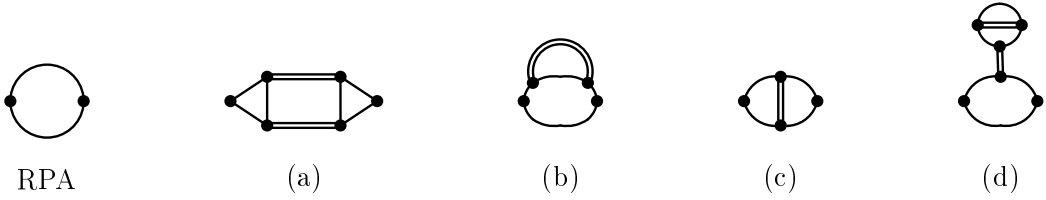,
     height=2.6cm, width=14.0cm}}\end{center}
\caption{\it Contributions to the meson polarization functions in leading
     (RPA) and next-to-leading order in $1/N_c$.}
\label{fig4} 
\end{figure}

We find four correction terms, $\delta \Pi_M^{(a)}$ to $\delta \Pi_M^{(d)}$,
to the meson polarization functions. These are displayed in \fig{fig4}
together with the leading order contribution. Solid lines denote again quark
propagators, double lines RPA meson propagators.  
This leads to the following form of the complete polarization function:
\beq
    {\tilde \Pi}_M(q) = \Pi_M(q) + \sum_{k=a,b,c,d}\; \delta \Pi_M^{(k)}(q)
\;.
\label{pol1}
\eeq
The correction terms, which contain either one RPA propagator and one quark
loop or two RPA propagators and two quark loops, are of order unity,
whereas the leading-order term is of order $N_c$.

The meson propagators are again, similar to the leading order, given as
solution of a Bethe-Salpeter equation. The only difference to the propagators
in RPA consists of taking into account the polarization functions up to
next-to-leading order. Thus we arrive at the following expressions for the
meson propagators, analogously to \eqs{dpisigma} and (\ref{drho}),
\beq
{\tilde D}_M(q) = \frac{-2 g_M}{1-2g_M {\tilde\Pi}_{M}(q)} \;.
\label{dm1}
\eeq
We should remark that these meson propagators contain arbitrary orders in
$1/N_c$ although we have determined the polarization functions by a strict
expansion in powers of $1/N_c$. This is due to the iteration of the
product of the polarization functions with the corresponding coupling constant.

This remark similarly concerns the meson masses. These are defined, in analogy
to \eq{mesonmass0}, as the location of the pole of the propagator,
\beq
   \tilde{D}_M^{-1}(q)|_{q^2 = m_M^2} = 0 \;.
\label{mesonmass1}
\eeq
Because of the implicit definition, $m_M$ also contains terms of arbritrary
orders in $1/N_c$. This definition is consistent with the Goldstone theorem,
but in the context of 
the Gell-Mann--Oakes--Renner relation we will encounter
difficulties caused by higher-order (beyond next-to-leading order)
contributions to the pion mass. This point will be discussed in detail in
\Sec{pionnc}.  

\begin{figure}[b!] 
\begin{center} 
\parbox{10cm}{ 
     \epsfig{file=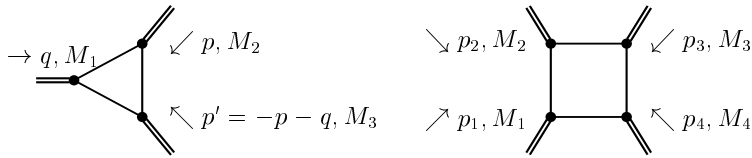,
     height=2.1cm, width=10.cm}}\end{center}
\caption{\it Quark triangle $ -i\Gamma_{M_1,M_2,M_3}(q,p)$ (left) and 
             quark box $-i\Gamma_{M_1,M_2,M_3,M_4}(p_1,p_2,p_3)$ (right).}
\label{fig5} 
\end{figure}

For the evaluation of the various contributions it is convenient to introduce
effective meson-meson vertices which consist of quark loops. We need two
different types of meson-meson vertices, a three-meson vertex, containing a
quark triangle, shown on the l.h.s.\@ of \fig{fig5} and a four-meson vertex,
containing a quark box, shown on the r.h.s.\@ of \fig{fig5}. 
For external mesons $M_1$, $M_2$ and $M_3$ the quark triangle has the
following form:
\bea
 -i \Gamma_{M_1,M_2,M_3}(q,p) &=& - \intk
  \Big\{ \Tr\,[\Gamma_{M_1}i S(k)\Gamma_{M_2} i S(k-p)\Gamma_{M_3}
 i  S(k+q)] \nonumber \\
& & \hspace{1.85cm}+ \Tr\,[\Gamma_{M_1}i S(k-q)\Gamma_{M_3}i 
    S(k+p)\Gamma_{M_2} i S(k)]\Big\}~. 
\label{trianglevertex}
\eea
The operators $\Gamma_M$ have already been defined below \eq{gap}. The above expression already
contains a sum over both possible orientations of the quark loop.

The four-meson vertices can be written in the following way:
\bea
&&  \hspace{-2.0cm}
-i\Gamma_{M_1,M_2,M_3,M_4}(p_1,p_2,p_3) \phantom{\intk}\nonumber\\
&&\hspace{-1.5cm}= -\intk \Big(
\Tr\,[\Gamma_{M_1}iS(k)\Gamma_{M_2}iS(k-p_2)\Gamma_{M_3}iS(k-p_2-p_3)
    \Gamma_{M_4}iS(k+p_1)]  
\nonumber\\ && \hspace{-1.5cm}\phantom{=\intk \Big(}
 +\Tr\,[\Gamma_{M_1}iS(k-p_1)\Gamma_{M_4}iS(k+p_2+p_3)\Gamma_{M_3}iS(k+p_2)
      \Gamma_{M_2}iS(k)]\Big)~.
\label{boxvertex}
\eea
Here again we have summed over both possible orientations of the quark loop.

With the help of the above definitions the $1/N_c$-correction terms to
the quark self-energy as well as to the meson polarization functions can be written in
a relatively compact form. 
For the momentum independent contribution  $\delta\Sigma^{(a)}$ to the quark
self-energy we obtain
\beq 
     \delta\Sigma^{(a)} = - \frac{1}{2} D_\sigma(0) \; 
     \sum_{M}\;\intp \;D_M(p)\,\Gamma_{M,M,\sigma}(p,-p) 
     = D_\sigma(0)\,\Delta \;.
     \label{deltasigmaa}
\eeq
Here we have introduced the constant $\Delta$,
\beq
    \Delta = \frac{1}{2}\intp\sum_M (-iD_M(p))
 (-i\Gamma_{M,M,\sigma}(p,-p))~, 
\label{Delta}
\eeq
which we will need later on for instance for the evaluation of diagram
$\delta\Pi_M^{(d)}$. The factor $1/2$ in the definition of $\Delta$ is
a symmetry factor, which is necessary in order to avoid double counting which
otherwise would arise from the sum over both possible orientations of the quark
loop in the definition of quark triangle, \eq{trianglevertex}. $\Delta$
consists of a quark triangle coupling to a meson loop and an external scalar
coupling. In principle we should have allowed also for other external
couplings, for instance a pseudoscalar one, and sum over all these couplings
and the corresponding meson propagators in \eq{deltasigmaa}. It turns out,
however, that only the scalar coupling leads to a nonvanishing contribution to
$\Delta$ because $\Gamma_{M,M,M_1}(p,-p)=0$ for $M_1 \neq \sigma$
(cf. App.~\ref{functions}).   

For the momentum dependent correction term 
$\delta\Sigma^{(b)}$ to the quark self-energy we arrive at the following
expression:  
\beq 
     \delta\Sigma^{(b)}(k) =  i \sum_{M}\; \intp \; D_{M}(p)\;
     \Gamma_M\,S(k-p)\,\Gamma_M \;.
     \label{deltasigmab}
\eeq
The $1/N_c$-correction to the quark condensate is closely related to the quark
self-energy (cf.~\eqs{qbq} and (\ref{deltaqbq})).
Inserting the expressions for $\delta\Sigma^{(a)}$
and $\delta\Sigma^{(b)}$ into \eq{deltaS}, and performing the trace in
\eq{deltaqbq} we obtain 
\beq
\delta\qq = -\Delta\Pi_\sigma(0)D_\sigma(0)+\Delta~.
\eeq
This expression can be further simplified with the help of the definition of
the RPA propagator $D_\sigma$ in terms of the corresponding polarization
function $\Pi_\sigma$, cf. \eq{dpisigma}. We finally obtain for the
$1/N_c$-correction to the quark condensate
\beq
    \delta\qq = - \frac{D_\sigma(0)\,\Delta}{2 g_s}\;.
\label{deltaqbqexp}
\eeq 

The evaluation of the various contributions to the polarization functions
give 
\bea
\delta\Pi^{(a)}_{M}(q) &=& \phantom{-} \frac{i}{2}\intp \sum_{M_1 M_2}
\Gamma_{M,M_1,M_2}(q,p)\, D_{M_1}(p)\,\Gamma_{M,M_1,M_2}(-q,-p)\,
D_{M_2}(-p-q)
\;, \nonumber\\
\delta\Pi^{(b)}_{M}(q)&=&- i\,\intp \hspace{3.0mm} \sum_{M_1} \;
\Gamma_{M,M_1,M_1,M}(q,p,-p)\,D_{M_1}(p)
\;, \nonumber\\
\delta\Pi^{(c)}_M(q)&=& -\frac{i}{2}\intp \hspace{2.5mm} \sum_{M_1} \; 
\Gamma_{M,M_1,M,M_1}(q,p,-q)\,D_{M_1}(p)
\;, \nonumber\\
\delta\Pi^{(d)}_M(q) &=& \frac{i}{2}\;\Gamma_{M,M,\sigma}(q,-q)\, D_\sigma(0)
\; \intp \hspace{3.0mm} \sum_{M_1}\; 
\Gamma_{M_1,M_1,\sigma}(p,-p) \, D_{M_1}(p)
\;, \nonumber\\
&=& -i\,\Gamma_{M,M,\sigma}(q,-q)\,D_\sigma(0)\,\Delta
\;.  
\label{deltapi}
\eea
The symmetry factor $1/2$ in $\delta\Pi^{(c)}_M$ and $\delta\Pi^{(d)}_M$ has
been introduced for the same reason as the factor $1/2$ in \eq{deltasigmaa}.
In $\delta\Pi^{(a)}_M$ it is necessary in order to account for the fact that
the exchange of $M_1$ and $M_2$ leads to the same diagram.

For the further evaluation of \eqs{deltasigmaa} to (\ref{deltapi}) we will 
proceed
in two steps. First the quark loops for the RPA meson propagators and the
effective three- and four-meson vertices will be calculated and then we will
evaluate the remaining meson loops.

In principle the sum in \eqs{deltasigmaa} to (\ref{deltapi}) is over all
possible intermediate states, i.e.\@ $\sigma$-,$\pi$-,$\rho-$ and
$a_1$-mesons.  However, pions, the lightest particles in the system, are
expected to yield the dominant contribution to most applications. For
instance, the behavior of the quark condensate at low temperatures is mainly
driven by thermally excited pions. The $\rho$-meson spectral function is also
primarily determined by a two-pion intermediate state. Other possible
contributions to $\delta\Pi^{(a)}$, like $\pi a_1$-, $\rho\sigma$-,
$\rho\rho$- or $a_1a_1$- intermediate states are expected to be much less
important since the corresponding decay channels open far above the
$\rho$-meson mass. In principle we have to keep in mind that the
physical $a_1$ as well as the $\rho$ are broad resonances which principally couple to
three- and two-pion intermediate states, respectively. Therefore the physical
threshold for the decay of the $\rho$ into $\pi a_1$ or $\rho\rho$ lies 
at $4 m_{\pi}\sim 560$ MeV, i.e.\@ below the $\rho$-meson mass of
about 770 MeV.  But of course considerable contributions will arise only above
$2 m_\rho\approx 1.6$~GeV and $m_{a_1} + m_\pi\approx 1.4$~GeV.  Besides,
experimentally the contribution of these intermediate states is negligible
anyway, one finds that the branching ratio of $\rho\rightarrow\pi\pi$ is about
100\%~\cite{PDG}.

From a phenomenological point of view it therefore seems well justified to
consider only pionic intermediate states. However, consistency with chiral
symmetry requires that also scalar states are considered, whereas vector and
axial intermediate states can be neglected without any difficulties.  Since
the evaluation of the various contributions can be simplified considerably by
neglecting vector and axial intermediate states we will exclusively take
scalar and pseudoscalar intermediate states into account for most of the
calculations. To describe a $\rho$-meson we have certainly to allow for an
external vector coupling in the diagrams shown in \fig{fig4}.

Another point which complicates the treatment of vector and axial
intermediate states is that they cause strong divergencies in the correction
terms to the polarization functions. On the one hand this has certainly to be
attributed to the non-renormalizability of the NJL model, on the other hand it
is related to a rather fundamental problem concerning the regularization of
the RPA vector and axial polarization function. This will be discussed in more
detail in \Sec{regularization} and \Sec{pionml}, respectively. 

\subsection{$\Phi$-derivable theory}
\label{Phi}
The main disadvantage of the approximation discussed in the previous section is
that the $1/N_c$-correction terms are treated perturbatively. For instance,
though we have considered corrections to the quark self-energy, these have not
been taken into account in the gap equation. Consequently all
quark propagators contributing to the $1/N_c$-corrected polarization functions
are determined in Hartree approximation. As long as the $1/N_c$-correction
terms are small compared with the leading-order contributions this approach
seems reasonable. But a description of the chiral phase transition will certainly not be possible within that scheme.

An approach which treats the mesons selfconsistently is highly desirable,
since e.g.\@ the thresholds for the decay $\rho\rightarrow\pi a_1$ would then
be 
located at $4 m_\pi$, but this is a very difficult task. For this reason we will be
content with a scheme which enables us at least to treat the quarks
selfconsistently. In the literature two functional methods can be found which,
in principle, allow to construct a selfconsistent scheme which is consistent
with chiral symmetry. One is the so-called
``$\Phi$-derivable''-method~\cite{luttinger,baym}, the other an expansion
of the ``effective action''~\cite{GSW,jona}. The latter will be discussed in
the next section. We will begin here by illustrating the $\Phi$-derivable
method on the Hartree + RPA scheme which emerges as the first approximation.
Subsequently we will go one step further which will generate a non-local
contribution to the gap equation.

The starting point of a $\Phi$-derivable theory is a functional $\Phi$, which
in principle contains all two-particle-irreducible skeleton diagrams which can
be constructed from the interactions and fields of the underlying Lagrangian.
Two-particle-irreducible diagrams are those which cannot be split into
parts by cutting two lines. In the diagrams all Greens functions have to
be full ones. The corresponding self-energies can be obtained by
functional derivatives of $\Phi$ with respect to the (full) Greens
functions. Generally a symmetry conserving approximation is generated by
taking any subset of diagrams~\cite{baym}. The thermodynamic potential per
volume, $\Omega$, which is identical to the energy density of the system in
vacuum, can be expressed as follows with the help of the functional $\Phi$, 
\beq 
\Omega = -i Tr \ln(i\;S^{-1}) + Tr
(\Sigma\, iS) + \Phi(S)~, 
\eeq
where $S^{-1}=S_0^{-1}+\Sigma$ represents the full Greens function, i.e.\@ in
our case the full quark propagator. The symbol $Tr$ denotes here a
trace over internal degrees of freedom, such as color-, spin- or isospin degrees
of freedom, as well as an integral over momentum or coordinate space. For instance,
in vacuum we have 
\[
Tr(\Sigma S) \equiv \intp \Tr(\Sigma(p) S(p))~.
\]
An expression for the self-energy can be derived by requiring stationarity of
$\Omega$ with respect to variations of the full propagator. One finds 
\beq
\Sigma = -\frac{\delta\Phi}{\delta\; (iS)}~.
\label{selbst}
\eeq
Evidently this means that the self-energy is given by the functional
derivative of $\Omega$ with respect to the full propagator. The scattering
kernel for the Bethe-Salpeter equation is in turn given by the functional
derivative of the self-energy with respect to the full propagator,
\beq
K = \frac{\delta\Sigma}{\delta iS}= - \frac{\delta^2\Phi}{\delta\; (iS)\delta\; (iS)}~.
\eeq

\begin{figure}[h!]
\parbox{16cm}{
\begin{center}
     \epsfig{file=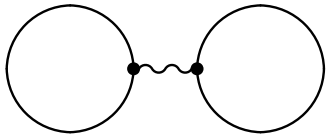}
\end{center}}
\caption{\it The ``glasses'', simplest contribution to the functional $\Phi$.}
\label{figbrille} 
\end{figure}

Let us now begin considering the simplest possible subset of
two-particle-irreducible diagrams. In the NJL model this is given by the
``glasses'', displayed in \fig{figbrille}. Again, solid lines represent quark
propagators. The wavy lines have been introduced to visualize the direction of
the interaction at the four-point vertices.
To that approximation $\Phi$ is explicitly given by 
\beq
\Phi^{(0)}[S^{(0)}] = -\sum_M g_M  \Big(-\intp\Tr(\Gamma_M \,i S^{(0)}(p))\Big)^2~.
\eeq
The superscript ${(0)}$ again denotes quantities to first approximation. 
Below we will demonstrate that the above expression for the
functional $\Phi$ generates the Hartree + RPA scheme for quarks and mesons,
respectively. 

According to \eq{selbst} the quark self-energy is a priori momentum
dependent and a matrix in color-, flavor- and Dirac-space. Performing the
functional derivative ($i$ and $j$ are multi-indices, referring to color-,
flavor- and Dirac-space),
\bea
\Sigma^{(0)}_{ij}(q_1,q_2) &\equiv& \delta(q_1+q_2)\Sigma^{(0)}_{ij}= \intf{1}\intf{2}
\frac{\delta\Phi^{(0)}[S^{(0)}]}{\delta\;
  (iS^{(0)}_{ij}(x_1,x_2))}\nonumber \\ &=&\delta(q_1+q_2)
\sum_M 2 i g_M \Gamma_{M,ji} \intp \Tr(\Gamma_M S^{(0)}(p))~, 
\label{selbstexp}
\eea 
we find, however, that the self-energy is at least momentum independent.
Remembering that only the scalar channel leads to a nonvanishing contribution
we conclude that it is in addition proportional to unity in all spaces.
After a comparison of the resulting expression for the self-energy with the 
gap equation,
\eq{gap}, it becomes obvious that $\Sigma^{(0)}_{ij}$ is identical to the
Hartree self-energy $\Sigma_H$ derived in \Sec{hartree}.

For the
scattering kernel $K^{(0)}$ we obtain
\bea
K^{(0)}_{ij,kl}(q_1,q_2,q_3,q_4)&=&
\int d^4x_1\int d^4x_2\int d^4x_3\int d^4x_4 {\mathrm e}^{-i
  (q_1\cdot x_1+q_2\cdot x_2+q_3\cdot x_3+q_4\cdot x_4)}\begin{array}{c}\frac{
    \delta\Sigma^{(0)}_{ij}(x_1,x_2)}{\delta\;
  (i S^{(0)}_{kl}(x_3,x_4))}\end{array}\nonumber \\
&=& \sum_M 2 g_M \Gamma_{M,ji}\Gamma_{M,lk} \delta(q_1+q_2+q_3+q_4)~.
\label{kern} 
\eea
As can be seen from a comparison with \eq{kkern} this expression meets our
expectations: It is indeed the scattering kernel for the $T$-matrix in RPA. We
therefore conclude that the Hartree + RPA approximation scheme can be derived
within the
$\Phi$-derivable method, if the functional $\Phi$ is restricted to the ``glasses''. 
\begin{figure}[h!]
\begin{center}
     \epsfig{file=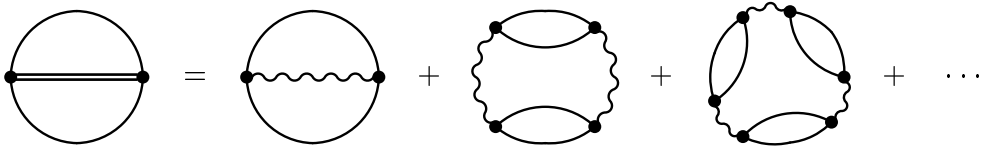}\end{center}
\caption{\it The `` ring sum'', contribution to the functional $\Phi$.}
\label{figring} 
\end{figure}      

\begin{figure}[h!]
\begin{center}
     \epsfig{file=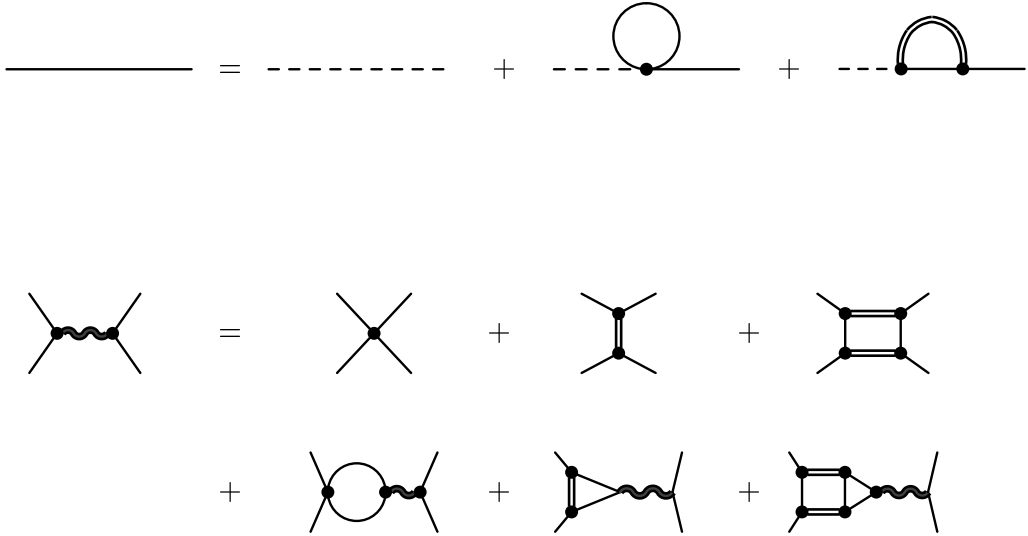}\end{center}
\caption{\it (upper panel) Gap equation as it results from $\Phi$ if the ring
  sum is included, 
  (lower panel) corresponding quark-antiquark scattering matrix.}
\label{figphi} 
\end{figure}

A suggestive extension for $\Phi$ is to take also the ``ring sum'', visualized
in \fig{figring}, into account. Evaluating the diagrams, ``glasses'' and ring
sum, we arrive at the following expression for $\Phi$: 
\bea
\Phi[S] &=& -\sum_M g_M  \Big(-\intp\Tr(\Gamma_M\, iS(p))\Big)^2\nonumber \\
&& -\frac{i}{2}\intp\{ \ln (1- 2 g_s\Pi_{\sigma}(p))+3 \ln (1- 2
g_s\Pi_{\pi}(p))\}~.
\eea
The functions $\Pi_{M}(p)$ were defined in \eq{pol0}. We have to keep in
mind that $S$ here denotes a full propagator and not a propagator in Hartree
approximation as it originally appeared in the definition of $\Pi_M$ in
\eq{pol0}.  Proceeding in the same way as above (cf. \eq{selbstexp}) we obtain
for the quark self-energy, $\Sigma_{ij}(q_1,q_2) =
\delta(q_1+q_2)\Sigma_{ij}(q_1)$,
\beq 
\Sigma_{ij}(q_1) = \Sigma_{H,ij} + \delta\Sigma^{(b)}_{ij}(q_1)~.
\label{sigmaring}
\eeq 
$\delta\Sigma^{(b)}(q)$ was defined in \eq{deltasigmab}. The same remark
of caution as above concerning the quark propagator $S$ is in order 
here: The expressions on the r.h.s.\@ of \eq{sigmaring} have to be
evaluated with the full propagator. The upper panel of \fig{figphi} displays
the corresponding gap equation. Here solid lines stand for quark propagators
as they emerge from the selfconsistent solution of \eq{sigmaring}. 

To determine the scattering kernel for the Bethe-Salpeter equation which
describes the mesons, the analogous procedure leading to \eq{kern} is
applied. One finds
\bea
K_{ij,kl}(q_1,q_2,q_3,q_4) 
&=& \delta(q_1+q_2+q_3+q_4) \Big\{\sum_M 2 g_M
\Gamma_{M,ji}\Gamma_{M,lk}+ D_M(q_1+q_4) \Gamma_{M,li}\Gamma_{M,jk}\nonumber\\
&&\hspace{-4cm} +2\; i \;\intk D_M(k+q_3)D_M(k-q_4) (\Gamma_M i
S(k-q_1-q_4)\Gamma_M)_{ji}(\Gamma_M i S(k)\Gamma_M)_{kl}\Big\}~. 
\label{kernring}
\eea
The corresponding Bethe-Salpeter equation for the quark-antiquark $T$-matrix
is displayed in the lower part of \fig{figphi}. In Ref.~\cite{OBW2} this
representation of the $T$-matrix was generated graphically by coupling an
external current to each quark line in the 
corresponding gap equation, which is shown in the upper part of
\fig{figphi}. Moreover, with the help of axial Ward identities one can show
generally the existence of a pole in $T$ for $(q_1+q_2)^2=0$. 

Because of the non-trivial momentum dependence of the self-energy it is a very
difficult task to obtain an explicit solution of the gap equation,
\eq{sigmaring}. Therefore we will put this scheme aside and concentrate in
explicit calculations on the expansion of the effective action which leads to
a local contribution to the gap equation and is much easier to handle. This
will be discussed in the next section.

\subsection{One-meson-loop expansion of the effective action}
\label{1ml}
The scheme we will derive within this section was first discussed by
Dmitra\v{s}inovi\'c et al.~\cite{dmitrasinovic}. The authors started from a
gap equation which in addition to the Hartree self-energy contains a local
meson-loop contribution. They found a consistent scheme to describe mesons and
proved various relations following from chiral symmetry explicitly , like
e.g.\@ the Goldstone theorem. 
The same scheme was later derived more systematically by Nikolov et
al.~\cite{nikolov} using the effective action formalism.  Here we will
follow the method of that reference.  Let us begin with a derivation
of an expression for the effective action. We will mainly follow the
presentation of Ref.~\cite{weinberg}, Chapter~16. The interested reader is
referred to that reference for further details.  The starting point is
a theory with an action $I[\phi]$. Throughout the derivation we will,
for simplicity, only regard a theory with a scalar field $\phi$, which
couples to an external classical source $j$. The generating functional
$W[j]$ is then given in path integral representation by
\beq
{\mathrm e}^{i W[j]} = \int {\mathcal D}(\phi) \exp\Big( i I[\phi]+ i \int
d^4x \phi(x) j(x)\Big)
\label{gfunct}
\eeq 
The functional $W[j]$ contains all connected vacuum diagrams,
which can be constructed from the underlying theory in the presence of the
source $j$. The vacuum expectation value of the fields $\phi$ can be obtained
from a derivative of $W[j]$ with respect to the external source $j$
\beq 
\ave{\phi(x)} = \frac{\delta W[j]}{\delta j(x)}\Big|_{j=0}~.
\eeq 
This expectation value can in general also be defined in the presence of the
source, we will call it $\varphi_J$, 
\beq 
\varphi_J(x) = \frac{\delta W[j]}{\delta j(x)}\Big|_{j=J}~.
\label{trafo}
\eeq 
For a given $\varphi_J$ this equation can in principle be solved for $J$.
Throughout the following derivation we will denote the current, which is related to $\varphi_J$
via \eq{trafo}, by $J$. A functional $\Gamma[\varphi_J]$, which no longer
depends on $j$ but on $\varphi_J$, can be obtained from $W[j]$ by performing a
Legendre transform,
\beq
\Gamma[\varphi_J] = W[J] - \int d^4 x\varphi_J(x) J(x)~.
\label{eadef}
\eeq This functional $\Gamma[\varphi_J]$ is called ``effective action''. It
contains all one-particle-irreducible diagrams (in the presence of the source)
of the theory~\cite{weinberg}. Similarly to two-particle-irreducible diagrams,
one-particle-irreducible ones are those which cannot be split into parts by
cutting one line. One important property of $\Gamma$ can be derived from the
derivative with respect to $\varphi_J$, 
\bea
\frac{\delta\Gamma[\varphi_J]}{\delta\varphi_J(y)} &=& \int d^4 x\frac{\delta
  W[j]}{\delta j(x)}\frac{\delta j(x)}{\delta \varphi_j(y)}\Big|_{j=J}-\int
d^4
x \varphi_J(x)\frac{\delta J(x)}{\delta \varphi(y)} - J(y)\nonumber \\
&=& -J(y)~.  
\eea 
Hence, if the source $J$ vanishes, only the fields
$\varphi_S$ at the stationary points of $\Gamma[\varphi_J]$, 
\beq
\frac{\delta\Gamma[\varphi_J]}{\delta\varphi_J(y)}\Big|_{\varphi=\varphi^S} =
0~, 
\eeq 
are possible fields $\varphi_J$. This means in particular that the
vacuum expectation values of the field $\phi$ are given as the values of
$\varphi_J$ at the stationary points of $\Gamma$, i.e.\@ $\varphi^S =
\ave{\phi}$. From this point of view the field $\varphi_J$ behaves similarly
to a classical field whose equations of motion can be generated from the
stationarity condition of the action $I[\phi]$. Thus the effective action can
be regarded as some sort of ``action with quantum fluctuations''.

Propagators can be obtained from the second derivative of
$W[j]$ with respect to the source,
\beq
\Delta(x,y) = \ave{\Phi(x)\Phi(y)} = \frac{\delta^2 W[j]}{\delta j(x)\delta
  j(y)}\Big|_{j=0}=\frac{\delta\varphi_J(x)}{\delta j(y)}\Big|_{j=0} ~.
\eeq
This can also be derived from the effective action,
\beq
\frac{\delta^2\Gamma[\varphi]}{\delta\varphi(x)\delta\varphi(y)}\Big|_{\varphi=\varphi^S} =
\frac{\delta j(x)}{\delta\varphi_j(y)}\Big|_{j=0}= \Delta(x,y)^{-1}~.
\label{tpoint}
\eeq

It is convenient to introduce another quantity, the {\it effective
  potential}. Let us assume that we wish to calculate the effective action
$\Gamma[\varphi]$ with constant, i.e.\@ space-time-independent, fields
$\varphi(x)\equiv \varphi_0$. Then $\Gamma[\varphi]$ will be proportional to
the volume of space-time ${\mathcal V}_4 = \int d^4 x$. Therefore we can write
the effective action in the form
\beq
\Gamma[\varphi] = -{\mathcal V}_4 V(\varphi_0)~,
\label{defep}
\eeq  
where $V(\varphi_0)$ is not a functional of $\varphi$ but an ordinary
function. This function is known as the effective potential~\cite{coleman}.
The effective potential can be interpreted in terms of the energy
density~\cite{weinberg}. $V(\varphi_0)$ can be
interpreted as the minimum of the energy density for all states with
expectation values $\varphi$ for the fields $\phi$. 

After having explained some general aspects of the effective action let us now
come to the determination of an explicit expression for $\Gamma[\varphi]$.
Certainly it will be impossible to obtain an exact result for the effective
action in an interacting theory because the path integral in \eq{eadef} can in
general not be carried out. This is only possible for constant, linear or
quadratic terms. Thus, expanding the argument of the exponential in \eq{gfunct}
around some fixed field $\varphi_0(x)$ which satisfies 
\beq 
\frac{\delta
  I[\phi]}{\delta \phi}\Big|_{\phi = \varphi_0} = -j 
\eeq 
up to quadratic
terms in the fields, we can evaluate the path integral and arrive at an
expression for the effective action $\Gamma[\varphi_0]$ up to
``one-loop'' (see~\cite{weinberg}, Chapter 16), 
\beq 
i\;\Gamma^{(1)}[\varphi] =
i I[\varphi] + \frac{1}{2} Tr\ln \Big(i \frac{\delta^2
  I[\varphi]}{\delta\varphi\delta\varphi}\Big)~.
\label{gamma1loop}
\eeq 
The first term in \eq{gamma1loop} is the ``zero-loop'' or ``tree-level'',
the second one the ``one-loop'' contribution. Strictly speaking the field
$\varphi$ which arises from the Legendre transform in \eq{eadef} not
necessarily needs to coincide with $\varphi_0$ which has been introduced for
the evaluation of the path integral. However, at tree-level it is obvious that
both fields are identical. Besides, it can be shown that in a perturbative
expansion this is true up to one-loop order~\cite{jackiw}.

It can be shown (see e.g.\@~\cite{weinberg}) that in most cases the effective action
$\Gamma[\varphi]$ has the same symmetries as the action
$I[\phi]$. Especially symmetries which are related to linear transformations
of the fields are conserved. This has an important consequence (see
e.g.\@~\cite{weinberg,ripka}): One can show that a spontaneously broken continuous
symmetry then generates a pole at $q^2=0$ in the two-point function
$\Delta(q)$. This corresponds to the existence of a Goldstone boson. 

We are now in a position to apply the effective action formalism to the NJL model. A detailed
discussion can be found
in Ref.~\cite{ripka}. We will begin our presentation with the lowest-order
contribution, i.e.\@ the tree-level approximation which will turn out to be
equivalent to Hartree approximation + RPA.

In the remaining part of this section we will drop the vector and axial vector
interaction and  start from a Lagrangian which contains only scalar and
pseudoscalar interaction terms, i.\@ e.
\beq
   {\cal L} \;=\; \pb ( i \partial{\hskip-2.0mm}/ - m_0) \psi
            \;+\; g_s\,[(\pb\psi)^2 + (\pb i\gamma_5{\vec\tau}\psi)^2]  
   \,.
\eeq
The generating functional of the system can be expressed in terms of the path
integral 
\beq
 Z[\eta,\bar{\eta}] = e^{iW[\eta,\bar{\eta}]} = \int {\mathcal D}(\pb){\mathcal D}(\psi)
 e^{iI[\pb,\psi]+ i\int d^4 x (\pb(x)\eta(x) + \bar{\eta}(x)\psi(x))}~,
\label{z}
\eeq
with the action 
\beq
I[\pb,\psi] = \int d^4 x {\cal L}(x)~.
\eeq
To further proceed it is convenient to ``bosonize'' the action. This is
achieved by introducing auxiliary hermitian fields $\Phi_a^{\prime}, a =
\{0,1,2,3\}$. These collective bosonic fields are chosen in such a way that
the action becomes bilinear in the quark fields, i.\@ e.
\bea
Z[\eta,\bar{\eta}] &=& \int {\mathcal D}(\pb){\mathcal D}(\psi){\mathcal D}(\Phi_a^{\prime})
\exp\Big\{iI(\pb,\psi)+ i \int d^4 x(\pb(x)\eta(x) +
\bar{\eta}(x)\psi(x)) \nonumber \\ &&\hspace{2cm}
-\frac{i}{4 g_s}\int d^4 x(\Phi_a^{\prime}+2 g_s\pb\Gamma_a\psi)^2 \Big\}~,
\eea
with $\Gamma_a = (1,i\gamma_5\vec{\tau})$. 
The quark fields can now be integrated out. 
After performing a shift of the auxiliary 
fields, $\Phi_a = \Phi^{\prime}_a + (m_0,\vec{0})$, one finally arrives at the
bosonized form of the generating functional 
\beq
Z[\eta,\bar{\eta}] = \int {\mathcal D}(\Phi_a)\exp\Big\{ i I[\Phi_a]+i\int d^4
x\int d^4 y \bar{\eta}(y)\,{\mathrm D}^{-1}(y,x)\eta(x)\Big\}
\eeq
with the action
\beq
 I[\Phi] =  -i\;Tr\ln i {\mathrm D} - \frac{1}{4 g_s}\int d^4x 
           (\Phi^2-2 m_0 \Phi_0+m_0^2)~.
\label{ba}
\eeq
D is the Dirac operator 
\beq
{\mathrm D}(x,y) = (i \dslash - \Gamma_a\Phi_a) \delta(x-y)~.
\eeq
This notation should not be confused with the meson
propagator $D_M(q)$ we introduced earlier.
The first term in \eq{ba} arises from the ``fermion determinant''. 
The symbol $Tr$ is therefore to be
understood as a functional trace and a trace over internal degrees of freedom
such as flavor, color and spin. This is the quark-loop contribution. Expanding the
fields around some constant field $\Phi_a= (M,\vec{0}) + \Delta \Phi_a$ we can
rewrite the Dirac operator as follows
\bea
{\mathrm D}(x,y) &=& (i \dslash-M)\delta(x-y)
-\Gamma_a\Delta\Phi_a\delta(x-y)\nonumber\\
&\equiv& {\mathrm D_0}(x,y) - V~,
\eea
where $V$ is of first order in the fluctuations $\Delta\Phi_a$. $D_0$
corresponds to a quark propagator with ``mass'' $M$, it can be written as
\beq
D_0(x,y) = \intp {\mathrm e}^{i p \cdot (x-y)} S(p)~.
\eeq
The logarithm
in the quark-loop term to the action can then be expressed in powers of the
fluctuations $\Delta\Phi_a$, 
\bea
\ln (i{\mathrm D}) &=& \ln (i{\mathrm D_0}) + \ln (1-{\mathrm D_0}^{-1}
V)\nonumber\\
&=&\ln (i{\mathrm D_0}) -\sum_{n=1}^{\infty}\frac{({\mathrm D_0}^{-1}
  V)^n}{n}~. 
\label{log}
\eea 
This expression will be helpful when we determine the derivatives of
the action $I$ with respect to the fields $\Phi$.
In which way are the expectation values of the auxiliary fields $\Phi$ related
to physical quantities we are interested in? This can most directly be seen
for the quark condensate:   
It can be expressed via the expectation value of $\Phi_0$ as 
\beq
\qq = \frac{\partial W}{\partial m_0} = -\frac{1}{2 g_s} (\ave{\Phi_0}-m_0)~.
\label{qqea}
\eeq
From \eq{tpoint} we can infer that the meson propagators can be obtained as
second-order derivatives from the effective action.
In ``zero-loop'' approximation the effective action is simply given by, 
\beq
\Gamma^{(0)}[\Phi] = I[\Phi]~,
\eeq
where
we have dropped the fermionic source terms $\eta$ and $\bar{\eta}$ for
simplicity. 
To determine the expectation values of the fields $\Phi$ we first have to build
the functional derivative with respect to the fluctuations
$\Delta\Phi$. To that end only the first order term in \eq{log} has to be
considered. We arrive at the following expression
\beq
\frac{\delta\Gamma^{(0)}}{\delta
  \Delta\Phi_b(z)}\Big|_{\Delta\Phi=0} = \intp
\Tr[i S(p)\Gamma_b]-\frac{1}{2 g_s}(M-m_0)\delta_{b0}~.
\eeq
Since the trace in the first term on the r.h.s.\@ vanishes for $b\neq 0$ there
is no contribution with $b\neq 0$. 
Comparing the first term on the r.h.s.\@ of the above expression with the
definition of the Hartree self-energy we conclude that the expectation value
of the fields $\Phi$ in the present approximation can be written in the form
$(M,\vec{0})$, where $M$ is given by the solution of
the following equation: 
\beq
M-m_0-\Sigma_H(M) = 0~,
\eeq
with $\Sigma_H$ as defined in \eq{gap}. If we identify the expectation value
of the zeroth component of the field $\Phi$, i.e.\@ $M$, with the constituent
quark mass $m_H$ in Hartree approximation, this equation is identical with the
gap equation (\eq{gap}). We have to emphasize here that the interpretation of
$\ave{\Phi_0}=M$ as a constituent quark mass is not clear a priori. In
principle the constituent quark mass should be determined from the pole of the
quark propagator. However, in Hartree approximation this pole coincides with
$\ave{\Phi_0}$. 

If we evaluate the second-order derivative of the effective action $\Gamma^{(0)}$
at the stationary point, we obtain the following result
\beq
\intf{1}\intf{2}\frac{\delta^2\Gamma^{(0)}}{\delta
  \Delta\Phi_a(x_1)\Delta\Phi_b(x_2)}\Big|_{\Phi=(M,\vec{0})} =
\delta(q_1+q_2)\delta_{bc}\Pi_{b}(q_1) -\frac{\delta(q_1+q_2)\delta_{bc}}{2 g_s}~,
\label{zloopmesons}
\eeq 
where we have already exploited the fact that in our case the polarization
function $\Pi_M$ is diagonal in flavor space. Thus, comparing this result
with the definition of the inverse meson propagators in \eq{tpoint}, we
conclude that we in this way exactly recover the inverse meson propagators in
RPA, cf. \eq{dpisigma}.  This allows us to draw the conclusion that the
effective action formalism in ``zero-loop'' approximation yields the same
results as the Hartree approximation + RPA. 

Extending the effective action to
``one-loop'', starting from a bosonized version of the NJL model, means that
we take into account mesonic fluctuations. The effective action is then given
by \eq{gamma1loop}.  The second term in \eq{gamma1loop} contains the mesonic
fluctuations.  As discussed in Ref.~\cite{ripka} the method is only meaningful
if the second-order functional derivative which enters into this term is
positive definite. Otherwise severe problems arise due to an ill-defined
logarithm, which would then be complex. We will come back to this point in
\Sec{solution}.

Although we introduced this approximation rather generally as ``one-loop''
approximation we will throughout this work adopt the notation of
Ref.~\cite{nikolov} and call this approximation scheme the ``one-meson-loop
approximation'' (MLA). This name is motivated by the fact that one-loop
contribution in fact contains mesonic fluctuations due to the preceding
bosonization procedure. In addition we have to mention that the 
''quark-loop'' approximation described above is not really a tree-level 
approximation which would be suggested by the introduction of $\Gamma^{(0)}$ 
in \eq{gamma1loop}. In order to arrive at the bosonized form of the action, we
have integrated out the quark fields, i.e.\@ the bosonized action in principle
contains all quark loops.  
\begin{figure}[b!]
\begin{center}
     \epsfig{file=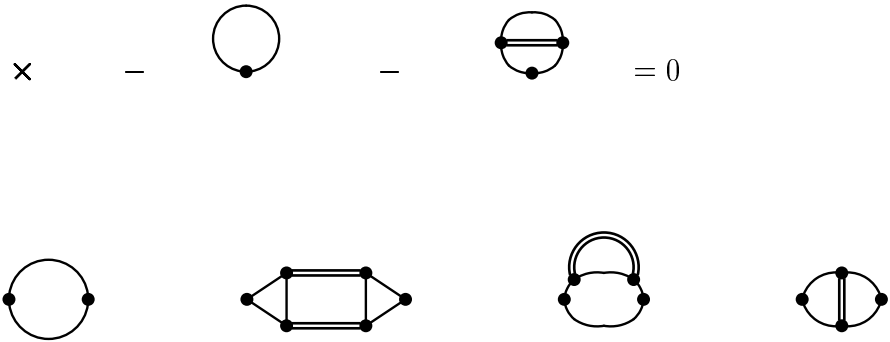}\end{center}
\caption{\it The ``one-meson-loop approximation scheme'': gap-equation (upper
  part), and the corresponding mesonic polarization functions (lower part). 
             The double line denotes an RPA meson propagator 
             (see \fig{fig2}), which is selfconsistently constructed from 
             the quark propagators dressed with the local self-energy of the
             present equation (solid line). The cross stands for the
             constant term $\ave{\Phi_0}-m_0$ in the gap equation
             (\eq{localgap}). }   
\label{figlocal} 
\end{figure}      
 
For the following derivation of explicit expressions in one-meson-loop 
approximation it is convenient to first establish relations between the
effective meson-vertices introduced in \Sec{1/N_c} and the third- and
fourth-order derivative of the action $I[\Phi]$,
\bea
\int d^4 x_1 d^4 x_2 d^4 x_3 {\mathrm e}^{-i (q_1\cdot x_1+
q_2\cdot x_2+q_3\cdot x_3)} \frac{\delta^3
  I[\Phi]}{\delta(\Phi_a(x_1))\delta(\Phi_b(x_2))
\delta(\Phi_c(x_3))}\Big|_{\Delta\Phi=0} &&\nonumber \\* =  
 \delta(q_1+q_2+q_3) \Gamma_{a,b,c}(q_1,q_2) && \nonumber\\*
\int d^4 x_1 d^4 x_2 d^4 x_3 d^4 x_4 {\mathrm e}^{-i (q_1\cdot
  x_1+q_2\cdot x_2+q_3\cdot x_3+q_4\cdot x_4)} \frac{\delta^4
  I[\Phi]}{\delta(\Phi_a(x_1))\delta(\Phi_b(x_2))
\delta(\Phi_c(x_3))\delta(\Phi_d(x_4))}\Big|_{\Delta\Phi=0}
&&\nonumber\\* = -
i \delta(q_1+q_2+q_3+q_4) \Big(\Gamma_{a,b,c,d}(q_1,q_2,q_3)+\Gamma_{a,b,d,c}(q_1,q_2,q_4)
+\Gamma_{a,c,b,d}(q_1,q_3,q_2) \Big)~.&&
\label{dreiviervertex}
\eea 
These relations enable us to straightforwardly derive an equation for the
stationary point of the effective action in \eq{gamma1loop}. The following 
``gap equation'' holds~\cite{nikolov}: 
\beq
\ave{\Phi_0}-m_0-\tilde{\Sigma}(\ave{\Phi_0}) =
\ave{\Phi_0}-m_0-\Sigma_H(\ave{\Phi_0})-\delta\tilde\Sigma(\ave{\Phi_0}) = 0~.
\label{localgap}
\eeq
Here $\Sigma_H$ is the Hartree contribution to $\ave{\Phi_0}$ as defined
in \eq{gap}. In the upper part of \fig{figlocal} this gap equation is
visualized graphically. The cross indicates the constant term
$\ave{\Phi_0}-m_0$, the second term corresponds to the Hartree contribution
and the correction term $\delta\tilde\Sigma$ corresponds to the third diagram.
It consists of a quark loop dressed by an RPA-meson loop.
It can be shown, that at the external vertex only the scalar interaction
contributes.  
Hence $\delta\tilde\Sigma$ is given by
\beq
    \delta\tilde\Sigma(\ave{\Phi_0}) = -2g_s\,\Delta(\ave{\Phi_0})   \;,
\label{deltatildesigma}
\eeq
where $\Delta$ is the constant defined in \eq{Delta}. 
We should emphasize that these diagrams have to be 
evaluated selfconsistently at $\ave{\Phi_0}$, which is a solution 
of \eq{localgap}. Thus all ``quark propagators'', also those which enter into
the calculation of the RPA meson propagators, have the form
$S(p)=(\pslash-\ave{\Phi_0})^{-1}$. Because of the new diagram
$\delta\tilde\Sigma$, these are in general different from the Hartree quark
propagators. 

We have to comment on the notations we use here. The utilization of the symbol
$\Sigma$ for the different contributions and the name ``quark propagator'' for
$S(p)$ suggests that we again, as in the
``zero-loop'' approximation, deal here with quark self-energies. But as already
pointed out, the interpretation of $\ave{\Phi_0}$ as a constituent quark mass
should be considered with great care. In the Hartree approximation the expectation
value of the field $\Phi_0$ is equal to the pole of the quark
propagator. Therefore within this approximation, the interpretation of
$\ave{\Phi_0}$ as a constituent quark mass is correct. The form of the quark
lines which enter the calculations in the MLA seems
to point in the
same direction. But this is not the case. In Ref.~\cite{nikolov} it has
already been emphasized that $\ave{\Phi_0}$ is not equal to the pole of the
``real'' quark propagator. 
This will become
clearer if we look at the quark condensate.
If we take $S(p)$ being a quark propagator literally, we expect to obtain the
quark condensate by evaluating \eq{qbq} with the ``quark propagator'' $S(p)$,
\beq
    \qq = - \frac{\Sigma_H(\ave{\Phi_0})}{2g_s} = - \frac{\ave{\Phi_0} - m_0}{2 g_s} \;-\; \Delta(\ave{\Phi_0})
    \;.
\label{qbqself1}
\eeq
This result does not agree with the result we obtain following the
prescription in \eq{qqea} to determine the quark condensate as
\beq
    \qq = - \frac{\tilde\Sigma(\ave{\Phi_0})}{2g_s} = - \frac{\ave{\Phi_0} -
      m_0}{2 g_s}~.
\label{qbqself2}
\eeq
This expression is reduced to the perturbative
result in a strict $1/N_c$ expansion, whereas by expanding \eq{qbqself1} one
only recovers the contribution of $\delta\Sigma^{(a)}$. 
Moreover, as we will discuss in
\Sec{piondstl}, $\qq$ according to \eq{qbqself2} is consistent with the Gell-Mann--Oakes--Renner relation.
Hence, within the MLA,
\eq{qbqself2} is the correct expression for the quark condensate, 
whereas the gap equation (\eq{localgap}) should not be interpreted as an
equation for the corresponding inverse quark propagator. For simplicity, 
however, we will still call $\ave{\Phi_0}\equiv m^{\prime}$ a ``constituent quark
mass'' and $S(p) = (\pslash - m^{\prime})^{-1}$ a ``quark propagator'', although this
is not entirely correct.

Because of the additional diagram in the gap equation,
the RPA is not the consistent scheme to describe mesons:
In the chiral limit, RPA pions are no longer massless. Consistent meson
propagators can be obtained in the same way from the effective action as the
RPA meson propagators resulted from the effective action in zero-loop
approximation.  
According to \eq{tpoint}, these are given as second-order derivatives of the
effective action, evaluated at the stationary point. With the help of 
\eq{dreiviervertex} this derivative can easily be evaluated. This leads to
three additional mesonic polarization diagrams which are displayed in the
lower part of \fig{figlocal} together with the RPA polarization loop. 
Obviously these diagrams are identical to $\delta\Pi^{(a)}_{M}$, 
$\delta\Pi^{(b)}_{M}$ and $\delta\Pi^{(c)}_{M}$, which we defined in
Sec.~\ref{1/N_c} (\fig{fig4}, \eq{deltapi}), i.e.\@ the new meson
propagators are given by 
\beq
{\tilde D}_M(q) = \frac{-2 g_M}{1-2g_M {\tilde\Pi}_{M}(q)} \;,
\label{dmtilde}
\eeq
with
\beq
    {\tilde \Pi}_M(q) = \Pi_M(q) + \sum_{k=a,b,c}\; \delta \Pi_M^{(k)}(q)
\;.
\label{poltilde}
\eeq
Here we have dropped the momentum conserving $\delta$-function (see
\eq{zloopmesons}). As mentioned above, all diagrams have to be evaluated at the
stationary point, i.e.\@ with a constituent quark mass $m^{\prime}$ which is a 
solution of \eq{localgap}. For simplicity we renounced to indicate this point 
by adding
an additional mass argument in the above expressions. 
Since the general structure of the expressions we have to
evaluate is the same in both schemes, we will introduce the convention that
unless otherwise stated all quantities within the Hartree + RPA and the
{\nce} will be calculated with the Hartree quark mass $m_H$ and within the MLA
with the selfconsistently determined quark mass $m^{\prime}$. 

The above discussion clearly reveals that the $1/N_c$-arguments, put forward
by the authors of Ref.~\cite{dmitrasinovic} in order to motivate their choice
of diagrams for the extended gap equation, \eq{localgap} are questionable. In
addition to the fact that in a perturbative $1/N_c$-expansion there would be
two corrections to the quark self-energy in next-to-leading order, a momentum
independent term and a momentum dependent term (cf. \fig{fig3}), whereas the
last one is missing within this non-perturbative scheme, one should stress
again, that the selfconsistent solution of the gap equation mixes all orders in
$1/N_c$ anyway. 
In addition to the systematic approach we presented here following
Ref.~\cite{nikolov}, a possibility to
engender this scheme diagrammatically was discussed in
Ref.~\cite{OBW2}. 

In fact, our discussion of the one-loop approximation to the effective action
shows that no reference to $1/N_c$-counting is needed, neither within the
diagrammatic derivation in Ref.~\cite{OBW2}.  
Besides, if one performs a strict $1/N_c$ expansion of the 
mesonic polarization diagrams up to next-to-leading order one exactly 
recovers the diagrams shown in \fig{fig4} \cite{dmitrasinovic}.
This is quite obvious for the diagrams $\delta\Pi_M^{(a)}$ to
$\delta\Pi_M^{(c)}$, which are explicitly contained in \eq{poltilde}.
Diagram $\delta\Pi_M^{(d)}$, which seems to be missing, is implicitly
contained in the quark-antiquark loop via the next-to-leading order
terms in the quark propagator, which arise from the extended gap equation. 
 
Finally, we would like to stress that this scheme is selfconsistent with
respect to the quarks, but not with respect to the mesons. This can be seen easily: The
quark propagator, obtained from the selfconsistent solution of \eq{localgap},
is used in all loops, whereas the intermediate meson propagators, i.e.\@ RPA
propagators calculated with the selfconsistent quark propagators, are not
identical to the improved meson propagators in \eq{dmtilde}. We will come back
to this point in \Sec{numerics}.

\section{Consistency with chiral symmetry}
\label{secpion}

It was mentioned in the previous section that the effective action
formalism provides approximations which are consistent with the Goldstone
theorem. This can be shown on a rather general
level~\cite{weinberg,ripka}. The only condition is that the approximation for
the effective action conserves the symmetries of the action and the Lagrangian
respectively, which is fulfilled by the loop expansion we performed in the
previous section~\cite{weinberg}. Hence, the Hartree approximation + RPA as
well as the MLA yield massless Goldstone bosons, the pions. 
Since the mesonic polarization functions of the MLA
contain all diagrams up to next-to-leading order of the
$1/N_c$-expansion scheme and the various contributions to the pion mass
have to cancel order by order in the chiral limit, this implies that the 
$1/N_c$ scheme discussed in \Sec{1/N_c} is also consistent with the 
Goldstone theorem. 

Nevertheless, for the numerical implementation it is instructive to show
the consistency of the different schemes with chiral symmetry on a
less formal level. Since most of the integrals which have to be evaluated 
are divergent and must be regularized one has to ensure that the 
various symmetry relations are not destroyed by the regularization. 
To this end, it is important to know how these relations emerge in detail.
This will also enable us to perform further approximations without violating
chiral symmetry. As we will see in \Sec{solution}, this is very important
for practical calculations within the MLA,
which cannot be applied as it stands.

For instructive reasons we begin by an explicit proof of the Goldstone
theorem and the Gell-Mann--Oakes--Renner (GOR) relation within the Hartree
approximation + RPA although this is completely standard and can be found in
many references (see e.g.\@~\cite{njl,vogl,klevansky,hatsuda}).
Afterwards this will be discussed within the MLA and the {\nce}. 
An explicit proof of the Goldstone theorem in both schemes was given first by
Dmitra\v{s}inovi\'{c} et al.~\cite{dmitrasinovic}.  
The GOR relation is of particular interest in the context of the proper
definition of the quark condensate in the MLA. 
(cf. Eqs.~(\ref{qbqself1}) and (\ref{qbqself2})).

To keep the formalism as simple as possible, we restrict ourselves in this
section to scalar and pseudoscalar interactions. The explicit proof of the
Goldstone theorem including vector and axial vector interactions can be found
in App.~\ref{vchiral}.


\subsection{Hartree approximation + RPA}
\label{pionrpa}
The Goldstone theorem states the existence of massless pions in the chiral
limit. Therefore one has to show that the inverse pion propagator vanishes in
the chiral limit for zero momentum,
\beq 
    2g_s\,\Pi_\pi(0) = 1 \qquad {\rm for} \quad m_0 = 0~.
    \label{goldstoneRPA}
\eeq
As before we use the notation $\Pi^{ab}_\pi = \delta_{ab}\Pi_\pi$.  
An evaluation of the RPA loop gives 
\beq
    2g_s\,\Pi_\pi(0) = \frac{\Sigma_H}{m}  \;.
\label{pi0}
\eeq
We designedly called the constituent quark mass in \eq{pi0} $m$ and not $m_H$
because the validity of this relation between the RPA polarization loop and the
Hartree quark self-energy is not restricted to the Hartree approximation + RPA
itself. If we take this relation at the Hartree mass $m_H$, together with the
Hartree gap equation $m_H = m_0 + \Sigma_H$, the validity of
\eq{goldstoneRPA} in the chiral limit is ensured.   

Going away from the chiral limit the pion receives a finite mass.
To lowest order in the current quark mass $m_0$ it is given by the 
Gell-Mann--Oakes--Renner (GOR) relation,
\beq
    m_\pi^2 \, f_\pi^2 = -m_0 \,\ave{\pb\psi} \;.
\label{GOR}
\eeq
To check the validity of this relation we have to expand $m_\pi^2$ to linear
order in $m_0$ whereas $f_\pi^2$ and the quark condensate have to be evaluated
in the chiral limit. 
The expression for the latter is taken from \eq{qbq0}. The pion
decay constant $f_\pi$ is calculated from the one-pion-to-vacuum axial matrix
element. In Hartree approximation + RPA we obtain
\beq
i\,\delta_{ab} f_{\pi}^{(0)} q_\mu =-\frac{g_{\pi qq}^{(0)}}{2} \intp
\;\Tr\,[\,i\gamma_\mu\gamma_5\tau^a \, iS(p+q)
                  \,i\gamma_5\tau^b\, iS(p)\,] \;.
\label{fpidef}
\eeq
This has to be taken for on-shell pions, i.e.\@ $q^2 = m_\pi^2$. The superscript
$(0)$ again denotes that we deal with a quantity in Hartree approximation +
RPA. 
The above expression resembles the definition of the quark-antiquark loop for the
mesonic polarization functions in RPA in \eq{pol0}. Basically we only replace
one vertex by $\gamma_\mu\gamma_5\tau^a/2$ and the other by $i \gamma_5\tau^b$
times the pion-quark coupling constant $g_{\pi qq}^{(0)}$ which has been
defined in \eq{mesonmass0}. 

We have to emphasize here that most of the following formal derivations are
independent of the actual value of the quark
mass. One should keep in mind, however, that observables, like $f_\pi,
m_\pi$ or the pion-quark coupling constant $g_{\pi qq}$ are only meaningful
when they are calculated with a quark mass consistent with the scheme
used. 

The pion decay constant can most
easily be evaluated with the help of the axial Ward identity
\beq
\gamma_5 \,\qslash = 2 m \gamma_5 \,+ \gamma_5 \,S^{-1}(q+p) 
                   \,+ S^{-1}(p)\, \gamma_5~.
\label{axialward}
\eeq
Contracting \eq{fpidef} with $q^\mu$, which means that we take the divergence
of the axial current, and inserting \eq{axialward} on the right hand side of
\eq{fpidef} we obtain
\beq
f_\pi^{(0)} =g_{\pi qq}^{(0)} m \Big(\frac{\Pi_\pi(q)-
  \Pi_\pi(0)}{q^2}\Big)\Big|_{q^2=m_\pi^2}~,
\label{fpirpa}
\eeq
where we have already divided out the common factors $i \delta_{ab}$ and have
used \eq{pi0} to cast the last term in the above form. In the chiral limit,
$q^2 =m_\pi^2\rightarrow 0$, we can use \eq{mesonmass0} to replace the
difference ratio on the right hand side of \eq{fpirpa} by a pion-quark
coupling constant. We then arrive at the Goldberger--Treiman
relation\footnote{The Goldberger-Treiman relation~\cite{goldberger} was originally 
  introduced for nucleons instead of quarks in the form $f_\pi g_{\pi NN} =
  g_A m_N$. If we had considered vector and axial vector states, we would have
  to deal with a $g_A\neq 1$ also in the case of quarks, see
  e.g.\@ Ref.~\cite{vogl}. },
\beq
f_\pi^{(0)} g_{\pi qq}^{(0)} = m_H~.
\label{gtr}
\eeq
Strictly speaking, at this point we encounter an
inconsistency if we use another quark mass than the Hartree mass $m_H$: We
assumed that the pion is massless in the chiral limit, which is not the case
for the RPA pion calculated with a mass $m\neq m_H$. Therefore, although we
till now at no stage of the derivation explicitly needed the RPA pion
propagator, here and in the following part of this section we will use $m_H$
instead of a general $m$. 

For the pion mass we start from \eqs{dpisigma} and (\ref{mesonmass0}) 
and expand the inverse pion propagator around $q^2=0$, i.\@ e.
\beq
    1 \;-\; 2g_s\,\Pi_\pi(0) \;-\; 2g_s\, 
    \Big(\frac{d}{dq^2}\,\Pi_\pi(q)\Big)\Big|_{q^2=0}\,m_\pi^{2\,(0)} 
    + {\cal O}(m_\pi^{(0)\,4}) = 0 \;.  
\label{mpi0}
\eeq
To find $m_\pi^{2\,(0)}$ in lowest non-vanishing order in $m_0$, we have to
look at the $m_0$ dependence of the different quantities entering
\eq{mpi0}. Using the relation given in \eq{pi0} and the Hartree gap equation
(\eq{gap}) we see that $1- 2g_s\Pi_\pi(0)$ contains a term linear in
$m_0$. 
The derivative only implicitly depends on $m_0$ via the
constituent quark mass $m_H$ and contains no term of linear order. We
have to take it therefore in the chiral limit and can then replace it by a
pion-quark coupling constant, \eq{mesonmass0}. Thus finally we obtain for
$m_\pi^{2\,(0)}$ in lowest (linear) order in $m_0$ 
\beq
 m_\pi^{2\,(0)} = \frac{m_0}{m_H}\frac{g_{\pi qq}^{2\,(0)}}{2 g_s}~.
\label{mpi0exp}
\eeq
Combining \eq{qbq0} for the quark condensate, \eq{gtr}, and \eq{mpi0exp} we
find that the Hartree and RPA quantities fulfill the GOR relation (\eq{GOR}). 

As already mentioned above, this proof in the Hartree approximation + RPA has
been listed for purely illustrative reasons. In the following section we will
proceed with the MLA because in this scheme the proof of the Goldstone theorem
as well as the GOR relation works in a similar way as in the Hartree + RPA
scheme. The main relations are formally the same except that we have to
replace the Hartree + RPA quantities by their MLA analogues.
In the {\nce}, in contrast, especially the GOR relation is more difficult
to show as we have to carefully expand the various quantities in orders of
$1/N_c$.

\subsection{One-meson-loop approximation}
\label{piondstl}

Being consistent with the Goldstone theorem
requires that, in the chiral limit, the inverse pion
propagator vanishes at zero momentum, 
\beq 
    2g_s\,{\tilde\Pi}_\pi(0) = 1 \qquad {\rm for} \quad m_0 = 0.
    \label{goldstone}
\eeq
The function ${\tilde\Pi_\pi}^{ab}$ was defined in \eq{poltilde}. It
consists of the RPA polarization loop $\Pi_\pi^{ab}$ and the three 
additional diagrams $\delta\Pi_\pi^{(k)\,ab}$, $k = a,b,c$. 
Restricting the calculation to the chiral limit and to zero momentum
simplifies the expressions considerably and Eq.~(\ref{goldstone}) can 
be proved analytically.

We should keep in mind that the constituent quark mass
is now given by the extended gap equation, \eq{localgap}.
Therefore, the r.h.s.\@ of \eq{pi0} is different from unity in the chiral limit
and RPA pions are not massless. This has important consequences for the
practical calculations within this scheme, which will be discussed in greater 
detail in section~\ref{solution}.

The correction terms $\delta\Pi_\pi^{(k)}$ are defined in \eq{deltapi}.
Let us begin by diagram $\delta\Pi_\pi^{(a)}$. As mentioned above, 
we neglect the $\rho$ and $a_1$ subspace for intermediate mesons.
Then one can easily see that the external pion can only couple to
a $\pi\sigma$ intermediate state, since a $\sigma\sigma$ intermediate state
would violate isospin conservation and  a $\pi\pi$ intermediate state is not
possible for parity reasons. Evaluating the trace in 
Eq.~(\ref{trianglevertex}) for zero external momentum one gets for
the corresponding $\pi\pi\sigma$-triangle diagram
\beq
  \Gamma_{\pi,\pi,\sigma}^{ab}(0,p) = -\delta_{ab}\ 4  N_c N_f\ 2 m\ I(p)~, 
  \label{gammapps}
\eeq
with $a$ and $b$ being isospin indices and the elementary integral
\beq
    I(p) = \intk \frac{1}{(k^2-m^2+i \eps)( (k+p)^2-m^2+i\eps)}\;.
\eeq
Inserting this into Eq.~(\ref{deltapi}) we find 
\beq
\delta\Pi_\pi^{(a)\,ab}(0) 
= i \delta_{ab}\intp  (4 N_c N_f I(p))^2 4 m^2\ D_\sigma(p)\ D_\pi(p) \;.
\label{pisigtri}
\eeq
Now the essential step is to realize that the product of the RPA sigma- 
and pion propagators can be converted into a difference \cite{dmitrasinovic}, 
\beq
D_\sigma(p)\ D_\pi(p) =
  i \,\frac{D_\sigma(p)-D_\pi(p)} {4 N_c N_f\ 2 m^2\ I(p)}\;,
\label{pisig}
\eeq
to finally obtain
\bea
\delta\Pi_\pi^{(a)\,ab}(0) =-\delta_{ab}\;4 N_c N_f\intp
2 I(p)\Big\{ D_\sigma(p)-D_\pi(p)\Big\} \;. 
\label{pisigend}
\eea 
The next two diagrams can be evaluated straightforwardly.
One finds
\bea
  \delta\Pi_\pi^{(b)\,ab}(0)&=&  -\delta_{ab}\;4 N_c N_f\intp\Big\{ 
  D_\sigma(p)\  \big(I(p)+I(0)-(p^2-4 m^2)\ K(p)\big) \nonumber \\
&&\hspace{3.5cm} +D_\pi(p)\ \big(3I(p)\hspace{0.2cm}+\;3I(0)\hspace{0.4cm}
-\;\;3p^2\ K(p)\big)\; \Big\} \;, 
\nonumber\\ 
  \delta\Pi_\pi^{(c)\,ab}(0)&=& -\delta_{ab} \;4 N_c N_f\intp I(p)\Big\{ 
  -D_\sigma(p) - D_\pi(p) \Big\} \;.
\label{pseudo} 
\eea
The elementary integral $K(p)$ is of the same type as the integral
$I(p)$ and is defined in App.~\ref{integrals}.

Recalling that in the Hartree + RPA scheme \eq{pi0} played an essential role
in the proof of the Goldstone theorem we presume that a similar relation is
valid for $\delta\Sigma$ and $\delta\Pi$. Therefore we look at the correction to
the quark self-energy which is given by \eq{deltatildesigma}. 
It is proportional to the constant $\Delta$, defined in
\eq{delta}, which is explicitly given by
\bea
 \Delta &=& 4 N_c N_f\ m \int\frac{d^4p}{(2\pi)^4}{\Big\{}
\ D_\sigma(p)\ (2\ I(p)+I(0)-(p^2-4 m^2)\ K(p)) \nonumber \\
&&\hspace{3.3cm} +  D_\pi(p)\  (\;3I(0)\;-\;3p^2\ K(p)\;) 
{\Big\}}~. 
\label{tdself}
\eea
Comparing \eqs{pisigend} and (\ref{pseudo}) with \eq{tdself} we obtain 
for the sum of the 
three correction terms to the pion polarization function
\beq
    \sum_{k=a,b,c}\; \delta \Pi_\pi^{(k)}(0) = - \,\frac{\Delta}{m} 
\label{sumabc}
\;.
\eeq
Now, \eq{deltatildesigma} can be employed to arrive at an expression similar
to the relation between the Hartree quark self energy and the RPA pion
polarization function for vanishing momentum as given in \eq{pi0}:
\beq
\sum_{k=a,b,c}\; \delta \Pi_\pi^{(k)}(0) = 
\frac{\delta\tilde\Sigma}{2g_s m}~.
\label{deltasum}
\eeq
It has 
already been emphasized in \Sec{pionrpa} that \eq{pi0} is valid for any 
quark mass. Thus we can combine it with \eq{deltasum} to finally arrive at the
following result:
\beq
2 g_s{\tilde\Pi}_\pi(0) = \frac{\Sigma_H}{2 g_s m} + \frac{\delta\Sigma}{2 g_s m} .
\eeq
Hence, together with the modified gap equation (\eq{localgap}) we obtain
\beq 
    2g_s\,{\tilde\Pi}_\pi(0) = 1 \;-\; \frac{m_0}{m^{\prime}}
\eeq
in agreement with Eq.~(\ref{goldstone}).

As already pointed out, most of the integrals we have to deal with  
are divergent and therefore have to be regularized. 
Therefore one has to make sure that all
steps which lead to Eq.~(\ref{deltasum}) remain valid in the 
regularized model.   
One important observation is that the desired result can be obtained
on the level of the $p$-integrand, i.e.\@ before performing the 
meson-loop integral. This means that the
regularization of this loop is not restricted. 
We did not need either to perform the various quark loop integrals explicitly
but we had to make use of several relations between them. For instance,
in order to arrive at Eq.~(\ref{pisigend}) we needed the similar structure
of the quark triangle $\Gamma_{\pi,\pi,\sigma}(0,p)$ and the RPA 
propagators $D_\sigma(p)$ and $D_\pi(p)$. Therefore all quark loops,
i.e.\@ RPA polarizations, triangles and box diagrams should consistently be 
regularized within the same scheme, whereas the meson loops can be 
regularized independently.

To linear order in the current quark mass $m_0$ the pion mass is given by the 
GOR relation (Eq.~(\ref{GOR})). 
This relation holds exactly in the MLA, if 
we choose the appropriate definition of the quark condensate. 
This will be demonstrated below.

For the pion decay constant $f_\pi$ we follow the same steps which led to
\eq{fpirpa} in the previous section. 
It is calculated from the one-pion to vacuum axial vector matrix element,
which basically corresponds to evaluating the mesonic polarization diagrams,
Fig.~\ref{fig4}, coupled to an external axial current and to a pion (cf. \eq{deltapi}), 
\bea
i\,\delta_{ab} f_{\pi}q_\mu &=&-\frac{g_{\pi qq}}{2} \intp
\;\Tr\,[\,i\gamma_\mu\gamma_5\tau^a \, iS(p+q)
                  \,i\gamma_5\tau^b\, iS(p)\,] \nonumber\\
&& + \frac{i g_{\pi qq}}{4}\intp \sum_{M_1 M_2}
\Gamma_{a_1,M_1,M_2}(q,p)\, D_{M_1}(p)\,\nonumber\\
&& \phantom{+\frac{i g_{\pi qq}}{4}\intp \sum_{M_1 M_2}} \Gamma_{\pi,M_1,M_2}(-q,-p)\,
D_{M_2}(-p-q)
\;, \nonumber\\
&&- \frac{i g_{\pi qq}}{2}\,\intp \hspace{3.0mm} \sum_{M_1} \;
\Gamma_{a_1,M_1,M_1,\pi}(q,p,-p)\,D_{M_1}(p)
\;, \nonumber\\
&& -\frac{i g_{\pi qq}}{4}\intp \hspace{2.5mm} \sum_{M_1} \; 
\Gamma_{a_1,M_1,\pi,M_1}(q,p,-q)\,D_{M_1}(p)
\;.
\label{deltafpi}
\eea
Here the modified pion-quark coupling constant is defined as
\beq
    g_{\pi qq}^{-2} =
\frac{d{\tilde\Pi}_\pi(q)}{dq^2}|_{q^2 = m_\pi^2} \;.
\label{gpiqqdstl}
\eeq
analogously to Eq.~(\ref{mesonmass0}). Now we take the divergence of the
axial current and then use the axial Ward-Takahashi identity (\eq{axialward})
to simplify the expressions~\cite{dmitrasinovic}. One finds
\bea
\delta_{ab} f_\pi &=& \frac{g_{\pi qq}}{q^2} \Big(\nonumber\\
&&m \Pi_\pi(q)-m
\Pi_\pi(0)\nonumber \\
&&+ m\delta\Pi^{(a)}(q) +
\frac{1}{2}\intp\Big\{
(D_\sigma(p)-D_\pi(p+q))\Gamma_{\pi,\sigma,\pi}^{ab}(-q,-p)\nonumber\\
&&\phantom{+ m\delta\Pi^{(a)}(q) +\frac{1}{2}\intp\Big\{}
(D_\pi(p)-D_\sigma(p+q))\Gamma_{\pi,\pi,\sigma}^{ab}(-q,-p)\Big\}\nonumber\\  
&&+ m \delta\Pi^{(b)}(q) 
+ \Delta \nonumber\\ &&\phantom{+ m \delta\Pi^{(b)}(q) }+ \frac{1}{2}\intp
\{
\phantom{-}D_\sigma(p)\Gamma_{\pi,\pi,\sigma}^{ab}(p+q,-q)
\nonumber\\ && \phantom{+ m \delta\Pi^{(b)}(q) + \frac{1}{2}\intp\{}-3
D_\pi(p)\Gamma_{\sigma,\pi,\pi}^{ab}(p+q,-q)\}\nonumber\\
&&+ m \delta\Pi^{(c)}(q) + \frac{1}{2}\intp
\{D_{\sigma}(p)\Gamma_{\pi,\pi,\sigma}^{ab}(p+q,-p)\nonumber\\ &&\phantom{+ m \delta\Pi^{(c)}(q) + \frac{1}{2}\intp}+
D_\pi(p)\Gamma_{\sigma,\pi,\pi}^{ab}(p+q,-q)\}\Big)~.\nonumber\\
\label{fpistep1}
\eea
To further proceed, let us first consider the terms which still contain an
integration over $p$. 
If one takes into account that the $\pi\pi\sigma$-vertex should be symmetric
with respect to exchange of the two pions and performs a shift of the
integration variable $p$ such that all meson propagators depend on $p$ and not
on $p+q$, one immediately sees that these terms cancel. 
Using \eq{sumabc} we then arrive at the following expression:
\beq
    f_{\pi} = g_{\pi qq} \; m\, \frac{{\tilde \Pi}_\pi(q) - 
    {\tilde \Pi}_\pi(0)}{q^2}\Big|_{q^2=m_\pi^2}\;. 
\label{fpidstl}
\eeq
In the chiral limit, $m_\pi^2\rightarrow 0$,
the difference ratio on the r.h.s.\@ of Eq.~(\ref{fpidstl})
can be replaced by the pion-quark coupling constant
(Eq.~(\ref{gpiqqdstl}). 
This leads to the Goldberger-Treiman relation
\beq
f_{\pi} g_{\pi qq} = m^{\prime}~,
\label{gtrmla}
\eeq
The same remark of caution as in the Hartree+RPA case concerning
the choice of the constituent quark mass $m$ should be considered
here. Although \eq{fpidstl} could formally be derived without any reference
to the extended gap equation (\eq{localgap}), for further proceeding we should
restrict the
discussion to the selfconsistently determined quark mass $m^{\prime}$ because
otherwise the pion mass would not vanish in the chiral limit. 

For the pion mass we start from Eqs.~(\ref{dm1}) and (\ref{mesonmass1}) 
and expand the inverse pion propagator around $q^2=0$
\beq
    1 \;-\; 2g_s\,{\tilde\Pi}_\pi(0) \;-\; 2g_s\, 
    \Big(\frac{d}{dq^2}\,{\tilde\Pi}_\pi(q)\Big)\Big|_{q^2=0}\,m_\pi^2 
    + {\cal O}(m_\pi^4) = 0 \;.  
\eeq
As explained in \Sec{pionrpa},
to find $m_\pi^2$ in lowest non-vanishing order in $m_0$,
we have to expand $1-2g_s {\tilde\Pi}_\pi(0)$ up
to linear order in $m_0$, 
while the derivative has to be calculated in the chiral limit, where it can
be identified with the inverse squared pion-quark coupling constant,
Eq.~(\ref{gpiqq}).
The result can be written in the form
\beq
m_{\pi}^2 = \frac{m_0}{m^\prime} \frac{g_{\pi qq}^2}{2 g_s}
            +  {\cal O}(m_0^2) \;.
\label{mpidstl}
\eeq
Multiplying this by $f_\pi^2$ as given by \eq{gtrmla} we get to linear order
in $m_0$
\beq
    m_\pi^2\,f_\pi^2 = m_0 \,\frac{m^\prime}{2g_s} \;.
\eeq
Obviously this is consistent with the GOR relation (\eq{GOR}) if the
quark condensate is given by \eq{qbqself2}, but not if it is given by 
\eq{qbqself1}. In \Sec{1ml} we saw that within the effective action
formalism the quark condensate is given by \eq{qbqself2}. 
Therefore at this point we clearly see that the interpretation of $m^{\prime}$ as a constituent
quark mass, which would mean that we have to calculate the quark condensate
according to \eq{qbqself1}, leads to a contradiction with the GOR
relation. 
Therefore, in the numerical part, we will calculate the quark condensate 
according to \eq{qbqself2}.


\subsection{$1/N_c$-expansion}
\label{pionnc}
Again we have to show the validity of \eq{goldstone}. In the {\nce}
the function $\tilde\Pi_\pi$ is given by \eq{pol1}, i.e.\@ it differs 
from the corresponding function in the MLA
(\eq{poltilde}) by the fact, that it contains the additional diagram $\delta
\Pi_\pi^{(d)}$ (As we already discussed in the MLA it is implicitly contained
in the RPA diagram.). The other diagrams have the same structure as
before and we can largely use the results of the previous subsection.

Since the gap equation is not changed in the perturbative $1/N_c$ expansion,
\eq{goldstoneRPA} remains true, i.e.\@ \eq{goldstone} is already fulfilled by
the RPA polarization loop alone. 
Therefore we have to show that the contributions of the correction terms 
add up to zero, i.\@ e.
\beq 
    \sum_{k=a,b,c,d}\;\delta\Pi_\pi^{(k)}(0) = 0
    \qquad {\rm for} \quad m_0 = 0.
    \label{deltapisum}
\eeq

In the previous subsection we calculated the contribution from diagrams
$(a)-(c)$. The sum of these correction diagrams is given by \eq{sumabc},
evaluated at the Hartree quark mass.
Thus, we are left with the calculation of $\delta\Pi_\pi^{(d)}(0)$.
According to Eq.~(\ref{deltapi}), it can be written in the form 
\beq
    \delta\Pi_\pi^{(d)\,ab}(0) = -i\Gamma^{ab}_{\pi,\pi,\sigma}(0,0) \,
    D_\sigma(0)\,\Delta  \;.
    \label{deltapid}
\eeq
The next step is to realize that $\Gamma_{\pi,\pi,\sigma}^{ab}(0,0)$ can be
written with the help of $\pi$- and $\sigma$-meson polarization functions in
the RPA,
\beq
\Gamma_{\pi,\pi,\sigma}^{ab}(0,0) =
\frac{i}{m}(\Pi_\pi(0)-\Pi_\sigma(0))~. 
\label{gammapps0}
\eeq
Till now the evaluation of the various correction terms has been independent
of the choice of the quark mass. Primarily the next step requires the use of
the Hartree quark mass $m_H$: 
In the chiral limit $\Pi_\pi(0,m_H)$ is equal to $1/2 g_s$ (see
\eq{goldstoneRPA}), and combining \eq{gammapps0} with the definition of the RPA
propagator of the $\sigma$-meson in \eq{dpisigma} one finds that the product
of the first two factors 
in Eq.~(\ref{deltapid}) is simply $\delta_{ab}/m_H$, i.e.\@ one gets   
\beq
 \delta\Pi_\pi^{(d)\,ab}(0) = \delta_{ab} \,\frac{\Delta}{m_H}~.
\label{pid}
\eeq
With these results it can easily be checked, together with \eq{deltasum} that Eq.~(\ref{deltapisum}) 
indeed holds in this scheme. 

The discussion concerning the regularization procedure can be repeated here. 
The structure of the proof again leads to the conclusion that we have to
regularize all quark loops in the same way, whereas we have the freedom to
choose the regularization for the meson loops independently.

Another important observation is that we, in both schemes, do not need the
explicit form of the
RPA propagators. $D_{\sigma}(p)$ and $D_{\pi}(p)$ only need to fulfill
Eq.~(\ref{pisig}). Thus, approximations to the RPA propagators can be made as
long as Eq.~(\ref{pisig}) remains valid.

In \Sec{pionrpa} we demonstrated that the Hartree + RPA
quantities, i.e.\@ the leading-order quantities in $1/N_c$, exactly fulfill the
GOR relation (\eq{GOR}) which determines the
behavior of the pion mass for small current quark masses $m_0$. This is
also true for the MLA. 
In contrary, in the $1/N_c$-expansion scheme we cannot expect that the GOR 
relation holds exactly. We already discussed in \Sec{1/N_c} that the pion
mass, in spite of the fact that we perform a strict $1/N_c$-expansion of the
mesonic polarization functions, contains arbitrary orders in $1/N_c$ because
of its implicit definition (see \eq{mesonmass1}). A second point is that we,
even if we carefully expand $m_\pi^2$ and $f_\pi^2$ in orders of $1/N_c$,
generate higher orders simply by multiplying both quantities on the left hand
side of \eq{GOR}. Thus, because we certainly expect the GOR relation to hold
in each order in $1/N_c$ separately, we have to expand both sides of this
relation up to a definite order.
In \Sec{1/N_c} we calculated the quark 
condensate in leading order and next-to-leading order in $1/N_c$.
Hence, to be consistent, we should also expand the l.h.s.\@ of the
GOR relation up to next-to-leading order in $1/N_c$  
\beq
m_{\pi}^{2(0)} f_{\pi}^{2(0)} + m_{\pi}^{2(0)}\delta f_{\pi}^2 
+ \delta m_{\pi}^2 f_{\pi}^{2(0)}
= -m_0 \,  \Big(\ave{\pb\psi}^{(0)} + \delta\ave{\pb\psi}\Big) \;.
\label{gor1}
\eeq
Here, similar to the notations we already introduced for the 
quark condensate, $m_{\pi}^{2(0)}$ and $f_{\pi}^{2(0)}$ denote the
leading-order and $\delta m_{\pi}^2$ and $\delta f_{\pi}^2$
the next-to-leading order contributions to the squared pion mass
and the squared pion decay constant, respectively. 
In contrast to the Hartree + RPA scheme and the MLA \eq{gor1} corresponds here
to a double expansion: Besides the usual expansion up to linear order in $m_0$ 
both sides are expanded in orders of $1/N_c$. 

The leading-order and next-to-leading-order expressions for the quark
condensate are given in Eqs.~(\ref{qbq0}) and (\ref{deltaqbqexp}).
For the pion decay constant $f_\pi$ we follow the same steps as in the
previous section. 
We have emphasized that the derivation of \eq{fpidstl}
is independent of the choice of the quark mass. Thus this equation also can be
employed in the present case to describe the contribution of the RPA
loop and the sum of correction terms $(a)$ to $(c)$. 
We therefore have to 
add only the contribution from diagram $(d)$ in \fig{fig4}, which is given by
\beq
i\,\delta_{ab} f_{\pi}^{(d)}q_\mu = -\frac{ i g_{\pi
    qq}}{2}\,\Gamma_{a_1,\pi,\sigma}(q,-q)\,D_\sigma(0)\,\Delta
\;.  
\label{deltafpid}
\eeq
Here the $1/N_c$-corrected pion-quark coupling constant is 
defined as
\beq
    g_{\pi qq}^{-2} = g_{\pi qq}^{-2(0)} + \dg =
\frac{d{\tilde\Pi}_\pi(q)}{dq^2}|_{q^2 = m_\pi^2} \;.
\label{gpiqq}
\eeq
With the help of the axial Ward-Takahashi identity (\eq{axialward}) and
\eq{gammapps0} we find 
\beq
f_\pi^{(d)} = \frac{g_{\pi qq}}{q^2} \Big(
+m_H \delta\Pi^{(d)}(q)-m_H\delta\Pi^{(d)}(0)+\Delta D_\sigma(0)(\Pi_\pi(q)-\Pi_\pi(0))\Big)~.
\label{fpistep1d}
\eeq
Adding up \eqs{fpidstl} and (\ref{fpistep1d}) we finally arrive at the
following expression for the pion decay constant in the {\nce}:
\beq
    f_{\pi} = g_{\pi qq} \; \Big(\; \frac{{\tilde \Pi}_\pi(q) - 
    {\tilde \Pi}_\pi(0)}{q^2}\,m_H + \frac{\Pi_\pi(q) - \Pi_\pi(0)}{q^2} 
    \, D_\sigma(0)\,\Delta\; \Big)\Big|_{q^2=m_\pi^2}\;. 
\label{fpi}
\eeq
In the chiral limit, $q^2 = m_\pi^2\rightarrow 0$,
Eqs.~(\ref{mesonmass0}) and (\ref{gpiqq}) can be employed to replace 
the difference ratios on the r.h.s.\@ by pion-quark coupling 
constants. 
If we square this result and only keep the leading order and the
next-to-leading order in $1/N_c$ we finally obtain
\beq
f_{\pi}^{2(0)} + \delta f_{\pi}^2 = 
m^2 \,g_{\pi qq}^{-2(0)} + \Big(\,m^2 \,\dg  +
2m\,D_\sigma(0)\,\Delta\,g_{\pi qq}^{-2(0)}\,\Big)~.
\label{fpi2}
\eeq

Following the analogous steps which led us to \eq{mpidstl}
we find for the pion mass
\beq
    m_\pi^2 = \frac{m_0}{m_H}\,\frac{g^2_{\pi qq}}{2g_s}\;
    \Big(\,1 \,-\,\frac{D_\sigma(0)\Delta}{m_H}\,\Big) 
    +  {\cal O}(m_0^2) \;.
\label{mpi}   
\eeq
Finally one has to expand this equation in powers of $1/N_c$.
This amounts to expanding $g_{\pi qq}^2$, which is the only term in 
Eq.~(\ref{mpi}) which is not of a definite order in $1/N_c$.
One gets
\beq
    m_\pi^{2(0)} + \delta m_\pi^2 = 
    m_0 \,\frac{m}{2g_s}\,\frac{g^{2(0)}_{\pi qq}}{m^2} \;-\;
    m_0 \,\frac{m}{2g_s}\,\frac{g^{2(0)}_{\pi qq}}{m^2} 
    \Big(\,g_{\pi qq}^{2(0)}\,\dg 
    \,+\,\frac{D_\sigma(0)\Delta}{m}\,\Big) 
    \;.
\label{mpinc}   
\eeq
Combining Eqs.~(\ref{deltaqbqexp}), (\ref{fpi2}), and 
(\ref{mpinc}) one finds that the GOR relation in next-to-leading order, 
Eq.~(\ref{gor1}), holds in this scheme.

However, one should emphasize that this result is obtained by a strict 
$1/N_c$-expans\-ion of the various properties which enter into the GOR
relation and of the GOR relation itself. If one takes $f_\pi$ and $m_\pi$
as they result from Eqs.~(\ref{fpi}) and (\ref{mpi}) and inserts them 
into the l.h.s.\@ of Eq.~(\ref{GOR}) one will in general find deviations from 
the r.h.s.\@ which are due to higher-order terms in $1/N_c$.  
As such one can take the  violation of the GOR relation as a measure 
for the importance of these higher-order terms \cite{oertel}.

\section{Transversality of the Rho-Meson}
\label{transverse}
In this section we want to show that the polarization function in the vector
channel fulfills the transversality condition, \eq{trans}, in the
Hartree + RPA scheme, and in the MLA as well as in the {\nce}. Our proof will
mainly rely on the Ward identity
\beq
\qslash = S^{-1}(p+q/2) - S^{-1}(p-q/2)~,
\label{ward}
\eeq
where $S(p)$ stands for a quark propagator. We will nowhere need the explicit
form of this propagator, $S^{-1}(p) = \pslash-m_H$ or $S^{-1}(p) =
\pslash-m^{\prime}$ respectively. Thus, concerning this proof, the only
difference between the approximation schemes will be the number of diagrams we
consider for the entire polarization function. This statement implies that the
RPA polarization loop and $\delta\Pi^{(d)}$ are tranverse themselves, whereas
only the sum of the three other contributions needs to be transverse. We will
explain the procedure in detail by proving the transversality of the
RPA polarization loop, which was defined in \eq{pol0}. For the other
terms we will be more brief since the procedure contains no essential
difference, only the expressions become more lengthy. This is also the reason
why we will neglect here again the contributions from intermediate vector and
axial vector states.

Contracting the RPA loop with $q_\mu$ and applying the Ward identity,
\eq{ward}, we obtain
\beq
q_\mu \Pi^{\mn,ab}(q) =-i\intp \;\Tr\,[(S^{-1}(p+\frac{q}{2})-S^{-1}(p-\frac{q}{2}))\tau^a \, iS(p+{q\over2})
                  \,\gamma^\nu\tau^b \, iS(p-{q\over2})\,] \;.
\eeq
This can be further simplified using the invariance of the trace under cyclic
permutations, 
\beq
q_\mu\Pi^{\mn,ab}(q) = -i \intp\Big\{\Tr[i\tau^a\gamma^\nu\tau^b i
S(p-\frac{q}{2})]-\Tr[i\tau^a\,
iS(p+\frac{q}{2})\gamma^\nu\tau^b]\Big\}~. 
\label{rpatrans}
\eeq
Performing a shift of the integration variable in the two addends,
$p'=p-\frac{q}{2}$ and $p' = p+\frac{q}{2}$ respectively, we immediately see
that \eq{trans} holds for the RPA polarization loop. As already
pointed out in \Sec{secpion} in connection with the proof of the Goldstone
theorem and the GOR relation within the various approximation schemes, 
we have to keep in mind that, most of the integrals have to be regularized
since they are divergent. 
Certainly the crucial point is here the shift in the integration variable,
whose realization is accompanied by a shift in the integration boundary -in
some regularization schemes- which would destroy transversality.

Let us proceed with $\delta\Pi_{\rho}^{(d)}$. Inspecting the definition
in \eq{deltapi} closely, we conclude that we only have to show the
transversality of the $\rho\rho\sigma$-triangle vertex. This will later on
be helpful in connection with diagram $(a)$ in \fig{fig4}. Again,
contracting the triangle vertex for one external $\rho$-meson and external
mesons $M_2$ and $M_3$, see \eq{trianglevertex}, with $q_\mu$ and
applying the Ward identity we obtain
\bea
 q_\mu\Gamma_{\rho,M_2,M_3}^{\mu,a}(q,p) &=&\nonumber\\
&&\hspace{-3.6cm}\intk
  \Big\{ \Tr\,[\tau^a i S(k)\Gamma_{M_2} i S(k-p)\Gamma_{M_3}]-\Tr[\tau^a i
  S(k+q)\Gamma_{M_2}i S(k-p)\Gamma_{M_3}] \nonumber \\
&&\hspace{-2.2cm} + \Tr\,[\tau^a i S(k-q)\Gamma_{M_3}i 
    S(k+p)\Gamma_{M_2}]-\Tr[\tau^a i S(k)\Gamma_{M_3}i S(k+p)\Gamma_{M_2}]\Big\}~. 
\label{triangleqmu}
\eea
Performing a shift of the integration variable, $k' = k+p$ in the first two
terms, $k'=k+q$ in the last two terms, and using the invariance of the trace
under cyclic permutations one immediately realizes that, if at least one of
the remaining mesons, i.e.\@ $M_2$ or $M_3$, is an isoscalar state, this triangle
vertex is tranverse. This is applicable especially for the case of the
$\rho\rho\sigma$-vertex. Hence, we can conclude that the correction term $\delta\Pi_\rho^{(d)}$ is
tranverse by itself. 

Let us now proceed with $\delta\Pi^{(a)}_\rho$. 
As mentioned above, we neglect here the $\rho$ and $a_1$ subspace for
intermediate mesons and therefore have to deal only with a two pion
intermediate state. The contraction of the $\rho\pi\pi$-vertex with $q_\mu$
can be written as a difference of RPA pion polarization loops, which in turn
can be written as a difference of inverse RPA propagators
\beq
q_\mu\Gamma_{\rho,\pi,\pi}^{\mu,abc}(q,p) = 2 i
\eps_{abc}(D_\pi^{-1}(p)-D_\pi^{-1}(p+q))~. 
\label{rhopipiward}
\eeq
This equation represents the Ward identity for the RPA pions.
Inserting this result into the definition of $\delta\Pi_\rho^{(a)}$ we obtain
\beq
q_\mu\delta\Pi_\rho^{(a),\mn,ab}(q) = -2
\eps_{acd}\intp(D_\pi(p+q)-D_\pi(p))\Gamma_{\rho,\pi,\pi}^{\nu,bcd}(-q,-p)~.
\label{pia}
\eeq
Following the same strategy we arrive at a similar result for the sum of
the two remaining diagrams, $\delta\Pi_\rho^{(b)}$ and $\delta\Pi_\rho^{(c)}$,
\beq
q_\mu(\delta\Pi_\rho^{(b),\mn,ab}(q)+\delta\Pi_\rho^{(c),\mn,ab}(q)) = -2
\eps_{acd}\intp
D_\pi(p)(\Gamma_{\rho,\pi,\pi}^{\nu,bcd}(-q,-p)-\Gamma_{\rho,\pi,\pi}^{\nu,bcd}(-q,p+q))
~.
\label{pibc}
\eeq
In principle there exists also a contribution to these two diagrams from the 
$\sigma$-meson. 
A straightforward calculation shows that their longitudinal part
vanishes. We only have to suppose that we can substitute the integration
variable $p'=-p$. If it is allowed to put $p'= -p-q$ in the last addend of
\eq{pibc}, we see that the sum of \eqs{pibc} and (\ref{pia}) vanishes,
i.e.\@ also the pionic contribution to the $\rho$-meson polarization function is
tranverse. If all the integrals were convergent, this step would certainly be
allowed and the $\rho$-meson polarization function would therefore be transverse.
Since this is not the case and we have to apply a regularization scheme, we
have to pay attention whether \eq{trans} still holds. In
\Sec{regularization} we will come back to this question.
\section{Relation to hadronic models}
\label{hadron}
This section discusses an approximation to our approximation schemes which
points out the relation to hadronic models. For instance, a possibility to
extract a hadronic description from the NJL model is to construct an effective
Lagrangian from its bosonized form, see e.g.\@ Ref.~\cite{osipov} and
references therein. We will follow here another strategy: In order to suppress
the quark effects in the present model, it is suggestive to assume that the
constituent quark mass is very large as compared with the relevant meson
momenta.  Certainly, in the various diagrams we have to evaluate we integrate
over meson momenta, i.e.\@ they can in principle be arbitrarily large. Thus we
assume, that the main contribution to the integrals comes from a region where
the momenta are small, i.e.\@ we perform an effective low-momentum
approximation for the quark-loop vertices. This amounts to expanding these
vertices up to the first non-vanishing order in the incoming momentum
(``static limit''). In fact, in most cases this corresponds to a zero-momentum
approximation.  In order to preserve chiral symmetry we then have to
approximate the RPA-meson propagators consistently. It turns out that a
straightforward generalization of the prescription we applied for the vertices
to the propagators would lead to a contradiction with the Goldstone theorem.
To make this obvious let us have a closer look at the way in which the latter
comes out within this approximation.  Since the essential step which
determines the form of the RPA-meson propagators within this approximation is
the same in both schemes, the static limit of the MLA and the {\nce}, we will
restrict the discussion here to the static limit of the MLA.

Our aim is now to find a form of the RPA propagators which ensures the
validity of \eq{goldstone} within the static limit of the MLA, i.e.\@ that the
pion polarization function vanishes at $q^2=0$ within this approximation, too.
For clarity of the expressions we will, as in \Sec{secpion}, neglect vector and
axial vector intermediate states for the proof. Before we look at the pion
polarization function we have to investigate the form of the extended gap
equation \eq{localgap} in the static limit in order to determine the
constituent quark mass. The completely momentum independent Hartree self
energy remains unchanged but the constant $\Delta$, see \eq{Delta}, occurring in the correction
term to the self-energy $\delta\tilde\Sigma$ is approximated by 
\bea 
\Delta
&=& 4 N_c N_f\ m \int\frac{d^4p}{(2\pi)^4}{\Big\{}
\ D_\sigma(p)\ (3\ I(0)+4 m^2\ K(0)) \nonumber \\
&&\hspace{3.3cm} + D_\pi(p)\ 3I(0) {\Big\}}~.
\label{tdselfsl}
\eea
For the pion polarization function let us begin by diagram
$\delta\Pi_\pi^{(a)}$. The $\pi\pi\sigma$-vertex is given by
\beq
  \Gamma_{\pi,\pi,\sigma}^{ab}(0,p=0) = -\delta_{ab}\ 4  N_c N_f\ 2 m\ I(0)~, 
  \label{gammappssl}
\eeq
with $a$ and $b$ again being isospin indices. The proof of the Goldstone
theorem in the full momentum dependent case relied mainly on the relation
between the $\pi$- and $\sigma$-propagator in the RPA, given in
\eq{pisig}. By a comparison of \eq{gammapps} and \eq{pisig}, we can write this
relation in the form
\beq
D_\sigma(p)\ D_\pi(p) =
  -i \,\frac{D_\sigma(p)-D_\pi(p)} {m\,\Gamma_{\pi,\pi,\sigma}(0,p)}\;, 
\label{pisigsl0}
\eeq
where we have used the notation $\Gamma_{\pi,\pi,\sigma}^{ab}\equiv
\delta_{ab} \Gamma_{\pi,\pi,\sigma}$. A straightforward generalization of
\eq{pisigsl0} to the static limit would give, explicitly,
\beq
D_\sigma(p)\ D_\pi(p) =
  i \,\frac{D_\sigma(p)-D_\pi(p)} {4 N_c N_f\ 2 m^2\ I(0)}\;.
\label{pisigsl}
\eeq
Exploiting this relation we arrive at the following expression for
$\delta\Pi_\pi^{(a)}$ in the static limit, similar to \eq{pisigend}:
\bea
\delta\Pi_\pi^{(a)\,ab}(0) =-\delta_{ab}\;4 N_c N_f\intp
2 I(0)\Big\{ D_\sigma(p)-D_\pi(p)\Big\} \;. 
\label{pisigendsl}
\eea 
Performing the
zero-momentum approximation for the vertices in the next two diagrams, we find
analogously to \eq{pseudo}, 
\bea
  \delta\Pi_\pi^{(b)\,ab}(0)&=&  -\delta_{ab}\;4 N_c N_f\intp\Big\{ 
  D_\sigma(p)\  \big(2\,I(0)+4 m^2\ K(0)\big) \nonumber \\
&&\hspace{3.5cm} +D_\pi(p)\ 6I(0)\Big\} \;, 
\nonumber\\ 
  \delta\Pi_\pi^{(c)\,ab}(0)&=& -\delta_{ab} \;4 N_c N_f\intp I(0)\Big\{ 
  -D_\sigma(p) - D_\pi(p) \Big\} \;.
\label{pseudosl} 
\eea
Adding up \eqs{pseudosl} and (\ref{pisigendsl}) and comparing the result with
\eq{tdselfsl} we conclude that \eq{sumabc} and thereby also \eq{goldstone}
holds within the static limit provided that the RPA propagators ascertain the
validity of \eq{pisigsl}. Thus, in order to be consistent with chiral symmetry
the RPA propagators have to fulfill \eq{pisigsl}. 
Including vector and axial vector intermediate states one can show that
the propagators of the RPA-$\rho$- and $a_1$-meson have to fulfill a similar
relationship.  
For all mesons this can be achieved by simply replacing the function $I(p)$ 
in the RPA polarization functions by $I(0)$. For instance, for the
inverse $\sigma$-propagator $-2 g_s D_\sigma^{-1}$ this corresponds to
replacing 
\beq
 1- 2 g_s \,2 i N_c N_f (2 I_1 - (p^2-4 m^2) \,I(p)) \quad{\rm by}\quad 
 1- 2 g_s \,2 i N_c N_f (2 I_1 - (p^2-4 m^2) \,I(0))~.
\eeq 
$I(0)$ can be related to the RPA pion decay constant $f_\pi^{(0)}$ 
in this approximation.
The RPA-meson propagators then take the form of free boson propagators,
\beq
D_M(q) = \frac{g^{2(0)}_{Mqq}}{q^2-m^{(0)2}_M}~,
\eeq
with
\beq
    m_\pi^{2(0)} = \frac{m_0 m}{2g_s f_\pi^{2(0)}}~, \qquad
    m_\rho^{2(0)} = \frac{3 m^2}{4g_v f_\pi^{2(0)}}~, \qquad
    m_\sigma^{2(0)} = m_\pi^{2(0)} + 4m^2~, \quad
    m_{a_1}^{2(0)} = m_\rho^{2(0)} + 6m^2~,
\label{mstat}
\eeq
and
\beq
g^{2(0)}_{\pi qq} = g^{2(0)}_{\sigma qq} = \frac{2}{3} g^{2(0)}_{\rho qq}
= \frac{2}{3} g^{2(0)}_{a_1 qq} = \frac{m^2}{f_\pi^{2(0)}} \;.
\label{gstat}
\eeq
We have to point out that, in a strict sense, this replacement does not correspond to an
expansion of the RPA polarization functions up to order momentum squared.
Consider, for instance, the inverse
$\sigma$-propagator. An expansion up to order $p^2$ 
would give
\beq
 -2 g_s D_\sigma^{-1} = 1- 2 g_s \,2 i N_c N_f (2 I_1 - p^2 \,(I(0)+4 m^2
 \frac{dI(p)}{p^2}\Big|_{p^2=0}))~, 
\eeq
whereas the above prescription leads to 
\beq
 -2 g_s D_\sigma^{-1} = 1- 2 g_s \,2 i N_c N_f (2 I_1 - (p^2-4 m^2) \,I(0))~. 
\eeq

\begin{figure}[t!]
\begin{center}
  \parbox{10cm}{\epsfig{file=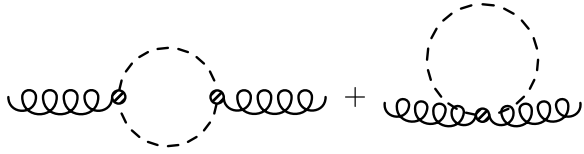}}
\end{center}
\caption{\it Contributions to the self-energy of the $\rho$-meson in standard
  hadronic models, pion loop (left) and pion tadpole (right). The dashed lines
  denote pions, the curly ones
  $\rho$-mesons.}
\label{figrhoselfh} 
\end{figure}

We proceed by expanding the remaining quark triangles and box diagrams to
first non-vanishing order in the external momenta, as it was demonstrated
above for the $\pi\pi\sigma$-triangle diagram (\eq{gammappssl}). 
In this way one obtains a hadronic model with effective meson-meson coupling
constants.  Let us, for example, look at the $\rho$-meson self-energy
which is generated as an approximation to the NJL model in MLA. 
Neglecting the $\rho$ and $a_1$ subspace for intermediate mesons, we are left
with a pion loop diagram, which is generated from the diagram shown in
Fig.~\ref{fig4}(a) and a pion and a sigma tadpole diagram
coupling directly to the $\rho$-meson, which
arise from the sum of the diagrams in Fig.~\ref{fig4}(b) and (c). In the
static limit of the {\nce} we would, from diagram $(d)$ in \fig{fig4}, in
addition get a pion and a sigma tadpole, coupling to the $\rho$-meson via an
intermediate sigma-meson. Assuming the contributions from intermediate
sigma-mesons to be negligible we 
exactly recover the diagrams which are calculated in standard hadronic
descriptions of the $\rho$-meson in vacuum, which are displayed in
\fig{figrhoselfh} (see e.g.\@ Refs.~\cite{herrmann, chanfray}). 
In fact, a simple reasoning reveals that in a consistent static
approximation the contributions from the $\sigma$-meson have to vanish
exactly. As mentioned above, the $\sigma$-mesons occur in the $\rho$-meson
polarization function within this approximation only in tadpole diagrams,
i.e.\@ their contribution is independent of the momentum of the
$\rho$-meson. The pionic contributions, however, preserve the consequences of
current conservation on
their own, in particular $\Pi_\rho(q=0) =0$ for $q^2=0$. Therefore the contributions from the $\sigma$-meson have to vanish
for vanishing momentum of the $\rho$-meson. Thus, since they are momentum
independent, they have to vanish identically. 

The comparison of the standard hadronic calculations with the static limit
provides us with a possibility to estimate the importance of dynamical quark
effects. For that purpose we will look at the pion loop in the $\rho$-meson self
energy $\Sigma^\mn$. We adopt here the usual notation for the self-energy, i.e.\@ the $\rho$-propagator is written as
\[
(q^2 - m_\rho^{(0)2}-\Sigma)^{-1} T^\mn + {\rm longitudinal~parts}~,
\]
with $\Sigma = 1/3 T_\mn\Sigma^\mn$. 
Straightforward evaluation of the corresponding diagram gives
\beq
g_\mn \Sigma^\mn_\rho(q) = 4 i g_{\rho\pi\pi}^2 \Big\{I_1(m_\pi)-(q^2- 4 m_\pi^2)
I(q,m_\pi) \Big\}~,
\label{rhohadron}
\eeq
with a $\rho\pi\pi$-coupling constant $g_{\rho\pi\pi}$. In the above expression we introduced an additional mass argument for the integrals $I_1$ and
$I(q)$ in order to indicate that they have to be evaluated at the pion mass 
$m_\pi$ and not, as usual, at the constituent quark mass $m$. 
If we want to compare
this with the static limit of an NJL model calculation we have to pay attention
on the fact that the polarization function we usually consider contains a
factor $g_{\rho qq}^2$ as compared with the usual definition of the
self-energy, 
\beq
1-2g_M\tilde\Pi_M(q)  = 1-2g_M(\Pi_M(q)+\delta\Pi_M(q)) =
-2 g_M g_{Mqq}^{-2(0)}(q^2-m_M^{2\,(0)}+g_{Mqq}^{2(0)} \delta\Pi_M(q))~.
\eeq
From the structure of that expression we conclude that $\Sigma$ is given 
by
\beq
\Sigma_M = -g_{Mqq}^{2(0)}\delta\Pi_M(q)~.
\eeq
For the contribution of the pion loop to the $\rho$-meson self-energy in the
static limit we find
\beq
g_\mn \Sigma^\mn_\rho(q) = 16 i  g_{\rho qq}^{2(0)} \Big\{ I_1(m_\pi)-(q^2- 4 m_\pi^2)I(q,m_\pi) \Big\}~.
\eeq
Comparing this result with \eq{rhohadron} and replacing the $\rho$-quark
coupling constant by the pion-quark coupling constant using \eq{gstat}, we
obtain for the effective $\rho\pi\pi$ coupling constant
\beq
g^2_{\rho\pi\pi} = 6 g^{2(0)}_{\pi qq} = 6 \frac{m^2}{f^{(0)2}_{\pi}}~.
\label{rhopipistat}
\eeq
If we take the commonly used value of about 6 for $g_{\rho\pi\pi}$ in 
hadronic models we obtain for the constituent quark mass
\beq
    m \approx \sqrt{6} f^{(0)}_{\pi}~.
\label{mapprox}
\eeq 
We should remark here that there is 
some uncertainty in the value we have to take for the pion decay constant in
RPA, $f_\pi^{(0)}$. In the next chapter we will perform a refit such that the corrected
quantities, i.e.\@  $f_\pi$ determined either in the MLA or in the {\nce}, are
equal to the empirical value of $f_\pi$. 
As will be seen in \Sec{pionfit}, $f_\pi^{(0)}$ is larger but not much larger
than the corrected quantities.
In any case, since we are only
interested in a rough estimate here we will
take a value of $f_{\pi}^{(0)} \sim$ 100-150 MeV. According to \eq{mapprox} 
this leads to a constituent 
quark mass of about 250-350 MeV. This is obviously in contradiction 
to our original assumption of very heavy quarks . Thus, if the main
contribution to the momentum dependent quark vertices does not arise from
momenta much smaller than 250-350 MeV, we would expect
that this approximation does not describe the full model very well. This 
point will be discussed in more detail in the next section. 

In hadronic models one is also interested in the coupling of a bare
$\rho$-meson to a photon, which is e.g.\@ needed to calculate the pion
electromagnetic form factor via vector-meson dominance. In the NJL model the 
bare $\rho$-meson corresponds to the RPA meson and its coupling to a photon
is basically given by the RPA polarization loop in the vector channel,
\beq
g_{\rho\gamma}^\mn(p) = -i \frac{e g_{\rho qq}^{(0)}}{2}\Pi_{\rho}(p)T^\mn(p)~.
\eeq
If we now perform the same low-momentum approximations as for the 
polarization functions we find that the vertex is given by
\beq
g_{\rho\gamma}^\mn(p) = 
i (e/g_{\rho\pi\pi}) p^2 T^{\mu\nu}(p)
\eeq
which exactly corresponds to the $\gamma\rho$ vertex in the vector
dominance model of Kroll, Lee and Zumino \cite{klz}. 

Similar to the proceeding for the $\rho$-meson, approximations to the self-energies of the other
mesons can be performed. For instance, this approximation to the
$1/N_c$-corrected NJL model generates an effective one loop approximation to
the linear sigma model in the $\pi$-$\sigma$ sector. The resulting effective
meson-meson coupling constants depend on the quark mass and $f_\pi^{(0)}$
which in turn depend on temperature and density. This dependence is used in
Ref.~\cite{davesne} within a linear sigma model calculation for in-medium pion
properties to simulate effects of (partial) chiral
symmetry restoration at nonzero temperature and density. This undertaking
should, of course, be interpreted with great care since all the effects which
can be modulated by their dependence on the quark mass are closely related to
unphysical quark effects present in the NJL model due to the lack of
confinement. We will discuss this question in more detail in
Chapters~\ref{numerics} and ~\ref{qqatt}. 

Another suggestive approximation to the full momentum dependent calculation is
to evaluate the vertices for on-shell intermediate mesons instead of
performing a low-momentum expansion. This approximation is discussed in
Ref.~\cite{he}. 
However, at least for processes dominated by intermediate pions
this gives very similar results to those obtained with the low-momentum
expansion.   

\chapter{Numerical results at zero temperature}
\label{numerics}

In this chapter we will discuss our numerical results at zero temperature. We
will begin by a brief description of the regularization scheme and then
discuss peculiarities related to the solution of the gap equation in the MLA.
After that we will study the influence of mesonic fluctuations on quantities
in the pion sector, thereby focusing on possible instabilities.  Finally we
will perform a refit of these quantities within the {\nce} and the MLA and
apply the model to observables in the $\rho$-meson sector.


\section{Regularization}
\label{regularization}

Before we begin the explicit calculation, we will have to fix our
regularization scheme.  As discussed in Sec.~\ref{secpion}, all quark loops,
i.e.\@ the RPA polarization diagrams, the quark triangles and the quark box
diagrams must be regularized in the same way in order to preserve chiral
symmetry. Besides Lorentz-covariance being a desirable feature for its own, we
prefer for computational convenience a regularization scheme for the quark
loops which does not destroy
Lorentz-covariance. There are several possible covariant schemes, e.g.\@
proper-time regularization, Pauli-Villars regularization, subtracted
dispersion relations or an $O(4)$-cutoff. Some of these have the great
disadvantage of displaying the wrong analytic behavior of the RPA meson
propagators.  For instance an investigation of Broniowski et
al.~\cite{broniowski} shows that the proper-time regularized RPA meson
propagators have poles in the complex plane near the imaginary axis.
Subtracted dispersion relations lead to poles for space-like momenta and an
$O(4)$-cutoff itself induces a non-analytic structure in the corresponding
meson propagator. Since, for the calculation of the mesonic fluctuations, we
need the RPA meson propagators for arbritrary four-momenta, we would like to
choose a regularization scheme which generates RPA propagators with the
correct analytic behavior. At first sight, Pauli-Villars-regularization
fulfills both requirements, the covariance and the analyticity of the RPA
meson propagators. We will therefore use this regularization scheme. Soon we
will become aware that Pauli-Villars regularization is not completely free of
peculiarities, either. We will, however, be able to circumvent most of the
occurring problems by slightly modifying the standard Pauli-Villars scheme.

To be consistent, all divergent terms should be regularized with the same
number of regulators, which is determined by the highest occurring degree of
divergence, i.e.\@ the quadratic divergence of the integral 
\beq
I_1 = \intk \frac{1}{k^2-m^2+i\eps},
\eeq
which we encounter, for instance, in the Hartree gap equation (\eq{gapexp}),
\beq
m_H = m_0 + 2 i g_s 4 N_c N_f m_H I_1(m_H)~.
\label{gapex}
\eeq
A quadratic divergence requires two regulators, thus we replace
\beq
    \intk f(k;m) \;\rrr\; \intk \sum_{j=0}^2 c_j\,f(k;\mu_j)~,
    \label{pv}
\eeq
by
\beq
    \mu_j^2 = m^2 + j\,\Lambda_q^2~;  \qquad
    c_0 = 1, \quad c_1 = -2, \quad c_2 = 1~.
\label{muj}
\eeq
Here $\Lambda_q$ is a cutoff parameter. Let us illustrate the standard way to
apply this prescription within the NJL model with the RPA polarization loop
in the $\sigma$-channel, defined in \eq{pol0}. A straightforward evaluation
gives
\beq
\Pi_\sigma(p) = 2 i N_c N_f \intk \Big(\frac{2}{k^2-m^2+ i\eps}- \frac{p^2-4 m^2}{(k^2-m^2+i\eps)((k+p)^2-m^2+i\eps)}\Big)~.
\label{sigma0}
\eeq
\begin{figure}[t] 
\begin{center}
\parbox{7.5cm}{
              \epsfig{file=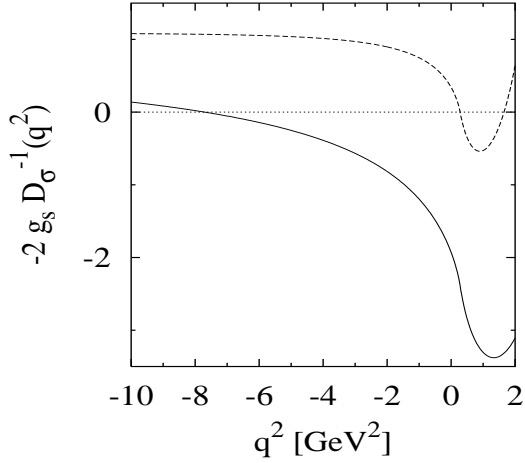,height=6.43cm,width=7.5cm}}
\caption{\it Inverse $\sigma$-propagator in RPA, $D_\sigma^{-1}(p^2)$, multiplied by
              $-2 g_s$ as a function of momentum squared, entire polarization
              loop regularized (solid line) and only $I_1$ and $I(p)$
              regularized (dashed line). The parameters used are listed in
              Table~\ref{tablence}, for a meson cutoff $\Lambda_M=0$ MeV.}
\end{center}
\label{figsigma}
\end{figure}
The most obvious procedure would be to replace the entire integrand in
\eq{sigma0} by the sum over the regulators, see \eq{pv}. This would be more in
the spirit of the original Pauli-Villars regularization~\cite{itzykson} than
the usual strategy~\cite{klevansky}, where the polarization loop is written
in terms of the already defined elementary integrals $I_1$ and $I(p)$,
\beq
\Pi_\sigma(p) = 2 i N_c N_f (2 I_1-(p^2-4 m^2) I(p))~.
\eeq
Here only the integrals $I_1$ and $I(p)$ are regularized, the factor $m^2$ in
front of $I(p)$ is not modified.   
The advantage of this strategy can be seen most easily from the inverse
$\sigma$-propagator which is displayed in \fig{figsigma} for both
methods. This calculation has been performed with the Hartree quark mass. The
naive application of \eq{pv} to the entire polarization loop obviously leads to
a zero of the inverse propagator, i.e.\@~a pole in the propagator, for space-like
momenta. The same observation can be made for the $a_1$-propagator.
This is the reason why we will adopt the usual form of
Pauli-Villars regularization in the NJL model and regularize only the
divergent elementary integrals throughout this investigation.

We should add a comment on a rather general problem which is related to
the regularization of the RPA polarization loops.
A straight-forward evaluation of the vector and axial vector polarization 
diagrams gives
\bea
    \Pi_\rho(p) &=&  -i {4\over3} N_c N_f\ (-2 I_1+(p^2+2 m^2)\ I(p))~,
    \label{pirhoex}\\
    \Pi_{a_1}(p) &=&  -i {4\over3} N_c N_f\ (-2 I_1+(p^2-4 m^2)\ I(p))~.
    \label{pia1ex}
\eea
Because of vector current conservation $\Pi_\rho$ should vanish
for $p^2$~=~0. This is only true if 
\beq
     m^2\,I(0) = I_1~,
\label{mi0}
\eeq
which is not the case if we regularize $I(p)$ and $I_1$ as described
above. Alternatively one could perform the replacement Eq.~(\ref{pv}) for 
the entire polarization loop. Then the factor 
$m^2$ in Eqs.~(\ref{pirhoex}) and (\ref{mi0}) should be replaced by a factor $\mu_j^2$ 
inside the sum over regulators and one can easily show that Eq.~(\ref{mi0}) 
holds (see~Eqs.~(\ref{i1}) and~(\ref{i0})). So it seems as if the strategy we
rejected above since it generates poles in the RPA $\sigma$- and $a_1$-meson
propagators for space-like momenta would be favorable to apply here.
However, this scheme leads to
another severe problem: 
From the gap equation (Eq.~\ref{gapex}) we conclude that $i I_1$ should 
be positive. On the other hand the pion decay constant in the chiral limit
and in leading order in $1/N_c$, i.e.\@ RPA, is given by, see \eq{fpirpa}
\beq
f_{\pi}^{2(0)} =  -2i N_c N_f\ m^2\ I(0)~,
\label{fpiex}
\eeq 
which implies that $i m^2 I(0)$ should be negative.  Thus, irrespective of
the regularization scheme Eq.~(\ref{mi0}) cannot be fulfilled if we want to
get reasonable results for $m_H$ and $f_\pi^{(0)}$ at the same time.
Therefore we will retain our choice of the standard form of Pauli-Villars
regularization in the NJL-model and replace the term $I_1$ in
Eq.~(\ref{pirhoex}) by hand by $m^2\ I(0)$. 
This leads to the following $\rho$-meson propagator 
\beq
D_\rho(p) = \frac{-2 g_v}{1+ 2i g_v\ {4\over3}
  N_c N_f\ 
  (-2 m^2\ I(0)+(p^2+2 m^2)\ I(p))}\label{rhoi}~.
\eeq
A priori it is not clear whether the symmetry properties of our approximation
schemes can be preserved if we perform the above subtraction for the vector
polarization function. We noted, however, that this is the case if we take care that
the relationship between the RPA $\rho$- and $a_1$-meson propagators,
\eq{vpisig}, is not affected. This can only be achieved if the $a_1$-meson
polarization function is treated analogously to the vector one. We then arrive
at the following RPA $a_1$-meson propagator
\beq
D_{a_1}(p) = \frac{-2 g_v}{1+ 2i g_v\ {4\over3} N_c N_f\ (-2 m^2\ 
  I(0)+(p^2-4 m^2)\ I(p))}~.\label{a1i} 
\eeq

After having explained the regularization scheme applied to the quark loops,
we will come to discuss the remaining meson-loop integration (integration over
$d^4 p$ in \eq{deltapi}). This integration
is not automatically rendered finite by regularizing the quark loops due to
the non-renormalizability of the model, so that we have to regularize it
separately. 
This regularization is not constrained by 
chiral symmetry and independent from the quark-loop regularization. 
For merely practical reasons we choose a three-dimensional cutoff $\Lambda_M$ 
in momentum space. In order to obtain a well-defined result we work in the 
rest frame of the ``improved''  meson.
The same regularization scheme was already used
in Refs.~\cite{oertel,OBW,OBW2}.

From an esthetic point of view, it would certainly be desirable to employ a
regularization scheme which allows to render both quark loops and meson
loops finite without the introduction of an additional parameter (or as in our
case, even another regularization scheme). This could be achieved in a NJL-type
model with a separable non-local
interaction~\cite{blaschke,bowler}. This non-local interaction
generates a form factor at the quark vertices. That this makes all integrals
finite can most directly be seen in the version of Ref.~\cite{blaschke}. The
authors of Ref.~\cite{blaschke} use a form factor which depends on the
three-momentum of the quarks. It can be chosen in such a way, that the
three-momentum of each quark is limited to absolute values below sum cutoff
parameter $\Lambda$. Then simple kinematical considerations lead to the
conclusion that the absolute value of the three-momentum of the mesons is
automatically restricted to lie below $2 \Lambda$. The disadvantage of this
model is that it is manifestly non-covariant. In
Refs.~\cite{bowler} similar models with form factors depending on the
four-momentum are considered. These are obviously covariant, but they have the
disadvantage of destroying the analytic properties of the RPA meson
propagators. Recently
mesonic fluctuations have been investigated~\cite{ripkapaper,plant} within the
model presented in Ref.~\cite{bowler}. 

\section{Evaluation of the various two-loop diagrams}
\label{expnumerics}
Within this section we will present some details concerning the rather involved
numerical evaluation of the two-loop diagrams 
  \footnote{Note that diagram (a)
  in \fig{fig4} is in principle a three-loop diagram. But, since the two quark
  loop vertices decouple, we actually have to deal only with a two-loop
  diagram. A similar remark concerns the RPA mesons which in principle contain
  an arbitrary number of loops.}  
contributing for instance to the mesonic polarization functions
or the modified gap equation in the MLA, \eq{localgap}. 

A general two-loop diagram contains a four-dimensional (momentum) space
integration for each loop, i.e.\@ altogether eight integrations. Some of these
can usually be performed analytically while the remaining ones have to
be performed numerically. Of course one is interested in performing as few as
possible integrations numerically since computing time rises 
and computing precision is reduced steeply the
more integrations one has. In our case all the quark loop integrals, i.e.\@
those for the RPA meson propagators and the effective meson-meson vertices,
can be written in terms of five ``elementary'' integrals defined in
App.~\ref{functions} which in turn, with one exception, can be evaluated
analytically. Only for the function 
\beq L(p_1,p_2,p_3) =
\intk\frac{1}{(k^2-m^2+i \eps) ( k_1^2-m^2+ i\eps) ( k_2^2-m^2+i
  \eps)(k_3^2-m^2 + i\eps)}~, 
\eeq 
with $k_i = k+p_i$, 
an analytic expression exists exclusively in kinematical regions where at least
one of the momenta lies above the corresponding (in our case quark-antiquark)
threshold, i.e.\@ $p_1^2, p_2^2,p_3^2, (p_1+p_2)^2, (p_1+p_3)^2$ or $(p_2+p_3)^2$
is larger than $4 m^2$. This function appears in the four-meson vertex
$\Gamma_{M_1,M_2,M_3,M_4}(p_1,p_2,p_3)$, ~\eq{boxvertex}.
Thus, we are at most left with a one-dimensional integration for the quark
loops. 

\begin{figure}[h!]
\begin{center}
\parbox{7.5cm}{
     \epsfig{file=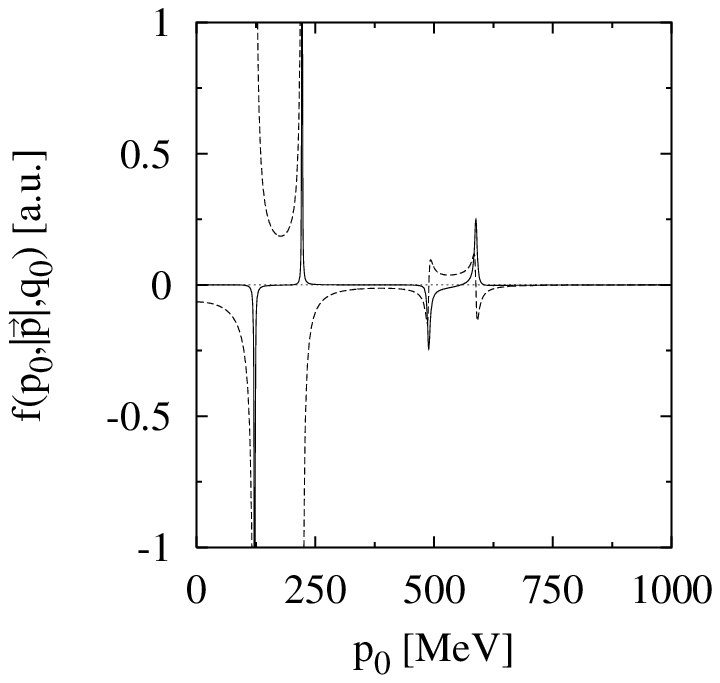,
     height=6.43cm, width=7.cm}\quad}
\parbox{7.5cm}{
     \epsfig{file=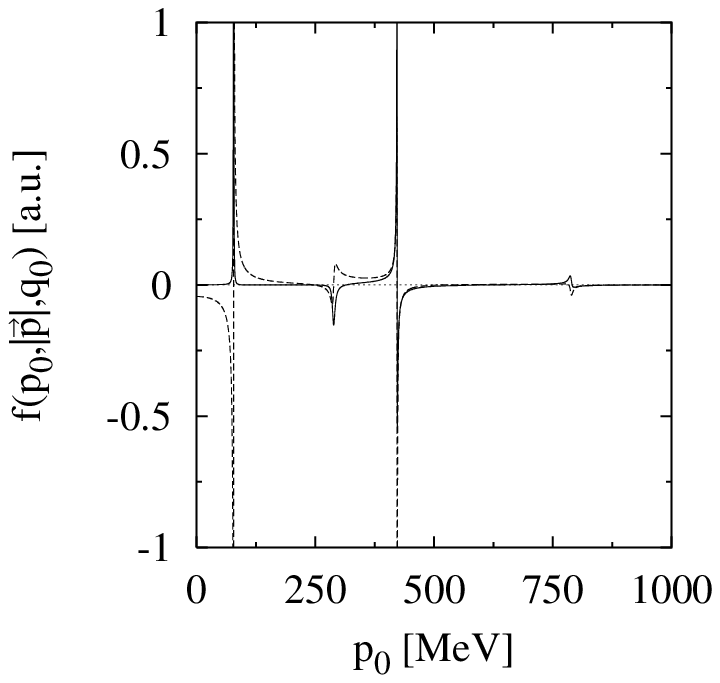,
     height=6.43cm, width=7.5cm}\quad}
\end{center}
\caption{\it Integrand for the $p_0$-integration, pionic part of the
     contribution $\delta\Pi^{(a)}$ to the $\sigma$-meson polarization
     function with $|\vec{p}|=100$~MeV, real (solid) and imaginary part
     (dashed), for $q_0=100$~MeV (left) and $q_0=500$~MeV (right).}
\label{figdpa} 
\end{figure}

Since we work in the rest frame of the mesons the angular part of
the meson-loop integration merely contributes a factor of $4 \pi$ and
therefore only an integration over energy and the absolute value of the
three-momentum remains. The former is most difficult to handle because we have
to take care of the various singularities and discontinuities arising from poles and thresholds of
the RPA meson propagators and the quark loop vertices. Let us consider for
instance the pionic part of the energy integrand for $\delta\Pi^{(a)}$ in the
scalar channel
\bea \frac{3\,i}{8 \pi^3}\int_0^{\Lambda_M}d|\vec{p}|\int_{-\infty}^{\infty} dp_0\ \vec{p}^2
\Gamma_{\sigma,\pi,\pi}(q,p-q/2)\,
D_{\pi}(p+q/2)\,\Gamma_{\sigma,\pi,\pi}(-q,-p+q/2)\, D_{\pi}(p-q/2)\nonumber\\ 
\equiv \frac{3}{8 \pi^3}\int_0^{\Lambda_M}d|\vec{p}|\int_{-\infty}^{\infty} dp_0 \ 
f(p_0,|\vec{p}|,q) ~,\hspace{6cm}
\label{sppint}
\eea 
where we have included a factor $3$ due to the sum over the isospin
indices of the intermediate pions.  We have performed a shift of the
integration variable $p \rightarrow p+q/2$ as compared with the definition of
$\delta\Pi^{(a)}$ in \eq{deltapi} in order to obtain a function
$f(p_0,|\vec{p}|,q)$ which is symmetric in $p_0$, enabling us to restrict the
energy integration to the interval $[0,\infty[$ by replacing
$f(p_0,|\vec{p}|,q)$ with $2 f(p_0,|\vec{p}|,q)$. This symmetry property
becomes obvious if we take into account that the RPA pion propagator
$D_\pi(p)$ is only a function of $p^2$ instead of $p$ and that the
$\sigma\pi\pi$-vertex function $\Gamma_{\sigma,\pi,\pi}(q,p)$ only depends on
the momentum of the incoming mesons squared, i.e.\@ $q^2, (p-q/2)^2$ and
$(p+q/2)^2$.  The real and imaginary parts of the symmetrized integrand are
shown in \fig{figdpa} for $|\vec{p}| = 100$~MeV and two values of $q_0$,
$100$~MeV and 500 MeV. We have used the parameters listed in Table~\ref{tablence}
for $\Lambda_M=0$~MeV.  The most pronounced peaks we observe are the pion
poles at $p_0 = |\pm q_0/2 +\sqrt{m_\pi^2+\vec{p}^2}|$. Note that we have
introduced by hand a small but nonzero imaginary part by replacing the quark
mass $m$ in the quark loop integrals with $m-i\eps$,
$\eps\approx1$~MeV. One consequence of this nonzero imaginary part is
the finite width of the pion poles. More or less steep
quark-antiquark thresholds appear if one of the momenta, i.e.\@ $q^2, (p+q/2)^2$
or $(p-q/2)^2$, equals $4 m^2$. These are invisible in \fig{figdpa} because of
the much larger scale of the singularities arising from the pion poles.
Additional thresholds exist due to the Pauli-Villars regulator masses
$m^2+\Lambda_q^2$ and $m^2 + 2\Lambda_q^2$, see \eq{muj}.

The values of $q_0=100$~MeV and $q_0=500$~MeV have been chosen for the
following reason: The former value for $q_0$ lies below any threshold and we
expect no imaginary part to be present at the end, the latter above the
two-pion threshold at $280$~MeV. The results we obtain for the imaginary part
below any thresholds are very sensitive to the details of the numerical
evaluation. Calculating this imaginary part, which we know to vanish, serves
as a sensitive measure for the precision of our delicate numerical integration
procedure.  An imaginary part is always related to a decay process, e.g.\@
$\sigma \rightarrow \pi\pi$. The three-dimensional cutoff we apply for the
$|\vec{p}|$-integration can at most restrict the phase space for that decay.
If for example $q_0 > 2 \sqrt{m_\pi^2+\Lambda_M^2}$, then no decay is
possible, thus the cutoff can suppress a decay process but in no way generate
one, and consequently no imaginary part. From this we conclude that the energy
integration alone has to be responsible for the imaginary part vanishing below
any threshold. Thus the integration over the imaginary part of
$f(p_0,|\vec{p}|,q)$ has to vanish for $q_0<280$~MeV, e.g.\@ for
$q_0=100$~MeV. Looking at \fig{figdpa} one realizes that this is a completely
non-trivial undertaking. Particular attention should be paid to three crucial
points: to proper integration over singularities and thresholds, to an
extrapolation to vanishing $\eps$, and to the inclusion of high-energy
contributions.  The first and the second point are connected, since the
quantity $\eps$ has been introduced by hand to soften the otherwise
numerically not tractable singularities and thresholds. This artificial width
enables us, by suitably adapting the location and number of mesh points in the
vicinity of singularities and thresholds, to perform the integration with
sufficient precision. The limit $\eps\rightarrow 0$ is approximated thereafter
by linear extrapolation. Including also contributions from very high energies
turns out to be difficult because the evaluation of the integrand becomes
numerically unstable for very large energies, i.e.\@ $p_0 \gsim$~15 GeV.
However, at least for the imaginary part these contributions cannot be
neglected since they are necessary for cancellations below any threshold. An
integration up to $p_0 \rightarrow \infty$ can be achieved by first applying
one of the standard substitutions of the integration variable: For large
energies, i.e.\@ above any threshold, we substitute $p_0 = \tan(x)$. The
resulting function of $x$ can very well be approximated by a quadratic
function and thus can easily be extrapolated up to the upper integration
boundary of $\pi/2$.  The real part remains almost unaltered by the operations
described above.

\begin{figure}[h!]
\begin{center}
\parbox{7.5cm}{
     \epsfig{file=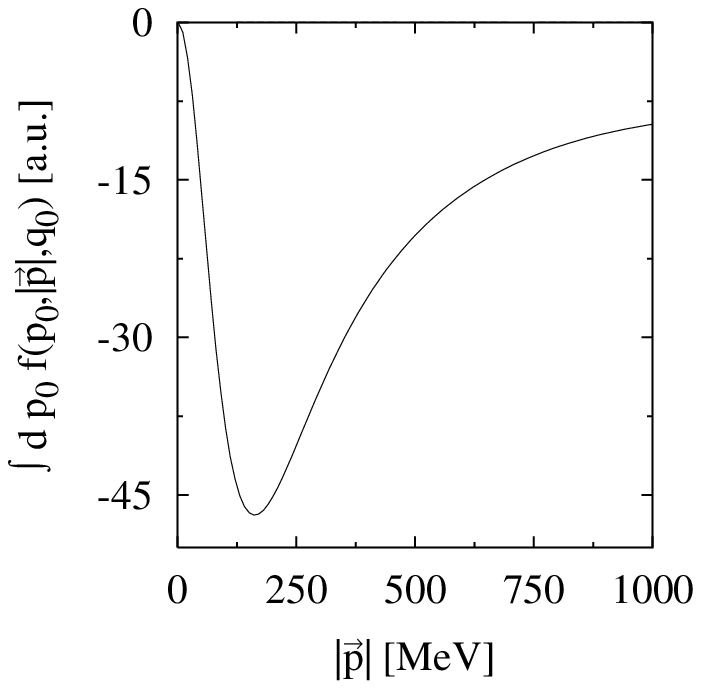,
     height=6.43cm, width=7.cm}\quad}
\parbox{7.5cm}{
     \epsfig{file=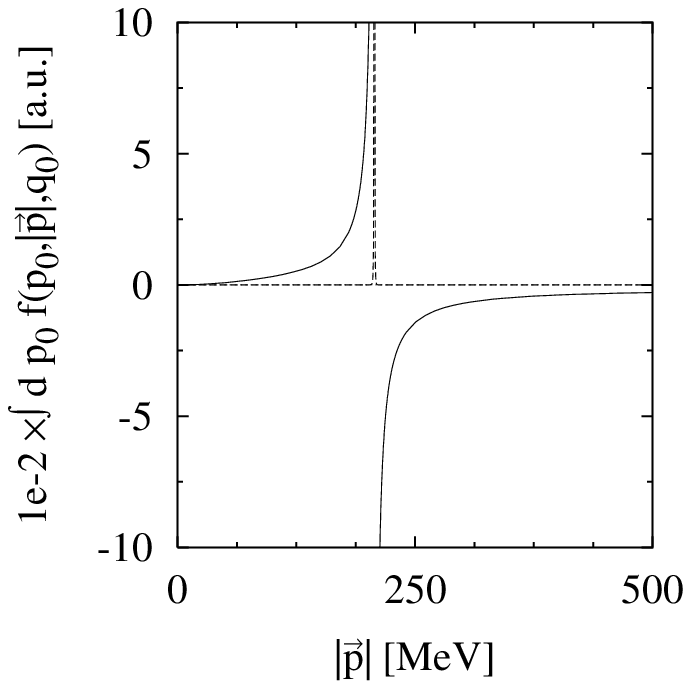,
     height=6.43cm, width=7.5cm}\quad}
\end{center}
\caption{\it Integrand for the $|\vec{p}|$-integration, pionic part of the
     contribution $\delta\Pi^{(a)}$ to the $\sigma$-meson polarization
     function, real (solid) and imaginary part (dashed), with (left)
     $q_0=100$~MeV and (right) $q_0=500$~MeV.}
\label{figpintgd} 
\end{figure}

The finally rather satisfying result can be seen in \fig{figpintgd} where we
have plotted the corresponding integrand for the $|\vec{p}|$-integration. The
left panel represents the result for $q_0=100$~MeV, the right panel that for
$q_0=500$~MeV. For $q_0=100$~MeV the imaginary part is suppressed by 5 orders
of magnitude as compared with the real part which reveals the efficiency of
our method of integration. The $|\vec{p}|$-integration is comparatively easy
to handle. For $q_0=100$~MeV the integrand is completely smooth, for
$q_0=500$~MeV only in the vicinity of $|\vec{p}|
= \sqrt{q_0^2/4-m_\pi^2}$, i.e.\@ $|\vec{p}| = 207$~MeV, we have to be a little
careful with placing the mesh points for the integration.  In most other
cases, more precisely in all cases where we do not have to deal with a
two-pion intermediate state, we also end up with a completely smooth function
for the $|\vec{p}|$-integration.

\section{Solution of the gap equation in the MLA}
\label{solution}
In the {\nce} all diagrams are calculated with the constituent quark mass in
Hartree approximation, i.e.\@ as it emerges from the Hartree gap equation,
\eq{gap}. In the MLA, on the contrary, the quark mass is determined by the
solution of the extended gap equation, \eq{localgap}, which was derived
in \Sec{1ml} from a one-meson-loop approximation to the effective action. 
The structure of this extended gap equation was indispensable in order to prove
the validity of 
various symmetry relations, namely the Goldstone theorem, the Goldberger
Treiman relation and the GOR relation, in \Sec{piondstl}. This is not
astonishing since the entire scheme is based on the extended gap equation. 
However, only the structure of this equation has been used but not the
explicit solution. This section is now devoted to a discussion of the explicit
solution of the extended gap equation. Unforeseen problems
occur which will force us to slightly modify the scheme.
   
\begin{figure}[b!]
\begin{center}
\parbox{7.5cm}{
     \epsfig{file=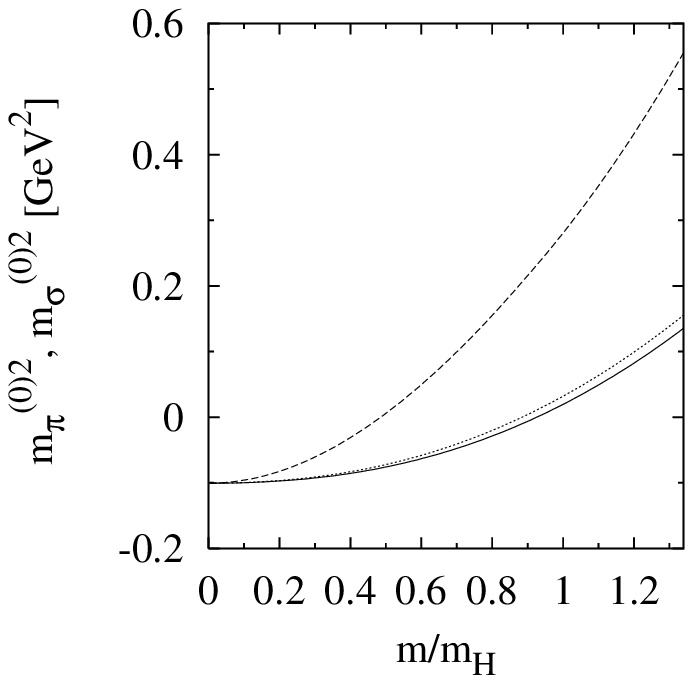,
     height=6.43cm, width=7.5cm}\quad}
\parbox{7.5cm}{
     \epsfig{file=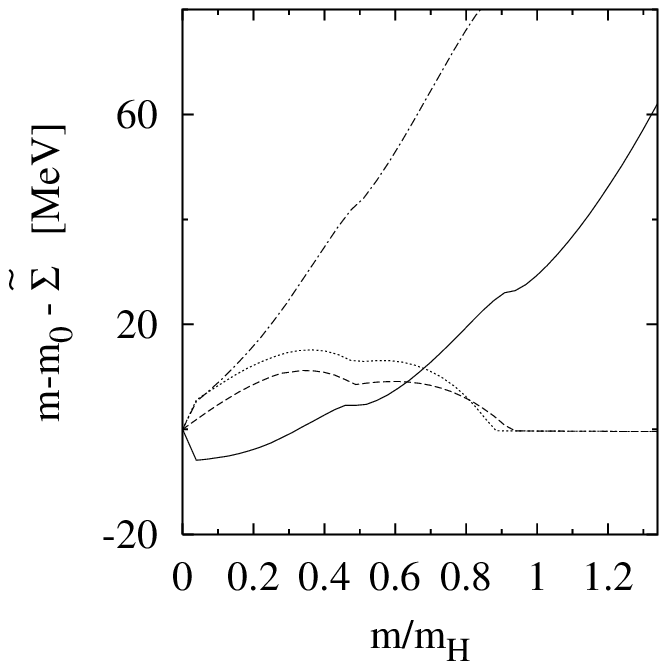,
     height=6.43cm, width=7.5cm}\quad}
\end{center}
\caption{\it (Left) Squared pole masses of the pion with $g_v = 0$ (solid),
  $g_v = 2 g_s$ (dotted) and the $\sigma$-meson (dashed) in RPA as functions
  of a trial constituent quark mass $m$ in units of the Hartree quark mass.
  (Right) Difference $m - m_0 - \tilde\Sigma(m)$ between the l.h.s.\@ and the
  r.h.s.\@ of the MLA gap equation, \eq{localgap}, as a function of the trial
  constituent quark mass $m$.  The real part is denoted by the solid line for
  $g_v=0$ and by the dashed-dotted line for $g_v=2 g_s$, the imaginary part by
  the dashed line for $g_v=0$ and by the dotted line for $g_v = 2 g_s$.}
\label{figqm} 
\end{figure}

First indications for these problems can already be inferred from the behavior
of the RPA mesons as a function of a trial constituent mass $m$. As already
pointed out, the RPA mesons, which enter the extended gap equation (see
\fig{figlocal}), have to be evaluated
with the selfconsistently determined quark mass $m^\prime$, which is in
general different from the Hartree quark mass $m_H$. Hence, the propagators of
these mesons differ from those in the Hartree + RPA scheme. To illustrate
this let us look at the masses of the pion and $\sigma$-meson in the RPA as a
function of a trial quark mass $m$ displayed on the l.h.s.\@ of \fig{figqm}. 
We have used the parameters
listed in Table~\ref{tablence} with $\Lambda_M=0$ MeV.
The effects of $\pi$-$a_1$-mixing can be governed by manipulating the vector
coupling constant $g_v$: With a vanishing $g_v$, $\pi$-$a_1$-mixing is turned
off, with an increasing $g_v$ it becomes more and more important.  For
comparison we have computed the pion mass with $g_v=0$ (solid line) and $g_v=
2 g_s$ (dotted line). The mass of the $\sigma$-meson (dashed line) does not
depend on $g_v$. The principal observation is that the pion becomes tachyonic,
i.e.\@ $m_\pi^{(0)2}$ becomes negative, for quark masses smaller than the
Hartree quark mass. Strictly speaking this is only true in the chiral limit,
for nonvanishing current quark masses we find negative
$m_\pi^{(0)2}$ at first for quark masses slightly below $m_H$.  

Qualitatively the results do not change if $\pi$-$a_1$-mixing is included.
If we keep all parameters fixed, it can be seen from the explicit form of the
pion propagator in the RPA (\eqs{pivv} and (\ref{pseudopiii})), that in the chiral
limit the results even do not change at all. Away from the chiral limit the
pion mass is slightly increased and hence the trial quark mass, where the pion mass
vanishes, is slightly decreased. In principle, the parameters should be
refitted with the pion mass fixed to its empirical value at $m=m_H$, which
diminishes the influence of $\pi$-$a_1$-mixing on the value of the quark mass
$m$ where the pion becomes tachyonic.

A similar observation can be made for $m_{\sigma}^{(0)2}$, but  
only for $m$ much smaller than the Hartree mass. This finding of tachyonic RPA
mesons for certain quark masses causes the meson-loop term in
the effective action (second term in \eq{gamma1loop}) to be no longer positive
definite in that range of quark masses. As discussed in \Sec{1ml} this reveals
that the effective action is ill-defined in that range. 

The latter point becomes obvious if we look at the quark self-energy which
becomes complex if it is calculated with tachyonic RPA mesons. Therefore we
have to hope that a solution of the extended gap equation can be found which
is larger than the Hartree mass. Any other solution necessarily must be
complex or negative.  To illustrate this point we consider the difference
between the l.h.s.\@ and the r.h.s.\@ of the extended gap equation, \eq{localgap},
as function of a (real) trial mass $m$. This difference is plotted in the
right panel of \fig{figqm}.  Let us begin the discussion by the results
obtained with $g_v=0$, i.e.\@ we turn off effects of $\pi$-$a_1$-mixing and the
contributions of vector and axial intermediate states in $\delta\tilde\Sigma$.
The solid line denotes the real part for this case, the dashed line the
imaginary part of $m - m_0 - \tilde\Sigma(m)$. Evidently the self-energy
indeed becomes complex for quark masses below some mass $m_c$. A comparison with
the pion mass reveals that $m_c$ coincides with the quark mass below which the
pion becomes tachyonic. A kink in the real as well as in the imaginary part
indicates the point where also the sigma becomes tachyonic.  Moreover, we see
that there is no solution of the gap equation for real constituent quark
masses besides the trivial solution at $m=0$. Hence, solutions, if there
exists any, lie somewhere in the complex plane off the positive real axis.
Consequently the RPA mesons would be built of quarks with either complex or
negative masses leading to peculiar properties. In any case this renders a
reasonable description of the mesons in MLA completely impossible.  This is
most clearly seen for the $\rho$-meson, whose properties are mainly determined
by intermediate RPA pions.  Including vector and axial vector intermediate
states as well as $\pi$-$a_1$-mixing does not alter this troubling result,
since, qualitatively, the function $m-m_0-\tilde\Sigma(m)$ shows the same
behavior in the case of $g_v=2 g_s$. It is displayed on the r.h.s.\@ of
\fig{figqm}, the dashed-dotted line corresponds to the real part, the dotted
line to the imaginary part.

A possible way out of this dilemma is to perform a modification, which has
been introduced in Ref.~\cite{nikolov}. Of course, any approximation we
perform has to preserve the symmetry properties of the MLA. As discussed in
\Sec{piondstl} this can be achieved if the validity of \eq{pisig} for the RPA
meson propagators is not affected. This remains true if we include vector and
axial vector intermediate states, see App.~\ref{vchiral}. In that case in
addition to \eq{pisig} for the pion and $\sigma$-meson propagators a similar
relationship for the RPA $\rho$- and $a_1$-meson propagators has to hold. Thus
the approximation suggested in Ref.~\cite{nikolov} for the purely scalar and
pseudoscalar case can straightforwardly be generalized to the case with vector
and axial vector interaction. This will be shown in more detail in
App.~\ref{vchiral}, here we will only illustrate the principle of this
approximation which can be done most easily if we set $g_v=0$.

The authors of Ref.~\cite{nikolov} simply replace the RPA pion propagator 
in the extended gap equation,
\beq
    D_{\pi}(p) = -2g_s\; 
    \Big [\; 1\;-\;2 i g_s\;4 N_c N_f \intk \frac{1}{k^2 - m^2 + i\varepsilon}
    + 2i g_s (2 N_c N_f)\, p^2 \,I(p) \,\Big ]^{-1}
\eeq
by 
\beq
    D_{\pi}(p) = -2g_s\; 
    \Big [\,\frac{m_0}{m} 
    + 2i g_s\;2 N_c N_f\, p^2 \,I(p) \, \Big ]^{-1}
\label{dpirepl}
\eeq 
and analogously for the $\sigma$-propagator. It is clear from the proof
of the Goldstone theorem in \Sec{piondstl} that the same replacements have to
be performed for the RPA meson propagators in the correction terms
$\delta\Pi_M^{(k)}$ to the mesonic polarization diagrams. In addition chiral
symmetry requires that the RPA contribution to the mesonic polarization
function, $\Pi_M$, itself is not changed. By performing this approximation we
not only obtain a real (positive) solution of the extended gap equation but
also the masses of the intermediate mesons remain real. Moreover, we
obtain massless intermediate pions in the chiral limit, which becomes obvious
by looking at \eq{dpirepl}. Besides, the various symmetry relations proved in
\Sec{piondstl} still hold.

We can certainly ask whether there is any other motivation for this
approximation in addition to the practical reasons discussed above. In the
Hartree approximation, i.e.\@ if the RPA mesons are calculated with quarks of
mass $m_H$, the above replacements would be exact. Following the
arguments put forward by the authors of Ref.~\cite{nikolov}, the correction
terms we neglect by performing the above replacements would be suppressed,
because they are of higher order in $1/N_c$. This is of course a questionable
argument in the MLA since already the quark mass $m^\prime$, the selfconsistent
solution of the extended gap equation, contains arbitrary orders in $1/N_c$.
In fact, we have seen that this approximation drastically changes the results.

We will keep the name ``one-meson-loop approximation'' for this scheme,
including the above replacements, although this is strictly speaking only
an approximation to the MLA as it was originally introduced in \Sec{1ml}.

\section{Meson-loop effects on quantities in the pion sector}
The additional cutoff $\Lambda_M$ we introduced in \Sec{regularization},
enables us to study the influence of mesonic fluctuations on several
quantities, e.g.\@ the quark condensate or the pion mass $m_\pi$ and the pion
decay constant $f_\pi$. This cutoff restricts the momentum of the intermediate
meson states. We can therefore, by slowly turning on this cutoff, control the
importance of the mesonic fluctuations. For $\Lambda_M=0$ we of course recover
the usual RPA result. It is advantageous to keep all other
parameters constant. We use $m_0=6.13$ MeV, $\Lambda_q=800 $MeV and
$g_s\Lambda_q^2=2.90$. These parameter values are chosen in such a way that
$m_\pi^{(0)}, f_\pi^{(0)}$, and $\qq^{(0)}$ reproduce the empirical values. We
obtain, for $g_v = 0$, $m_\pi^{(0)}=140.0$ MeV, $f_\pi^{(0)} = 93.6$ MeV and
$\qq^{(0)} = -2 (241.1~{\rm MeV})^3$. For the constituent quark mass in
the Hartree approximation we get $m_H=260$ MeV. These values are slightly changed
if we include vector mesons, i.e.\@ $g_v \neq 0$.  In the following section we
will discuss our results for the quark condensate, in \Sec{pionml} we will
consider the properties of the pion, among other things $m_\pi$ and $f_\pi$
which are related to the quark condensate via the GOR relation.

Of course, an application of the approximation schemes, the {\nce} and the
MLA, to physical processes will only be meaningful if a refit of the
parameters is performed to reproduce the empirical values of $m_\pi, f_\pi$
and $\qq$. This will be discussed in \Sec{pionfit}.

\subsection{Quark condensate}
\label{qqattz}
\begin{figure}[b!]
\begin{center}
\parbox{7.5cm}{
     \epsfig{file=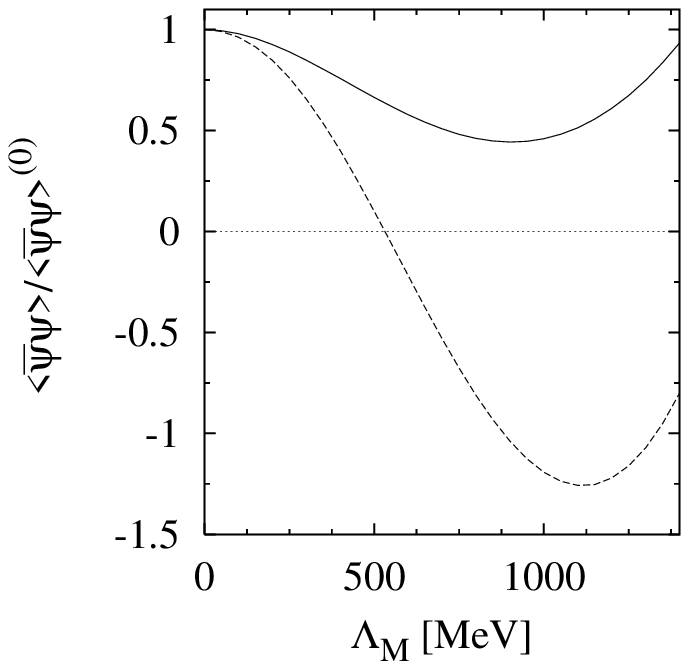,
     height=6.43cm, width=7.5cm}\quad}
\parbox{7.5cm}{
     \epsfig{file=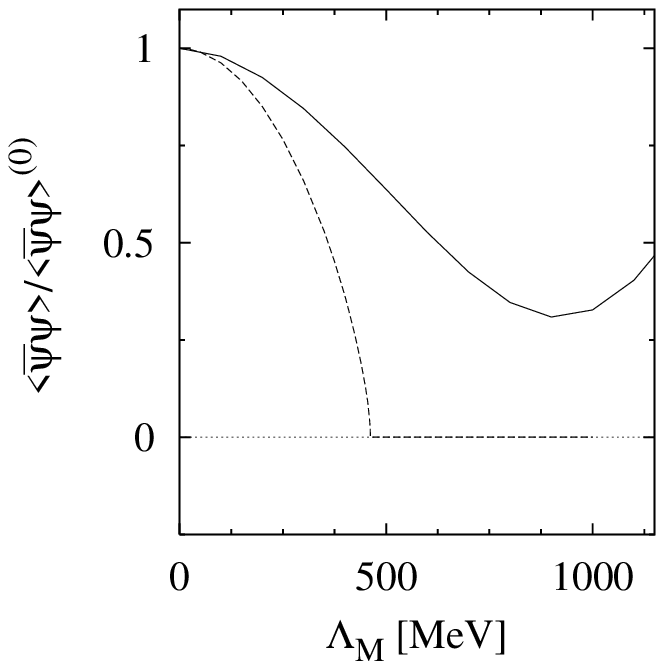,
     height=6.43cm, width=7.5cm}\quad}\\
\parbox{7.5cm}{
     \epsfig{file=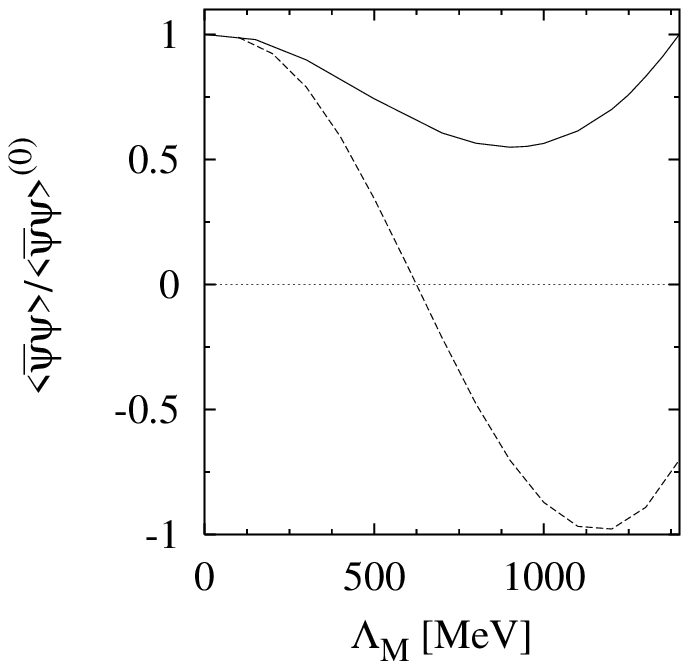,
     height=6.43cm, width=7.5cm}\quad}
\parbox{7.5cm}{
     \epsfig{file=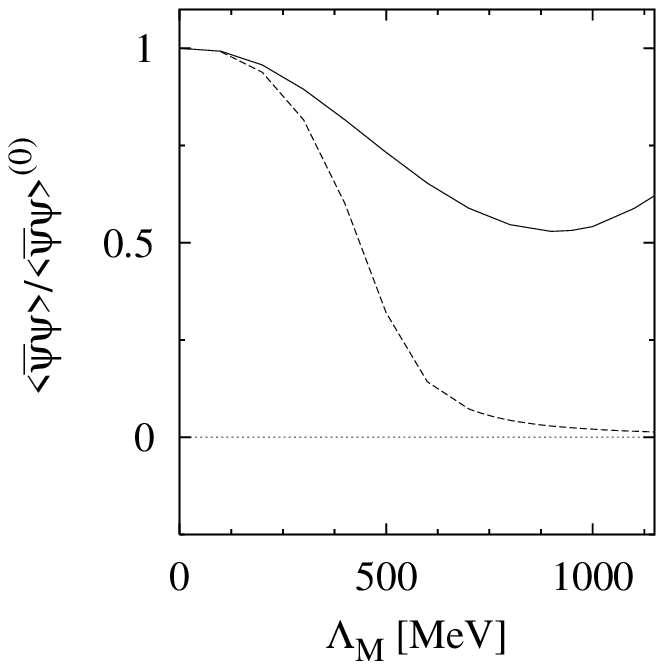,
     height=6.43cm, width=7.5cm}\quad}
\end{center}
\caption{\it The ratio $\qq/\qq^{(0)}$ as a function of the meson cutoff
     $\Lambda_M$. The solid line corresponds to $g_v=0$, i.e.\@ the result
     without vector and axial intermediate states, the dashed line to $g_v=2
     g_s$, (left) \nce, (right) MLA. The curves in the upper panel are
     calculated with a vanishing current quark mass,  in the lower panel
     the corresponding results with a realistic value of $m_0=6.13$~MeV are
     shown.} 
\label{figqqattz} 
\end{figure}

The behavior of the quark condensate as a function of the mesonic cutoff
$\Lambda_M$ is displayed in \fig{figqqattz}. The upper panel shows the results
in the chiral limit and the lower panel those with $m_0=6.13$~MeV, i.e.\@ with a
realistic pion mass. The solid lines correspond to $g_v=0$, i.e.\@ excluding
vector interactions, and the dashed lines to $g_v=2 g_s$. We will begin the
discussion by the results obtained without vector and axial vector
intermediate states which were already investigated in Ref.~\cite{OBW2}.
We observe that in the chiral limit as well as with a nonvanishing current
quark mass the absolute value of the quark condensate decreases first with an
increasing cutoff and then, if the cutoff exceeds a certain value, goes up
again. This feature is common to the {\nce} (left panel) and the MLA (right
panel).  This effect is related to an unphysical pole in the pion propagator
we detect for large values of the cutoff, see \Sec{instabilities}.  Studying
the effects of vector and axial vector fluctuations is an interesting issue in
the context of a suggestion made by Plant and Birse~\cite{plant}, saying that these
instabilities do not survive if one takes into account fluctuations of vector
and axial vector mesons and especially the $\pi$-$a_1$-mixing. They
corroborate this suggestion by a calculation of the quark self-energy in a
generalized, non-local, version of the NJL model (cf. \Sec{regularization}).
Without vector meson intermediate states they encounter in next-to-leading
order in $1/N_c$ a minimum in the self-energy for large momenta, whereas the
vector and axial vector states seem to suppress this effect.

Considering the results with $g_v= 2 g_s$ we should primarily state that the influence of the vector and axial vector fluctuations is
remarkably big. In the {\nce} the minimum does not
vanish, neither in the chiral limit nor with a nonzero current quark mass. 
It is only relocated at higher values of the cutoff. It becomes even more
steep compared with the case without vector and axial mesons. In the MLA the minimum
has entirely disappeared. Looking at the results in the chiral limit we instead
observe a behavior typically for a 
phase transition: 
For small values of the cutoff the quark condensate
rapidly approaches zero, at $\Lambda_M\gsim$450~MeV it vanishes. The slope of
the condensate at the ``critical cutoff'' diverges, indicating indeed a phase
transition of second or first order.  
In \Sec{instabilities} we will establish a conclusive answer to the nature of
this phase transition from the
effective potential. In that section we will also explain how
the calculations in the chiral limit, numerically a difficult task,
are actually performed. 
For $m_0=6.13$~MeV the condensate shows the typical cross-over behavior: For
small values of $\Lambda_M$ it rapidly approaches zero, at $\Lambda_M  
\approx 550$~MeV the slope rather abruptly changes such that the quark
condensate seems to reach zero asymptotically. 

The authors of Ref.~\cite{plant} state that mainly the longitudinal
parts of the vector mesons are responsible for the disappearing of the minimum. In
our case, as shown in \Sec{transverse}, 
the $\rho\rho\sigma$-vertex is transverse and hence the longitudinal part of
the $\rho$-meson gives no contribution at all, whereas the longitudinal part
of the axial polarization, absorbed in the pion propagator via
$\pi$-$a_1$-mixing, can very well be important. 

\begin{figure}[b!]
\begin{center}
\parbox{7.5cm}{
     \epsfig{file=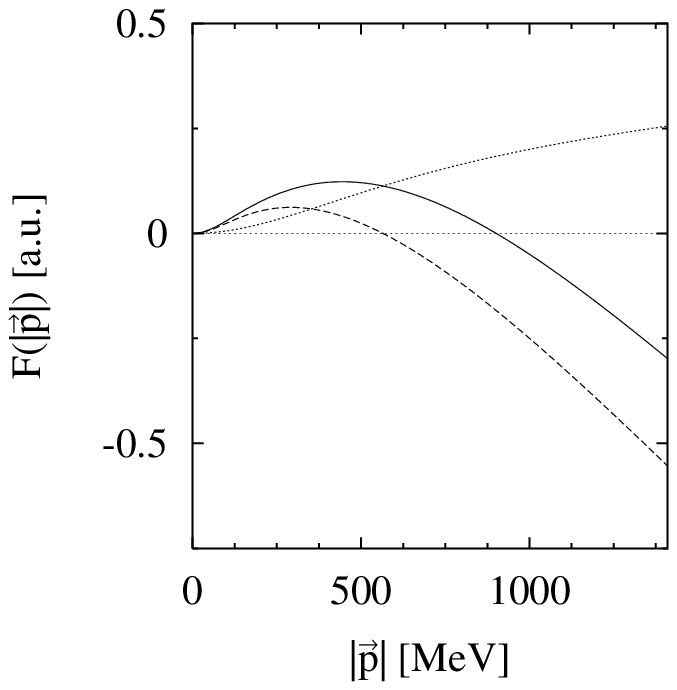,
     height=6.43cm, width=7.5cm}\quad}
\parbox{7.5cm}{
     \epsfig{file=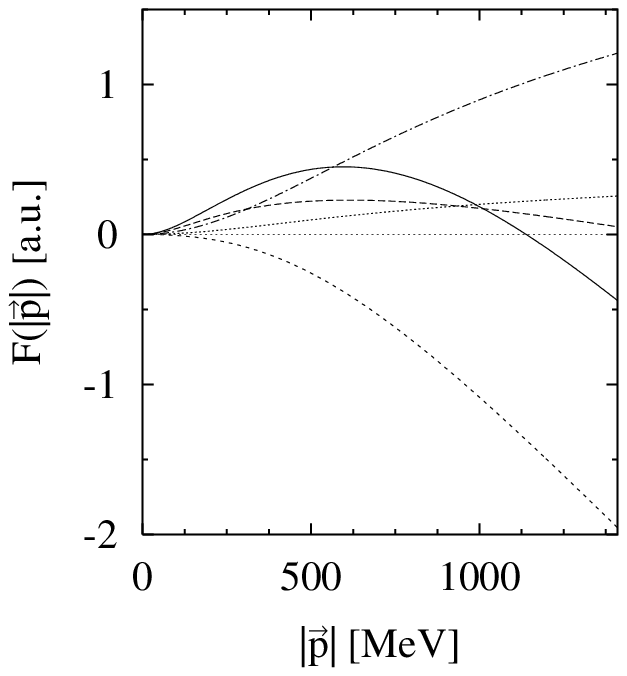,
     height=6.43cm, width=7.5cm}\quad}
\end{center}
\caption{\it Integrands for $\Delta$ as a function of $|\vec{p}|$. (Left:)
     $g_v=0$, contribution from the pion (dashed line), $\sigma$-meson
     (dotted), and the sum of both contributions (solid). (Right:) $g_v= 2
     g_s$, contribution from the pion (dashed), $\sigma$-meson (dotted),
     $\rho$-meson (short-dashed), $a_1$-meson (dashed-dotted), and the sum of
     all contributions (solid). }
\label{figdelta} 
\end{figure}

To investigate this question we will look at the constant $\Delta$ which is
the main element of the corrections to the quark condensate, both in
the {\nce} and in the MLA. In the {\nce} this can be seen directly (cf.
\eqs{qbq0} and (\ref{deltaqbqexp})): As a function of $\Lambda_M$ the Hartree contribution and the RPA $\sigma$-propagator remain
constant. Thus the mesonic fluctuations can only have an effect on $\Delta$. In
the MLA this connection is not as obvious since we have to solve \eq{localgap}
selfconsistently. Nevertheless the qualitative behavior of the quark
condensate can be deduced from that of $\Delta$.  In order to understand the
peculiar ``turn-around'' as a function of the meson cutoff $\Lambda_M$, it is
convenient to look at the integrand instead of $\Delta$ itself. The function
$F(|\vec{p}|)$, displayed in \fig{figdelta}, has to be integrated over
$|\vec{p}|$ to obtain $\Delta$. This representation allows us to estimate the
influence of the cutoff value on the result for $\Delta$.  We show the
contributions of the different mesonic intermediate states and the
corresponding sum, on the left hand side we neglected the $\rho$- and
$a_1$-meson subspace, whereas on the right hand side it is included (together
with $\pi$-$a_1$-mixing).

Let us begin our discussion by the purely scalar-pseudoscalar case. The sum
of both contributions first rises, then, at $|\vec{p}| \sim 445 $ MeV, it
starts to decrease. At $|\vec{p}| \sim 900$ MeV it even crosses the axis and
its slope indicates that it will become more and more negative for larger
values of $|\vec{p}|$.  We see that the pionic contribution is much more
important than the contribution from $\sigma$-meson intermediate states. It
also mainly produces the negative contributions which are responsible for the
``turn-around'' of the quark condensate as a function of the cutoff. 
The contribution of the $\sigma$-meson of course
remains completely unchanged if we include vector and axial intermediate states. The pionic one, however, is affected by
$\pi$-$a_1$-mixing. As can be seen on the r.h.s.\@ of \fig{figdelta} the
decrease of the pionic contribution is weakened and shifted to higher momenta.
The (transverse) vector contribution is mainly responsible for the decrease of
the integrand, whereas the tranverse part of the axial one, i.e.\@ the $a_1$,
tends to increase it. The maximum and the zero of the overall sum is
relocated at higher absolute values of the three-momentum compared with the
purely scalar-pseudoscalar case, but qualitatively the behavior is not
altered. This observation is directly reflected in the behavior of the quark
condensate in the {\nce}, where vector interactions only lead to a
shift of the observed minimum as a function of the mesonic cutoff $\Lambda_M$
to higher energies. In the MLA, however, the relation between $\Delta$ and the
quark condensate is less direct mainly because of the selfconsistent
determination of quark condensate. In any case also in the MLA the 
general behavior remains unaltered.   

\subsection{Problems with vector intermediate states}
Before we begin our discussion of pionic properties let us explain the
difficulties we encounter if we try to include vector and axial vector meson
intermediate states for the calculation of the corrections to the mesonic
polarization functions. The problem is related to the rather general problem
discussed in \Sec{regularization} of preserving the consequences of current
conservation for the vector propagator in RPA, i.e.\@ transversality and
$\Pi_\rho(0)=0$, and simultaneously not inducing wrong analytical properties
for $\sigma$- and $a_1$-mesons. As an example we will consider the diagram
$(b)$ in \fig{fig4}. It contains among other contributions one four-meson vertex where two
external pions with momentum $q$ couple to an axial state with momentum $p$.
Contracting this vertex with $p_\mu p_\nu/p^2$ we receive contributions to the
integral for $\delta\Pi_\pi^{(b)}(q)$, defined in \eq{deltapi}, from an
$a_1$-meson as well as from the axial part of the pion. In fact, the resulting
scalar vertex function, 
\bea
-i\frac{p_\mu
  p_\nu}{p^2}\delta_{bc}\Gamma_{\pi,a_1,a_1,\pi}^{\mn,abcd}(q,p,-q)
&=&\nonumber\\
&&\hspace{-4.5cm} -12 N_c
N_f\Big(\phantom{+} I(p+q) (p+q)^2 - I(q) (q^2+ p\cdot q)-I(0) p\cdot q\nonumber\\
&&\hspace{-4.5cm}\phantom{-12 N_cN_f\Big(} +K(q) q^2
(p\cdot q-p^2) + 4 m^2 p^2 K(p)-4 m^2\, p^2\, q^2 L(p,-q,0)\Big)
\label{pia1vertex}
\eea
is
multiplied by the difference of the RPA $a_1$-meson propagator and the axial
part of the pion propagator, $D_{a_1}(p)-D_{\pi_v}(p)$.

The problem arises from the combination of two types of terms: The first type
are addends proportional to $1/p^2$, e.g.\@ the term $q^2/p^2 I(q)$ in
\eq{pia1vertex}, the second type are terms depending only on the momentum $q$,
i.e.\@ being constant with respect to the integration variable $p$, e.g.\@ the
term $q^2 K(q)$ in \eq{pia1vertex}. Thus, on the one hand the difference of
the RPA $a_1$-propagator and the axial part of the pion propagator should
vanish as $p^2$ for $p^2 \rightarrow 0$ in order not to produce unphysical
poles at $p^2 = 0$, and on the other hand it should vanish at least as
$1/p_0$ for large energies to guarantee the convergence of the energy
integration. Using the explicit expressions for the $a_1$-propagator and the
axial part of the pion propagator given in \eqs{a1i} and (\ref{pivv}),
respectively, we see that the difference vanishes at $p^2=0$. Hence we
can conclude that the first condition is fulfilled for the RPA propagators in
their present form. Since the three-dimensional cutoff $\Lambda_M$ restricts
the absolute value of the three-momentum $|\vec{p}|$, we can gain insight into
the behavior of the difference for large energies by looking at it for large
$p^2$. Using the properties of the Pauli-Villars regulators introduced in
\eq{pv} we infer that the integral $I(p)$ in leading order behaves like 
\beq
I(p) \rightarrow 2 \frac{I_1}{p^2}~;\quad {\rm for}\quad p^2 \gg m^2~.  
\eeq
This enables us to conclude that the difference of the RPA $a_1$-propagator and
the axial part of the pion propagator contains besides terms vanishing at
least with $1/p^2$ a constant term proportional to
\[
- m^2 I(0)+ I_1~.
\]
This term is reminiscent of the $a_1$ polarization function. Note that the
subtraction we performed in \Sec{regularization} in order to preserve gauge
invariance for the vector polarization function here induces this problem. On
the other hand a closer examination shows that the difference would not vanish
at $p^2=0$ if we had not performed this subtraction. Thus the only solution
seems to be providing for \eq{mi0} to hold. This is a dilemma since we saw in
\Sec{regularization} that this leads to even more severe problems. We
therefore decided to put the vector and axial vector meson intermediate states
aside and will only treat scalar and pseudoscalar states throughout the
remaining part of this paper. 

\subsection{Pion Properties} 
\label{pionml}

In this subsection we want to study the influence of mesonic fluctuations
on the pion propagator, which will enable us to study especially the effects
on the pion mass and the pion decay constant, both within the {\nce} and within
the MLA. Major parts of this analysis can be found in Refs.~\cite{oertel,
  OBW2}.
We proceed in the same way as in the previous subsection: 
Since the strength of the fluctuations is controlled by the meson cutoff
$\Lambda_M$, we first keep all other parameters fixed and investigate how
the above quantities change, when $\Lambda_M$ is varied.

\begin{figure}[t!]
\begin{center}
\parbox{7.5cm}{
     \epsfig{file=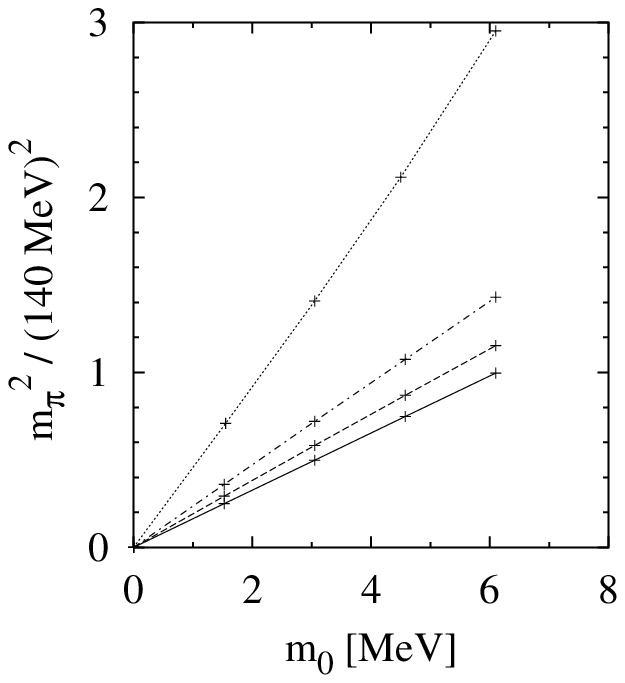,
     height=6.43cm, width=7.5cm}\quad}
\parbox{7.5cm}{
     \epsfig{file=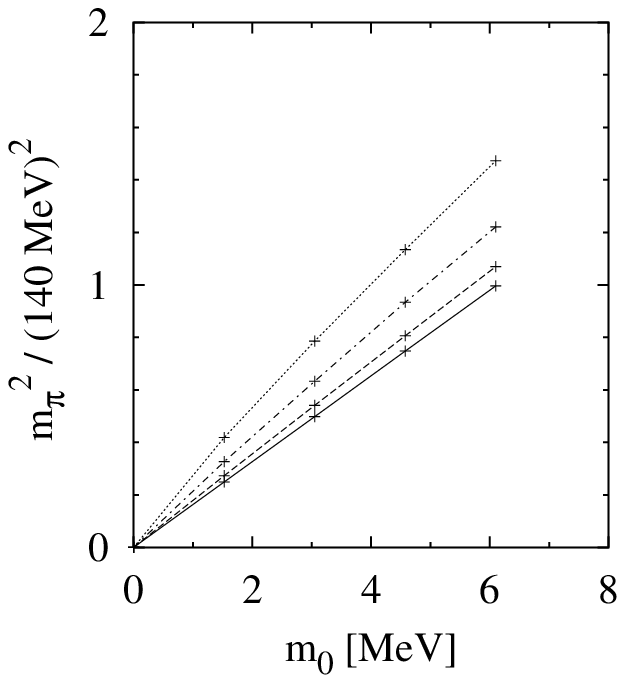,
     height=6.43cm, width=7.5cm}\quad}
\end{center}
\caption{\it Squared pion mass as a function of the current quark
             mass $m_0$ for different meson-loop cutoffs; (left) {\nce}:
             $\Lambda_M$~=~0~MeV (solid), 500~MeV (dashed), 
             900~MeV (dashed-dotted) and 1300~MeV (dotted); (right) MLA:
             $\Lambda_M = 0$~MeV (solid), 300~MeV (dashed), 500~MeV
             (dashed-dotted) and 700~MeV (dotted). The calculated
             points are explicitly 
             marked.}
\label{figmpi}
\end{figure}
Fortunately, the pion mass provides us with a decisive test for the stability
of the rather involved numerics.  The Goldstone theorem, which was proved
analytically in \Sec{secpion} for the {\nce} and the MLA, clearly states that
the pion mass has to vanish for a vanishing current quark mass $m_0$. For
numerical reasons we cannot compute the pion mass exactly in the chiral limit,
but its value can be extrapolated from the behavior of the squared pion mass
as a function of the current quark mass $m_0$. This is displayed in
\fig{figmpi} for different values of $\Lambda_M$, on the l.h.s.\@ evaluated
within the {\nce}, on the r.h.s.\@ in the MLA. Obviously the dependence in both
schemes is almost linear, i.e.\@ the analytically calculated point at $m_0=0,
m_\pi^2 = 0$ and the numerically computed points, corresponding to the same
cutoff, lie almost on a straight line. This result is suited for convincing us
of the consistency of our calculations with chiral symmetry and of the
stability of the numerics. 

\begin{figure}[t] 
\parbox{16cm}{\begin{center}
              \epsfig{file=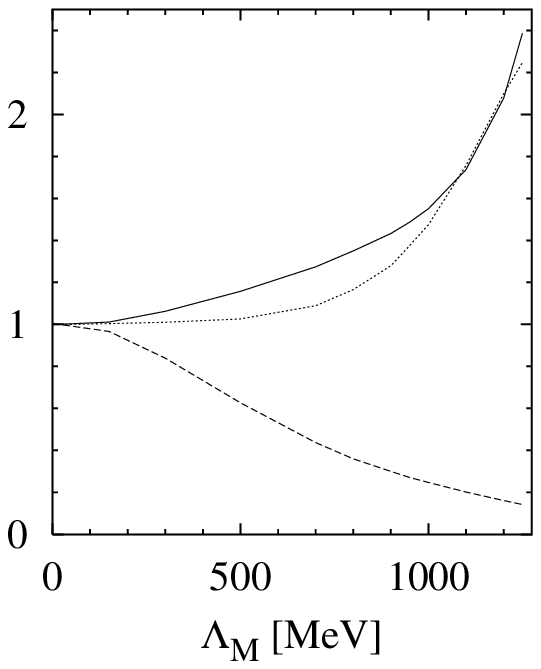,height=6.43cm,width=7.5cm}
              \epsfig{file=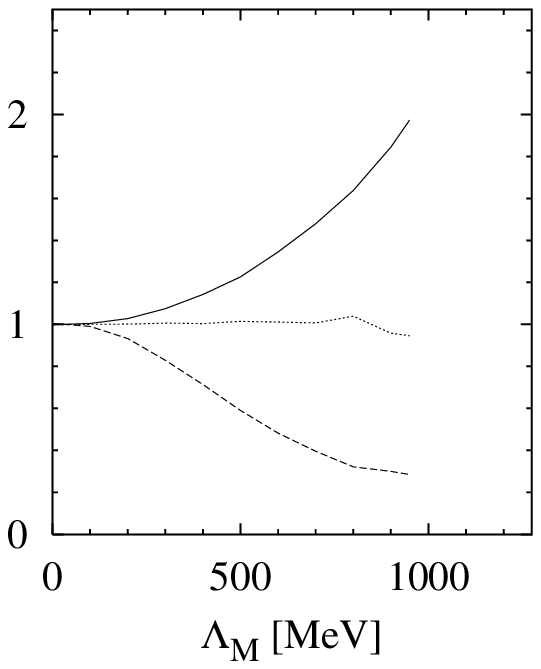,height=6.43cm,width=7.5cm}\end{center}}
\caption{\it The ratios $m_\pi^2/{m_\pi^2}^{(0)}$ (solid),
         $f_\pi^2/{f_\pi^2}^{(0)}$ (dashed),
         and the combination
         $-m_0\ave{\pb\psi}/m_{\pi}^2f_{\pi}^2$ (dotted)  
         as a function of the meson loop cutoff $\Lambda_M$ in the {\nce}
         (left) and the MLA (right).} 
\label{figpion}
\end{figure}

Now we fix the current quark mass again to $m_0=6.13$ MeV and slowly turn on
the mesonic cutoff
$\Lambda_M$ and with it the strength of the mesonic fluctuations.
The resulting behavior of $m_\pi^2$ and $f_\pi^2$
as a function of $\Lambda_M$ is displayed in Fig.~\ref{figpion}.
The left panel shows the results for the \nce, the right panel for the MLA.
It is clearly visible, that the qualitative behavior of $f_\pi$ and $m_\pi$
turns out to be the same in both schemes: The first (dashed lines) is reduced
whereas the latter (solid) is increased. If only
$\sigma$ and $\pi$ intermediate states are considered it was observed in the
previous section that both schemes show similar results for the quark
condensate.  

In \Sec{pionrpa} we showed that in the Hartree approximation + RPA the
quantities ${m_\pi^2}^{(0)}$, ${f_\pi^2}^{(0)}$ 
and $\ave{\pb\psi}^{(0)}$, should be in good agreement with the GOR 
relation, Eq.~(\ref{GOR}). This should also hold for the corresponding
quantities $m_\pi, f_\pi$, and $\qq$ in the MLA, see \Sec{piondstl}. The
{\nce}, on the contrary, is consistent with the GOR relation only up to
next-to-leading order and, as discussed in \Sec{pionnc}, the ``corrected''
quantities violate the relation by higher-order terms.  
For this reason we expect a less perfect agreement in this
scheme, becoming worse with increasing values of $\Lambda_M$. 

These expectations are more or less confirmed by the results.  In the Hartree
+ RPA scheme we find only deviations -due to corrections in higher orders in $m_0$- of about
0.1\%. In the MLA the deviations are slightly larger, due to the numerics
being much more difficult to handle. One reaches up to about 3\% for a cutoff of
$\Lambda_M = 900$ MeV. These deviations can be inferred from the ratio of the
r.h.s.\@ and the l.h.s.\@ of the GOR relation, \eq{GOR}, displayed in \fig{figpion}
by the dotted lines. As can be seen, in the MLA (right panel) it almost
coincides with the desired value of $1$, whereas in the {\nce} (left panel) we
indeed observe considerable discrepancies. Within this scheme the GOR relation
holds at least within 30\% for $\Lambda_M\leq$~900~MeV, tolerable for 
higher-order corrections in 
this perturbative scheme. However, when the meson
cutoff is further increased the agreement with the GOR relation rapidly
deteriorates, indicating that in this regime higher-order corrections become
important. Our perturbative scheme should not be trusted there any more.

One may wonder why we do not plot the various curves for cutoffs larger than
$\Lambda_M=1250$~MeV in the {\nce} and $\Lambda_M = 900$~MeV in the MLA. The
reason is that upon further increasing $\Lambda_M$ we detect a second,
unphysical, pole in the pion propagator~\cite{oertel,OBW2}. This phenomenon
will be discussed in more detail in the next section.

\section{Instabilities due to strong mesonic fluctuations?}
\label{instabilities}
As already mentioned it has recently been claimed by Kleinert and van den
Bossche~\cite{kleinert} that in the NJL model chiral symmetry, which is
spontaneously broken in the mean field ground state, gets restored due to
strong mesonic fluctuations. One principal objection one can raise against 
this supposition is the non-renormalizability of the NJL model. Including
fluctuations beyond mean-field we cannot avoid encountering additional 
divergencies. Thus a priori an additional cutoff parameter has to be introduced
as discussed previously. This additional cutoff parameter then controls the 
strength of mesonic fluctuations~\cite{oertel}. Nevertheless it
might be possible to observe some kind of ``chiral symmetry restoration'' at a
certain value of the cutoff parameter.

\begin{figure}[t] 
\parbox{16cm}{\begin{center}
              \epsfig{file=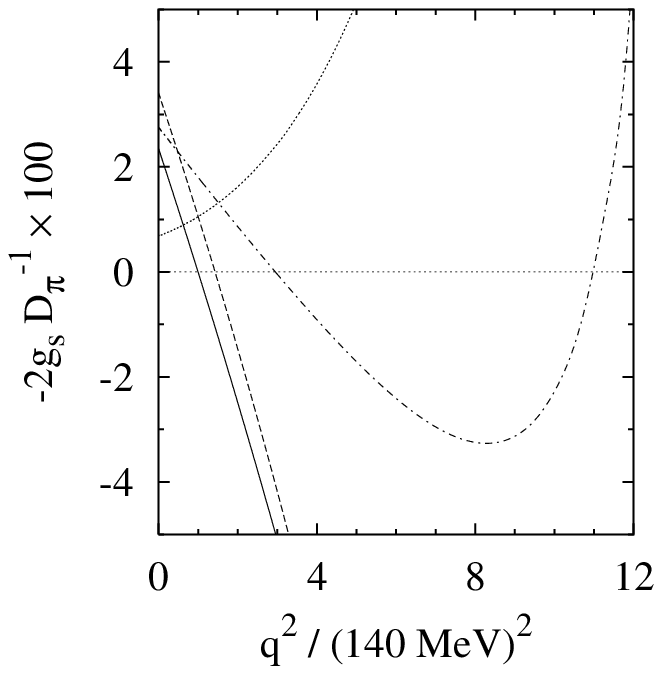,height=6.43cm,width=7.5cm}\hspace{.5cm}
              \epsfig{file=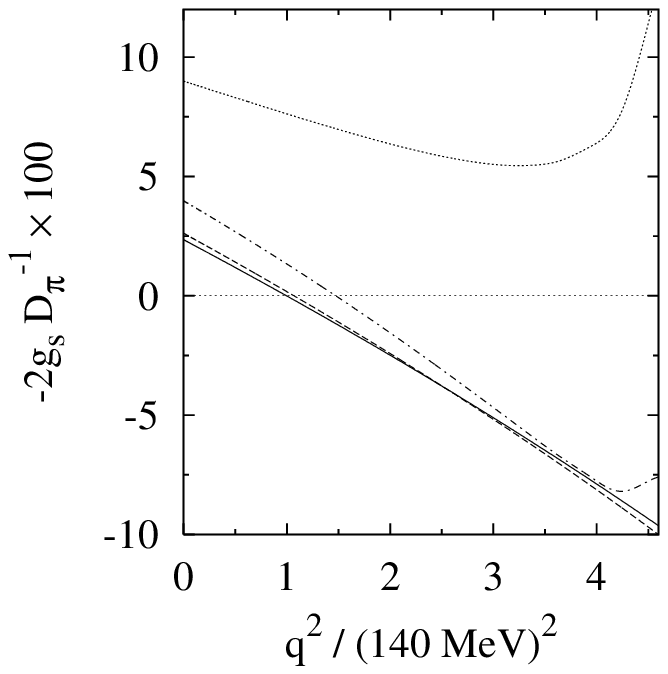,height=6.43cm,width=7.5cm}\end{center}}
\caption{\it Inverse pion propagator, $D_\pi^{-1}$, multiplied by
         $-2g_s$ as a function of the 4-momentum squared for various
         meson cutoffs; in the {\nce} (left): $\Lambda_M$~=~0 (solid), 900~MeV
         (dashed), 1300~MeV (dashed-dotted) and 1500~MeV (dotted) and the MLA
         (right): $\Lambda_M=0$~MeV (solid), 500 MeV (dashed),
         700~MeV(dashed-dotted) and 1100~MeV (dotted).}
\label{figpionprop}
\end{figure}
In the previous section we stated that a second, unphysical, pole
emerges in the pion propagator if the cutoff exceeds a certain value. This
pole has a residue of the ``wrong'' sign, leading to an imaginary pion-quark
coupling constant and a negative value for the squared pion decay constant
$f_\pi^2$. This becomes clear from \fig{figpionprop}, where the inverse pion
propagator $D_\pi^{-1}$ is plotted as a function of the momentum squared $q^2$
for different values of the cutoff $\Lambda_M$. The left panel shows the
results in the {\nce}, the right panel those in MLA. In the chiral limit all
curves would go through $-2 g_s D_\pi^{-1}=0$ at $q^2 = 0$ with a negative
slope. In the Hartree approximation + RPA the effect of a non-zero current quark
mass is to shift this value up to $m_0/m_H$, i.e.\@ the zero of the inverse
propagator, corresponding to the pole of the propagator, is then located at
some positive (time-like) value of $q^2$. In the {\nce} the main effect of the
meson-loop contributions for $\Lambda_M\lsim 900$ MeV is to further increase
the value of $D_\pi^{-1}$ at $q^2=0$ and with it the value of $q^2$ at the
zero, i.e.\@ the pion mass. Upon further increasing $\Lambda_M$ the value of
$D_\pi^{-1}(0)$ is again reduced. From \eqs{sumabc} and (\ref{pid}) we
conclude that the $1/N_c$-correction terms to $D_\pi^{-1}(0)$ are proportional
to $\Delta$. Thus, the behavior of $D_\pi^{-1}(0)$ reflects that of the
constant $\Delta$, discussed in \Sec{qqattz}. At the same time the curve
flattens considerably, leading to a strongly rising pion mass. For $\Lambda_M
\gsim 1250$ MeV we observe that it turns around below the quark-antiquark
threshold, causing the second pole in the pion propagator. Obviously the slope
of $D_\pi^{-1}$ at this second zero is inverse as compared with the first zero.
This reveals the above mentioned fact of a residue with the ``wrong'' sign.
When the cutoff is further increased the two zeros move towards each other,
merging at $\Lambda_M \approx 1350$ MeV and disappearing thereafter from the
real positive $q^2$-axis. Qualitatively the same behavior is observed within
the MLA. The only difference is that the second unphysical zero emerges
already for $\Lambda_M\gsim 910$~MeV and the two poles merge already for
$\Lambda_M \approx 1095$~MeV. 
\begin{figure}[p!]
\begin{center}
\parbox{7.5cm}{
     \epsfig{file=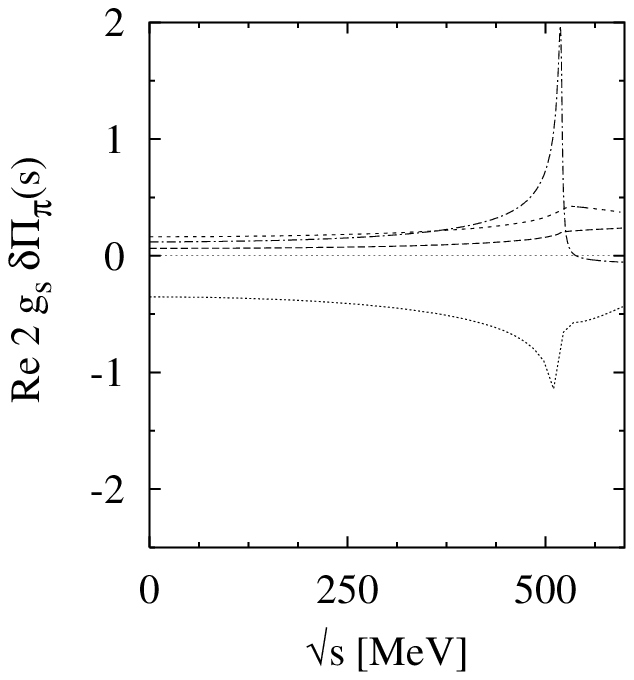,
     height=6.43cm, width=7.5cm}}
\parbox{7.5cm}{
     \epsfig{file=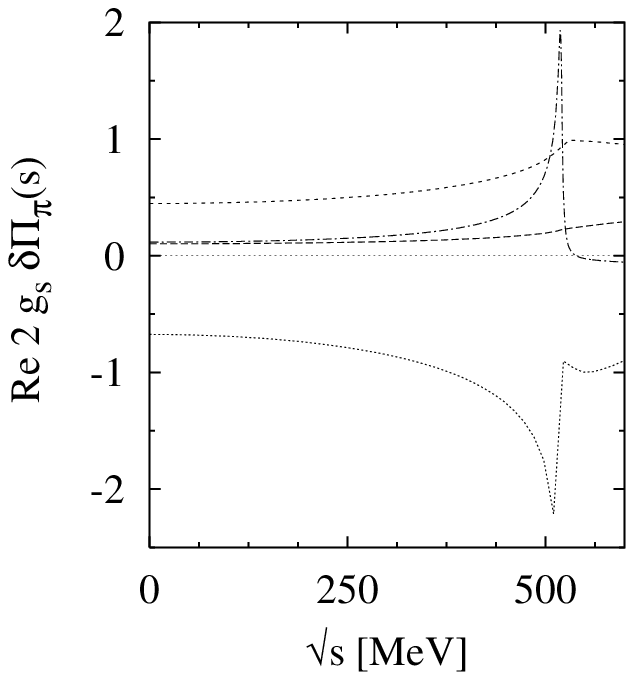,
     height=6.43cm, width=7.5cm}}\\
\parbox{7.5cm}{
     \epsfig{file=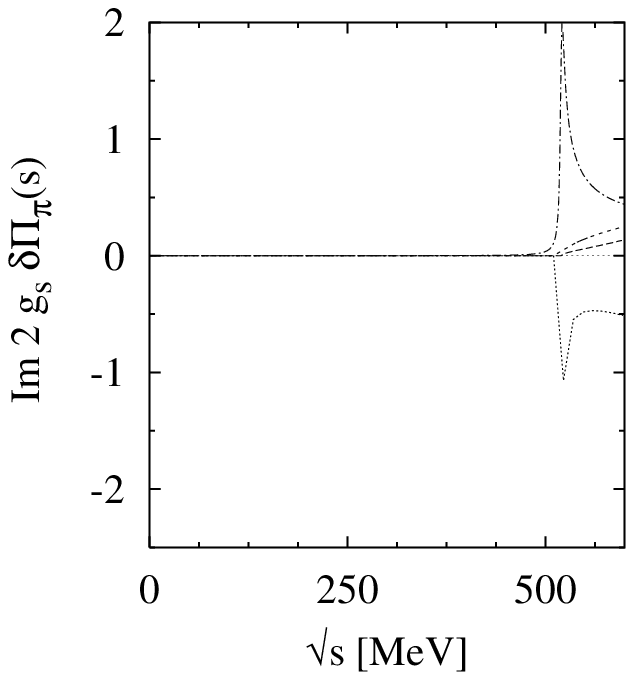,
     height=6.43cm, width=7.5cm}}
\parbox{7.5cm}{
     \epsfig{file=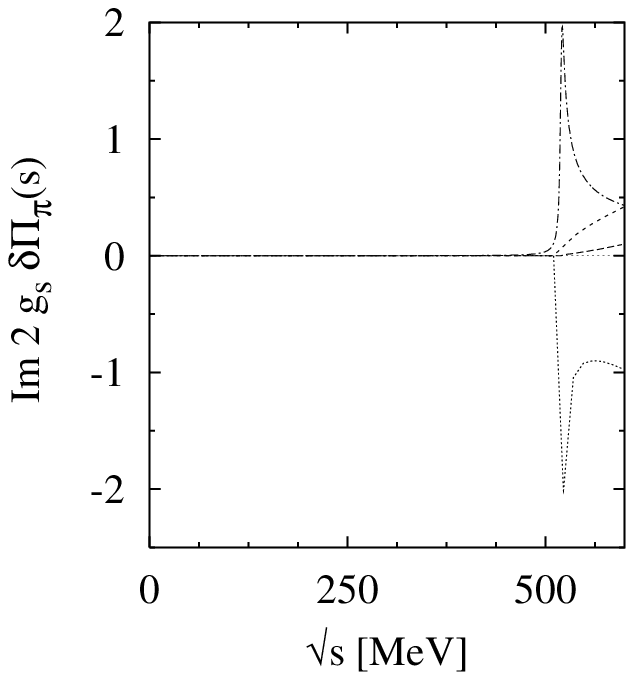,
     height=6.43cm, width=7.5cm}}\\
\parbox{7.5cm}{
     \epsfig{file=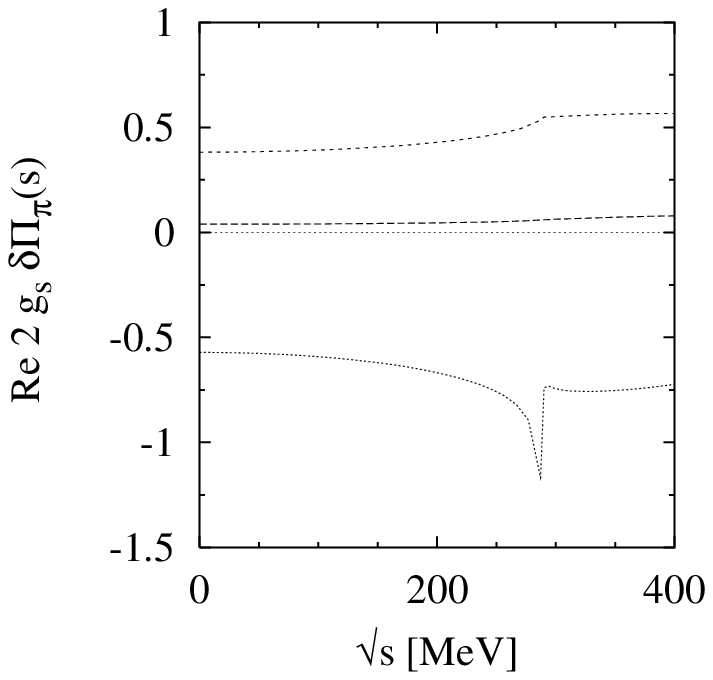,
     height=6.43cm, width=7.5cm}}
\parbox{7.5cm}{
     \epsfig{file=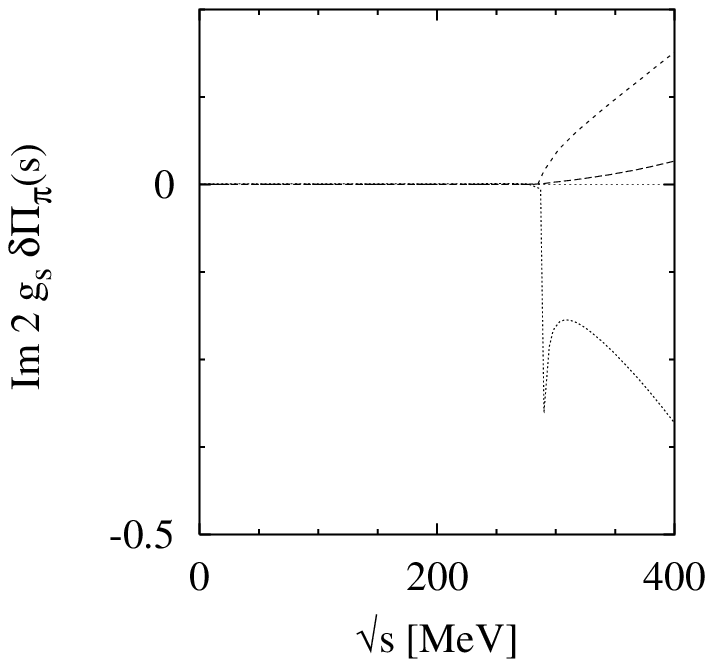,
     height=6.43cm, width=7.5cm}}
\end{center}
\caption{\it Real (upper panel) and imaginary part (middle) of the
     different $1/N_c$-correction terms to the pion polarization
     function, $\delta\Pi_\pi^{(a)}$ (long-dashed), $\delta\Pi_\pi^{(b)}$
     (dotted),  
     $\delta\Pi_\pi^{(c)}$ (short-dashed), and $\delta\Pi_\pi^{(d)}$
     (dashed-dotted), for $\Lambda_M=700$~MeV (left), and 1100~MeV
     (right). Different corrections term to the pion polarization function in
     the MLA for $\Lambda_M=1000$~MeV (lower panel), $\delta\Pi^{(a)}$
     (long-dashed), $\delta\Pi^{(b)}$ (dotted), and $\delta\Pi^{(c)}$
     (short-dashed), real part (left) and imaginary part (right).}
\label{figdpiabcd} 
\end{figure}

Within both schemes this unphysical behavior can be traced back mainly to the
contributions from $\delta\Pi_\pi^{(b)}$. This term has a peak in the
imaginary part above the
$\bar{q}q$-threshold with the ``wrong'' sign. This can be seen from
\fig{figdpiabcd}, where the various contributions to the pion polarization
function are displayed. For low values of the cutoff the sum of all other
contributions cancels this effect such that the imaginary part of the entire
polarization function behaves reasonably. In the {\nce} especially
$\delta\Pi_\pi^{(d)}$, proportional to $\Delta$, adds to this
cancellation. Here the growing of the peak in $\delta\Pi_\pi^{(b)}$ together
with the diminishing value of $\Delta$ for $\Lambda_M\gsim 900$ MeV lead to
the imaginary part of the polarization function changing sign if the cutoff
exceeds a certain value. This imaginary part with the ``wrong'' sign can then
be related via dispersion relations to the ``turn-around'' in the real part below
the $\bar{q}q$-threshold. 

The appearance of the unphysical second pole already indicates that the pion
becomes unstable if the strength of the mesonic fluctuations is increased too
much. 
In Ref.~\cite{oertel} we suggested that these instabilities of the pion
propagator might indicate an instability of the underlying ground state.
This was motivated by referring to similar situations in many-body systems,
where e.g.\@ complex energy eigenvalues in the excitation spectrum are a hint
at an instability of the ground state~\cite{thouless}. In that reference we
exclusively dealt with the {\nce}, where it is evidently impossible to
investigate the question whether the ground state develops an instability
against mesonic fluctuations, since these are built perturbatively on the
Hartree ground state. 
However, the MLA offers a possibility to examine the influence of mesonic
fluctuations on the structure of the ground state more closely~\cite{OBW2}. 
Since we
encounter the same type of instabilities within this scheme as in the {\nce},
we can probably gain deeper insight into the nature of these
instabilities. The principal tool to study the structure of the ground state
is certainly the effective action, \eq{gamma1loop}. In our case it can be
written in terms of the effective potential, defined in \eq{defep}, which 
describes the energy density of the system. It is explicitly given by  
\bea
V(m) &=& 2 \,i N_c N_f \intp \ln(\frac{m^{2}-p^2}{m_0^2-p^2}) 
+ \frac{(m-m_0)^2}{4 g_s} \nonumber\\ &&-
\frac{i}{2}\intp\{ \ln (1- 2 g_s\Pi_{\sigma}(p))+3 \ln (1- 2
g_s\Pi_{\pi}(p))\} + const.~.
\label{tp}
\eea 
The irrelevant constant can be chosen in such a way that $V(0) = 0$.
Here we only considered $\pi$- and $\sigma$-meson as intermediate states. Of
course this can be generalized to include also $\rho$- and $a_1$-meson
states. The resulting explicit expression can be found in App.~\ref{veffp}.

Since we plan to study a possible ``chiral symmetry restoration'' the only
sensible procedure would be to perform the calculations in the chiral limit.
We have already mentioned in connection with the pion mass that it is
numerically an extremely difficult task to evaluate the expressions in vacuum
exactly in the chiral limit. Thus we are forced to build this limit either
by extrapolating from nonzero current quark masses or from nonzero
temperatures. We have checked that both methods lead to almost the same
results. To be precise we should point out that we will proceed in the following
way to build the limit: Starting from the parameter set given above, we keep
the Hartree constituent quark mass, $m_H =$~260~MeV, fixed, while $m_0$ is
reduced from 6.1~MeV to zero. The consequence is a slightly enhanced coupling
constant, $g_s\Lambda_q^2 = 2.96$ as compared with $g_s\Lambda_q^2= 2.90$.
The same procedure was also applied to the calculations of meson loop
effects on the quark condensate described in \Sec{qqattz}.

In which way can we now identify a possible ``chiral symmetry restoration'' in
the behavior of the effective potential? First let us remind ourselves that
the positions of the extrema of $V(m)$ correspond to the solutions of the gap
equation~(\ref{localgap}). As the system always tends to minimize its energy, the
vacuum expectation value $m^{\prime}$ is therefore given by the value of $m$
at the absolute minimum of $V$. For a vanishing current quark mass
$m^{\prime}$ is, according to \eq{qbqself2}, proportional to the quark
condensate and consequently to the order parameter of chiral symmetry.  Hence,
for a given value of $\Lambda_M$, a non-zero value of $m^{\prime}$ indicates a
spontaneously broken ground state whereas chiral symmetry is restored if the
absolute minimum of $V$ is located at $m=0$.

\begin{figure}[t]
\begin{center} 
\parbox{16cm}{
              \epsfig{file=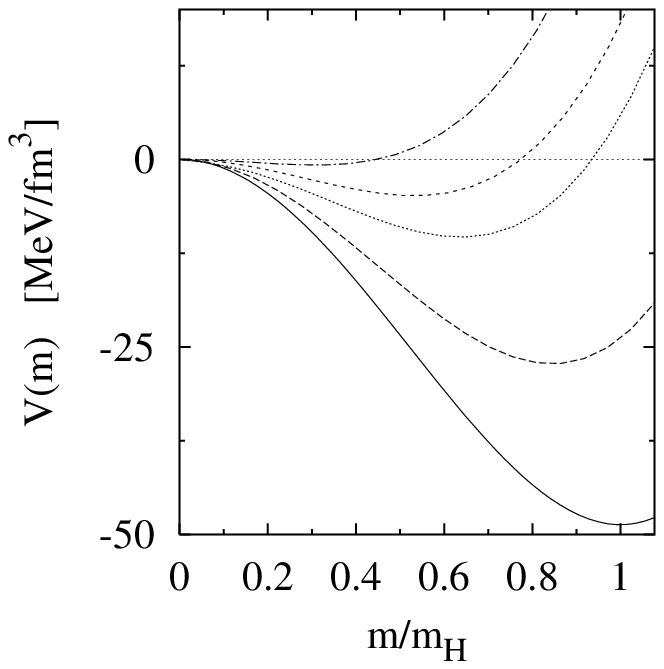,height=6.43cm,width=7.5cm}
              \hspace{.5cm}
              \epsfig{file=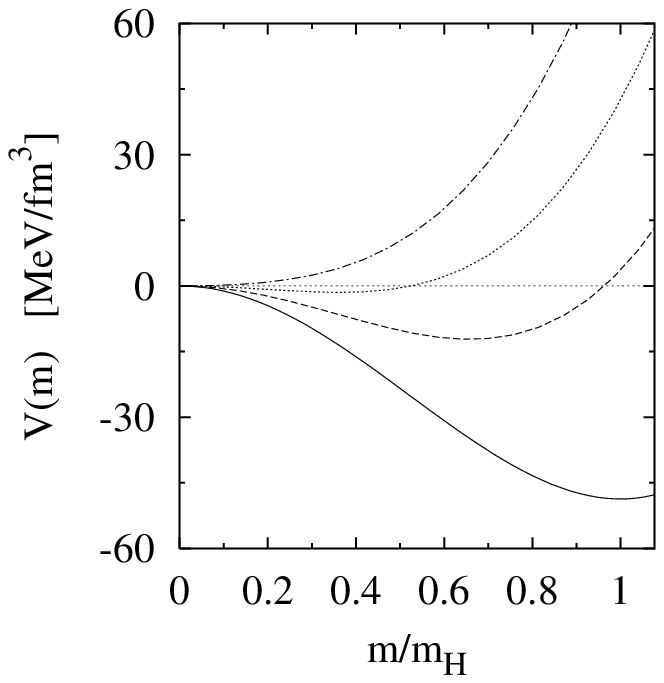,height=6.43cm,width=7.5cm}}\end{center}
\caption{\it  Effective potential $V(m)$ as a function of 
              $m/m_H$ in the MLA for different values of
              the meson cutoff $\Lambda_M$, on the l.h.s.\@ without vector meson
              intermediate states, i.e.\@ $g_v=0$, on the r.h.s.\@ with $g_v = 2
              g_s$. (left) 0~MeV (solid), 300~MeV
              (long-dashed), 500~MeV (dotted), 900~MeV (dashed-dotted) and
              1200 MeV (short-dashed), (right) 0~MeV (solid), 300~MeV
              (dashed), 400~MeV (dotted) and 500~MeV (dashed-dotted). }
\label{figtp}
\end{figure}

$V(m)$ as a function of $m/m_H$ 
for different values of $\Lambda_M$ is displayed in \fig{figtp}. Taking into
account only fluctuations of $\pi$- and $\sigma$-mesons we obtain the
results shown on the l.h.s.\@ in the other case those on the r.h.s.\@ 
For $\Lambda_M = 0$ we of course encounter the usual picture: A local maximum
is located at $m=0$ and we find a minimum at $m=m_H=260$~MeV. This is the well
established result in the Hartree approximation. We have already explained that we
would expect the maximum at $m=0$ to convert into a minimum upon increasing
$\Lambda_M$ if there was
indeed some kind of ``phase transition''.
It is obvious from \fig{figtp} that in the purely scalar and pseudoscalar case
no phase transition occurs, although the results for $\Lambda_M\lsim 900$~MeV
seem to point in this direction. In this regime not only the vacuum
expectation value $m^{\prime}$ is considerably reduced -to about 30\% of the
Hartree mass for $\Lambda_M=900$~MeV- but also the ``bag constant'', which is
given as the difference in energy between the maximum at $m=0$ and the minimum at
$m= m^{\prime}$, i.e.\@ $B=V(0)-V(m^{\prime})$, moves towards zero. It decreases
from 48.7~MeV/fm$^3$ at $\Lambda_M$~=~0 to 0.8~MeV/fm$^3$ at
$\Lambda_M$~=~900~MeV. However, when $\Lambda_M$ is further increased we
also observe the traces of the ``turn-around'' of $\Delta$ (see
\Sec{qqattz}): Both, $m^{\prime}$ and $B$, go up again. This means in particular
that $V(0)$ always remains a local maximum leading to the conclusion that a
``phase transition'' is ruled out as an explanation for the instabilities we
found. 

Note that the results do not completely exclude the possibility of a ``phase
transition'' due to strong mesonic fluctuations. The small bag constant at
$\Lambda_M=900$~MeV indicates that probably a careful parameter study brings
to light 
a ``phase transition'' in a certain parameter range. This supposition is
corroborated by the results including the $\rho$- and
$a_1$-meson subspace for intermediate states. As we already suspected
from the results for the quark condensate in \Sec{qqattz}, we observe a second
order phase transition. The existence of a phase transition is clearly
discernible from the behavior of the effective potential for different
cutoffs displayed on the r.h.s.\@ of \fig{figtp}. For $\Lambda_M=0$~MeV the curve
certainly does not differ from that in the purely scalar and pseudoscalar
case. However, as already stated in the previous sections, the vector and axial
intermediate states amplify the effects of mesonic fluctuations, leading to
an accelerated decrease of $m^{\prime}$ and $B$ with an
increasing cutoff. For $\Lambda_M=500$~MeV the maximum has disappeared and the
only extremum we find is a minimum at $m = 0$. A closer examination of the
development strongly suggests that we deal here with a second order phase
transition with a ``critical'' $\Lambda_{M,c} = 461.82$~MeV. The exact value of
this ``critical'' meson cutoff should not be taken too seriously since we have
chosen the value of $g_v=2 g_s$ rather deliberately. For instance,  if we take
$g_v=g_s$ the critical cutoff is augmented by almost 100 MeV.
 
We have to remark here that our investigations are not suitable to give a
conclusive answer as to whether for a certain choice of parameters ``chiral
symmetry restoration'' according to the conjecture by Kleinert and van den
Bossche~\cite{kleinert} is observed. The phenomenon the authors of
Ref.~\cite{kleinert} describe is closely related to one from strong-coupling
superconductors. In addition to the superconducting phase in strong-coupling
superconductors there exists a so-called pseudo-gap phase above
$T_c$~\cite{nozieres,sademelo,babaev}, where Cooper pairs are still formed but do not
condense.  Analogously in such a phase the quarks would still acquire a
non-vanishing constituent mass if the mass was identified with the modulus
of the field $\Phi$ introduced in \Sec{1ml}. Since the phase of the
$\Phi$-field is strongly fluctuating the expectation value of $\qq$ would
nevertheless vanish, indicating that chiral symmetry would not be broken
within this phase. Our approach implicitly assumes a uniform phase factor and
we can therefore not exclude a phase transition into a pseudo-gap phase of
that type.
  
In any case we would like to understand the reason for the instabilities we
found. A hint is probably the discussion of another type of vacuum instability
which has been performed recently by Ripka~\cite{ripkapaper}. In
\Sec{regularization} we pointed out that several regularization schemes
produce unphysical poles of the RPA propagators, i.e.\@ poles in the complex
plane where they are forbidden by microcausality. Ripka states that the
instabilities he finds by examinating the effective potential are caused by
these unphysical poles of the RPA propagators, which are induced by the
regularization procedure. His analysis is performed with a 4-momentum cutoff
and a Gaussian form factor. As discussed in \Sec{regularization} at the first
sight the Pauli-Villars scheme does not generate such difficulties. However,
the regulators lead to an ``overshooting'' of the imaginary part of the RPA
propagators, i.e.\@ in some kinematical regions a $\bar{q}q$-continuum is
created with the wrong sign. Although the above discussion, in connection with
the observation that the instabilities in the pion propagator can be traced
back mainly to an imaginary part with the wrong sign in $\delta\Pi_\pi^{(b)}$,
strongly corroborates the supposition that the instabilities are related to
the analytical properties of the RPA propagators. Certainly further
investigations are necessary in order to give a conclusive answer.  From that
point of view it is also not very astonishing that recently a second
(unphysical) pole in the pion propagator has also been found in a non-local
generalization of the NJL model~\cite{plant} using a Gaussian form factor
similar to that discussed by Ripka~\cite{ripkapaper}.  The above reflections
render it implausible that these instabilities could principally be removed by
including vector and axial vector intermediate states, as suggested by Plant
and Birse~\cite{plant}, since the analytical properties of the RPA propagators
are thereby not altered. Besides, we observed in \Sec{qqattz} that the
behavior of the quark condensate, strongly related to the instabilities we
found in the pion propagator, is not changed qualitatively if we include
vector interactions. Probably vector and axial vector intermediate states
could enlarge the region in parameter space, where we are far off the region
where instabilities occur and at the same time obtain a reasonable fit of
observables. We will enter into this question in the next section.
\section{Parameter fit}
\label{secpi2}

Up to now we have investigated the influence of mesonic fluctuations on various
quantities in the pion sector. To that end all parameters have been kept fixed, which have
been determined by fitting $f_\pi^{(0)}, m_\pi^{(0)}$, and $\qq^{(0)}$, except
the meson cutoff $\Lambda_M$. It is clear that, if the model should be
applicable to describe physical processes, a refit of the parameters to
reproduce the empirical values of $f_\pi, m_\pi$, and $\qq$ should be
performed. For the {\nce} this has been done in Ref.~\cite{OBW}. In the MLA,
however, we did not achieve such a fit~\cite{OBW2} for reasons explained
below. Of course, the three observables in the pion sector do not suffice to
conclusively determine the five model parameters $m_0, g_s, g_v, \Lambda_q$, 
and $\Lambda_M$. The fact that we are compelled to exclude vector and axial
vector intermediate states, see \Sec{pionml}, considerably simplifies the
fitting procedure because in this case pionic properties do not depend on the
vector coupling constant $g_v$. We will therefore follow the strategy
established in Refs.~\cite{OBW,OBW2}: For various values of $\Lambda_M$ we
first fix the current quark mass $m_0$, the quark-loop cutoff $\Lambda_q$, and
the scalar coupling constant to fit the quantities in the pion sector,
i.e.\@ $m_\pi, f_\pi$, and $\qq$. With the two remaining parameters, $g_v$ and
$\Lambda_M$, we will then try to reproduce the data for the pion
electromagnetic form factor in the time-like region which is, assuming
vector-meson dominance, primarily affected by the properties of the
$\rho$-meson. The latter is very well suited for this purpose since it,
besides being a vector state, cannot be described reasonably without including
mesonic fluctuations. Since the phenomenologically important two pion
intermediate state in our model consists of RPA pions, we are forced to fit
$m_\pi^{(0)}$ and not $m_\pi$ to the empirical pion mass if we desire to get
the correct threshold behavior for the $\rho$-meson. This is certainly a
slight inconsistency and will restrict the applicability of our model to
regions where the deviations are not too large. 

Another point additionally restricts the regime of possible parameter sets: The
unphysical $q\bar{q}$-threshold must lie well above the peak in the
$\rho$-meson spectral function in order to obtain a realistic description of
the $\rho$-meson. Consequently the constituent quark mass, $m_H$ in the {\nce}
and $m^{\prime}$ in the MLA, has to be larger than about 400 MeV. We have to
emphasize here that the constituent quark mass is entirely determined by
the particular parameter set fixed by fitting observables. Nevertheless
we have some freedom: The empirical value of the quark condensate is not known
very precisely because it is not a directly measurable quantity. We have to
rely on sum rule analyses or results from the lattice. From the former an
upper limit for the absolute value of $\qq$ of about 2(260~MeV)$^3$ at a 
renormalization scale of 1~GeV can be
extracted~\cite{dosch}. The value obtained on the lattice lies in the same
range, recent results give $\qq$~=~{\mbox -2($(231 \pm 4 \pm 8 \pm
  6)$~MeV)$^3$} \cite{giusti}. Thus, as long as the value of the quark
condensate stays within the above-given boundaries, we can try to increase the
constituent quark mass as much as possible. In the MLA it will be much more
difficult to obtain a quark mass which is large enough than in the {\nce}. The
reason is obvious: The meson-loop
corrections, which in the MLA affect the constituent quark mass via the
solution of the extended gap equation, \eq{localgap}, tend to diminish it. In
the {\nce}, on the contrary, the determining quantity is the
Hartree quark mass $m_H$, which is not influenced by any mesonic fluctuations. 
In addition, in the {\nce}, the mesonic fluctuations lower the absolute value of
the quark condensate, whereas in the MLA the direct relation between the quark
condensate and the constituent quark mass (\eq{qbqself2}) at the same time
leads to a decreasing quark mass.
\subsection{Pion sector}
\label{pionfit}
Our results~\cite{OBW,OBW2} in the pion sector are listed in
Table~\ref{tablence} for the {\nce} and in Table~\ref{tableMLA} for the MLA. 
We show five parameter sets, corresponding to meson-loop cutoffs of 0 MeV
(RPA), 300 MeV, 500 MeV, 600 MeV, and 700 MeV, together with the respective
constituent quark mass and the resulting values of $m_\pi, f_\pi$, and
$\qq$. Since the properties of the intermediate pion states, in particular the
pion mass $m_\pi^{(0)}$, are important in connection with the $\rho$-meson, we
also display the RPA quantities $m_\pi^{(0)}, f_\pi^{(0)}$, and $\qq^{(0)}$. In
the MLA, however, these are strictly speaking no RPA quantities since within
that scheme the
intermediate ``RPA'' states consist of quarks with the constituent quark mass
$m^{\prime}$. This, among other consequences, results in the quark
condensate $\qq$ coinciding with the ``RPA'' value $\qq^{(0)}$,
\begin{table}[h!]
\begin{center}
\begin{tabular}{|c|c|c|c|c|c|}
\hline
$\Lambda_M$~/~MeV   &   0.  & 300.  & 500.  & 600.  & 700.  \\ \hline
$\Lambda_q$~/~MeV   & 800.  & 800.  & 800.  & 820.  & 852.  \\ \hline
$m_0$~/~MeV         & 6.13  & 6.40  & 6.77  & 6.70  & 6.54  \\ \hline
$g_s\Lambda_q^2$    & 2.90  & 3.07  & 3.49  & 3.70  & 4.16  \\ \hline
$m_H$~/~MeV           & 260.  & 304.  & 396.  & 446.  & 550.  \\ \hline
$m_\pi^{(0)}$~/~MeV & 140.0 & 140.0 & 140.0 & 140.0 & 140.0 \\ \hline
$m_\pi$~/~MeV       & 140.0 & 143.8 & 149.6 & 153.2 & 158.1 \\ \hline
$f_\pi^{(0)}$~/~MeV &  93.6 & 100.6 & 111.1 & 117.0 & 126.0 \\ \hline
$f_\pi$~/~MeV       &  93.6 &  93.1 &  93.0 &  93.1 & 93.4  \\ \hline
$\qq^{(0)}$~/~MeV$^3$ & -2(241.1)$^3$ & -2(249.3)$^3$ & -2(261.2)$^3$ 
                      & -2(271.3)$^3$ & -2(287.2)$^3$ \\ \hline
$\qq$~/~MeV$^3$       & -2(241.1)$^3$ & -2(241.7)$^3$ & -2(244.1)$^3$ 
                      & -2(249.5)$^3$ & -2(261.4)$^3$ \\ \hline
-$m_0 \qq / m_\pi^2 f_\pi^2$ & 1.001 & 1.007 & 1.018 & 1.023 & 1.072
\\ \hline
\end{tabular}
\end{center}
\caption{{\it The model parameters ($\Lambda_M$, $\Lambda_q$, $m_0$ and 
$g_s$) and the resulting values of $m_\pi$, $f_\pi$ and
$\qq$ (together with the corresponding leading-order quantities), 
the constituent quark mass $m_H$ in the \nce. The ratio $-m_0 \qq / m_\pi^2 f_\pi^2$,
is also given.
}}
\label{tablence}
\end{table}  
\begin{table}[h!]
\begin{center}
\begin{tabular}{|c|c|c|c|c|c|}
\hline
$\Lambda_M$~/~MeV   &   0.  & 300.  & 500.  & 600.  & 700.  \\ \hline
$\Lambda_q$~/~MeV   & 800.  & 800.  & 810.  &820. & 835.  \\ \hline
$m_0$~/~MeV         & 6.13  & 6.47 & 7.02  & 7.30 & 7.90  \\ \hline
$g_s\Lambda_q^2$    & 2.90  & 3.08 & 3.44  & 3.71 & 4.52  \\ \hline
$m_H$~/~MeV         & 260.  & 305. & 390.  & 450. & 600.  \\ \hline
$m^{\prime}$~/~MeV  & 260.  & 278.2 & 320.0 & 355.7& 468.4  \\ \hline
$m_\pi^{(0)}$~/~MeV & 140.0 & 139.9 & 140.0 & 139.7& 140.0 \\ \hline
$m_\pi$~/~MeV       & 140.0 & 145.1 & 156.3 & 164.5& 182.7 \\ \hline
$f_\pi^{(0)}$~/~MeV &  93.6 & 96.7 & 103.6 & 108.4& 120.0\\ \hline
$f_\pi$~/~MeV       &  93.6 & 93.2  &  92.9 & 92.9 & 92.8  \\ \hline
$\qq$~/~MeV$^3$     & -2(241.1)$^3$ & -2(241.7)$^3$ & -2(246.2)$^3$ 
                       &- 2(250.8)$^3$ & -2(260.9)$^3$\\ \hline
-$m_0 \qq / m_\pi^2 f_\pi^2$ & 1.001 & 1.001& 1.006 &1.01& 1.02
\\ \hline
\end{tabular}
\end{center}
\caption{\it The same as in Table~\ref{tablence} for the MLA. $\qq^{(0)}$ is
                       not shown since it agrees with $\qq$, see text.}
\label{tableMLA}
\end{table}  
cf. \eqs{qbq0} and (\ref{qbqself2}). We therefore renounced to list both values
in Table~\ref{tableMLA}. For comparison we also give the corresponding
Hartree quark mass $m_H$ in Table~\ref{tableMLA}. The ratio of the right hand
side of the GOR relation, \eq{GOR}, divided by the left hand side, cf. the
last line in Tables~\ref{tablence} and \ref{tableMLA}, provides us on the one
hand with a measure for higher order (beyond next-to-leading order)
corrections in the {\nce}, and on the other hand with the possibility to
estimate numerical uncertainties. 
The latter can be deduced from the
discrepancies between the exact value of unity and the obtained value in the
MLA. 
The almost perfect agreement of the RPA 
quantities with the GOR relation suggest that we deal here indeed mainly with 
numerical uncertainties and not with higher-order corrections in $m_0$. 
We obtain deviations of at most 2\%, assuring our confidence in the
numerics. With the above parameter sets we also stay in a region where
higher order corrections in the {\nce} remain small: the deviations in the
{\nce} are less than $10\%$, for $\Lambda_M\leq600$~MeV even less than 3\%.

Although we find in both schemes that the constituent quark mass, $m_H$ in the
{\nce} and $m^{\prime}$ in the MLA, increases with an increasing cutoff, the
values in the MLA are considerably smaller than those in the {\nce}. Within
the latter we find a region of meson cutoffs, 500 MeV $\leq\Lambda_M\leq$ 700
MeV, where on the one hand the constituent quark mass is large enough to shift the
$q\bar{q}$-threshold above the peak in the $\rho$-meson spectral function, and
where on the other hand we can reproduce the empirical values of $f_\pi, m_\pi$ and
$\qq$. The upper boundary for $\Lambda_M$ is caused by the impossibility to
simultaneously stay below the limit for $\qq$ and reproduce the correct value
for $f_\pi$ upon further increasing $\Lambda_M$. If one remembers the above
discussion concerning the constituent quark mass, it is not astonishing that
the region of meson cutoffs where we obtain sensible results is much more
narrow in the MLA. From the values listed in Table~\ref{tableMLA} it can be
seen that for $\Lambda_M=600$~MeV $m^{\prime}$ is still too small whereas for
$\Lambda_M=700$~MeV the value of $\qq$ already slightly exceeds the limit. The main
reason is that simultaneously with the constituent quark mass the absolute
value of the quark condensate rises, rendering it much more difficult to stay
below the limit for $\qq$.

\subsection{Description of the $\rho$-meson}
\label{secrho}

In the previous section we started the parameter fit by determining parameter
sets for various values of the meson cutoff $\Lambda_M$, which reproduce the
empirical values of $m_\pi, f_\pi$, and $\qq$. In both schemes we established a
lower and an upper boundary for the meson cutoff $\Lambda_M$. The lower 
boundary was chosen such that the quark threshold lies above the 
$\rho$-meson mass, whereas the upper boundary mainly resulted from the 
impossibility to reproduce the empirical values of $\qq$ and $f_\pi$ at 
the same time. In this section
we will attempt to decisively fix $\Lambda_M$ by looking at quantities related
to the $\rho$-meson. This will allow us as well to fix the vector coupling
constant $g_v$. In the {\nce} we will succeed~\cite{OBW}, whereas in the MLA we
will enter into a region where instabilities occur~\cite{OBW2}, similar to
those observed in the pion propagator (see \Sec{instabilities}). This will
become more clear from the following discussion.

Including meson-loop effects the polarization function of the $\rho$-meson
reads (cf. \eqs{pol1} and (\ref{poltilde})) 
\beq
    {\tilde \Pi}_\rho^{\mu\nu, ab}(q) = \Pi_\rho^{\mu\nu, ab}(q)
+ \sum_{k}\; \delta \Pi_\rho^{(k)\;\mu\nu, ab}(q)
\;.
\label{rhopol1}
\eeq
Here $k$ runs over $\{a,b,c,d\}$ in the {\nce} and over $\{a,b,c\}$
in the MLA. It is understood that the functions have to be evaluated at the
Hartree quark mass $m_H$ in the {\nce} and at $m=m^{\prime}$ in the MLA. 
In \Sec{transverse} we proved, with the help of Ward identities, that
this polarization function fulfills the transversality condition, \eq{trans},
provided we apply a regularization scheme which preserves this
property. The delicate point was the shift in the integration variable
needed to establish for instance the cancellations in \eq{rpatrans}, which is 
accompanied by a shift in the integration boundary in some regularization
schemes, destroying transversality. This is obviously not the case for the
Pauli-Villars scheme we employed to regularize the RPA polarization loop. With
the three-dimensional cutoff $\Lambda_M$ we use to regularize the meson loops
we have to be more careful: In principle not only Lorentz
covariance is violated but also the transverse structure of the correction
terms $\delta\Pi$ to the polarization function is in general not
retained. However, we are working in the rest
frame of the $\rho$-meson, i.e.\@ $\vec{q}=0$, so at least transversality
can be preserved. Thus, if we assume Lorentz invariance to be fulfilled, the
ansatz for the tensor structure of the polarization function in
\eq{tensortrans} is also applicable, enabling us to evaluate a scalar function
$\tilde\Pi_\rho(q)=-1/3 g_\mn \tilde\Pi_\rho^\mn(q)$ (cf. \eq{scalartrans})
instead of all tensor components. 

A second consequence of vector current conservation is, that the 
polarization function should vanish for $q^2=0$. In \Sec{regularization} we
saw that already in the Hartree approximation + RPA this can only be
achieved by a subtraction. The reason should not be searched for in a
particular 
regularization procedure we have applied, in this case Pauli-Villars, but to a
greater extent in the rather general problem we noted in \Sec{regularization}
concerning the inconsistency of several regularized RPA quantities. 
The three-dimensional cutoff we apply to render the meson-loop
integrals finite violates current conservation directly, i.e.\@ the correction
terms to the $\rho$-meson polarization function do not vanish a priori.   
We cure this problem again by performing a subtraction,
\beq
    \sum_{k}\; \delta \Pi_\rho^{(k)}(q) \;\rrr\;
    \sum_{k}\; \Big(\delta \Pi_\rho^{(k)}(q) \;-\;
    \delta \Pi_\rho^{(k)}(0) \Big) \;.
\label{sub}
\eeq 

\begin{figure}[t]
\hspace{1cm}
\parbox{14cm}{
     \epsfig{file=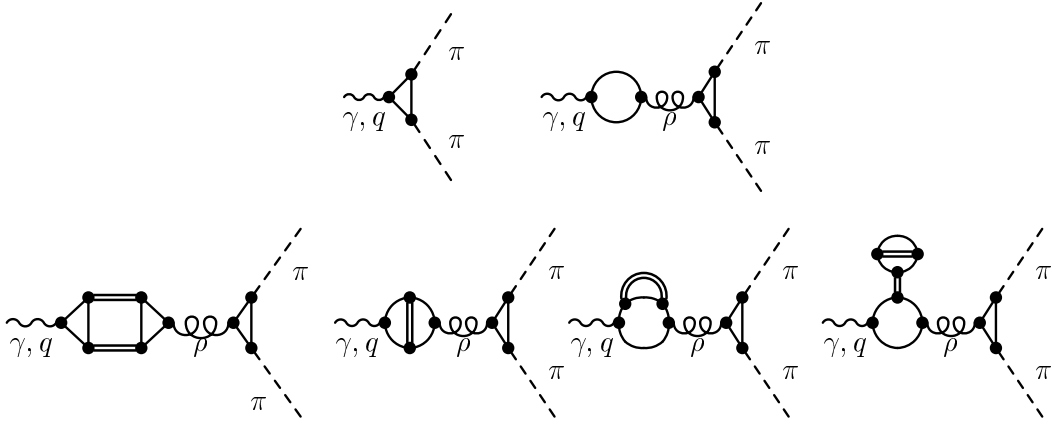,
     height=5.6cm, width=14.cm}}
\caption{\it Contributions to the pion electromagnetic form factor
         in the {\nce}. The propagator denoted by the curly line corresponds
         to the $1/N_c$-corrected $rho$-meson, while the double lines
         indicate RPA pions and sigmas.}
\label{figfeynpefm} 
\end{figure}

We are now in a position to attack the evaluation of the pion electromagnetic
form factor, enabling us to determine the two remaining parameters, $g_v$ and
$\Lambda_M$. This form factor is (in the time-like region) defined as the ratio of the cross section for
$e^+e^- \rightarrow \pi^+\pi^-$ divided by the cross section for the
production of point-like pions. The amplitude for this process is determined by the
exchange of a (virtual) photon. Assuming vector meson dominance the
coupling of the exchanged photon to the pions is mediated by a
$\rho$-meson. This ansatz exhibits great success in describing the experimental
data. It can also be simulated within the NJL model. 
The standard Hartree + RPA calculation~\cite{lutz} for the corresponding
amplitude, however, suffers from the completely unphysical nature of the intermediate
$\rho$-meson, containing only quark-antiquark loops. We consider now two types
of improvements. The first one is obvious: The RPA $\rho$-meson is replaced by
a full one, containing mesonic, in particular pionic, fluctuations. The second
one is more or less a question of consistency. In the standard scheme the
photon couples to the $\rho$-meson via a quark-antiquark loop similar to the
RPA polarization loop. Therefore we modify the coupling of the
photon to the $\rho$-meson in the same way as we did for the polarization
functions. The resulting diagrams for the {\nce} are displayed in
\fig{figfeynpefm}. In the MLA the diagram corresponding to $\delta\Pi^{(d)}$
(fourth diagram in the lower line) is certainly missing. Note
that the diagrams in the upper line would correspond to those in the standard
scheme if we replaced the ``improved'' $\rho$-meson propagator (curly line) by
an RPA one. 

Another type of possible improvement has been discussed in Ref.~\cite{lemmer}:
modifications of the quark triangle responsible for the coupling of the
$\rho$-meson or the photon, respectively, to the pions. These would be
suggestive if we considered full pions, i.e.\@ including meson-loop effects. We
decided, however, to take the external pions to be RPA ones. This is motivated
by the following fact: it is more consistent with dressing the $\rho$-meson by
RPA pions and in addition, the mass $m_\pi^{(0)}$ has been fitted to the
empirical pion mass.  
 
Evaluating the diagrams of Fig.~\ref{figfeynpefm} for on-shell pions we notice
that the resulting expression for the amplitude displays the same tensorial
and isospin structure as that for point-like
particles, i.e.\@ it is proportional to $(p^\mu -
p^{\prime\mu}) \eps_{abc}$ times the electric charge $e$. Here $p$ and
$p^\prime$ are the four-momenta of the external pions, $a, b$, and $c$ are
isospin indices of the pions and the photon. 
Calculating the ratio of the cross sections we become aware that this common
factor and the common phase space factor cancel.
We finally obtain for the form factor, $q=-p-p^\prime$,\footnote{Note that 
  in the corresponding formula in Ref.~\cite{OBW} a factor of 1/2 is missing.} 
\beq
|F_{\pi}(q)|^2 = \frac{1}{4}\,\Big|\,g_{\pi qq}^{(0)2} 
f(p,p^{\prime}) \, \Big(1-{\tilde\Pi}_{\rho}(q) 
{\tilde D}_{\rho}(q)\Big)\,\Big|^2~,
\label{piform}
\eeq  
where $f(p,p^{\prime})$ is a scalar function appearing in the
$\rho\pi\pi$-vertex function (see App.~\ref{functions}). Of course, the
external pions are to be taken on-shell, i.e.\@ 
$p^2 = p^{\prime2} = m_{\pi}^2$ and $p\cdot p^{\prime} = q^2/2-m_{\pi}^2$.  
Since a real photon ($q^2=0$) should ``see'' the actual charge of the pion,
the form factor should equal unity at $q^2 = 0$. 
The subtraction performed in \eq{sub} guarantees that the $\rho$-meson self
energy ${\tilde\Pi}_{\rho}$ vanishes at this point. Therefore we have to check
whether $|g_{\pi qq}^{(0)2} f(p,p^\prime)|^2=4$ holds for $q^2=0$. This can
most easily be shown with the help of the Ward-Takahashi identity
derived in \Sec{transverse} for the $\rho\pi\pi$-vertex (see
\eq{rhopipiward}). This identity is not directly applicable to our problem
since the vertex is not contracted with $q_\mu$ in our case. However, this can
be handled by using the following relation,
\beq
\Gamma^{\mu, abc}_{\rho,\pi,\pi}(q,p) =
\frac{d\,(q_\nu\Gamma^{\nu,abc}_{\rho,\pi,\pi}(q,p))}{d\, q_\mu}
- q_\nu \frac{d\,\Gamma^{\nu,abc}_{\rho,\pi,\pi}(q,p))}{d\, q_\mu}~.
\eeq
Using the Ward-Takahashi identity, \eq{rhopipiward}, we can write the first term on the
r.h.s.\@ of this expression as the derivative of a
difference of pion polarization functions in the RPA. Performing the limit $q
\rightarrow 0$ and comparing the result for the vertex with the definition of
the pion quark coupling constant $g_{\pi qq}^{(0)}$ we deduce
\beq
g_{\pi qq}^{(0)\, 2} f(p,p^\prime) \Big|_{q^2=0}
= 2~,
\eeq
leading to the desired value for the form factor at $q^2 = 0$.
\begin{figure}[t]
\begin{center}
\parbox{7.5cm}{
     \epsfig{file=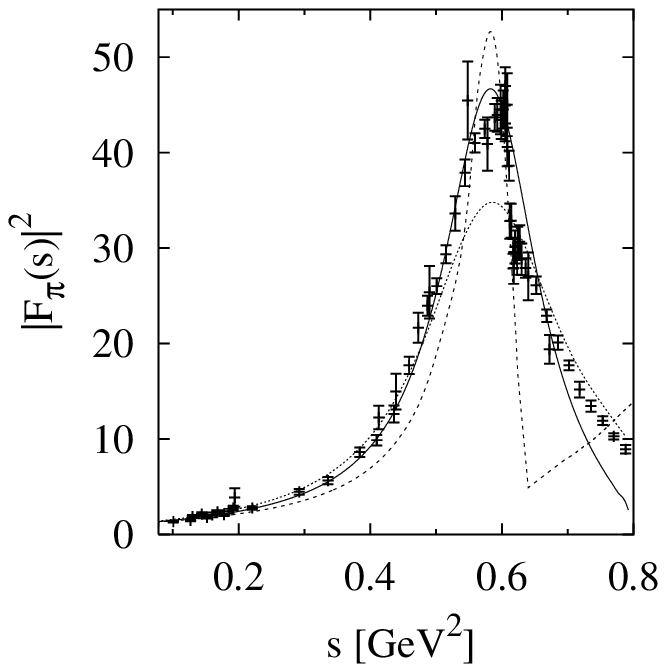,
     height=6.43cm, width=7.5cm}\quad}
\parbox{7.5cm}{
     \epsfig{file=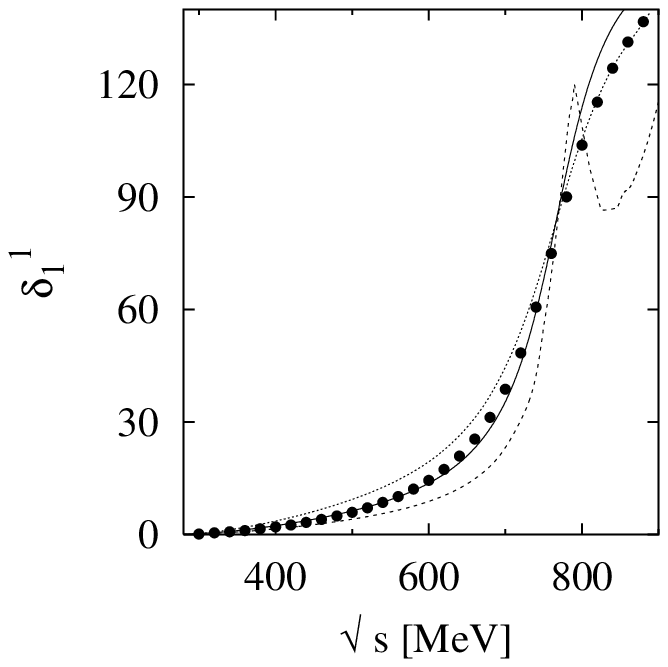,
     height=6.43cm, width=7.5cm}\quad}
\end{center}
\caption{\it Pion electromagnetic form factor (left panel) and 
             the $\pi\pi$-phase 
             shifts in the vector-isovector channel (right panel) for
             $\Lambda_M=500$~MeV and $g_v=1.0 g_s$ (dashed),
             $\Lambda_M=600$~MeV and $g_v = 1.6 g_s$ (solid) as well as
             $\Lambda_M=700$~MeV and $g_v=2.4 g_s$ (dotted) in the {\nce}. The
             other parameter values are taken from Table~\ref{tablence}. 
             The data points are taken from refs.~\cite{barkov} and
             \cite{frogatt}, respectively.}
\label{figpefm} 
\end{figure}

Our results for $|F_\pi|^2$ in the {\nce} as a function of the center-of-mass
energy squared are displayed in the left panel of Fig.~\ref{figpefm}, 
together with the experimental data \cite{barkov}. The three theoretical
curves are calculated with different values of the meson cutoff,
$\Lambda_M=500$~MeV (dashed line), $\Lambda_M=600$~MeV (solid), and
$\Lambda_M=700$~MeV (dotted). 
We have already
emphasized that these values for the meson cutoff lie within the small window
of cutoff values where we can hope to produce reasonable results. 
The vector coupling constant $g_v$ is chosen such that the maximum of the form
factor lies at the right position. The other parameters are taken from
Table~\ref{tablence}. For $\Lambda_M=500$~MeV we observe a kink in the form
factor slightly above the maximum ascribable to the quark-antiquark threshold
at $s=0.63~{\rm GeV}^2$. Below the kink the form factor drops very steeply due to
the sub-threshold attraction in the vector channel. Thus the poor description
of the data above the maximum can be attributed to quark-antiquark threshold
effects. Therefore we would conclude that our lower boundary for the meson
cutoff, determined by the prescription that the threshold lies above the
maximum, has not been chosen restrictively enough. However, the behavior of the
form factor below the maximum, where threshold effects are less important,
indicates that a higher value of the cutoff presumably produces better
agreement with the data. Since increasing the cutoff reinforces the mesonic,
in this case pionic, fluctuations we expect that with a larger $\Lambda_M$ the width of the form
factor will be less underestimated. In fact, with a cutoff of
$\Lambda_M=700$~MeV the width is already noticeably overestimated coming along
with a diminished height of the maximum. Below the maximum the description of
the data is almost perfect with a cutoff of $\Lambda_M=600$~MeV. Above the
maximum we certainly miss the detailed structure of the form
factor around $s=0.61~{\rm GeV}^2$ which is due to $\rho-\omega$-mixing,
not included in our model. Approaching the
threshold at $s=0.80~{\rm GeV}^2$, effects induced by quarks become more
and more important resulting in an underestimation of the data in this
region. Probably the overall agreement with the data can be somewhat improved
by taking a slightly larger value for $\Lambda_M$ but since we are not
interested in fine-tuning here, we keep the fit with $\Lambda_M=600$~MeV.  

Two quantities closely related to the pion electromagnetic form factor are
well suited for confirming the consistency of our calculations with
experimental data. The first one is the charge radius of the pion,
which is defined as
\beq
     \ave{r_\pi^2} = 6\,\frac{d F_\pi}{d q^2}\Big|_{q^2 = 0} \;.
\eeq 
Obviously at $q^2=0$ the details of the $\rho$-meson peak are less important
than in the higher energy part. A simple pole ansatz for the form
factor leads to $\ave{r_\pi^2}^{1/2}=\sqrt{6}/m_\rho=0.63$~fm~\cite{bhaduri},
a value which is already close to the experimental value of  
$\ave{r_\pi^2}^{1/2}$~= (0.663~$\pm$~0.006)~fm~\cite{amendolia}. 
Our results, $\ave{r_\pi^2}^{1/2} = 0.59$~fm for
$\Lambda_M=500$~MeV, $\ave{r_\pi^2}^{1/2} = 0.61$~fm for
$\Lambda_M=600$~MeV and $\ave{r_\pi^2}^{1/2} = 0.66$~fm for
$\Lambda_M=700$~MeV, also satisfactorily agree with the experimental value.  
Since this quantity is not very well suited to fix the model parameters, which
is clear from the above reasoning, we shall not worry about the fact that for
$\Lambda_M=700$~MeV, which has been found to be too large to reproduce the
data in the region of the maximum, the agreement with the data is almost
perfect whereas for lower values of the cutoff the charge radius turns out to
be slightly too small. 

\begin{figure}[h!]\begin{center}
     \epsfig{file=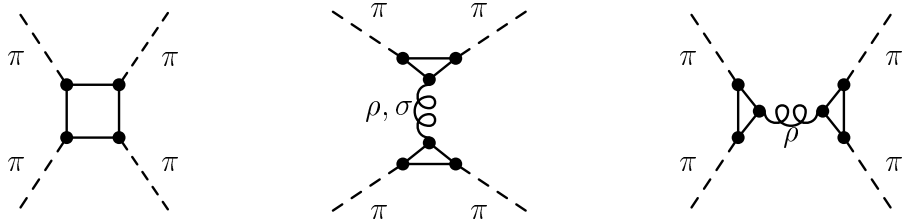}\end{center}
\caption{\it Diagrams contributing to the $\pi\pi$-scattering 
             amplitude: Quark box diagram (left), t- and u-channel $\rho$- or
             $\sigma$-meson exchange (middle) and 
             s-channel $\rho$-meson exchange (right). Only the latter is taken
             into account in our calculations.}
\label{figfeynphase} 
\end{figure}
Other observables are the $\pi\pi$-phase shifts in the vector-isovector channel.
We assume here that the dominant contribution to these phase-shifts arise from
the s-channel $\rho$-meson exchange, visualized on the r.h.s.\@ of
\fig{figfeynphase}. The other processes shown in \fig{figfeynphase}, the
direct scattering via quark box or the t- and u-channel exchange of a $\rho$-
or a $\sigma$-meson, respectively, are neglected. 
The result for different values of $\Lambda_M$, together with the empirical
data~\cite{frogatt}, is displayed in the right panel of \fig{figpefm}. Again,
below the $\rho$-meson peak, which is to be understood as the location where
the phase shifts equal $90\deg$, the best description of the data is
obtained with a cutoff of $\Lambda_M=600$~MeV, whereas for higher energies,
where threshold effects begin to influence the results for the latter cutoff
substantially, $\Lambda_M=700$~MeV produces better agreement
with the data.

The choice of contributions to the phase shifts has been made on purely
phenomenological grounds, motivated by the decisive role of the s-channel
$\rho$-meson exchange. In addition to the s-channel $\rho$-meson exchange the
direct scattering process via a quark box seems to be a suggestive
contribution since it is of leading order in $1/N_c$, i.e.\@ Hartree +
RPA~\cite{bernard}.  Obviously it destroys unitarity of the $S$-matrix if it
is not iterated. Because of the non-trivial momentum dependence of the quark
box diagram, this would be a very difficult task. On the other hand phase
shifts are only properly defined if the $S$-matrix fulfills unitarity. Our
evaluation of the phase shifts in the vector-isovector benefits from the
dominant role of the s-channel $\rho$-meson exchange which, when taken apart, is not
in contradiction to unitarity. In any case, it can be checked that including
the quark box diagram leaves the results for the scattering amplitude almost
unchanged, thus justifying our assumption.  In the scalar-isoscalar channel, on the
contrary, we do not expect this pragmatic approach to be as successful.
In that case the $t$- and $u$-channel $\sigma$-meson exchange is essential in
order to fulfill some of the low-energy theorems. Anyway,
the $\sigma$-meson within our model suffers from severe shortcomings which
render it senseless to attack the problem of calculating phase shifts. This
will be discussed in \Sec{secsigma}.


\begin{figure}[h!]
\hspace{3.5cm}
     \epsfig{file=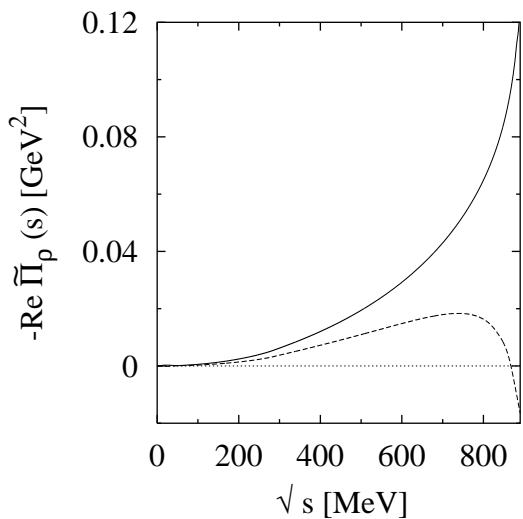}
\caption{\it Real part of the $\rho$-meson polarization function 
             $\tilde\Pi_\rho$ 
             as a function of the energy $\sqrt{s}$ in the rest frame
             of the meson. The solid line corresponds to the {\nce}
             with $\Lambda_M$~=~600~MeV, the dashed line to the MLA
             with $\Lambda_M$~=~700~MeV. The other parameters are given
             in Table~\ref{tablence} and Table~\ref{tableMLA}, respectively.}
\label{figrhoselfsum} 
\end{figure}

The preceding presentation of the results in the {\nce} offered the
possibility to decisively fix a consistent parameter set within this scheme.
However, the much more narrow range for variations of the meson cutoff we
found for the MLA in the previous section, renders it unlikely that such a fit
is possible also in the MLA. In the {\nce} we found threshold effects to cause
a steep drop in the pion electromagnetic form factor just below this
threshold, leading to the conclusion that a constituent quark mass of about 400 MeV, shifting
the threshold only slightly above the $\rho$-meson peak in the form factor, is
not sufficiently large. This further restricts the possible values for the
meson cutoff to a small range just below $\Lambda_M=700$~MeV. Hence the only
variable parameter to fit the form factor is the vector coupling constant
$g_v$.  In fact, if we take $\Lambda_M=700$~MeV, it turns out that we
unfortunately run into instabilities in the $\rho$-meson propagator similar to
those found in the propagator of the pion (see
\Sec{instabilities})~\cite{OBW2}. 
This observation is of course independent of
the value of $g_v$ since it can be detected already in the polarization
function $\tilde\Pi_\rho$. 

To this end the real part of the 
of the $\rho$-meson polarization function $\tilde\Pi_\rho$ is plotted in
\fig{figrhoselfsum} as a function of the energy $\sqrt{s}$ in the rest 
frame of the meson. The MLA result corresponds to the dashed line.
For comparison we also show this function in the {\nce}, using the `best-fit 
parameters' given above (solid line).

The ``proper'' behavior of the polarization function can be derived
from what we expect of the inverse propagator. It has to develop a zero for
$\sqrt{s} \simeq m_\rho$ to produce the
well pronounced peak in the form factor. Thus $2 g_v Re\tilde\Pi_\rho$ has to
become equal to unity for an energy of about 770 MeV. In the {\nce} this can
easily be achieved by an appropriate choice of the coupling constant since the
polarization function is (at least below the quark-antiquark threshold) a
steadily rising function. In the MLA, on the contrary, this condition can by
no means be fulfilled due to the maximum at $\sqrt{s}\sim 740$~MeV and the
subsequent steep drop the polarization function displays. Consequently for a weak
coupling, i.e.\@ small $g_v$, we observe no zero at all in the inverse propagator
whereas for stronger couplings we find two solutions, one at lower energies and
one above the maximum. Obviously within those scenarios we would obtain only a
very unrealistic description of the $\rho$-meson. 
Thus we conclude that we did not succeed in performing a fit of the
observables in the pion sector, $m_\pi, f_\pi$, and $\qq$, simultaneously with
properties of the $\rho$-meson in the MLA.
  

\begin{figure}[h!]
\begin{center}
\parbox{7.5cm}{\epsfig{file=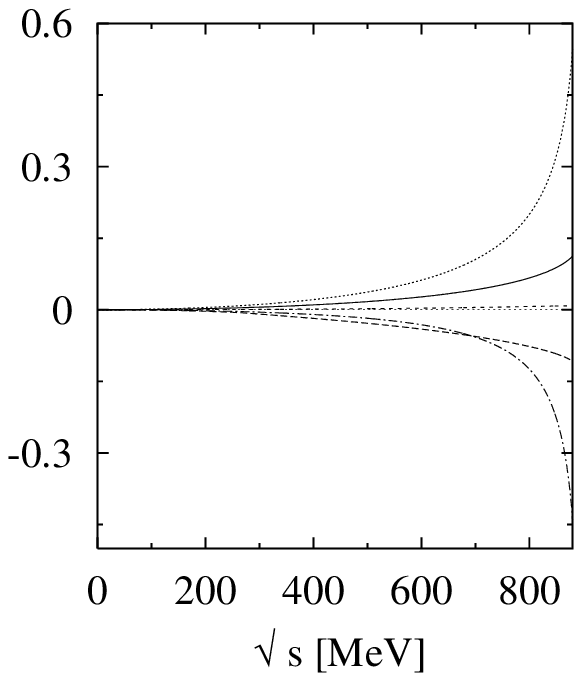,height=6.43cm,width=7.5cm}}
\parbox{7.5cm}{\epsfig{file=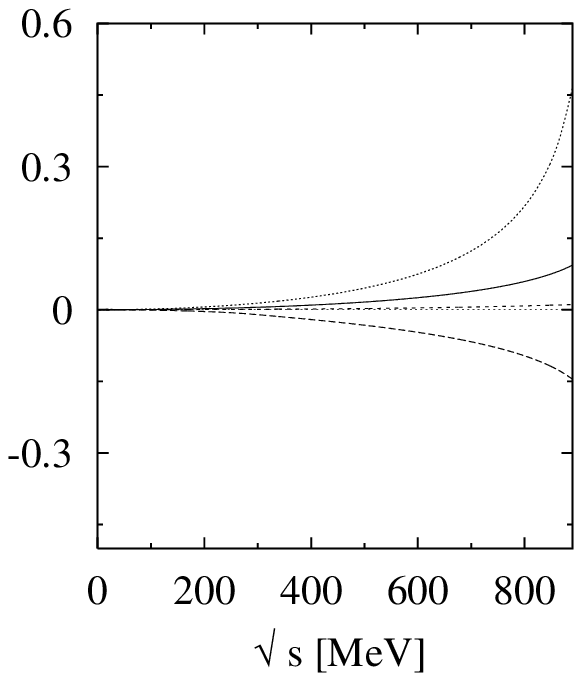,height=6.43cm,width=7.5cm}}
\end{center}
\caption{\it Real parts of the RPA contribution $\Pi_\rho$ (solid) and the various
             correction terms to $\tilde\Pi_\rho$:
             $\delta\Pi_\rho^{(a)}$ (long-dashed),
             $\delta\Pi_\rho^{(b)}$ (dotted), $\delta\Pi_\rho^{(c)})$
             (short-dashed), 
             and $\delta\Pi_\rho^{(d)}$ (dashed-dotted).
             For all contributions we performed a subtraction, such
             that they vanish at $\sqrt{s}$~=~0.
             The left panel corresponds to the {\nce}, the right panel
             to the MLA. The model parameters are the same as in 
             \fig{figrhoselfsum}.} 
\label{figrhoself} 
\end{figure}


The cusp in the $\rho$-meson polarization function responsible for the
peculiar behavior of the inverse $\rho$-propagator in the MLA looks very
similar to that emerging in the inverse pion propagator for large values of
the mesonic cutoff, see the discussion in \Sec{instabilities}. It has also
the same origin:  the strong momentum dependence of the only negative
contribution, diagram $\delta\Pi^{(b)}_\rho$. This can be seen from the right panel
of \fig{figrhoself}, where we have separately plotted the various contributions
to the polarization function of the $\rho$-meson. On the l.h.s.\@ of
\fig{figrhoself} the same is shown for the {\nce}. One might wonder, why the
results in the {\nce} permit a reasonable description of the $\rho$-meson,
although qualitatively the behavior of diagram $\delta\Pi_\rho^{(b)}$,
responsible for the unphysical behavior in the MLA, agrees with that in the
MLA. However, in the {\nce} its contribution is almost cancelled by that of
diagram $\delta\Pi^{(d)}_\rho$, which is not present in the MLA where the
corresponding quark self-energy insertion is selfconsistently absorbed in the
modified gap equation (\eq{localgap}). Thus probably the purely perturbative
treatment of both next-to-leading order quark self-energy insertions,
$\delta\Pi^{(b)}$ {\it and} $\delta\Pi^{(d)}$, within the {\nce} helps to
suppress unphysical effects. However, we should emphasize that any
unphysical behavior we detect here, although observed below the
quark-antiquark threshold, is mainly induced by quark effects. In
\Sec{instabilities} we already noted that in particular the momentum
dependence of the real part of $\delta\Pi^{(b)}_\pi$ below the threshold is
related via dispersion relations to an imaginary part with the ``wrong'' sign
above the threshold. In the same way the real part of all other diagrams below
the threshold is influenced by the quark-antiquark continuum. Only digram
$\delta\Pi^{(a)}$ is not exclusively governed by effects which can be traced
back to the quark-antiquark continuum: Here the imaginary part due to the
two-meson intermediate state in a large range of momenta dominates. 
In the
following section we will carry on this discussion in connection with a comparison of
these results with those obtained in the static limit.

In any case the results suggest that the unphysical behavior of the
polarization function can be at least weakened in energy regions relevant for
the description of hadrons by pushing the constituent quark mass further up.
This presumption is mainly supported by the steep drop in $\delta\Pi^{(b)}$
occurring only just below the threshold. One might suggest that the constituent
quark mass is increased if further intermediate states are included in the
model. However, we have seen that it is an extremely difficult task to
implement $\rho$- and $a_1$-mesons within the present model (see
\Sec{pionml}). In addition the observations made in \Sec{instabilities}
indicate that we have to cope with an even more severe problem when including
vector and axial intermediate states: We observed ``restoration'' of chiral
symmetry if the mesonic fluctuations became too strong. Thus, it seems as if 
the inclusion of the $\rho$- and $a_1$-subspace for the
intermediate states would not help without further modifications (or
extensions) of the model. 
\section{Comparison with the static limit}
\label{secstatic}
This section is devoted to a comparison of the results obtained in the full
model with those in the static limit, i.e.\@ the approximation introduced in
\Sec{hadron}. This approximation is (in vacuum)
equivalent to a purely hadronic description, i.e.\@ all quark effects are
suppressed. The only reminiscence of the full model can be found in the quark
mass dependence of the effective coupling constants.  Since the approximations
to the MLA and the {\nce} do not differ qualitatively from each other, it is
completely sufficient for our purpose to restrict the discussion within this
section to one of the schemes. We have decided to take the {\nce} because only
within that scheme we are able to consistently describe the data in the pion
sector as well as for the $\rho$-meson.

From the very beginning, we are for two reasons very distrustful of the static
limit being a good approximation to the full calculation. First, the estimate
of a realistic value of the constituent quark mass within this approximation
(see \eq{mapprox})
seems to be in contradiction with the motivation of large quark masses as
compared with the relevant momenta, with which we started in \Sec{hadron}. In
addition indications were found in the previous sections that quark effects are
important for the mesonic polarization functions. The estimate performed in
\Sec{hadron} can also be applied vice versa: If we insert the parameters of
our best fit to the pion electromagnetic form factor in the {\nce} (those
listed in Table~\ref{tablence} for $\Lambda_M=600$~MeV) into \eq{rhopipistat},
we obtain a value of $g_{\rho\pi\pi} =9.4$ for the effective $\rho\pi\pi$
coupling constant, which is considerably larger than the value of
$g_{\rho\pi\pi}\approx 6$ usually needed in hadronic models to describe the
data. Already with $\Lambda_M=500$~MeV, which in the full calculation
noticeably underestimates the width of the $\rho$-meson, the value of
$g_{\rho\pi\pi}\approx 6$ is exceeded by about 40\%.

\begin{figure}[t]
\hspace{3.75cm}
\parbox{7.5cm}{
     \epsfig{file=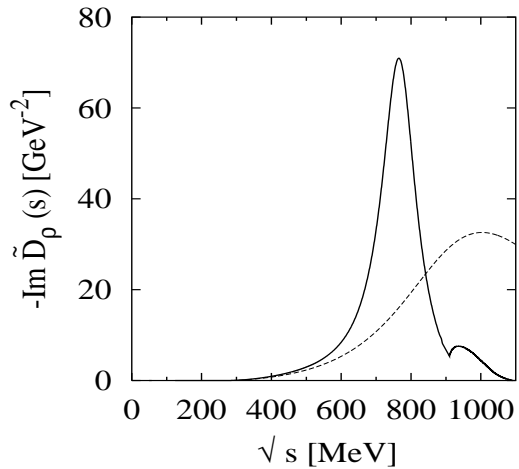,
     height=6.43cm, width=7.5cm}\quad}
\hspace{3.75cm}
\caption{\it The imaginary part of the $1/N_c$-corrected $\rho$-meson
             propagator. The solid line represents the full calculation 
             while the dashed line denotes the static limit.}
\label{figspectral} 
\end{figure}
To compare the static limit with an exact treatment of the {\nce} we will look at
the imaginary part of the $\rho$-meson propagator which is displayed in
\fig{figspectral}. The solid line represents a full calculation with
$\Lambda_M = 600$~MeV, the dashed line the corresponding static limit. Since we
have fitted the parameters used in the full calculation to the pion
electromagnetic form factor, the maximum of the corresponding imaginary part is close to
the $\rho$-meson mass of 770~MeV whereas this does not necessarily need to be
the case in the static limit. In fact, the peak in the static limit is much
broader and shifted to higher energies. Another obvious difference is the
existence of the quark-antiquark threshold at $\sqrt{s}=892$~MeV in the full
model which is not present in the static limit since within that approximation
all quark effects are suppressed.

\begin{figure}[t!]
\begin{center}
\parbox{7.5cm}{
     \epsfig{file=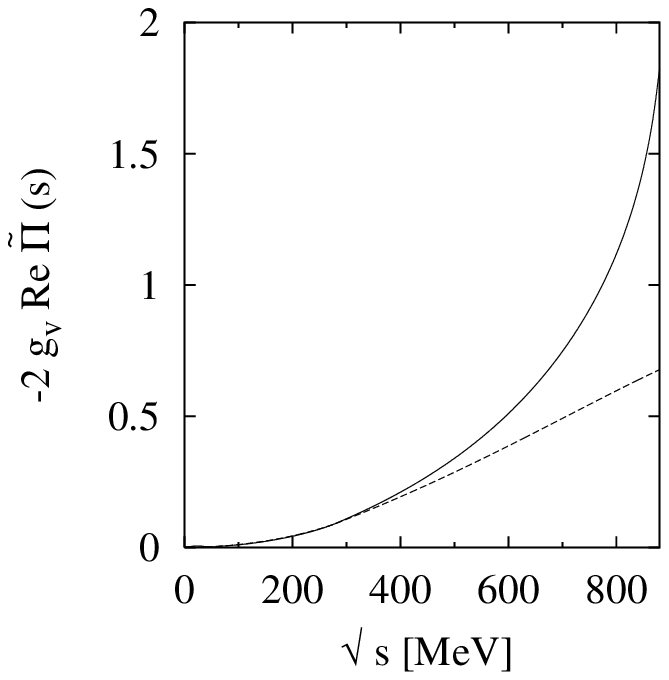,
     height=6.43cm, width=7.5cm}}
\parbox{7.5cm}{
     \epsfig{file=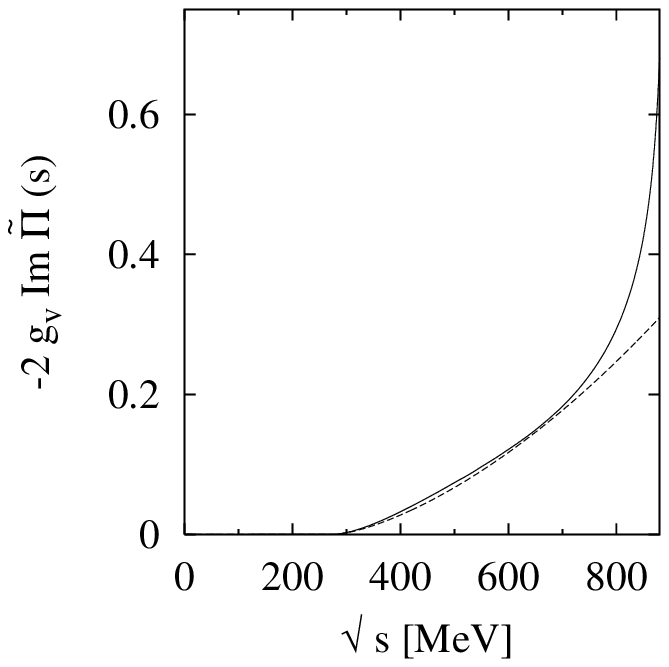,
     height=6.43cm, width=7.5cm}\quad}
\end{center}
\caption{\it The real part (left) and the imaginary part (right) of the
             $1/N_c$-corrected $\rho$-meson self-energy, multiplied
             by $-2g_v$. The solid line represents the full 
             calculation, while the dashed line indicates the static limit.}
\label{figrhostat} 
\end{figure}

For a better understanding of this intriguingly different behavior already
below the unphysical threshold we also compare the real and imaginary parts
of the corresponding self-energies (\fig{figrhostat}). Note that the imaginary
part in this region is exclusively generated by the decay process
$\rho\rightarrow \pi\pi$. Differences in the imaginary part therefore directly
reflect the non-trivial momentum dependence of the quark loops in the
meson-meson vertices.
They become important only
near the threshold. For $\sqrt{s} \lsim 800$~MeV the two curves almost
coincide. In the real part the differences are much larger. Coincidence
of the two curves can hardly be acknowledged up to $\sqrt{s} \sim 400$ MeV. 
The real part in the static limit is obviously much less attractive than
that in the full calculation. Besides, the slope of the curve is much steeper
in the full calculation, which explains the broadening and the shift
to higher energies of the peak in the $\rho$-meson propagator as shown in
\fig{figspectral}. As already emphasized several times, the real part below
the threshold is, via dispersion relations, related to the entire imaginary
part including that above the threshold. Hence the discrepancies between the
static limit and the full model here not only reflect the momentum dependence
of the quark loops contained in the vertices but 
additionally indirect effects arising from quark decay channels. 

The only conclusion we can draw from the preceeding discussion is that quark
effects are indeed important. At this point one might argue that one has to
distinguish between ``physical quark effects'', for instance those modifying
the point-like structure of the meson-meson vertices, and unphysical quark
effects leading to unconfined quarks. However, these two types of effects
cannot be considered separately since they are linked to each other via
dispersion relations.  This is related to a rather fundamental question,
namely whether ``physical quark effects'' exist at all. Of course, at
sufficiently high energies quarks should be the relevant degrees of freedom
which in one way or another influence hadronic properties. An approach like QCD
sum rules~\cite{shifman} in principle offers the possibility to map at least
different energy regions. But in any case an explanation of this influence
requires first the understanding of quark confinement itself.

Our scheme can only exemplify how a hadronic model
emerges naturally from the underlying quark structure provided that the
necessary degrees of freedom are included. This can be seen nicely from the
low energy region where the static limit almost coincides with the exact
treatment of our model. 

\section{Results for the sigma-meson}
\label{secsigma}
\begin{figure}[t!]
\begin{center}
\parbox{7.5cm}{
     \epsfig{file=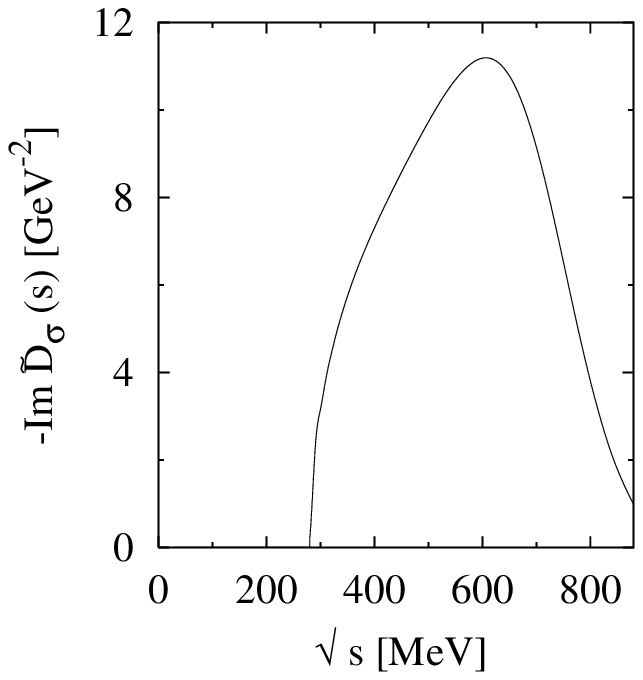,
     height=6.43cm, width=7.5cm}}
\parbox{7.5cm}{
     \epsfig{file=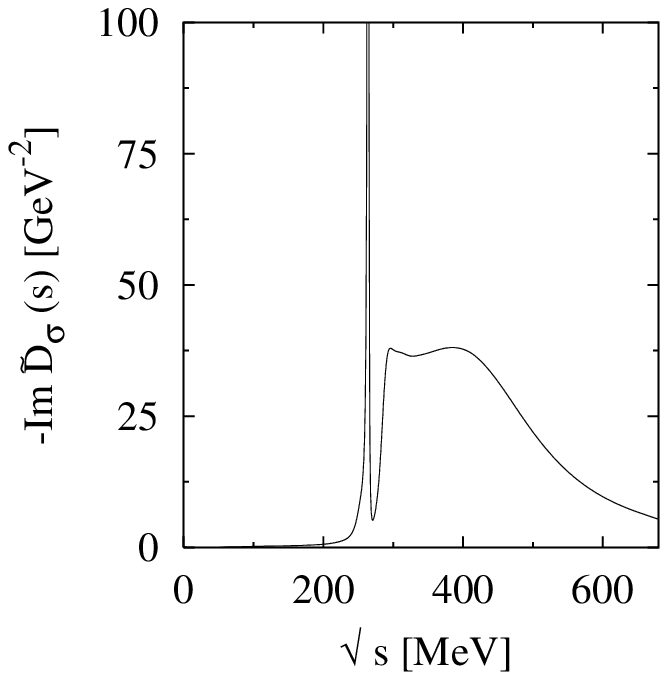,
     height=6.43cm, width=7.5cm}}
\end{center}
\caption{\it The imaginary part of the ``improved '' $\sigma$-meson
     propagator, $\tilde{D}_\sigma$ in the {\nce} (left) and in the MLA
     (right) for $\Lambda_M=600$~MeV.}
\label{figsigim} 
\end{figure}
The sigma-meson is not a very well established particle in nature. In the
``Review of Particle Physics''~\cite{PDG} we read concerning the $\sigma$:
``The interpretation of this entry as a particle is controversial.'' The mass
of this ``particle'' lies between 400 and 1200 MeV with a width of the same
magnitude, 600-1000 MeV.  However, in chiral models where chiral symmetry is
linearly realized, like the NJL model, it inevitably accompanies the pion as
its chiral partner. Most of the models deal with a sharp
$\sigma$-particle on tree-level. However, 
due to its strong coupling to a two-pion state the $\sigma$-meson is
significantly broadened. Within a linear sigma model approach this notion has
been successfully applied to predict not only a broad resonance in vacuum but
also its sharpening due to partial restoration of chiral symmetry in medium
(see Schuck et al.~\cite{schuck} and references therein).  The description of
the $\sigma$-meson in the NJL model in the standard Hartree + RPA scheme is
certainly not suitable to obviate this problem. Within that approximation the
$\sigma$-meson is an (almost) sharp particle, located at the
$q\bar{q}$-threshold or just above. Thus, the inclusion of mesonic
fluctuations, in particular pionic ones, in the {\nce} or the MLA seems to be
a promising starting point to improve the phenomenology of the $\sigma$-meson
within the NJL model.
\begin{figure}[t!]
\begin{center}
\parbox{7cm}{
     \epsfig{file=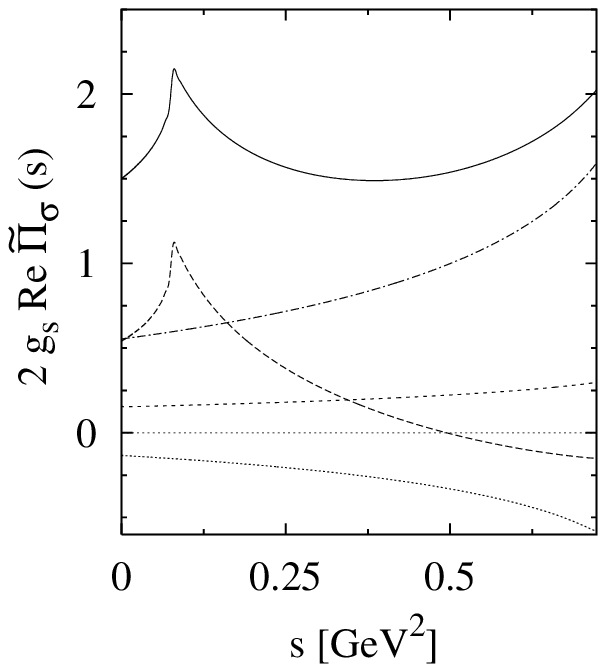,
     height=6.43cm, width=7.5cm}\quad}
\hspace{1cm}
\parbox{7cm}{
     \epsfig{file=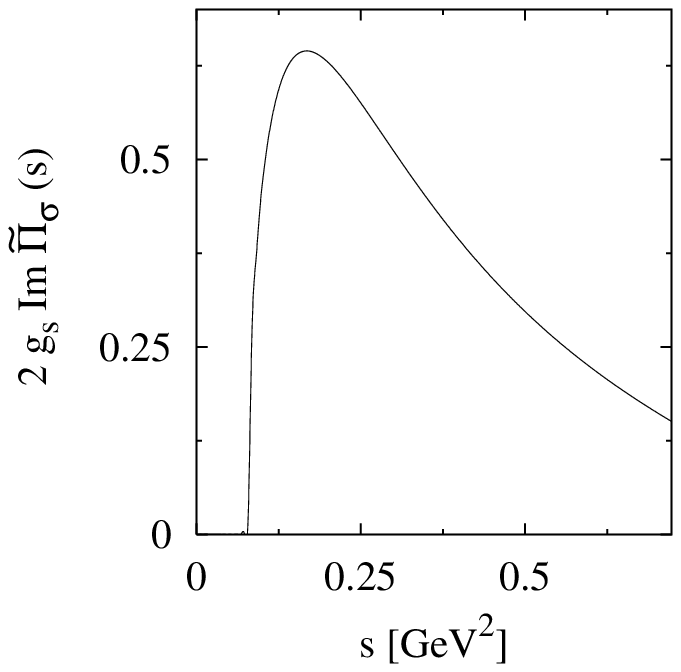,
     height=6.43cm, width=7.5cm}\quad}
\\
\parbox{7cm}{
     \epsfig{file=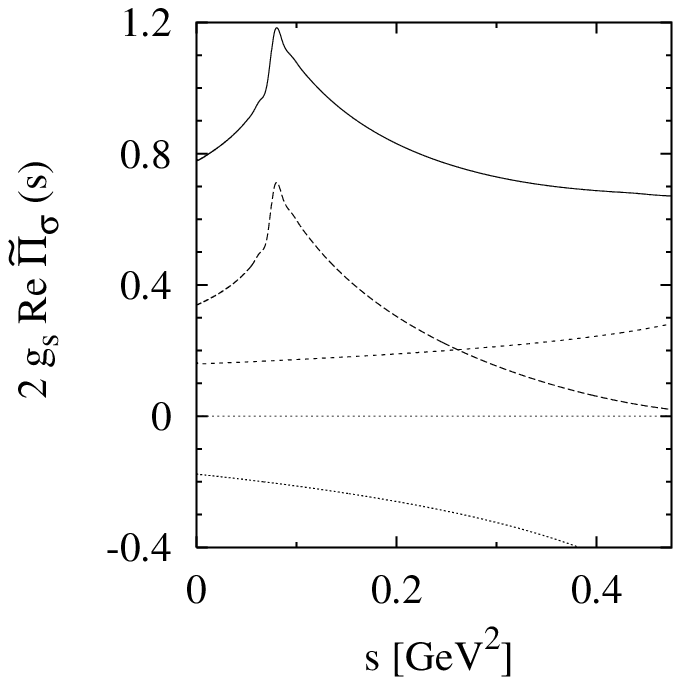,
     height=6.43cm, width=7.5cm}\quad}
\hspace{1cm}
\parbox{7cm}{
     \epsfig{file=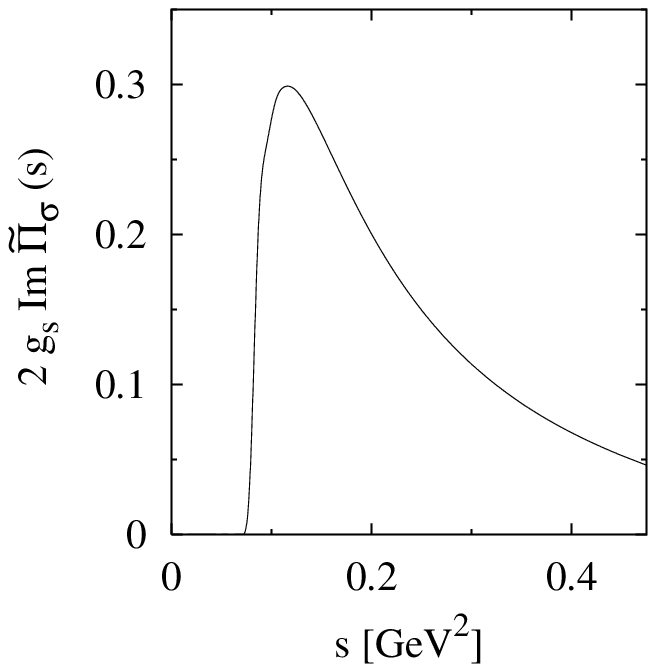,
     height=6.43cm, width=7.5cm}\quad}
\end{center}
\caption{\it The real part (left) and the imaginary part (right) of the
     various corrections terms to the $\sigma$-meson polarization function,
             multiplied by $2g_s$, in the {\nce} (upper panel) and in the MLA
             (lower panel): total polarization function 
             $\tilde\Pi_\sigma$ (solid line), $\delta\Pi_\sigma^{(a)}$ (long
             dashed line), $\delta\Pi_\sigma^{(b)}$ (dotted line),
             $\delta\Pi_\sigma^{(c)}$ (short dashed line), and
             $\delta\Pi_\sigma^{(d)}$ (dashed-dotted). Note that the
             contribution of $\delta\Pi^{(d)}_\sigma$ is missing
             in the MLA. }
\label{figsig} 
\end{figure}

Actually, in both schemes, the {\nce} and the MLA, the $\sigma$-meson is
substantially broadened above the two-pion threshold as can be inferred
from the left and right panel of \fig{figsigim}, respectively, where the
imaginary part of the $\sigma$-propagator $\tilde{D}_\sigma(s)$ as a function
of the energy $\sqrt{s}$ in the rest frame of the meson is displayed. For the
{\nce} the calculation has been performed with $\Lambda_M=600$~MeV and the
corresponding ``best fit'' parameter set listed in Table~\ref{tablence}. In
the MLA, where we did not succeed in performing a consistent fit, we also took
$\Lambda_M=600$~MeV together with the parameters shown for that value of the
cutoff in Table~\ref{tableMLA}. In the MLA we find, in addition to the
broadening above the two-pion threshold, a sharp peak just below that
threshold which would be a $\delta$-function if we had not introduced by hand
some small but nonzero imaginary part. The origin of this peak will become
clear from the behavior of the polarization function itself.

The left panels of \fig{figsig} show the real parts of the different
contributions to the $\sigma$-meson polarization function as a function of
$s$. The momentum dependence of the real part is mainly caused by the
contribution arising from $\delta\Pi_\sigma^{(a)}$ (long-dashed line), i.e.\@
the correction term containing the two-pion intermediate state. Below the
$q\bar{q}$-threshold this term is in fact the only one producing a
nonvanishing imaginary part. This imaginary part is shown on the r.h.s.\@ of
\fig{figsig}.  These features are common for the {\nce} and the MLA.

\begin{figure}[b!]
\begin{center}
\parbox{7.5cm}{
     \epsfig{file=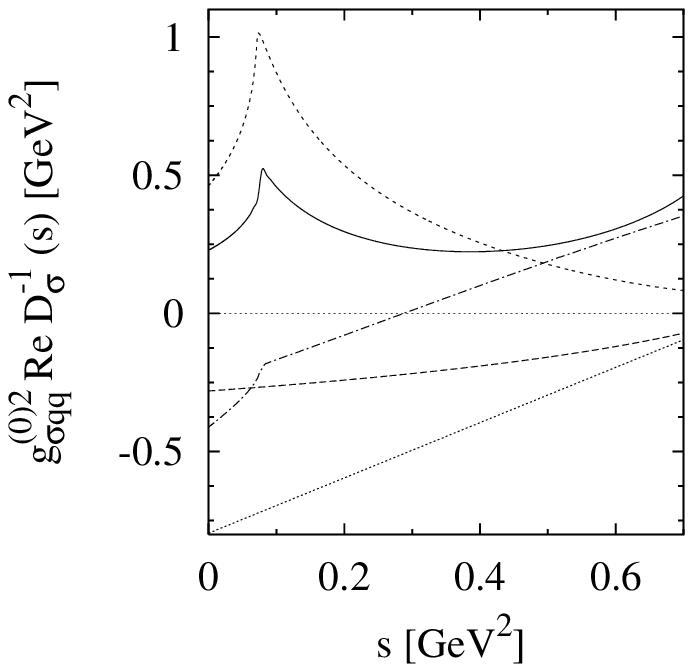,
     height=6.43cm, width=7.5cm}}
\parbox{7.5cm}{
     \epsfig{file=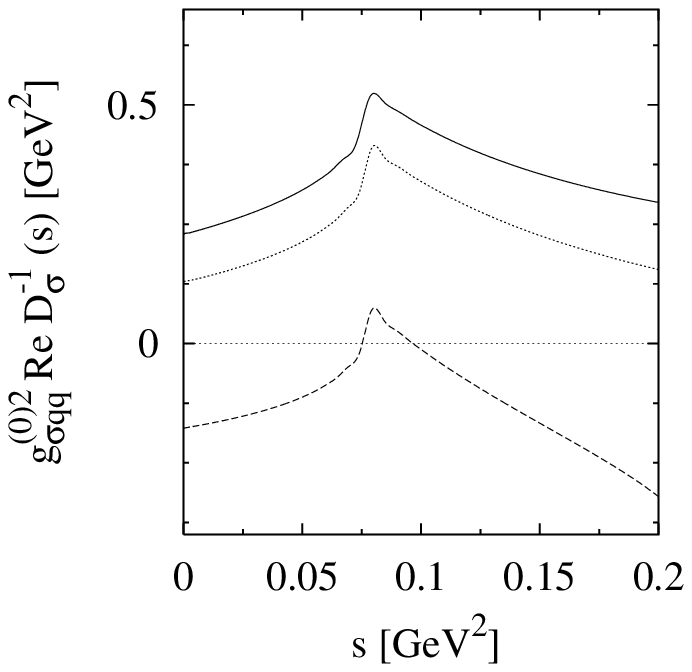,
     height=6.43cm, width=7.5cm}}
\end{center}
\caption{\it Real part of the inverse $\sigma$-meson propagator. (Left): 
  $g_{\sigma qq}^{(0)2} \tilde{D}_\sigma^{-1}(s)$ in the {\nce} with
  $\Lambda_M=600$~MeV (solid line) and in the static limit (short-dashed
  line), $g_{\sigma qq}^{(0)2} D_\sigma^{-1}(s)$ in RPA (long-dashed line),
  $s-4 m_H^2$ (dotted line), and the corresponding linear sigma-model result
  by Aouissat et al.~\cite{aouissat} (dashed-dotted line). (Right): {\nce}
  with $\Lambda_M=600$ MeV (dashed), $\Lambda_M=600$ MeV (solid), and
  $\Lambda_M=900$ MeV (dotted).}
\label{figsigprp} 
\end{figure}
This rather encouraging result, at least for the {\nce}, is spoiled, however,
if we investigate the analytic properties of the propagator more closely.
Recalling that the propagator acquires a pole if $2 g_s \tilde\Pi_\sigma(s)$
becomes equal to unity, one understands the existence of the sharp peak in the
$\sigma$-meson propagator in MLA. In the {\nce}, on the other hand, the existence
of a pole is not as obvious. In the time-like region we find no pole at all,
explaining the smaller scale of the imaginary part of the propagator as
compared with that in the MLA (\fig{figsigim}).  We only guess from the slope
of the polarization function at $s=0$ that this might very well occur for
negative, i.e.\@ space-like, $s$. This would of course be a very disturbing
unphysical feature of the $\sigma$-propagator.  Unfortunately we cannot
support this guess by explicit results: Since we work in the rest frame of the
mesons, we are not able to perform calculations for space-like momenta.

However, another fact which can already be deduced from the value of the real
part of the polarization function at $s=0$ corroborates the doubts concerning
our results for the $\sigma$-meson: The sign of the inverse propagator at
$s=0$ differs from what we except from a comparison with a free propagator
$1/(s-m_\sigma^2)$.  On the l.h.s.\@ of \fig{figsigprp} we have displayed the
inverse $\sigma$-propagator as it emerges from the {\nce} (solid line), in the
static limit (short-dashed line), the free one with an arbitrarily chosen mass
of $m_\sigma = 2 m_H$ (dotted line) and for comparison the RPA one
(long-dashed line) and the result obtained by Aouissat et al.~\cite{aouissat}
(dashed-dotted line). Note that we have normalized the results in RPA and in
the {\nce} to the RPA $\sigma$-quark coupling constant $g_{\sigma qq}^{(0)2}$.
In the vicinity of $s$=0 the slope of the various inverse propagators agrees
at least concerning the sign, while the function itself has positive values in
the {\nce} and negative ones in the two other cases.  This behavior is
basically not altered if we vary the mesonic cutoff. This can be inferred from
the right panel of \fig{figsigprp} where we have plotted the inverse
$\sigma$-propagator in the {\nce} using different values of $\Lambda_M$ while
all other parameters are kept constant. The effect of varying the cutoff is
mainly to shift the entire curve up or down. The shape remains almost the
same. For small values of $\Lambda_M$, cf. the dashed line with
$\Lambda_M=200$~MeV, we encounter the even more peculiar situation, which has
already been observed in the MLA (see \fig{figsig}), where we have two zeros
in the real part of the inverse $\sigma$-propagator in the time-like region,
one below the two-pion threshold at about 280 MeV and one above.  With
increased cutoff the curve is first shifted upwards, then, if the cutoff
exceeds $\Lambda_M \gsim 900 $~MeV it begins to move downwards again.
Nevertheless it seems to be hopeless that we can obtain a more realistic
description of the $\sigma$-propagator for very large values of the cutoff. On
the one hand, if the shape of the curve is not altered, we cannot avoid
finding finally two poles in the time-like region instead of one.  On the
other hand the investigations in the previous sections, concerning the pion
and the $\rho$-meson, render it plausible that we run into instabilities when
applying such large values for the cutoff, besides completely failing to
describe the data in the pion and $\rho$-meson sector.
\begin{figure}[t!]
\begin{center}
     \epsfig{file=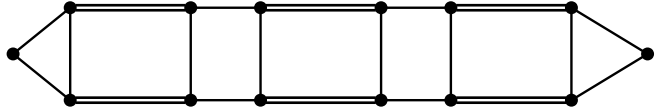}
\end{center}
\caption{\it Example for contributions to the $\sigma$-meson polarization
  function similar to those included in Ref.~\cite{aouissat} (see text).}
\label{figsigscat} 
\end{figure}

Till now each unphysical phenomenon we detected could be ascribed to strong
quark effects. This is not true for the present case. The left panel of
\fig{figsigprp} also displays the result in the static limit (short-dashed line), where all quark effects are suppressed, which shows essentially the
same behavior, at least in the low-energy region, as the calculation in the
{\nce}. This leads us to the conclusion that the strong attraction of the
two-pion state in the scalar-isoscalar channel is responsible for this
behavior. It is therefore suggestive that it can be mended by imitating the
terms which produce reasonable results in a linear sigma-model calculation by
Aouissat et al.~\cite{aouissat} (see l.h.s.\@ of \fig{figsigprp},
dashed-dotted line).  The RPA used by the authors of Ref.~\cite{aouissat}
consists of iterating the entire pion-pion scattering amplitude in
the scalar-isoscalar channel for the self-energy of the $\sigma$-meson. In our
case this would mainly correspond to terms like those presented in
\fig{figsigscat}. Terms of that type would have been included in the
``improved'' polarization function if we had applied the $\Phi$-derivable
method discussed in \Sec{1ml}. The scattering kernel of the quark-antiquark
Bethe-Salpeter equation visualized in \fig{figphi} precisely contains those
non-local contributions necessary to generate the desired $\sigma$-meson
polarization function. In any case, we resigned to pursue that scheme further
because it is very difficult to handle due to the non-local structure, which
already shows up in the modified gap equation (cf.~\fig{figphi}).
\chapter{Results at nonzero temperature}
\label{temperatur}
It is commonly believed that chiral symmetry, which is spontaneously broken in
vacuum, gets restored at high temperatures as well as at high densities. For a
recent review on the present view of the QCD phase diagram see
Ref.~\cite{rajagopal}. In wide ranges of densities and temperatures this
notion is, however, based on model calculations for lack of fundamental
knowledge. Only at low temperatures and low densities model independent
results can be obtained by considering a gas of pions or a gas of pions and
nucleons, respectively. Pions are the dominant degrees of freedom in that
range because of their small mass as compared with other possible degrees of
freedom. At zero chemical potential, i.e.\@ zero density, lattice calculations
provide us with results also in the vicinity of the phase transition, but at
nonzero chemical potential no realistic lattice results exist. Thus
in that region we have to rely entirely on model calculations.

One of the models which has been applied by many authors to study the effects
of (partial) chiral symmetry restoration at nonzero temperature and density
on various quantities, in particular the quark condensate as the order
parameter of chiral symmetry and properties of mesons, is the NJL model. Most
of these investigations have been performed in the standard approximation
scheme, i.e.\@ Hartree + RPA (see e.g.\@
\cite{klevansky,hatsuda,sklimt,mlutz}).  The main drawback of these
examinations is certainly the unphysical nature of the degrees of
freedom, deconfined quarks, taken into account to describe the thermodynamical
properties of the system. The presence of unphysical degrees of freedom is not
the only weak point of these studies. In addition, the relevant degrees of
freedom, at low temperatures and densities mainly pions, are completely
missing such that, for instance, the model independent results within that
range cannot be reproduced by these calculations. Hence, although the
fundamental problem of lack of confinement cannot be overcome, we can hope to
improve the results by introducing mesonic degrees of freedom within an
approximation beyond Hartree + RPA, i.e.\@  in the {\nce} or the MLA. Since both
schemes do not contain any nucleons, essential ingredients at nonzero
densities, we will restrict our examinations in the {\nce} and the MLA to
nonzero temperatures but zero density.
\section{Results in Hartree approximation + RPA}

Before we present first results at nonzero temperature in the {\nce} and
the MLA, we will briefly review the results in the standard approximation
scheme, namely Hartree + RPA. To this end we first have to generalize the
formalism presented in Chapter~\ref{NJLModell} to nonzero temperature and
chemical potential. We will keep the latter since it enters into the
expressions in almost the same way as nonzero temperature.
In principle we can choose between
two different methods to deal with thermodynamic properties of the system:
the imaginary time or Matsubara formalism (see e.g.\@ \cite{fetter}), and the
real time formalism (see e.g.\@ \cite{landsbrecht}). Within this paper we will
adopt the Matsubara formalism for simplicity. We should remark that it is not
applicable to determine the temperature dependence of the mesonic properties in
the {\nce} or the MLA. This requires the analytic continuation of
functions only known numerically for a finite number of points, which cannot be
done uniquely. In any case we will restrict the later discussion on the quark
condensate which is a static quantity and can therefore be treated within the
Matsubara formalism. 

The basics of the Matsubara formalism can be found in textbooks, e.g.\@ in
Ref.~\cite{fetter}. In App.~\ref{aptemp} we summarize the Feynman rules we
obtain starting from the Lagrangian in \eq{lagrange}. The resulting
expressions at nonzero temperature and chemical potential $\mu$ can be
written formally in almost the same way as in vacuum if 
we replace the integration
over energy in the vacuum  expressions by a sum over fermionic
or bosonic Matsubara
frequencies $i \omega_n$, i.\@e.
\beq
i \intk f(k)\rightarrow -T\sum_n\int\frac{d^3 k}{(2\pi)^3}
f(i\omega_n,\vec{k})~.
\label{rpt}
\eeq
If we have a nonvanishing chemical potential for the quarks, $\mu$ should be
added to the fermionic Matsubara frequency in all expressions, i.e.\@ $i
\omega_n \rightarrow i \omega_n + \mu$.  
The replacement prescription in \eq{rpt} can be applied to define the analogue to the
elementary integrals in vacuum, e.g.\@ 
\bea
I(p) &=& \intk\frac{1}{(k^2-m^2+i \eps)( (k+p)^2-m^2+
  i\eps)}\rightarrow\nonumber\\
I(i\omega_l,\vec{p})&=&
iT\sum_n\int\frac{d^3k}{(2\pi)^3} \frac{1}
{( (i\omega_n+\mu)^2-\vec{k}^2-m^2)
  ( (i\omega_n+\mu+i\omega_l)^2-(\vec{k}+\vec{p})^2-m^2)}~.\nonumber\\
\eea
Note that the expressions at nonzero $T$ and $\mu$ depend separately on
energy and three-momentum because a preferred frame of
reference exists in the heat bath. This simplifies our notation: the second
argument enables us to clearly distinguish vacuum quantities from their medium
analogues. Functions with one argument, e.g.\@ $p$, denote vacuum quantities,
whereas functions with two arguments, e.g.\@ $i\omega_l,\vec{p}$, are the
corresponding in-medium quantities, such
that we can use the same symbols to mark analogous quantities. The only
exception are static quantities which do not depend on momentum at all. There
we will add a suffix $T$, e.g.\@ the quark condensate in medium will be denoted
by $\qq_T$. 
\begin{figure}[t!]
\begin{center}
\parbox{7.5cm}{
     \epsfig{file=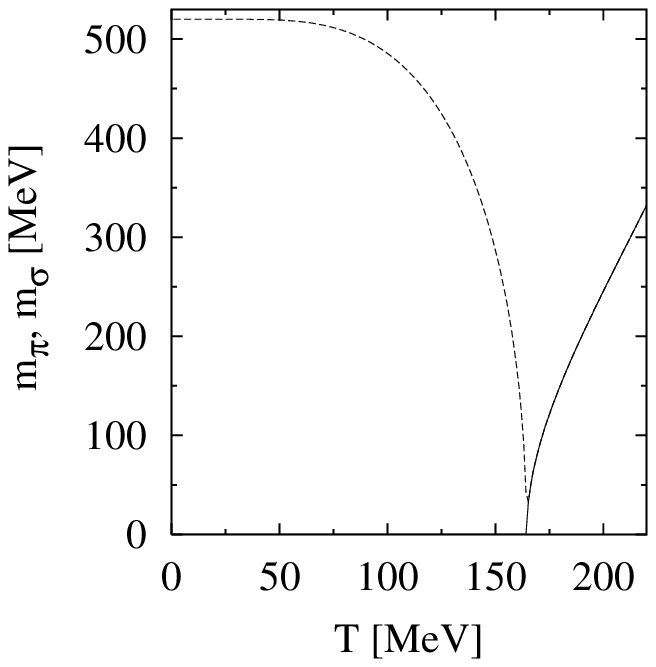,
     height=6.43cm, width=7.5cm}}
\parbox{7.5cm}{
     \epsfig{file=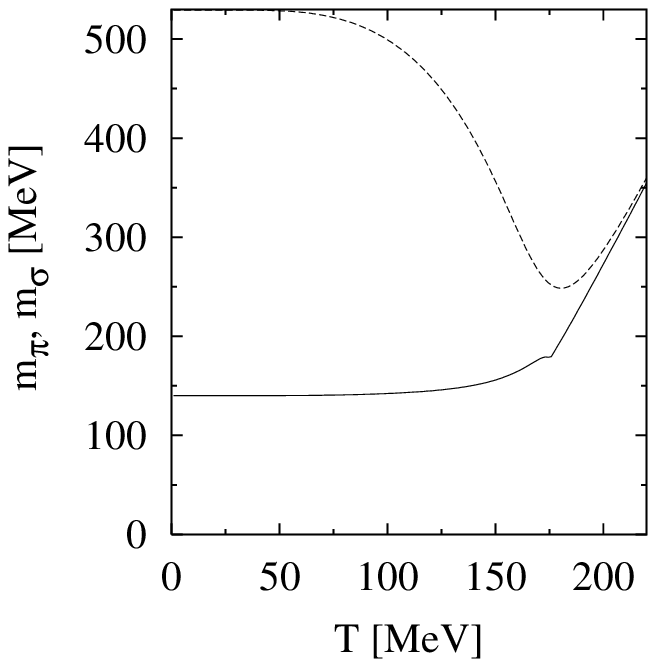,
     height=6.43cm, width=7.5cm}}
\end{center}
\caption{\it $m_\pi^{(0)}$ (solid line) and $m_\sigma^{(0)}$ (dashed) as a
     function of temperature
     in the chiral limit (left panel) and with $m_0=6.13$~MeV (right panel). For the
     remaining parameters see Table~\ref{tablence} for $\Lambda_M=0$.}
\label{figmpisig} 
\end{figure}
\begin{figure}[t!]
\begin{center}
\parbox{9cm}{\epsfig{file=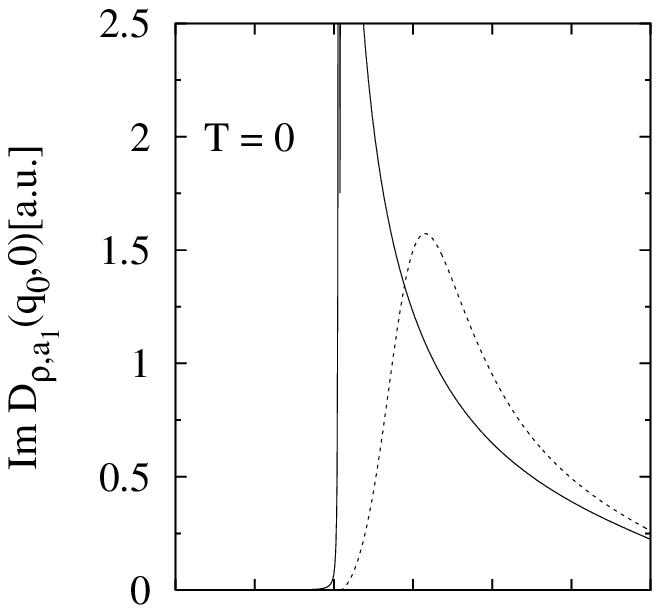, height=5.38cm, width=7.cm}}
\vskip-10.1mm
\parbox{9cm}{\epsfig{file=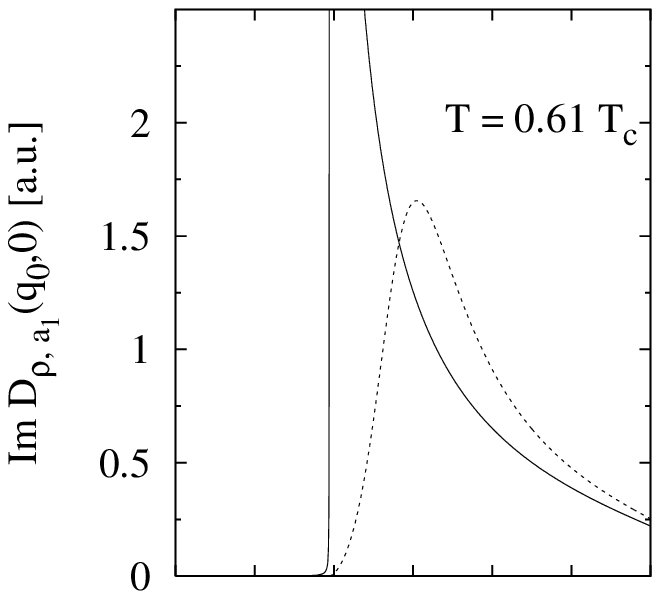, height=5.38cm, width=7.cm}}
\vskip-10.6mm
\parbox{9cm}{\epsfig{file=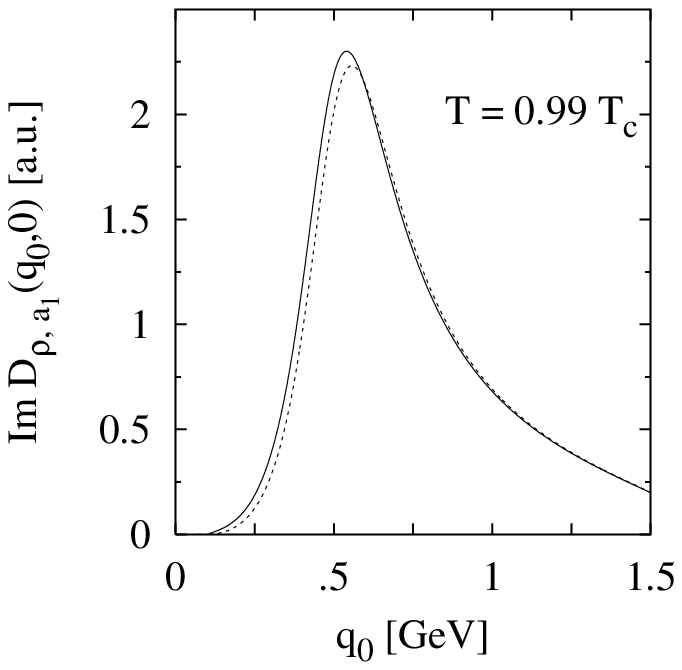, height=6.21cm, width=7.cm}}
\end{center}
\caption{\it Imaginary part of the $\rho$-meson (solid line) and $a_1$-meson
  (dashed line) propagator in RPA for different temperatures.}  
\label{figrhoa1t}
\end{figure}

The proportionality of the constituent quark mass to the quark condensate in
the Hartree approximation, see \eq{qbq0}, suggests that this quark mass is
strongly influenced by medium effects in the vicinity of the phase transition. In fact, the medium analogue of the
gap-equation, \eq{gap},  
\bea
m_{HT} &=& m_0 - 2 g_s 4 N_c N_f m_{HT} T\sum_n\int\frac{d^3 k}{(2 \pi)^3}
 \frac{1}{(i\omega_n+\mu)^2-E^2}\nonumber\\
 &=& m_0 + 2 g_s 4 N_c N_f m_{HT} I_{1T}~,
\label{gapt}
\eea  
with $E=\sqrt{\vec{k}^2+m_{HT}^2}$ and $\omega_n = (2n+ 1)\pi T $ being
fermionic Matsubara frequencies, reveals the $T$ as well as the $\mu$ dependence
of the constituent quark mass $m_{HT}$. The self-consistent solutions of this
equation correspond to extrema of the thermodynamic potential 
\beq
\Omega^{(0)}(m_T,T,\mu) =  -2 N_c N_f \intkt \ln\Big(\frac{E^2-(i \omega_n+\mu)^2}{m_0^2+\vec{k}^2-(i\omega_n+\mu)^2}\Big) 
+ \frac{(m_T-m_0)^2}{4 g_s} + const.~,
\label{tpt0}
\eeq
which can be employed to study the phase structure of the system. In the above
formula we set $g_v=0$ for simplicity. For details how to include vector
interactions see e.g.\@ Ref.~\cite{azakawa,mbuballa}. In this
approximation the NJL model exhibits a second order phase transition at $\mu=0, T\neq 0$
with a critical temperature of $T_c\approx 150-200$~MeV and, depending
strongly on the choice of parameters~\cite{mbuballa}, a second or a first order
phase transition at $T=0, \mu\neq 0$ with a critical chemical potential of the
same order of magnitude as the constituent quark mass in vacuum, $\mu_c
\approx 300-400$~MeV.

In addition, the properties of the mesons in the RPA are modified by medium effects. 
The polarization functions for the RPA pion and the $\sigma$-meson read 
\bea 
\Pi_{\sigma}(i \omega_l,\vec{p})& =& 4i N_c N_f I_{1T} - 2i N_c N_f
((i\omega_l)^2-\vec{p}^2-4 m_T^2) I(i \omega_l,\vec{p})\nonumber \\
\Pi_{\pi}(i\omega_l,\vec{p})& =& 4i N_c N_f I_{1T} - 2i N_c N_f
((i\omega_l)^2-\vec{p}^2) I(i \omega_l,\vec{p})~,
\eea
with $\omega_l = 2 l \pi T$ being bosonic Matsubara
frequencies. Below the phase transition the integral $I_{1T}$ can
again be replaced with the help of the gap equation \eq{gapt}
(cf. \eqs{sigmai} and (\ref{pseudopi})). The $\rho$-meson and the $a_1$-meson
are a little more difficult to handle since we have to keep in mind, that,
although the four-dimensional tranverse structure of the corresponding
polarization functions is preserved in medium, we have to distinguish
between a three-dimensional longitudinal and a three-dimensional transverse
part. This will be explained in more detail in App.~\ref{pia1}. 

Although the thermodynamics within this approximation is exclusively driven by
unphysical degrees of freedom, deconfined quarks, some features of chiral
symmetry restoration can nicely be illustrated. One example is the degeneracy
of the spectral functions of the chiral partners, i.e.\@ $\pi$ and $\sigma$ as
well as $\rho$ and $a_1$, above the phase transition can be mentioned here.
For the $\pi$-$\sigma$-system this can be inferred from \fig{figmpisig} where
the masses of pion and $\sigma$-meson
\footnote{We define masses here as the
  zero in the real part of the inverse propagator of the meson at rest.}  
as a
function of temperature for $\mu=0$ are displayed. For parameters we chose
our standard RPA parameter set. This leads to a critical temperature of
$T_c=164.4$~MeV. In the chiral limit (left panel) the mass of the pion,
denoted by the solid line, vanishes below the phase transition.  Away from the
chiral limit the pion mass, although non-vanishing, remains almost constant up
to $T\approx T_c$. In both cases the mass of the $\sigma$-meson (dashed line)
decreases with temperature, approaching zero at $T_c$ in the chiral limit, and
increasing again above the phase transition. Away from the chiral limit pion
and $\sigma$ mass do not exactly coincide at (and above) $T_c$. For the
$\rho$-$a_1$-system it is not sufficient to consider masses alone since both,
the $\rho$- and the $a_1$-meson, lie well above the $q\bar{q}$-threshold already for
temperatures $T \ll T_c$. In \fig{figrhoa1t} we have therefore displayed the
imaginary part of the propagators for $T=0$, $T= .61 T_c$ and $T=.99 T_c$.
Obviously these imaginary parts coincide for $T=T_c$ in the chiral limit. We
have to emphasize again that these results can only serve as an illustration.
Among other points the observation that the $\rho$- and the $a_1$-meson lie above the
$q\bar{q}$-threshold over wide temperature ranges reveals the unphysical nature
of these results.

\section{Quark condensate at $T\neq 0$ beyond mean field}
\label{qqatt}

Within this section we will study the behavior of the quark
condensate at nonzero temperature (and $\mu=0$) in the {\nce} and the MLA.
At the beginning of chapter~\ref{temperatur} we expressed the hope that at least the 
low-temperature behavior can be improved considerably as compared with the
Hartree + RPA scheme by including mesonic fluctuations within the
{\nce} or the MLA. To support this we will first compare the low-temperature
behavior we obtain with the expectations from model independent
considerations. Within this section we will again neglect fluctuations of
$\rho$- or $a_1$-mesons. 

\subsection{Low-temperature behavior}
\label{lowtemp}

Model independent results for the quark condensate can be obtained by a strict
low-temperature expansion in chiral perturbation theory. In the chiral limit
and at vanishing baryon density this reads~\cite{gasser}
\beq
\qq_T = \qq \Big(1 - \frac{T^2}{8 f_{\pi}^2}-\frac{T^4}{384 f_{\pi}^4} +
\dots\Big)~.
\label{ChiPT}
\eeq 
Here $\qq$ denotes the quark condensate at zero temperature.
The term proportional to $T^2$ arises from a pure (massless) pion gas. It can be derived
by considering the pressure of a free pion gas, 
\beq
p_\pi = - 3 T \int\frac{d^3k}{ (2\pi)^3} \ln (1-e^{-E_\pi/T})~,
\label{piongas}
\eeq
where $E_\pi = \sqrt{\vec{k}^2+m_\pi^2}$. The quark condensate is in general
given as the negative of the derivative of the total pressure with respect to the current
quark mass. Assuming that for small $T$ the change of the quark condensate with
temperature is determined by the pion gas and using the GOR
relation, \eq{GOR}, we can express the quark condensate as follows: 
\bea
\qq_T = \qq + \delta\qq_T &=& \qq +\frac{\qq}{f_\pi^2}\frac{\partial
  p_\pi}{\partial m_\pi^2} \nonumber\\
&=& \qq -\frac{\qq}{f_\pi^2}\frac{3}{2} \int\frac{d^3k}{ (2\pi)^3}
\frac{1}{E_\pi}\frac{1}{1-e^{-E_\pi/T}}~. 
\label{qqtm}
\eea 
In the chiral limit the integral in the last line of the above expression can
be evaluated analytically to give the $T^2$-term in
\eq{ChiPT}. The higher-order
terms in addition contain interactions between the pions. In the chiral limit
it is not astonishing that the massless Goldstone bosons, the pions, are the
dominant degrees of freedom at low temperatures. Thus we expect at least the
$T^2$-term to reflect only the underlying chiral symmetry and not any model
specific contributions. However, it can be shown~\cite{gasser} that aside from
the $T^2$- also the $T^4$-term of this expansion follows from chiral symmetry
alone. This implies that every model which exhibits chiral symmetry, among
others the NJL model, should in principle show the same behavior up to
$T^4$. In contrast an expansion of the quark condensate in Hartree
approximation~\cite{mlutz,ripka},
\beq
\qq^{(0)}_T = \qq^{(0)} \Big(1- \frac{(2 m_H T)^{3/2}}{\pi^{3/2} \qq^{(0)}}
e^{-\frac{m_H}{T}} + \dots\Big)~,
\label{mf}
\eeq
obviously does not reproduce these terms. There exists a very simple
argument~\cite{florkowski} why the
Hartree approximation is not suited to reproduce these terms: From the fact
that $f_\pi$ is of the order $\sqrt{N_c}$ we gather that the $T^2$-term is of
the order $1/N_c$ and the $T^4$-term of the order $1/N_c^2$. Remembering that
the Hartree approximation corresponds to the leading order in an
$1/N_c$-expansion we immediately see that this approximation must fail in the
desired low-temperature behavior. This reasoning also reveals that by extending
the calculations to next-to-leading order in $1/N_c$ we will at least be able
to reproduce the $T^2$-term~\cite{OBW2}. Although the MLA does not correspond
to an expansion in orders of $1/N_c$ the $T^2$-term can also be reproduced
within that approximation scheme~\cite{florkowski} because it contains all
$1/N_c$-correction terms in next-to-leading order. 
We will now proceed with
demonstrating this first for the {\nce}, and then for the MLA. 

\begin{figure}[t!]
\begin{center}
\parbox{7.5cm}{
     \epsfig{file=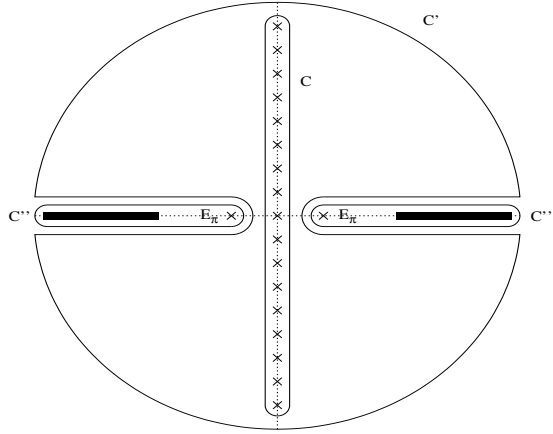,
     height=6.43cm, width=7.5cm}}
\end{center}
\caption{\it Contours $C$, $C'$ and $C''$ for the evaluation of
     $\Delta_T$. The poles at the Matsubara frequencies and at the pion energy
     are explicitly marked as well as the quark-antiquark cuts.}
\label{figcontour}
\end{figure}

Using the notations introduced in the previous section for the temperature
dependent quantities, we can write the quark condensate in next-to-leading order in $1/N_c$ in almost the same way as in vacuum (see \eqs{qbq0} and
(\ref{deltaqbqexp}),
\beq
\qq_T = \qq^{(0)}_T + \delta\qq_T = -\frac{m_{HT}-m_0}{2 g_s}-\frac{D_{\sigma}(0,0)
  \Delta_T}{2 g_s}~.
\label{qq}
\eeq
From \eq{mf} we conclude that to order $T^2$ thermal effects in the
leading-order term $\qq^{(0)}$ can be neglected. In the same way one can
reason that the temperature dependence of the $\sigma$-propagator does not
contribute. Thus we are left with an expansion of $\Delta_T$ at low temperatures.
$\Delta_T$ is
explicitly given by
\bea
\Delta_T 
&=& 4i N_c N_f\ m_T T \int\frac{d^3p}{(2\pi)^3}\sum_l{\Big\{}
\nonumber\\ &&\hspace{2.3cm}
D_\sigma(i\omega_l,\vec{p})(2I(i\omega_l,\vec{p})+I(0,0)-((i\omega_l)^2-\vec{p^2}-4
m_T^2) K(i\omega_l,\vec{p})) \nonumber \\
&&\hspace{2.cm} +  D_{\pi}(i\omega_l,\vec{p})\  (\;3I(0,0)\;-\;3 ((i\omega_l)^2-\vec{p}^2)\ K(i\omega_l,\vec{p})\;){\Big\}}~, 
\label{delta}
\eea
where $\omega_l$ are bosonic Matsubara frequencies.
If standard techniques are used the sum over the Matsubara frequencies in
Eq.~(\ref{delta}) can be converted into a contour integral~\cite{fetter},
where $C$ is a contour encircling the imaginary axis in the positive sense,
see \fig{figcontour},
\bea
 \Delta_T 
&=& 4i N_c N_f\ m_T \frac{1}{2\pi i}
\int\frac{d^3p}{(2\pi)^3}\int_C\frac{dz}{e^{z/T}-1}{\Big\{}
D_{\pi}(z,\vec{p})\  (3I(0,0)-3 (z^2-\vec{p}^2)\
K(z,\vec{p}))\nonumber \\
&&\hspace{1.8cm} +  D_\sigma(z,\vec{p})\ (2\ 
I(z,\vec{p})+I(0,0)-(z^2-\vec{p^2}-4 m_T^2) \ K(z,\vec{p})) {\Big\}}~. 
\label{contour}
\eea
One proceeds by deforming the contour $C$, first to $C'$, then to
$C''$, in such a way that only the
poles of the addends in \eq{delta} contribute. In our case  we encounter poles at 
\[
z
= \pm \sqrt{\vec{p}^2+m_\pi^{2(0)}}\equiv \pm E_\pi
\]
 and cuts for 
\[
z <-\sqrt{4
  m_{HT}^2+\vec{p}^2}\quad {\rm and}\quad z>\sqrt{4 m_{HT}^2+\vec{p}^2}~.
\]
The resulting
contour $C''$ is also shown in \fig{figcontour}. In the evaluation this contour
integration each pole at $z = z_p$ is weighted with a factor
$1/(e^{z_p/T}-1)$. Thus at low temperatures the lowest-lying pion pole
at $z = \pm E_\pi$ gives the main contribution. Subtracting the vacuum
part it can be expressed in the following way:
\beq
 (\Delta_T -\Delta)_\pi
= -4i N_c N_f\ m 
\int\frac{d^3p}{(2\pi)^3}\frac{2}{e^{E_\pi/T}-1}{\rm Res}{\Big\{}
D_{\pi}(E_\pi,\vec{p})\  (3I(0,0)-3 (E_\pi^2-\vec{p}^2)\
K(E_\pi,\vec{p}))\Big\}~.
\eeq
The exponential suppression of the
other contributions is certainly the stronger the lower the pion mass is as
compared with two times the constituent quark mass $m_{HT}$. In the
chiral limit we can therefore approximate $\Delta_T$ for low temperatures by
the pionic contribution 
\beq
 \Delta_T - \Delta
= 4 N_c N_f\ m 
\int\frac{d^3p}{(2\pi)^3}\frac{2}{e^{|\vec{p}|/T}-1}{\Big\{}
\frac{3}{2|\vec{p}| 2 N_c N_f} {\Big\}}~.
\eeq
This integral can be evaluated analytically and we obtain
\beq
 \Delta_T-\Delta = m\,{\frac{T}{2}}^2~.
\label{deltat}
\eeq
Up to now we have shown that the changes of the quark condensate for low
temperatures are indeed proportional to $T^2$ in the {\nce}. We are left with
a comparison of the coefficients in \eqs{ChiPT} and (\ref{qq}), respectively. 
In the chiral limit the vacuum
$\sigma$-meson propagator can be expressed with the help of the leading-order
pion decay constant as, (cf. \eqs{sigmaii} and (\ref{fpiex})),
\beq
    D_{\sigma}(0) = -\frac{1}{4f_{\pi}^{2(0)}}~.
\eeq
Inserting this expression together with \eq{qbq0} for the leading order quark
condensate and \eq{deltat}, evaluated at the Hartree mass, into \eq{qq} we
finally arrive at the following expression for the quark condensate in
next-to-leading order in $1/N_c$ at low temperatures:
\beq
\qq_T = \qq- \qq^{(0)}\frac{T^2}{8f_{\pi}^{2(0)}}~.  
\label{qqt}
\eeq
A comparison with \eq{ChiPT} shows that the coefficients in front of the
$T^2$-term are similar but not identical. In the {\nce} the leading-order
quantities, $\qq^{(0)}$ and $f_\pi^{(0)}$, determine the coefficient in
contrast to $\qq$ and $f_\pi$ in \eq{ChiPT}. On the one hand this simply
reflects the fact that we only get the lowest order contribution to
this coefficient by expanding the quark condensate up to next-to-leading order
in $1/N_c$. 
On the other hand it can be understood from physical
arguments: The $1/N_c$-corrections to the quark condensate consist of
fluctuating RPA mesons. Hence the low temperature behavior within the {\nce} is driven by
excited RPA pions and consequently quantities corresponding to the RPA pions.

For the MLA, a similar result has been derived in Ref.~\cite{florkowski}.
With the preceding results this derivation is easily comprehensible. In the
chiral limit the quark condensate in the MLA, cf. \eq{qbqself2}, is proportional
to the quark mass $m^\prime_T$,
\beq
\qq = -\frac{m^\prime_T-m_0}{2 g_s}~,
\eeq
where $m^\prime_T$ is a solution of the modified gap equation, 
cf.\eq{localgap},
\beq
m^\prime_T = m_0 + \Sigma_{HT}(m^\prime_T)-2 g_s \Delta_T(m^\prime_T)~.
\label{mqt}
\eeq  
This equation implicitly defines the temperature dependence of $m^\prime_T$
and therefore also that of the quark condensate. To obtain the
$T^2$-dependence we
will now take the derivative of \eq{mqt} with respect to $T^2$ at $T=0$ and
solve this for $d m^\prime_T/d T^2$, i.\@ e.  
\bea
\frac{d m^\prime_T}{ d T^2}\Big|_{T=0} &=& \frac{\partial (\Sigma_{HT}-2
  g_s\Delta_T)}{\partial T^2}\Big|_{T=0} (1-\frac{\partial (\Sigma_H-2
    g_s\Delta)}{\partial m^\prime})^{-1}\nonumber\\
&=& -2 g_s \frac{m^\prime}{2}(1-\frac{\partial
    (\Sigma_H-2g_s\Delta)}{\partial m^\prime})^{-1}~,
\eea
where in the last step we have employed \eq{deltat} and used the fact that the
derivative of $\Sigma_{HT}$ with respect to $T^2$ vanishes for $T=0$. 
From the definition of the inverse meson propagators in the MLA in
\Sec{1ml} it is obvious that the derivative of the quark self energy with
respect to the quark mass is
related to the $\sigma$-meson polarization function
$\tilde\Pi_\sigma(0)$. Thus the 
above expression can be written in terms of
the $\sigma$-meson propagator for vanishing momentum
\beq
\frac{d m^\prime_T}{ d T^2}\Big|_{T=0} =
\tilde{D}_\sigma(0)\frac{m^\prime}{2}~. 
\label{qqtmla0}
\eeq
The authors of Ref.~\cite{florkowski} argue that the leading order piece in
$1/N_c$ of the above expression is the RPA $\sigma$-meson propagator which in
turn can be expressed with the pion decay constant to that order. Consequently
the authors find 
\beq
\qq_T = \qq \Big(\,1 \;-\; \frac{T^2}{8f_{\pi}^{2(0)}}\,\Big)~.  
\label{qqtMLA}
\eeq
We have to keep in mind that $f_{\pi}^{2(0)}$ has to be understood as the
RPA-pion decay constant, 
\eq{fpiex}, evaluated at the quark mass $m^\prime$. This result is consistent
with the fact that in the MLA the pionic excitations, which determine the
thermal corrections to the quark condensate at low temperatures, consist of
RPA pions built of quarks with the constituent quark mass $m^\prime$. Since
$\qq^{(0)}$, evaluated at
the new mass $m^\prime$, agrees with $\qq$, $\qq_T$ to order $T^2$
is here, in
contrast to the {\nce}, proportional to the ``improved'' quark
condensate $\qq$. 

Of course, the $1/N_c$-expansion argument, leading from \eq{qqtmla0} to
\eq{qqtMLA}, is questionable because already the constituent quark mass
contains arbitrary orders in $1/N_c$. Nevertheless the result, \eq{qqtMLA},
seems well justified on physical grounds, since the dominant degrees of
freedom are in any case the excited RPA pions. Together with the discussion of
our numerical results in \Sec{tMLA} we will come back to the question whether
this approximation is well justified. 

 
\subsection{Numerical results within the $1/N_c$-expansion scheme}
\label{tnc}

\begin{figure}[b!]
\begin{center}
\parbox{7.5cm}{
     \epsfig{file=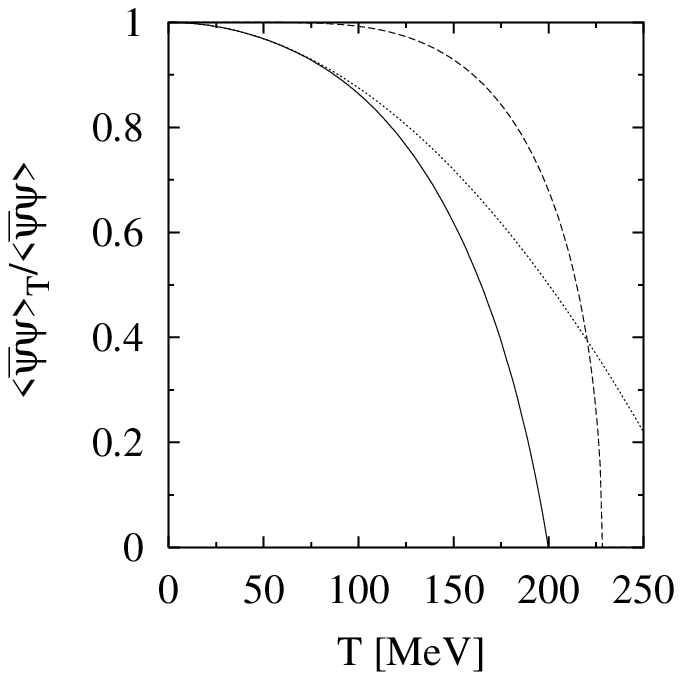,
     height=6.43cm, width=7.5cm}}
\parbox{7.5cm}{
     \epsfig{file=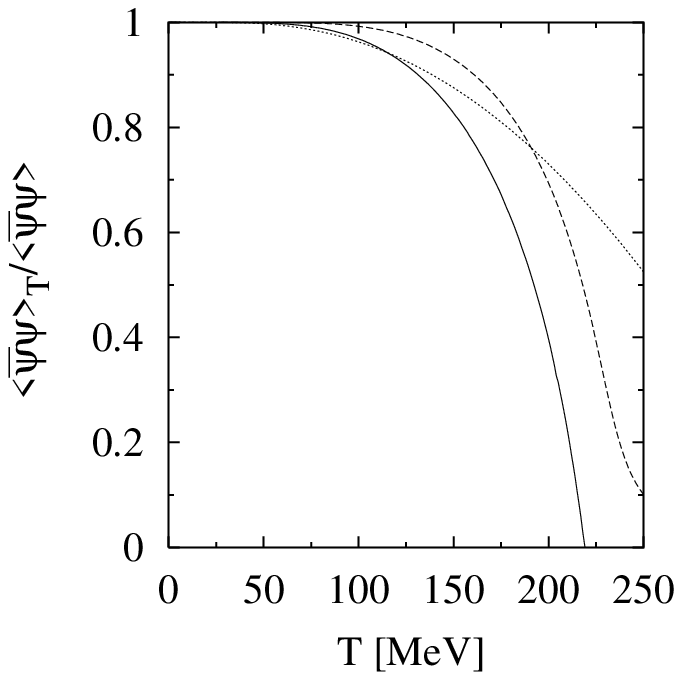,
     height=6.43cm, width=7.5cm}}
\end{center}
\caption{\it Quark condensate as a function of temperature, normalized to the
     vacuum value, in the chiral limit (left) and with $m_{\pi}^{(0)} = 140$
     MeV (right). Leading order in
     $1/N_c$ (dashed line), next-to-leading order (solid line), and free
     pion gas (dotted line).}
\label{fig10} 
\end{figure}

Our numerical results for the temperature behavior of the quark condensate
within the {\nce} are displayed in Fig.~\ref{fig10}. For the r.h.s.\@ we have used
the parameters listed in Table~\ref{tablence} for $\Lambda_M=600$~MeV which
allows a reasonable description of data in vacuum. In particular the RPA pion
mass reproduces the empirical value of $m_\pi^{(0)}=140$~MeV. The l.h.s.\@ of
\fig{fig10} corresponds to the chiral limit which has been built as described
in \Sec{instabilities} starting from the above parameter set. 

We will begin the discussion by the results in the chiral limit. Up to
temperatures of $T \approx100$~MeV the next-to-leading order result almost
perfectly agrees with the pure pion gas result whereas the leading order
result remains almost constant. We have to admit, that the denotation ``pure
pion gas'' result has to be considered carefully. Motivated by the
low-temperature expansion in the previous subsection, see \eq{qqt}, we compare
our result here with a gas of free leading-order pions. The applied parameter
set, however, has been fixed by fitting the $1/N_c$-corrected quantities
(except the pion mass, cf. \Sec{pionfit}), hence the value of $f_\pi^{(0)}$,
determining the behavior of the quark condensate at low temperatures, does not
coincide with the empirical value. The deviation is of about 25\%.
Nevertheless we can state that the $1/N_c$-corrections lead to a considerable
improvement as compared with the leading order result since our comparison
reveals the dominant role of pionic excitations at least in the low
temperature regime, where quark degrees of freedom are exponentially
suppressed.

\begin{figure}[t!]
\begin{center}
\parbox{7.5cm}{
     \epsfig{file=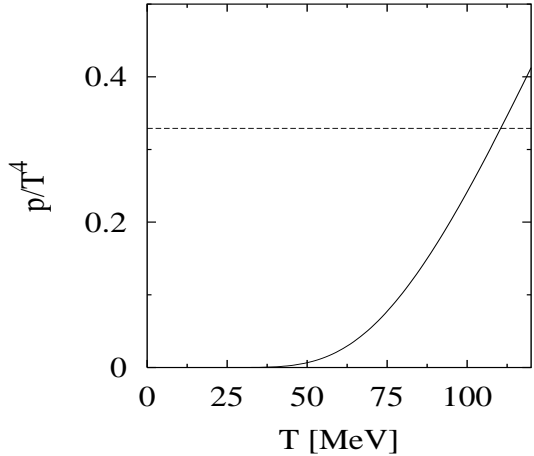,
     height=6.43cm, width=7.5cm}}
\end{center}
\caption{\it Pressure of a free gas of quarks with a mass of $m=446 MeV$
     (solid line) compared with that of a gas of massless pions (dashed line)
     as a function of temperature.}
\label{figdruck}
\end{figure}

However, we cannot by-pass the fundamental problem of lack of confinement in
the NJL model. The larger degeneracy factor of the quarks (24 as compared with
3) will consequently favor thermally excited quarks at higher
temperatures. To obtain a rough estimate of the temperature range where quark
effects become important let us compare the pressure of an ideal quark gas, 
\beq
p_q = 4 N_c N_f T \int \frac{d^3 k}{(2 \pi)^3} \ln(1+ e^{-E_q/T})~,
\eeq 
$E_q = \sqrt{\vec{k}^2+ m^2}$, 
with that of pions, see \eq{piongas}. For the case of massless pions,
i.e.\@ in the chiral limit, the latter can be evaluated analytically to obtain
\beq
p_\pi = T^4 \frac{\pi^2}{30}~.
\eeq
In \fig{figdruck} the solid line represents the pressure of a gas of free
quarks with mass $m = 446$~MeV, the dashed line that of a gas of massless
pions. 
The temperature of roughly 100 MeV where first deviations of the results in
the {\nce} from the pure pion gas result become visible lies in the same range
where the quark pressure becomes compatible with the pion pressure in the free
gas approximation.

At this point we enter into the question about the physical meaning of quark
effects at these temperatures. The excitation of quark degrees of freedom is
not completely excluded by nature. At temperatures above the deconfinement
phase transition quarks are, besides gluons, the principal degrees of freedom.
In the NJL model, however, there is no confinement and consequently no
deconfinement transition. 
On the other hand lattice
calculations~\cite{laermann} indicate that the deconfinement transition coincides with the chiral phase
transition which can be examined within the framework of the NJL model. Thus
the critical temperature for the chiral phase transition determines the scale
for the judgement on the relevance of quark effects. In principle one might
take the position that quark effects acquire a physical meaning only above the
phase transition. But as long as we are ignorant about the detailed
dynamics of the deconfinement transition we could allow for quark degrees
of freedom already in the vicinity of the phase transition. From a
phenomenological point of view this is consistent with the finding that a
resonance gas with many degrees of freedom can be, close to the phase
transition, effectively modeled by a quark gas (``quark-hadron duality''). 

Unfortunately, in the {\nce} the perturbative treatment of the mesonic
fluctuations prevents us from investigating a possible phase transition and
consequently from estimating whether thermally excited quarks become important
near the phase transition or much below.  Anyway, probably the applicability
of the perturbative expansion scheme has to be considered with great care
already for much lower temperatures than $T_c$. We will come back to the
question about the relevance of thermally excited quarks in the next section
in connection with the MLA.

This section will be completed by a discussion of the results away from the
chiral limit, i.e.\@ with $m_0 \neq 0$, shown on the r.h.s.\@ of \fig{fig10}. The
fundamental difference to the chiral limit is the presence of massive pions
which are now, like all other particles, exponentially
suppressed. This leads to the much more flat behavior of the quark condensate
at low temperatures. Although the pions are massive they are still the
lightest particles and can therefore be excited most easily. This is
confirmed by comparing the pure pion gas result (dashed line) with the
next-to-leading order result (solid): For temperatures below $T \lsim 100$~MeV
the two curves satisfactorily agree, though less perfectly than in the
chiral limit. The former has been calculated according to \eq{qqtm} with a
nonzero pion mass. Stronger deviations of the next-to-leading order result
from the pure pion gas result above $T\approx 100$~MeV indicate the increasing
importance of quark effects, which become visible at almost the same
temperature as in the chiral limit. 

 
\subsection{Meson loop approximation}
\label{tMLA}

\begin{figure}[t!]
\begin{center}
\parbox{7.5cm}{
     \epsfig{file=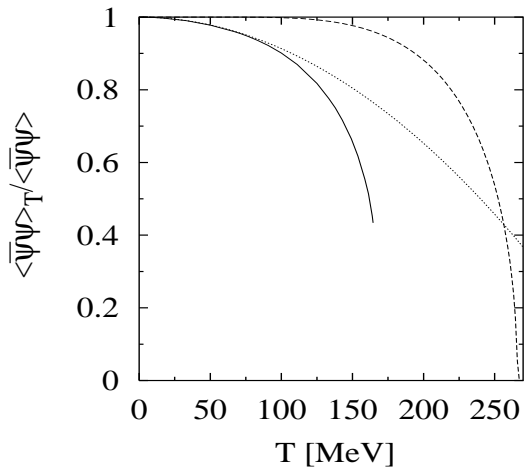,
     height=6.43cm, width=7.5cm}}
\end{center}
\caption{\it Quark condensate in the chiral limit as a function of
     temperature, normalized to the vacuum value, Hartree approximation 
     (dashed), meson loop approximation (solid) and free
     pion gas (dotted).}
\label{fig11} 
\end{figure}
\begin{figure}[t!]
\begin{center}
\parbox{7.5cm}{
     \epsfig{file=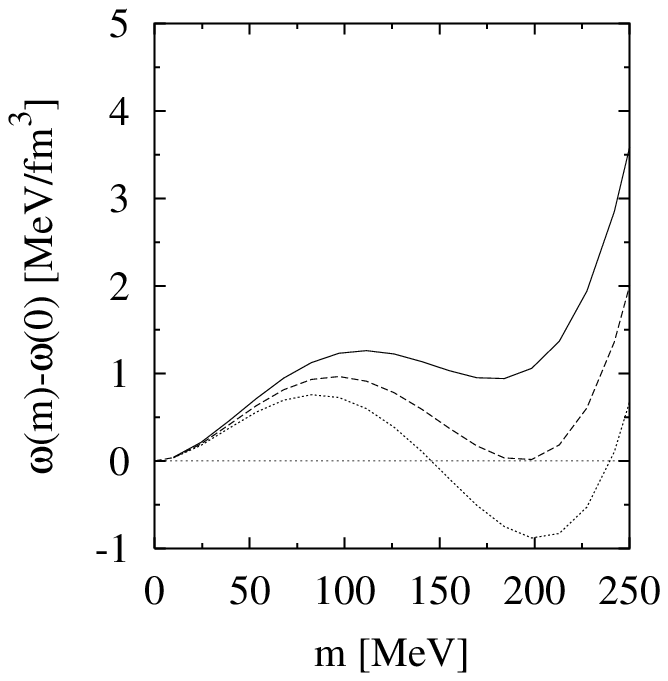,
     height=6.43cm, width=7.5cm}\quad}
\parbox{7.5cm}{
     \epsfig{file=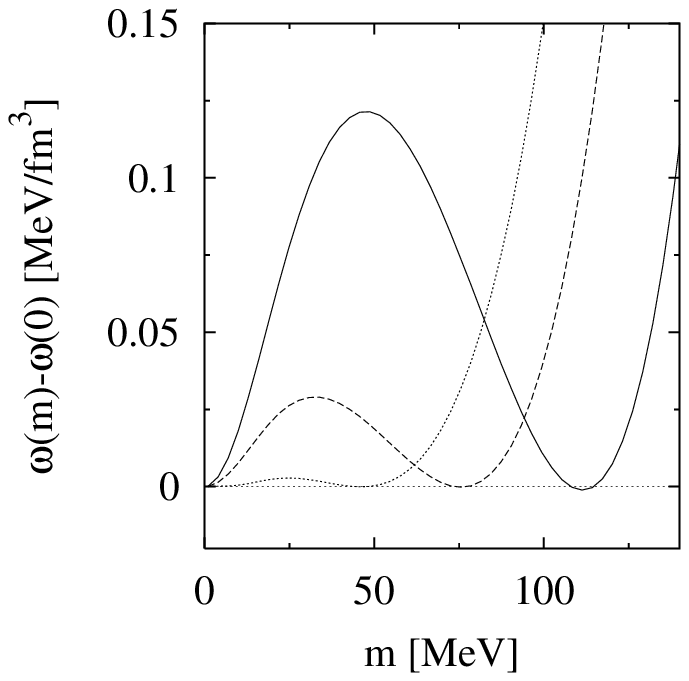,
     height=6.43cm, width=7.5cm}}
\end{center}
\caption{\it Thermodynamic potential per volume as a function of the
     constituent quark mass in the MLA with $\Lambda_M=700$~MeV (left panel)
     at $T=163.9$ MeV (dotted), 
     $T=164.5$ MeV (dashed), and $165.3$ MeV (solid) and (right panel) with
     $\Lambda_M=10$~MeV (dotted), 50 MeV (solid), and 100 MeV (dashed) at the
     corresponding critical temperature.}
\label{figomega} 
\end{figure}

Let us now compare the results of the previous subsection with the 
analogous calculations in the MLA. 
A study of the temperature dependence of the quark 
condensate within the MLA was first performed in Ref.~\cite{florkowski}
and can, for the chiral limit, also be found in Ref.~\cite{OBW2}.  
Here we likewise restrict ourselves to the chiral limit because our main aims
are on the one hand to be able to judge the relevance of quark effects on more
sensible grounds than in the {\nce} and, on the other hand, to test the quality
of the approximation which led us to \eq{qqtMLA}.  

Our results are shown in \fig{fig11}.  The calculations have
been performed using the parameters of Table~\ref{tableMLA} for $\Lambda_M =
700$~MeV, but $m_0 = 0$.  As discussed in \Sec{lowtemp}, at low temperatures
the model behaves again like a free pion gas (dotted line) decreasing with
$T^2$ for low temperatures. Had we assumed \eq{qqtMLA} to be exact we should
take the coefficient determining the $T^2$-behavior to be equal to $1/8
f_\pi^{2(0)}$, where $f_\pi^{(0)}$ corresponds to the pion decay constant of
RPA pions evaluated at the selfconsistently determined quark mass $m^\prime$.
From Table~\ref{tableMLA} we read off a value of $f_\pi^{(0)} = 120.0$~MeV for
the present parameter set with $m_0=7.9$~MeV. In the chiral limit, the value
of $f_\pi^{(0)}$ is slightly smaller, we obtain $f_\pi^{(0)}=118.6$~MeV. The
best agreement of the MLA result with the pion gas result can be achieved by
taking $f_\pi^{(0)}=119.0$~MeV supporting in hindsight \eq{qqtMLA}. At first
sight the approximation which has been performed to arrive at \eq{qqtMLA}
seemed a little deliberate although suggested by intuition.

Deviations from the pure pion gas behavior become visible at $T\approx
100$~MeV, which is quite similar to our observations in the {\nce}.  These
deviations arise from quark effects which could be tolerated only close to the
phase transition. In contrast to the {\nce} an examination of the phase
transition is possible in the MLA. The present parameter set leads to a
critical temperature of $T_c = 164.5$~MeV. Hence quark effects cease to be
negligible at a temperature of about $.6 T_c$, where one cannot avoid
admitting that they are completely unphysical.

Despite this obvious shortcoming of the model let us now have a closer look at
the effect of the mesonic degrees of freedom on the phase transition itself.
The critical temperature is considerably lowered as compared with the
corresponding value in Hartree approximation, $T_c=266.1$~MeV. This reduction
is caused on the one hand by the smaller constituent quark mass in vacuum,
$m^\prime=447.1$~MeV whereas $m_H=600$~MeV, and on the other hand by the
direct influence of mesonic fluctuations on the temperature dependence of the
quark condensate. It is worth noting that the critical temperature almost
agrees with the value we obtain in Hartree approximation with a parameter set
fitted to the RPA quantities. Had we succeeded in determining a parameter set in
the MLA which fits the MLA quantities this would of course be the only
sensible comparison. But, since it is to be expected that the properties of
the $\rho$-meson do not influence the quark condensate and therefore the
critical temperature considerably, $T_c$ presumably remains in the same range as
long as the values of $\qq$ and $f_\pi$ are refitted.

In accordance with the findings in Ref.~\cite{florkowski} the phase transition
is of first order whereas it is of second order in Hartree approximation. This
is already indicated by the jump in the order parameter at $T=T_c$. This
result can be assured by considering the behavior of 
the thermodynamic potential $\Omega$, the nonzero temperature analogue to
the effective potential (cf. \eq{tp}),
\bea
\Omega(m_T,T)&=& \Omega^{(0)}(m_T,T,\mu=0) \nonumber\\ &&+
\frac{1}{2}\intpt\{ \ln (1- 2 g_s\Pi_{\sigma}(i\omega_n,\vec{p}))+3 \ln (1- 2
g_s\Pi_{\pi}(i\omega_n,\vec{p}))\} + const.~.\nonumber\\
\label{tpt}
\eea 
The irrelevant constant is chosen in such a way that $\Omega(0,T)=0$. Though
this choice hinders the direct comparison of the absolute values of $\Omega$
for different temperatures, it facilitates the identification of the
order of the phase transition. That we indeed observe a first order phase
transition can be inferred from the l.h.s.\@ of \fig{figomega}, where we have
displayed $\Omega$ for different temperatures as a function of the constituent
quark mass $m_T$. At $T=T_c=164.5$~MeV one can doubtlessly identify two
degenerate minima at $m=0$ and $m\neq 0$, infallible evidence for a first
order phase transition at that temperature.  
One might ask whether this finding depends on the choice of the parameters, in
particular on the choice of $\Lambda_M$ which controls the strength of the
mesonic fluctuations. One indication that this is not the case is the fact
that the authors of Ref.~\cite{florkowski} report the same although they use a
different parameter set. In fact, varying the meson cutoff $\Lambda_M$ we find
that the discontinuity decreases with decreasing $\Lambda_M$ but nevertheless
the order of the phase transition remains the same. The results for $\Omega$
at the corresponding critical temperature are shown on the r.h.s.\@ of \fig{figomega} for
different values of the cutoff. All other parameters are kept constant.
We conclude that a first order phase transition seems to be a property of the
present approximation. 

This is probably, besides the unphysical quark degrees of freedom, another
shortcoming of the approximation, because universality arguments suggest that the
finite-temperature chiral phase transition in QCD with two massless quarks is
of second order~\cite{pisarski,rajagopalwilzcek}. This conjecture is based on
the assumption that the critical behavior is completely determined by the four
(at $T_c$ massless) bosonic degrees of freedom and that QCD therefore lies in
the same universality class as the $O(4)$ model which is known to exhibit a
second order phase transition. The same arguments can be applied to the NJL
model, which has the same underlying symmetry as QCD with two massless
flavors. However, one of the objections one might raise against the above
hypothesis is that a theory with composite boson fields not necessarily
belongs to the same universality class as the $O(4)$ model~\cite{kocic}. This
objection is among others valid for the NJL model.

In the same context we should mention that in the linear sigma model including
bosonic fluctuations in one-loop calculation similarly induces a first order
phase transition~\cite{BJpirner1}.  This model is usually chosen as prototype
of a model being in the same universality class as $O(4)$.  Applying the RPA
to this model also leads to a first order phase transition~\cite{zoheir}.
Here the application of renormalization group techniques led to a second order
phase transition~\cite{BJpirner2}. Thus it seems interesting to apply these
techniques also to the NJL model.
\chapter*{Summary}
\addcontentsline{toc}{chapter}{Summary}
\label{summary}

We investigated quark and meson properties within the Nambu--Jona-Lasinio
model, including meson-loop corrections. Here an approximation scheme beyond
the usual Hartree + Random Phase approximation which is consistent with
prescriptions following from the underlying chiral symmetry of the Lagrangian 
is required.
One possibility to construct a consistent scheme beyond Hartree + RPA is to
include the ring sum in the functional $\Phi$ within the framework of the
``$\Phi$-derivable method''. This scheme generates a non-local contribution to
the gap equation. We decided not to perform explicit calculations within this
scheme because of the difficulties which arise due to the non-trivial momentum
dependence of the non-local quark self-energy term.  The two following
schemes, an expansion of the self-energies in powers of $1/N_c$ up to
next-to-leading order, established in
Refs.~\cite{dmitrasinovic,oertel,OBW,OBW2}, and a one-meson-loop approximation
(MLA) to the effective action~\cite{nikolov}, were applied to examine quark
and meson properties.  The former is a perturbative scheme, where one starts
from the usual Hartree quark self-energy. The latter generates a local
correction term to the self-energy.  We explicitly showed that both schemes
are in accordance with the Goldstone theorem, i.e.\@ in the chiral limit the
pions emerge as massless particles. In the MLA in addition the validity of the
Gell-Mann--Oakes--Renner relation, which describes the behavior of the pion
mass for small current quark masses, can be proved. In
the {\nce} the GOR relation only holds if one carefully expands both sides
of the relation up to next-to-leading order in $1/N_c$.

The non-renormalizability of the NJL model leads to additional divergencies if
mesonic fluctuations are included. We therefore decided to introduce an
additional cutoff parameter $\Lambda_M$, which a priori is not related to the
regularization of the quark-loop contributions. This cutoff parameter controls
the strength of the mesonic fluctuations. Thus, varying this cutoff, we are
able to study the influence of mesonic fluctuations on various quantities.
This is interesting especially in the context of the question whether chiral
symmetry gets restored due to strong mesonic fluctuations~\cite{kleinert}.  In
fact, for large values of the cutoff $\Lambda_M$ we found instabilities in the
pion propagator as well in the {\nce} as in the MLA. However, due to the
perturbative treatment of the mesonic fluctuations a closer examination of the
question whether these instabilities are related to an unstable ground state,
as suggested in Ref.~\cite{oertel}, leading to a ``chiral restoration phase
transition'', was not possible within the {\nce}.  In the MLA an examination
of the effective potential as a function of $\Lambda_M$ enabled us to
establish a more decisive answer. We observed a ``phase transition'' only if
vector interactions are included.  
One
consequence of this observation is that the instabilities we find in the pion
propagator are not caused by a ``phase transition''. A plausible explanation
for these instabilities could be found in some peculiarities in the analytic
structure of the RPA meson propagators.

For the quark condensate in MLA we as well found a ``phase transition'' or a
``cross-over'' as a function of the cutoff $\Lambda_M$, depending on the value
of the current quark mass only if we include vector and axial intermediate
states. In the other cases, i.e.\@ in
the {\nce} and in the MLA without vector interactions, 
the quark condensate
exhibits a minimum as a function of $\Lambda_M$ in the same region of cutoffs
where the instabilities in the pion propagator emerge. Thus this observation 
might be related to the
instabilities in the pion propagator. 
In the {\nce} the
effect of vector interactions was simply that the minimum is then relocated at
higher values of the cutoff.

For the investigations described above all parameters except the mesonic
cutoff $\Lambda_M$ were kept constant. The corresponding values have been
chosen in such a way that the empirical values of the RPA quantities,
$\qq^{(0)}, m_\pi^{(0)}$, and $f_\pi^{(0)}$ could be reproduced.  However, at
the end the values of the parameters, including $\Lambda_M$, should be
determined by fitting physical observables. In addition to the quantities in
the pion sector, the quark condensate, the pion mass, and the pion decay constant, the properties of the
$\rho$-meson are very well suited for that purpose since in particular pionic
fluctuations are absolutely crucial if one wants to include the dominant
$\rho\rightarrow \pi\pi$-decay channel.  The standard Hartree + RPA scheme
contains only unphysical $q\bar{q}$-decay channels. A related problem, which
additionally constrains the possible choice of parameters is the following:
The presence of those unphysical decay modes can by no means be avoided
because they are a consequence of the lack of confinement in the NJL model.
One way in which this problem can be circumvented is to increase the
constituent quark mass sufficiently, such that this decay channel opens far
above the relevant region for the $\rho$-meson, i.e.\@ far above its mass.

In the {\nce} a set of parameter values which fulfills all these requirements
could be found. With these parameter values a reasonable fit of the data for
the pion electromagnetic form factor, corresponding roughly to a fit of the
mass and the width of the $\rho$-meson, together with reasonable values for
$\qq, m_\pi^{(0)}$ and $f_\pi$ can be obtained. With the corresponding value
of the constituent quark mass, $m_H=446$~MeV, the $q\bar{q}$-threshold lies
at $\approx 892$~MeV, i.e.\@ about 120 MeV above the maximum of the $\rho$-meson
spectral function. Moreover, with these parameter values we are far away from
the region where we encountered instabilities in the pion propagator. In fact,
from the moderate changes of the quantities in the pion sector - $f_\pi$ and
$\qq$ are lowered by about 20 \% and $m_\pi$ is increased by about 10\% - we
can conclude that higher-order terms in $1/N_c$ are small and that the
convergence of our perturbative {\nce} is satisfying.

In the MLA, on the contrary, we did not succeed in finding a suitable
parameter set. The main reason for this failure is the fact that in this
scheme the meson-loop effects lower the constituent quark mass as compared
with the Hartree mass $m_H$. Therefore it is much more difficult to prevent
the $\rho$-meson from decaying considerably into $q\bar{q}$ pairs. A
relatively large cutoff of $\Lambda_M \approx 700$~MeV was necessary in order
to shift the corresponding threshold above the $\rho$-meson peak and 
simultaneously to obtain reasonable values for other physical observables, in particular
$f_\pi$.  It turned out that, applying these parameter values, we encounter
instabilities in the $\rho$-meson propagator similar to those which emerge in
the pion propagator in certain regions of parameter space. Conclusively it can
be stated that we are unable to describe $\rho$-meson and pion properties
reasonably at the same time within the MLA. At this point one might ask
whether the inclusion of further intermediate states, e.g.\@ vector and axial
vector mesons, could improve the situation. We can give no conclusive answer
to that question at this stage. We can only remark that the results for the
quark condensate and the effective potential on the one hand reveal additional
problems, in particular a possible ``phase transition'', and on the other hand
probably enlarge the parameter space where we do not encounter any
instabilities. Besides, a different way of regularizing the occurring
divergencies is worth being investigated within this context.

Not only for the description of the $\rho$-meson but also for that of the
$\sigma$-meson it is absolutely necessary to include the two-pion intermediate
state which can be achieved in the {\nce} or the MLA. Our results for the
$\sigma$-meson, however, suffered from another disease: the 
attraction in the two-pion channel turned out to be too strong, leading to the
absence of one physical, time-like, pole in the propagator. This problem is
not an artifact of the NJL model but is found in the same way in similar
hadronic calculations. The similarity to difficulties arising in hadronic
calculations provided us with a hint, in which way this problem probably could
be cured: The $\sigma$-meson self-energy as it would come out if we applied
the ``$\Phi$-derivable-method'' has a promising structure and should therefore
be examined more closely.  

As well for the description of the $\rho$- as for that of the $\sigma$-meson
the two-pion intermediate state gives the dominant contribution which is
certainly an improvement compared with standard calculations in the NJL model,
which incorporate only quark effects. As such it is interesting to see how a
hadronic description emerges from an underlying quark structure. In this
context we performed a comparison of our results with that obtained in the
``static limit''. The static limit is an approximation to the full schemes
where all quark effects are suppressed by performing an expansion in 
the momenta of the incoming mesons up to the first non-vanishing order. This 
approximation was motivated by the   
assumption that the relevant momenta
are much larger than the constituent quark mass.  
This comparison shows that quark effects in our
calculations should be regarded with care: 
Although we were able to exclude the presence of unphysical $q\bar{q}$-decay
channels in the energy region relevant for the description of $\rho$- or
$\sigma$-meson
indirect quark effects considerably influence also
the propagator of the $\rho$-meson near its mass.

In the last part of the present paper we examined the temperature
dependence of several quantities. First we reviewed known results for
the temperature dependence of the RPA quantities, in particular the masses of
pion and $\sigma$-meson and the spectral functions of $\rho$- and $a_1$-meson.
In that way one of the principal consequences of chiral symmetry restoration
at nonzero temperature could be illustrated: Above the phase transition the
chiral partners, i.e.\@ $\pi$ and $\sigma$ or $\rho$ and $a_1$, respectively,
degenerate. Then we came to address the temperature dependence of the quark
condensate including mesonic fluctuations.  We find that as well in the
{\nce} as in the MLA the low-temperature behavior is almost exclusively driven
by thermally excited pions. Furthermore, it could be shown that in the chiral
limit it is consistent with the lowest-order chiral perturbation theory
result, i.e.\@ the contributions from a free pion gas. This is certainly a
considerable improvement as compared with the usual Hartree approximation
which completely fails to describe this behavior because pionic degrees of freedom, which
should be the dominant ones in this temperature region, are missing. Within
that scheme thermally excited quarks are the only possible degrees of freedom.
At higher temperatures, however, we cannot avoid having effects arising from
thermally excited quarks also in the two extended schemes.  One could argue
that the existence of these unphysical degrees of freedom is not necessarily
depreciable, because they could be tolerable in the vicinity of the
deconfinement transition, which coincides, according to lattice results, with
the chiral phase transition.

Of course, in the {\nce} we cannot judge whether quark effects become visible
near the chiral phase transition or much below, because the perturbative
treatment of the mesonic fluctuations does not allow an examination of the
phase transition. In principle this is possible within the MLA. Here the only
difficulty is to suitably choose the parameter values since we did not succeed
in determining a conclusive parameter set. However, the estimate we obtained
by, rather deliberately, taking the parameter set for $\Lambda_M=700$~MeV
seems to be robust: We find a critical temperature of 164.5~MeV, and on the
other hand, quark effects are visible already at a temperature of
$\sim$~100~MeV, i.e.\@ at about $0.6\ T_c$. This is much too early to be
realistic, i.e.\@ at this point the model obviously fails. The strong
influence of vector interactions on the results for the quark condensate at
zero temperatures suggests that this point should be reexamined including
additional intermediate states.

We should mention that, in agreement with Ref.~\cite{florkowski}, we
find a first-order phase transition. This point should be investigated further
since it is generally believed that the chiral phase transition
at nonzero temperature in a model with two light flavors is of second
order. In this context the application of renormalization group techniques
seems promising.   
\begin{appendix}

\chapter{Definition of elementary integrals}
\label{integrals}

It is possible to reduce the expressions for the quark loops to some
elementary integrals~\cite{passarinoveltman}, see App.~\ref{correlators}
and~\ref{functions}. In this section we give the definitions of these 
integrals. 
\bea
&&\hspace{-12mm} I_1 = \intk \frac{1}{k^2-m^2+i \eps}~, \label{onepoint}\\
&&\hspace{-12mm} I(p) = \intk \frac{1}{(k^2-m^2+i \eps)( (k+p)^2-m^2+
  i\eps)}~,\label{twopoint}\\
&&\hspace{-12mm} K(p) = \intk \frac{1}{(k^2-m^2+i \eps)^2( (k+p)^2-m^2+
  i\eps)}~,\label{threepoint1} \\
&&\hspace{-12mm} M(p_1,p_2) = \intk \frac{1}{(k^2-m^2+i \eps)
                 ( k_1^2-m^2+ i\eps)( k_2^2-m^2+i \eps)}~,\label{threepoint}\\
&&\hspace{-12mm} L(p_1,p_2,p_3) = \intk\frac{1}{(k^2-m^2+i \eps)
  ( k_1^2-m^2+ i\eps)
  ( k_2^2-m^2+i \eps)( k_3^2-m^2 + i\eps)}~,\label{fourpoint}\\
&&\hspace{-12mm} p_1^{\mu} M_1(p_1,p_2)+p_2^{\mu} M_1(p_2,p_1)
  =\hspace{-1mm} \intk \frac{k^{\mu}}{(k^2-m^2+i \eps)( k_1^2-m^2+ i\eps)
  (k_2^2-m^2+i\eps)},\label{tensor}
\eea
with $ k_i = k+p_i$. 
The function $M_1(p_1,p_2)$ can be expressed in terms of the other integrals,
\beq
M_1(p_1,p_2)
= \frac{p_1\!\cdot\!p_2\ I(p_1) - p_2^2\ I(p_2)
   + (p_2^2-p_1\!\cdot\!p_2)\  I(p_1-p_2)  
  + p_2^2\ (p_1^2-p_1\!\cdot\!p_2)\ M(p_1,p_2)}
  {2 \; ((p_1\!\cdot\!p_2)^2 \,-\, p_1^2\,p_2^2)}~,
\eeq
All integrals in Eqs.~(\ref{onepoint}) to (\ref{tensor}), are understood 
to be regularized. As described in \Sec{regularization} we use 
Pauli-Villars regularization with two regulators, i.e. we replace
\beq
    \intk f(k;m) \;\rrr\; \intk \sum_{j=0}^2 c_j\,f(k;\mu_j)~,
\eeq
by
\beq
    \mu_j^2 = m^2 + j\,\Lambda_q^2~;  \qquad
    c_0 = 1, \quad c_1 = -2, \quad c_2 = 1~.
\eeq
One then gets the following relatively simple analytic expressions for 
the integrals $I_1$, $I(p)$ and $K(p)$: 
\bea
&&\hspace{-12mm} 
  I_1=\frac{-i}{16\pi^2}\sum_j c_j\, \mu_j^2 \ln(\mu_j^2)\label{i1}\\
&&\hspace{-12mm}
  I(p) = \frac{-i}{16\pi^2}\sum_j c_j\, \Big(x_{j1} \ln(x_{j1})
  +x_{j2} \ln(-x_{j2})+x_{j1}\ln(-p^2x_{j1})+x_{j2}\ln(p^2x_{j2})\Big)\\ 
&&\hspace{-12mm}
  I(p=0) = \frac{-i}{16\pi^2}\sum_j c_j\, \ln(\mu_j^2)\label{i0}\\
&&\hspace{-12mm}
  K(p) = \frac{-i}{16\pi^2}\sum_j c_j\,\frac{1}{2 p^2(x_{j1}-x_{j2})} 
\Big(-\ln(x_{j1})-\ln(-x_{j1})+ \ln(x_{j2})+\ln(-x_{j2})\Big)~,
\eea
with 
\beq
x_{j1,2} = {1\over2}\pm{1\over2}\sqrt{1-{4 \mu_j^2\over p^2}} \;.
\eeq
An analytic expression for the three-point function (Eq.~\ref{threepoint})
can be found in Refs.~\cite{vanoldenborgh} and~\cite{veltman}. 
In certain kinematical regions the four-point function (eq.~\ref{fourpoint}) 
is also known analytically \cite{vanoldenborgh,veltman}.


\chapter{RPA propagators}
\label{correlators}

Using the definitions given in the previous section, we can write the gap equation
(Eq.~(\ref{gap})) in the following the form
\beq
    m = m_0 + 2ig_s\ 4 N_c N_f\ m\ I_1 \;.
\label{gapapp}
\eeq
Similarly one can evaluate the quark-antiquark polarization diagrams 
(Eq.~(\ref{pol0})) and calculate the RPA meson propagators.
The results for $\sigma$-meson and pion read 
\bea
  D_\sigma(p) &=& \frac{-2 g_s}{1-2i g_s\ 2 N_c N_f\ 
 (2 I_1 - (p^2-4m^2)\ I(p))}~,\label{sigmai}\\   
  D_{\pi}(p) &=& \frac{-2 g_s}{1- 2i g_s\ 2 N_c N_f\ (2 I_1-p^2\ I(p))}~.
 \label{pseudopi} \   
\eea
If we evaluate these propagators with the constituent
quark mass in Hartree approximation we can simplify the above expressions with
the help of the gap equation (Eq. \ref{gapapp}) to obtain
\bea   
  D_\sigma(p) &=& \frac{-2 g_s}{{m_0\over m} +2i g_s\ 2 N_c N_f\ 
 (p^2-4m^2)\ I(p)}~,\label{sigmaii}\\   
  D_{\pi}(p) &=& \frac{-2 g_s}{\frac{m_0}{m}+2i g_s\ 2 N_c N_f\ p^2\ I(p)}~.
 \label{pseudopiii}
\eea
As discussed in \Sec{solution}, this form is also used for the internal
meson propagators in the MLA.

A straight-forward evaluation of the vector and axial vector polarization 
diagrams gives
\bea
    \Pi_\rho(p) &=&  -i {4\over3} N_c N_f\ (-2 I_1+(p^2+2 m^2)\ I(p))~,
    \\
    \Pi_{a_1}(p) &=&  -i {4\over3} N_c N_f\ (-2 I_1+(p^2-4 m^2)\ I(p))~.
\eea
Because of vector current conservation $\Pi_\rho$ should vanish
for $p^2$~=~0. in \Sec{regularization} we discussed that in order to achieve
this we have to perform a subtraction. We then obtain for the $\rho$- and 
$a_1$-meson propagator 
\bea
  D_\rho(p) &=& \frac{-2 g_v}{1+ 2i g_v\ {4\over3} N_c N_f\ 
 (-2 m^2\ I(0)+(p^2+2 m^2)\ I(p))}\label{rhoii}~,\\   
  D_{a_1}(p) &=& \frac{-2  g_v}{1+ 2i g_v\ {4\over3} N_c N_f\ (-2 m^2\
    I(0)+(p^2-4 m^2)\ I(p))}~.\label{a1ii}
\eea   

\chapter{Details concerning vector and axial vector intermediate states}
\label{pia1}
\section{RPA meson propagators}
\label{pia1rpa}
One direct consequence of the presence of possible axial vector
excitations is $\pi$-$a_1$-mixing which occurs already on the level of the Hartree
approximation + RPA. As mentioned in \Sec{hartree} the quark-antiquark
$T$-matrix is no longer diagonal in the pion channel. Therefore we have to
solve a matrix equation to obtain the pion propagator in RPA. A generalized
ansatz for the $T$-matrix,
\bea
T_{\pi, ijkl}(q) &=& -D_{\pi_s}(q) \Gamma_{\pi,ij}\Gamma_{\pi,kl}
-D_{\pi_{sv}}(q)
\frac{q_\mu}{\sqrt{q^2}}\Gamma_{\pi,ij}\Gamma_{a_1,kl}^\mu\nonumber\\ 
&&-D_{\pi_{vs}}(q)
  \frac{q_\mu}{\sqrt{q^2}}
    \Gamma_{a_1,ij}^\mu\Gamma_{\pi,kl}-D_{\pi_v}(q)\frac{q_\mu
    q_\mu}{q^2}\Gamma_{a_1,ij}^\mu\Gamma_{a_1,kl}^\nu~,
\label{tpi}
\eea
leads to the following equation
\beq
\left(\begin{array}{cc}D_{\pi_s}&D_{\pi_{sv}}\\
    D_{\pi_{vs}}&D_{\pi_v}\end{array}\right) = 
\left(\begin{array}{cc}2 g_s&0\\
    0&-2 g_v\end{array}\right) +
\left(\begin{array}{cc}2 g_s&0\\
    0&-2 g_v\end{array}\right) 
\left(\begin{array}{cc}\Pi_{\pi \pi}&\Pi_{\pi a_1}\\
    \Pi_{a_1 \pi}&\Pi_{a_1 a_1}\end{array}\right) 
\left(\begin{array}{cc}D_{\pi_s}&D_{\pi_{sv}}\\
    D_{\pi_{vs}}&D_{\pi_v}\end{array}\right)~.
\label{dpii}
\eeq 
For simplicity we neglected the isospin indices and in the latter expression
also the momentum arguments. We have generalized here the definition for the
polarization functions, cf. \eq{pol0}, to
\beq
   \Pi_{MN}(q)=-i\intp \;\Tr\,[\,\Gamma_{M} \, iS(p+{q\over2})
                  \,\Gamma_{N} \, iS(p-{q\over2})\,] \;.
\label{pol0app}
\eeq
At this stage we should mention that our notation
$\Gamma_{a_1}^\mu$ for the axial coupling $\gamma^\mu\gamma_5\vec{\tau}$, introduced in \eq{gammas} is here
somewhat misleading because this vertex is not only related to the coupling of
an $a_1$-meson, but also to that of several parts of the pion
propagator. Nevertheless we will keep that notation.
Note that only the longitudinal part of the axial
polarization function contributes to the pion channel. The solution of this
matrix equation can be written in the following way:
\bea
D_{\pi_s}(q) &=& \frac{-2 g_s (1- 2 g_v 8 N_c N_f\, m^2\, i I(q))}{D(q)}\nonumber\\
D_{\pi_{sv}}(q) &=& \frac{-2 g_s 2 g_v 4 N_c N_f\, m \sqrt{q^2}\, i
  I(q)}{D(q)}\nonumber\\
D_{\pi_{vs}}(q) &=& -D_{\pi_{sv}}(q)\nonumber\\
D_{\pi_{v}}(q) &=& \frac{2 g_v (1-2g_s\Pi_{\pi_s}(q))}{D(q)}~,
\label{pivv}
\eea
with the determinant
\beq
D(q) = \big(1-2 g_s 2 i N_c N_f (2 I_1-p^2 I(p))\big)\big(1+ 2g_v 8 i N_c N_f
m^2 I(q)\big)-4 g_s g_v m^2 q^2\big(4 N_c N_f I(q)\big)^2~.
\label{detpi}
\eeq
With the help of the Hartree gap equation, \eq{gapexp}, the integral $I_1$,
occurring several times in the above expressions, can be replaced analogously to
the case without $\pi$-$a_1$-mixing.
 
The structure of the pion propagator, \eqs{tpi} and (\ref{pivv}), shows that in the present case, including
$\pi$-$a_1$-mixing, it is not sufficient to define a single pion-quark coupling constant analogously to \eq{mesonmass0}. One has to define a pseudoscalar and
an axial pion-quark coupling constant by taking the residue of the propagator
at the pion pole. For more details see e.g.\@ Ref.~\cite{vogl}. 
\section{Consistency with chiral symmetry}
\label{vchiral}
In \Sec{secpion} we neglected possible vector and axial vector intermediate
states while proving the consistency of our scheme with chiral symmetry. In
this section we will make up leeway. However, since there is conceptually no
difference to the MLA we restrict the discussion to the {\nce}.

One has to show that, in the chiral limit, the inverse pion propagator
vanishes at zero momentum, 
\beq 
    2g_s\,{\tilde\Pi}_\pi(0) = 1 \qquad {\rm for} \quad m_0 = 0.
    \label{vgoldstone}
\eeq
As before we use the notation
${\tilde\Pi}^{ab}_\pi = \delta_{ab}{\tilde\Pi}_\pi$.  
Similarly to \Sec{pionnc} we only need to show
that the contributions of the correction terms add up to zero,
\beq 
    \sum_{k=a,b,c,d}\;\delta\Pi_\pi^{(k)}(0) = 0
    \qquad {\rm for} \quad m_0 = 0.
    \label{vdeltapisum}
\eeq
Let us begin by diagram $\delta\Pi_\pi^{(a)}$. The external
pion can now couple to a $\pi\sigma$, a $\pi\rho$, a $\sigma a_1$, as well as
to a $\rho a_1$ intermediate state.
Evaluating the trace in 
Eq.~(\ref{trianglevertex}) for zero external momentum one gets for
the corresponding triangle diagrams
\bea
  \Gamma_{\pi,\pi,\sigma}^{ab}(0,p) &=& -\delta_{ab}\ 4  N_c N_f\ 2 m\ I(p)~,\nonumber\\ 
  \Gamma_{\pi,a_1,\sigma}^{\mu,ab}(0,p) &=& \delta_{ab}\ 4  i N_c N_f\ p^{\mu}\ I(p)~, \nonumber\\
  \Gamma_{\pi,\pi,\rho}^{\mu,abc}(0,p) &=& \eps_{abc}\ 4 i  N_c N_f\ p^{\mu}\
  I(p)~,\nonumber\\
  \Gamma_{\pi,\rho,a_1}^{\mu\nu,abc}(0,p) &=& \eps_{abc}\ 4  N_c N_f\ 2 m\ g^{\mu\nu}\ I(p)~.
  \label{vgammapps}
\eea 
The vertices for the coupling of the mixed or axial part of the pion propagator, cf.
\Sec{pia1rpa}, can be obtained by multiplying the corresponding $a_1$-vertex
with the momentum of the pion, see comment on the notation below \eq{dpii}.  We can see that, as the $\pi,\pi,\rho$-
and the $\pi,\sigma,a_1$-vertices are proportional to the momentum of the
$\rho$- and $a_1$-meson, the transverse parts of the propagators do not
contribute.  Inserting this into Eq.~(\ref{deltapi}) we find 
\bea
\delta\Pi^{(a)\,ab}_{M}(0) &=& i \intp\Big\{
  \Gamma^{ac}_{\pi,\pi,\sigma}(0,p)\,
  D_{\pi_s}(p)\,\Gamma^{bc}_{\pi,\pi,\sigma}(0,-p)\,
  D_{\sigma}(p)\nonumber\\
  & & \phantom{\intp} + \Gamma^{\mu,ac}_{\pi,a_1,\sigma}(0,p)\,
  \frac{p_{\mu}p_{\nu}}{p^2}D_{\pi_v}(p)\,\Gamma^{\nu,bc}_{\pi,a_1,\sigma}(0,-p)\,
  D_{\sigma}(p)\nonumber\\
  & & \phantom{\intp} + \Gamma^{\mu,ac}_{\pi,a_1,\sigma}(0,p)\, (-i p_{\mu})
  D_{\pi_{sv}}(p)\,\Gamma^{bc}_{\pi,\pi,\sigma}(0,-p)\,
  D_{\sigma}(p)\nonumber\\
  & & \phantom{\intp} + \Gamma^{ac}_{\pi,\pi,\sigma}(0,p)\, (i p_{\nu})
  D_{\pi_{sv}}(p)\,\Gamma^{\nu,bc}_{\pi,a_1,\sigma}(0,-p)\,
  D_{\sigma}(p)\nonumber\\
  & & \phantom{\intp} + \Gamma^{\mu\nu,acd}_{\pi,a_1,\rho}(0,p)\,
  T_{\mu\kappa} D_{a_1}(p)\,\Gamma^{\kappa
    \lambda,bcd}_{\pi,a_1,\rho}(0,-p)\,T_{\lambda\nu}
  D_{\rho}(p)\nonumber\\
  & &  \phantom{\intp} +
  \Gamma^{\mu,acd}_{\pi,\pi,\rho}(0,p)\, 
  D_{\pi_s}(p)\,\Gamma^{\nu,bcd}_{\pi,\pi,\rho}(0,-p)\,L_{\mu\nu}
  D_{\rho}(0)\nonumber\\
  & &  \phantom{\intp} +
  \Gamma^{\mu\nu,acd}_{\pi,a_1,\rho}(0,p)\, L_{\mu\kappa}
  D_{\pi_v}(p)\,\Gamma^{\kappa
    \lambda,bcd}_{\pi,a_1,\rho}(0,-p)\,L_{\lambda\nu}
  D_{\rho}(0)\nonumber\\
  & &  \phantom{\intp} +
  \Gamma^{\mu\nu,acd}_{\pi,a_1,\rho}(0,p)\, (-ip_\mu)
  D_{\pi_{sv}}(p)\,\Gamma^{
    \lambda,bcd}_{\pi,\pi,\rho}(0,-p)\,L_{\lambda\nu}
  D_{\rho}(0)\nonumber\\
  & &  \phantom{\intp} +
  \Gamma^{\mu,acd}_{\pi,\pi,\rho}(0,p)\, (ip_\kappa)
  D_{\pi_{sv}}(p)\,\Gamma^{
    \kappa\lambda,bcd}_{\pi,a_1,\rho}(0,-p)\,L_{\lambda\mu}
  D_{\rho}(0)\Big\}~.
  \eea
Inserting the expressions for the vertices (Eq.~\ref{gammapps}) one obtains
\bea
  \delta\Pi^{(a)\,ab}_{M}(0) 
  &=&\nonumber\\ && \hspace{-2.5cm}
i \delta_{ab} \intp (4 N_c N_f I(p))^2 \Big\{\phantom{+}D_{\sigma}(p) (D_{\pi_s}(p)\,
  4 m^2 + p^2\, D_{\pi_v}-4 m p^2 D_{\pi_{sv}}(p))\nonumber \\
  & &\hspace{-2.5cm} \phantom{{i \delta_{ab} \intp (4 N_c N_f I(p))^2}\Big\{}
  + D_{a_1}(p) D_{\rho}(p)
  24 m^2 \nonumber\\ 
  & &\hspace{-2.5cm} \phantom{{i \delta_{ab} \intp (4 N_c N_f I(p))^2}\Big\{} 
  - D_{\rho}(0) 2\,(p^2 D_{\pi_s}(p)+4 m^2 D_{\pi_v}(p)-4 m
  p^2D_{\pi_{sv}}(p))\Big\}
\eea
The essential step in the proof in \Sec{pionnc} was the validity of
\eq{pisig} which enabled us to write the product of the RPA pion and
$\sigma$-meson propagator as a difference. Of course this expression cannot be
directly applied to the pion propagator in the present case since it has been
modified by $\pi$-$a_1$-mixing. Nevertheless we can derive a similar helpful
expression. Besides, for the transverse part of the RPA $\rho$-meson
propagator and the $a_1$-meson propagator \eq{pisig} can be adopted almost
unchanged. We find  
\bea
D_\sigma(p)\ D_{\pi_s}(p) &=&
  i \,\frac{D_\sigma(p)-D_{\pi_s}(p)} {4 N_c N_f\ 2 m^2\ I(p)} -\frac{4
    m^2}{p^2} 2 g_v (\frac{i}{4 N_c N_f\ 2 m^2\ I(p)}+D_{\sigma}(p))\;,\nonumber\\
D_{a_1}(p)\ D_{\rho}(p) &=&
  i \,\frac{D_{\rho}(p)-D_{a_1}(p)} {4 N_c N_f\ 2 m^2\ I(p)}
\label{vpisig}
\eea
to finally obtain for $\delta\Pi_\pi^{(a)}$
\bea
\delta\Pi_\pi^{(a)\,ab}(0) &=&-\delta_{ab}\;4 N_c N_f\intp
2 I(p)\Big\{ D_\sigma(p)-D_{\pi_s}(p)+ 6 D_{\rho}(p)-6 D_{a_1}(p)\nonumber\\
& & \phantom{-\delta_{ab} 4 N_c N_f\intp 2 I(p)} - 2 G_v+2 D_{\rho}(0)\Big\} \;. 
\label{vpisigend}
\eea 
The next two diagrams can be evaluated straightforwardly.
One finds
\bea
  \delta\Pi_\pi^{(b)\,ab}(0)&=&  -\delta_{ab}\;4 N_c N_f\intp\Big\{ 
  D_\sigma(p)\  \big(I(p)+I(0)-(p^2-4 m^2)\ K(p)\big) \nonumber \\
&&\hspace{3.5cm} +D_{\pi_s}(p)\ \big(3I(p)\hspace{0.2cm}+\;3I(0)\hspace{0.4cm}
-\;\;3p^2\ K(p)\big)\nonumber \\
&&\hspace{3.5cm} +D_{\pi_{sv}}(p)\ 3\frac{p^2}{m} 4 m^2 K(p)\nonumber \\
&&\hspace{3.5cm} +D_{\pi_v}(p)\ \big(-3I(p)\hspace{0.4cm}
-\;\;12 m^2\ K(p)\big)\nonumber \\
&&\hspace{3.5cm} +D_{\rho}(p)\ \big(-3I(p)\hspace{0.4cm}-6 I(0)+6 (p^2+ 2 m^2)
\,K(p)\big)\nonumber \\
&&\hspace{3.5cm} +D_{\rho}(0)\ \big(-3I(p)\big)\nonumber \\
&&\hspace{3.5cm} +D_{a_1}(p)\ \big(-3I(p)\hspace{0.4cm}-6 I(0)+6 (p^2- 4 m^2)
\,K(p)\big)\; \Big\} \;, \nonumber\\ 
  \delta\Pi_\pi^{(c)\,ab}(0)&=& -\delta_{ab} \;4 N_c N_f\intp I(p)\Big\{ 
  3 D_{a_1}(p)-3 D_{\rho}(p)-D_{\rho}(0)\nonumber\\
  & &\hspace{3.5cm} -D_\sigma(p) - D_{\pi_s}(p)+D_{\pi_v}(p) \Big\} \;.
\label{vpseudo} 
\eea
Finally we have to calculate $\delta\Pi_\pi^{(d)}(0)$.
According to Eq.~(\ref{deltapi}), it can be written in the form 
\beq
    \delta\Pi_\pi^{(d)\,ab}(0) = -i\Gamma^{ab}_{\pi,\pi,\sigma}(0,0) \,
    D_\sigma(0)\,\Delta  \;,
    \label{vdeltapid}
\eeq
with 
\bea
 \Delta &=& \frac{1}{2}\intp\sum_M (-iD_M(p))
 (-i\Gamma_{M,M,\sigma}(p,-p)) 
   \nonumber \\
&=&\hspace{-2mm} 4 N_c N_f\ m \int\frac{d^4p}{(2\pi)^4}{\Big\{}
\ D_\sigma(p)\ (2\ I(p)+I(0)-(p^2-4 m^2)\ K(p)) \nonumber \\
&&\hspace{3.5cm} +  D_{\pi_s}(p)\  (\;3I(0)\;-\;3p^2\ K(p)\;) \nonumber\\
&&\hspace{3.5cm} +  D_{\pi_{sv}}(p)\ \frac{p^2}{m} (\;3I(p)\;+\;12m^2\ K(p)\;) \nonumber\\
&&\hspace{3.5cm} +  D_{\pi_v}(p)\  (\;-6I(p)\;-\;12m^2\ K(p)\;) \nonumber\\
&&\hspace{3.5cm} +  D_{\rho}(p)\  (\;6I(p)\;-6 I(0)+ 6 (p^2+2 m^2)\ K(p)\;) \nonumber\\
&&\hspace{3.5cm} +  D_{a_1}(p)\  (\;-12I(p)\;-6 I(0)+ 6 (p^2-4 m^2)\ K(p)\;){\Big\}}. 
\label{vtdself}
\eea
Evaluating $D_\sigma(0)$ in the chiral limit at the Hartree mass $m_H$ and comparing the result with
Eq.~(\ref{vgammapps}) one finds that the product of the first two factors 
in Eq.~(\ref{vdeltapid}) is simply $\delta_{ab}/m_H$, i.e. one gets   
\beq
 \delta\Pi_\pi^{(d)\,ab}(0) = \delta_{ab} \,\frac{\Delta}{m_H}~.
\label{vpid}
\eeq
With these results one can easily check that Eq.~(\ref{vdeltapisum}) 
indeed holds in our scheme. 

It now remains to check whether the replacement for the RPA meson propagators in
the MLA, see \Sec{solution}, does preserve the symmetry properties of the
schemes also if vector and axial vector intermediate states are included. The
question is whether \eq{vpisig} still holds for the RPA pion and $\sigma$-meson
propagators. The analogue of the replacement in
\eq{dpirepl} would be to replace the
determinant in the denominator of the RPA pion propagator, \eq{detpi}, by 
\beq
D(q) = \big(\frac{m_0}{m}+2 g_s 2 i N_c N_f p^2 I(p))\big)\big(1+ 2g_v 8 i N_c N_f
m^2 I(q)\big)-4 g_s g_v m^2 q^2\big(4 N_c N_f I(q)\big)^2~.
\eeq
Inserting this into \eq{pivv} for the pion propagator one can easily 
verify that \eq{vpisig} still holds for pion and sigma.
Because of gauge invariance we had to perform a subtraction for
the vector and axial vector polarization function in RPA anyway, such that no
more replacement is needed here. This subtraction is consistent with
\eq{vpisig}. 
\section{Effective potential including Rho- and $a_1$-mesons}
\label{veffp}
The explicit formula for the effective potential in vacuum including vector
interactions reads
\bea
V(m) &=& -4i N_c N_f \intp \ln(\frac{m^{2}-p^2}{m_0^2-p^2}) 
+ \frac{(m-m_0)^2}{4 g_s} \nonumber\\ &&-
\frac{i}{2}\intp\{ \phantom{+3}\ln (-D_{\sigma}^{-1}(p)/2g_s)+3 {\rm tr}\ln (-D_\pi^{-1} /2g_s)\nonumber \\
&&\phantom{-\frac{i}{2}\intp\{}\hspace{-4mm} +3 \ln (-D_{\rho}^{-1}(p)/2 g_v)+3 \ln
(-D_{a_1}^{-1}(p)/2 g_v) \} + const.~,
\eea 
where the pion propagator has to be taken from \eq{dpii}. The trace is here to
be understood to sum only the components of the $2\times 2$matrix in \eq{dpii}.

\chapter{Explicit expressions for the meson-meson vertices}
\label{functions}

In this section we list the explicit formulae for the meson-meson vertices. We
restrict ourselves to those combinations which are needed for the calculations
presented in this paper. We introduce the convention that a vertex labelled
with ``$\pi$'' is to be understood to contain a pseudoscalar coupling, whereas
a vertex labelled with ``$a_1$'' is to be understood to be calculated with an
axial coupling although the latter, due to $\pi$-$a_1$-mixing, couples also
pionic states. 

We begin with the three-meson vertices $\Gamma_{M_1,M_2,M_3}(q,p)$
(see~Fig.~\ref{fig4}): 
\bea
-i\Gamma_{\sigma,\sigma,\sigma}(q,p) &=& i 2 m N \Big(I(p^{\prime})
          +I(q)+I(p)+(4m^2-\frac{1}{2}(p^{\prime2}+p^2+q^2)) 
          M (p,-q)\Big) ~,
\nonumber\\
-i\Gamma^{ab}_{\pi,\pi,\sigma}(q,p) &=& i 2 m N\delta_{ab} 
\Big(I(p^{\prime})+p\!\cdot\!q M(p,-q)\Big) ~,
\nonumber\\
-i\Gamma^{\mu\lambda,ab}_{\rho,\rho,\sigma}(q,p)
&=&\delta_{ab}h(q,p)\Big(g^{\mu\lambda} 
- \frac{p^2\,q^{\mu}q^{\lambda} + q^2\,p^{\mu}p^{\lambda}
        \;-\;p\!\cdot\!q\,(p^{\mu}q^{\lambda}+q^{\mu}p^{\lambda})}
        {p^2q^2 \;-\; (p\!\cdot\!q)^2} \Big)~,
\nonumber\\ 
h(q,p) &=& i m N\Big(I(q)+I(p)-2 I(p^{\prime})
+(4 m^2-2 p\!\cdot\!q-p^{\prime2}) M(p,-q)\Big)~,
\nonumber\\
-i\Gamma^{\mu\lambda,ab}_{a_1,a_1,\sigma}(q,p)
&=&\delta_{ab}h(q,p)\Big(g^{\mu\lambda} 
- \frac{p^2\,q^{\mu}q^{\lambda} + q^2\,p^{\mu}p^{\lambda}
        \;-\;p\!\cdot\!q\,(p^{\mu}q^{\lambda}+q^{\mu}p^{\lambda})}
        {p^2q^2 \;-\; (p\!\cdot\!q)^2} \Big)
\nonumber\\ 
&& -2 i m N\delta_{ab} g^{\mu\lambda} \Big(I(p)+I(q)-(p^{\prime 2}-4 m^2)
M(p,-q)\Big)~, 
\nonumber\\
-i\Gamma_{a_1,\pi,\sigma}^{\mu,ab}(q,p) &=&  \delta_{ab} \Big(q^{\mu}
f_1(q,p)+p^{\mu} f_2(q,p)\Big)~, \nonumber\\
f_1(q,p) &=& -N \Big(I(p^\prime)-p^2 M(p,-q)+2 p\!\cdot\!p^\prime
M_1(q,-p)\Big)~,\nonumber\\ 
f_2(q,p) &=& -N \Big(I(p^\prime)+ I(p)-(q^2-4 m^2) M(p,-q)-2 p\!\cdot\!p^\prime
M_1(-p,q)\Big)~,\nonumber\\ 
-i\Gamma_{\pi,\pi,\rho}^{\mu,abc}(q,p) &=&  \eps_{abc} \Big(q^{\mu}
f(q,p)-p^{\mu} f(p,q)\Big)~, \nonumber\\
f(q,p) &=& N \Big(-I(q)+p^2 M(p,-q)+2 p\!\cdot\!q M_1(q,-p)\Big)~,
\eea
with $p^{\prime}= -p-q$ and $N=4 N_c N_f$. To calculate the constant $\Delta$
we need the vertices coupling to $\sigma$ for the special case where the
momentum of the $\sigma$ is vanishing. Since this limit is in several cases
difficult to build, we will also list here the corresponding expressions,
\bea
-i\Gamma_{\sigma,\sigma,\sigma}(-p,p) &=& i 2 m N \Big(I(0)
          +2 I(p)+(4m^2-p^{2}) 
          K (p)\Big) ~,
\nonumber\\
-i\Gamma^{ab}_{\pi,\pi,\sigma}(-p,p) &=& i 2 m N\delta_{ab} 
\Big(I(0)-p^2 K(p)\Big) ~,
\nonumber\\
-i\Gamma^{\mu\lambda,ab}_{\rho,\rho,\sigma}(-p,p)
&=&\delta_{ab} \frac{4i N}{3} T^{\mu\lambda}\Big(I(p)-I(0)+(p^2+2 m^2) K(p)\Big)~,
\nonumber\\ 
-i\Gamma^{\mu\lambda,ab}_{a_1,a_1,\sigma}(-p,p)
&=&\delta_{ab}\frac{-4 i N}{3 p^2}\Big(\phantom{+}
g^{\mu\lambda} p^2\,(I(0)+2 I(p)-(p^2-4
m^2) K (p)) \nonumber\\
&& \phantom{\delta_{ab}\frac{-4 i N}{3 p^2}\Big(}+ p^\mu p^\lambda (I(p)-I(0)+(p^2 + 2 m^2) K(p))\Big)~, 
\nonumber\\
-i\Gamma_{a_1,\pi,\sigma}^{\mu,ab}(-p,p) &=&  -N \delta_{ab} p^\mu\Big(I(p)+4
m^2 K(p)\Big)~. \nonumber\\
\eea

For the four-meson vertices we only need to consider the special cases 
needed for the diagrams (b) and (c) in Fig.~\ref{fig3}:
\bea
-i\Gamma_{\sigma,\sigma,\sigma,\sigma}(q,p,-q)&=& -N \Big\{
I(p-q)+I(p+q) + 8 m^2\ (M(p,q)+M(p,-q))
\nonumber\\ && \hspace{1cm} 
+ 4 \big(m^2\ (4 m^2-p^2-q^2)-\frac{p^2q^2}{4}\big)\
L(p,-q,p-q)\Big\}
\nonumber\\
-i\Gamma_{\sigma,\sigma,\sigma,\sigma}(q,p,-p) &=& -N \Big\{ I(p+q) + I(0)
+ 4m^2 \big( K(p)+K(q)+2 M(p,-q)\big) 
\nonumber\\ && \hspace{1cm} 
+2 p\!\cdot\!q M(p,-q) -q^2 K(q)-p^2 K(p) 
\nonumber\\ && \hspace{1cm} 
+ m^2\big( 16 m^2-4 p^2-4 q^2 + \frac{p^2 q^2}{m^2}\big) L(p,-q,0)\Big\}
\nonumber\\
-i \Gamma_{\sigma,\pi,\sigma,\pi}^{ab}(q,p,-q) 
&=&\delta_{ab} N\Big\{I(p+q)+I(p-q)+ p^2 (4m^2-q^2) L(p,-q,p-q)\Big\}
\nonumber\\
-i \Gamma_{\sigma,\pi,\pi,\sigma}^{ab}(q,p,-p) 
&=&\delta_{ab} N\Big\{-I(p+q)-I(0)- (4m^2-q^2) (K(q)-p^2 L(p,-q,0))
\nonumber\\ && \hspace{1cm}
+p^2 K(p)-2 p\!\cdot\!q\  M(p,-q)\Big\}
\nonumber\\
-i\Gamma_{\pi,\pi,\pi,\pi}^{abcd}(q,p,-q) 
&=& -N\kappa_{abcd}\Big\{I(p+q)+I(p-q)-p^2q^2 L(p,-q,p-q)\Big\}
\nonumber\\
-i\Gamma_{\pi,\pi,\pi,\pi}^{abcd}(q,p,-p) 
&=& -N\kappa_{abcd}\Big\{I(p+q)+I(0)-p^2 K(p)
\nonumber\\ && \hspace{1.7cm}
-q^2 K(q)+2 p\!\cdot\!q M(p,-q)+p^2 q^2 L(p,-q,0)\Big\}
\nonumber\\
-i\Gamma_{\rho,\sigma,\rho,\sigma}^{ab}(q,p,-q) 
&=& -2\delta_{ab}N\Big\{I(p+q)+I(p-q)+2 I(q) -p\!\cdot\!q (M(p,-q)-M(p,q))
\nonumber\\ && \hspace{1.5cm} 
+(4 m^2-2 p^2)(M(p,q)+M(p,-q))
\nonumber\\ && \hspace{1.5cm}
+m^2 (8 m^2-6 p^2+4 q^2+\frac{p^4-(p\!\cdot\!q)^2}{m^2}) L(p,-q,p-q)\Big\}
\nonumber\\
-i\Gamma_{\rho,\sigma,\sigma,\rho}^{ab}(q,p,-q) 
&=& -2\delta_{ab}N\Big\{-I(p+q)-I(0)+(p^2-4 m^2) K(p)
\nonumber\\ &&\hspace{1.5cm}
+(q^2+2 m^2) K(q) +(4 m^2-2 p\!\cdot\!q) M(p,-q)
\nonumber\\ &&\hspace{1.5cm}
+m^2 (8 m^2-2 p^2+4 q^2-\frac{p^2 q^2}{m^2}) L(p,-q,0)\Big\}
\nonumber\\ 
-i\Gamma_{\rho,\pi,\rho,\pi}^{abcd}(q,p,-q) 
&=& 2N\kappa_{abcd}\Big\{-I(p+q)-I(p-q)-2 I(q)
\nonumber\\ &&\hspace{1.5cm}
+2 p^2 (M(p,q)+M(p,-q)) +p\!\cdot\!q(M(p,-q)-M(p,q)) 
\nonumber\\ &&\hspace{1.5cm}
+(2 m^2 p^2- p^4+(p\!\cdot\!q)^2) L(p,-q,p-q)\Big\}
\nonumber\\
-i\Gamma_{\rho,\pi,\pi,\rho}^{abcd}(q,p,-q) 
&=& 2 N\kappa_{abcd}\Big\{I(p+q)+I(0)-p^2 K(p)-(q^2+2 m^2) K(q)
\nonumber\\ &&\hspace{1.5cm}
+2p\!\cdot\!q M(p,-q)+p^2 (2 m^2+q^2) L(p,-q,0)\Big\}~,
\eea
with $\Gamma_{\rho,M,M,\rho}(q,p,-q) =
g_{\mu\nu}\Gamma_{\rho,M,M,\rho}^{\mu\nu}(q,p,-q),
\Gamma_{\rho,M,\rho,M}(q,p,-p) = g_{\mu\nu}\Gamma_{\rho,M,\rho,M}^{\mu\nu}(q,p,-p)$ and $\kappa_{abcd} = \delta_{ab}\delta_{cd}+\delta_{ad}\delta_{bc}-\delta_{ac}\delta_{bd}$.


\chapter{Expressions at nonzero temperature}
\label{aptemp}
\section{Feyman rules at nonzero temperature and chemical potential}
\begin{itemize}
\item{Quark propagator:
\beq
\parbox{60pt}
  {\begin{fmfgraph*}(60,40)
     \fmfpen{thick}
     \fmfleft{l} \fmfright{r}
     \fmf{plain_arrow,label=$i\omega_n,,\vec{p}$}{l,r}
   \end{fmfgraph*}
  } = S(i\omega_n,\vec{p}) = \frac{(i \omega_n +
  \mu)\gamma_0-\vec{p}\vec{\gamma}-m}{(i \omega_n+\mu)^2-E^2}~,
\eeq
with $\omega_n = (2n+1) \pi T$ fermionic Matsubara frequency, $\mu$ the
chemical potential and $E = \sqrt{\vec{p}^2+m^2}$.}
\item{Vertices:
\beq
\parbox{60pt}{
   \begin{fmfgraph}(60,40)
     \fmfpen{thick}
      \fmfleft{l1,l2}\fmfright{r1,r2}
      \fmf{plain}{l1,v1,r2}
      \fmf{plain}{l2,v1,r1}
      \fmfdot{v1}
   \end{fmfgraph}}
= -2g_M\Gamma_M\otimes \Gamma_M
\eeq}
\item{A factor (-1) and a trace for closed fermion loops}
\item{The four-dimensional momentum space integration is converted into a
    three-dimensional one and a sum over the corresponding Matsubara
    frequencies, i.\@ e. 
\beq
\intp \longrightarrow T \sum_n \int \frac{d^3 p}{(2\pi)^3}~.
\eeq }
\end{itemize} 
\section{Rho- and $a_1$-meson propagators in the medium}
In this section we will explain the structure of the $\rho$- and $a_1$-meson
propagators in medium. Denoting the four-vector describing the motion of the
heat bath with $u^\mu$, we have in contrast to the vacuum here two independent
four-vectors, $u^\mu$ and the momentum $q^\mu$ of the meson, from which we can construct the tensor structure of the
polarization functions. Generally we can write
\beq
\Pi^\mn = g^\mn f_1 + \frac{q^\mu q^\nu}{q^2} f_2 + u^\mu u^\nu
f_3 + \frac{q^\mu u^\nu + u^\mu q^\nu}{q\cdot u} f_4~,
\eeq   
where we have already used the symmetry of the polarization function with
respect to an exchange of $\mu$ and $\nu$. The functions $f_1$ to $f_4$ are, a
priori independent, scalar functions, depending on the Lorentz scalars $q^2$, and $q\cdot u$. The most simple choice for the four-vector $u^\mu$ is
$u^\mu = (1,0,0,0)$, i.e.\@ we assume to be in the rest frame of the heat
bath. With that choice the general polarization function can be splited into a
four-dimensional transverse and three-dimensional transverse, a
four-dimensional transverse and three-dimensional longitudinal as well as two
four-dimensional longitudinal components. Here three-dimensional transverse or
longitudinal, respectively, means that the polarization function is transverse
(longitudinal) with respect to $\vec{q}$. Since the vector polarization
function is in our case four-dimensional transverse and the $a_1$ is
constructed exclusively from the four-dimensional transverse part of the axial
polarization function we can restrict the discussion to that part. The two
projectors on the three-dimensional transverse, $P_{TT}^\mn$, and
longitudinal, $P_{TL}^\mn$, 
part are,
\bea
P_{TT}^\mn &=& \left\{\begin{array}[c]{ll}\delta_{ij}-\frac{q_i q_j}{\vec{q}^2},
    \quad {\rm if} \quad &i,j = 1,2,3\\
 0 \quad    & {\rm otherwise}\end{array}\right .~, \nonumber\\
P_{TL}^\mn &=& P_T^\mn - P_{TT}^\mn
\eea
In the rest frame of the meson, i.e.\@ for $\vec{q}=0$, the two parts of the
polarization function are equal and can be extracted in the same way as in
vacuum with the help of the $P_T^\mn$. For instance, for the propagators
of $\rho$- and $a_1$-meson in RPA we then arrive at
\bea
  D_\rho(q_0, 0) &=& \frac{-2 g_v}{1+ 2i g_v\ {4\over3} N_c N_f\ 
 (-2 m^2\ I(0,0)+(q_0^2+2 m^2)\ I(q_0,0))}~,\\   
  D_{a_1}(q_0,0) &=& \frac{-2  g_v}{1+ 2i g_v\ {4\over3} N_c N_f\ (-2 m^2\
    I(0,0)+(q_0^2-4 m^2)\ I(q_0,0))}~.
\eea   
\end{appendix}
\end{fmffile}

\newpage
\centerline{\underline{\bfseries \large Danksagung}}

\vspace{2.5cm}
\noindent
Zuerst m{\"o}chte ich mich bei Herrn Prof. Dr. J. Wambach f{\"u}r die Anregung zu
dieser Arbeit und sein st{\"a}ndiges Interesse bedanken.
\vspace{1cm}\\
Bei Herrn Prof.~Dr.~A. Richter m\"ochte ich mich f\"ur die \"Ubernahme des
Korreferats und seine F\"orderung w\"ahrend der gesamten Studienzeit
bedanken.\\[1cm] 
Allen Mitgliedern der NHC-Gruppe danke ich f\"ur die angenehme Atmosph\"are. 
Mein besonderer Dank gilt
dabei Dr.~Michael~Buballa f{\"u}r seine st{\"andige}
Diskussionsbereitschaft. Au{\ss}erdem habe ich auch von vielen Diskussionen mit
Michael Urban und Dr.~Bernd-Jochen~Schaefer profitiert. Auch Carsten
Isselhorst und Dr. Zoheir Aouissat m\"ochte ich erw\"ahnen, von denen ich
insbesondere \"uber das $\sigma$-Meson viel gelernt habe.   
\\[1cm]
Herrn Prof.~G.~Ripka bin ich f\"ur kritische Diskussionen und fruchtbare
Anregungen zu Dank verpflichtet. Auch bei den
Profs. Drs. G. Chanfray, M. Ericson und P. Schuck und bei Dr. Dany Davesne m\"ochte ich mich f\"ur ihr
Interesse an dieser Arbeit und einige wichtige Hinweise bedanken.\\[1cm]
Ein ganz besonderer Dank geb\"uhrt auch Geert Jan van Oldenborgh, der mir wichtige
Hilfestellungen zur Benutzung seines FF-Programmpaketes gegeben hat, das in Teilen dieser Arbeit benutzt
wurde. \\[1cm]
Auch bei meinen Eltern m\"ochte ich mich f\"ur ihre Unterst\"utzung
bedanken, vor allem bei meinem Vater f\"ur seine kritischen Hinweise zu
englischen Formulierungen.
\end{document}